\documentclass{phd}

\begin{document}
	\logotypes{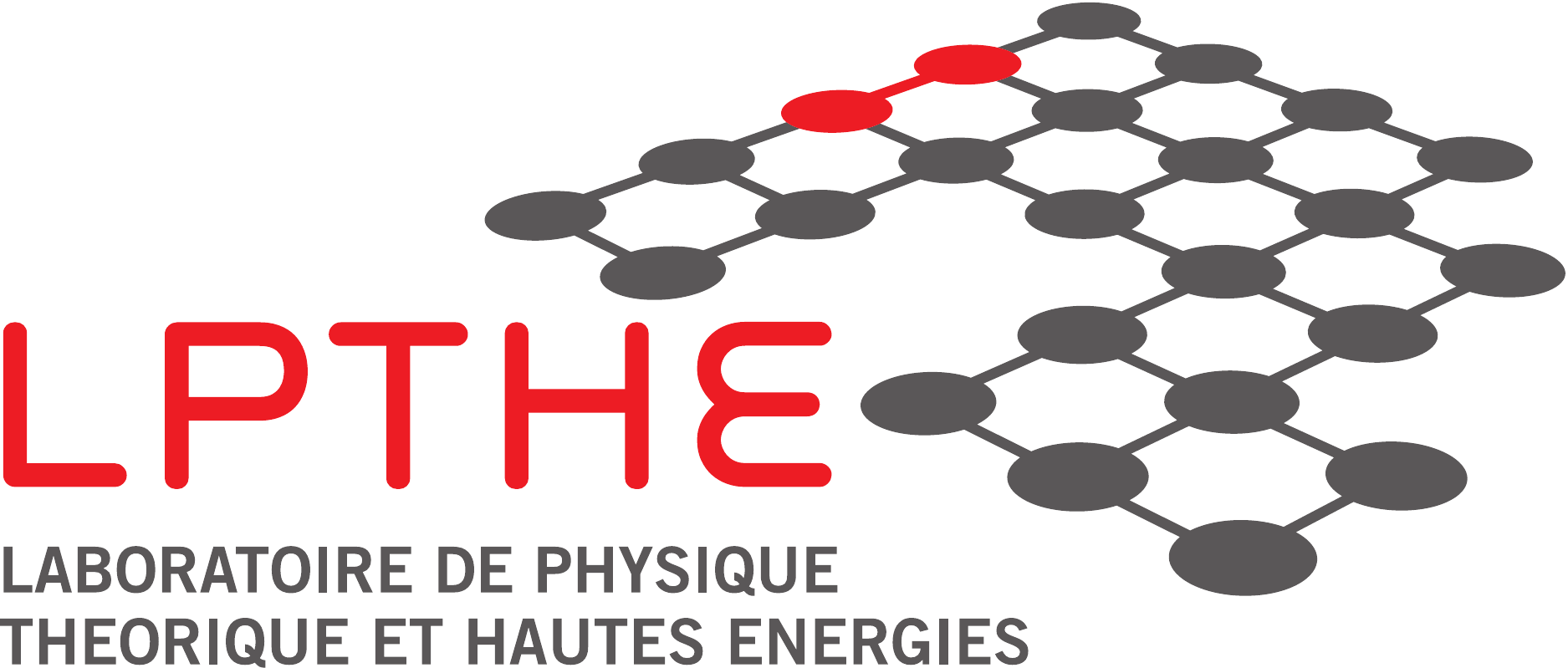}{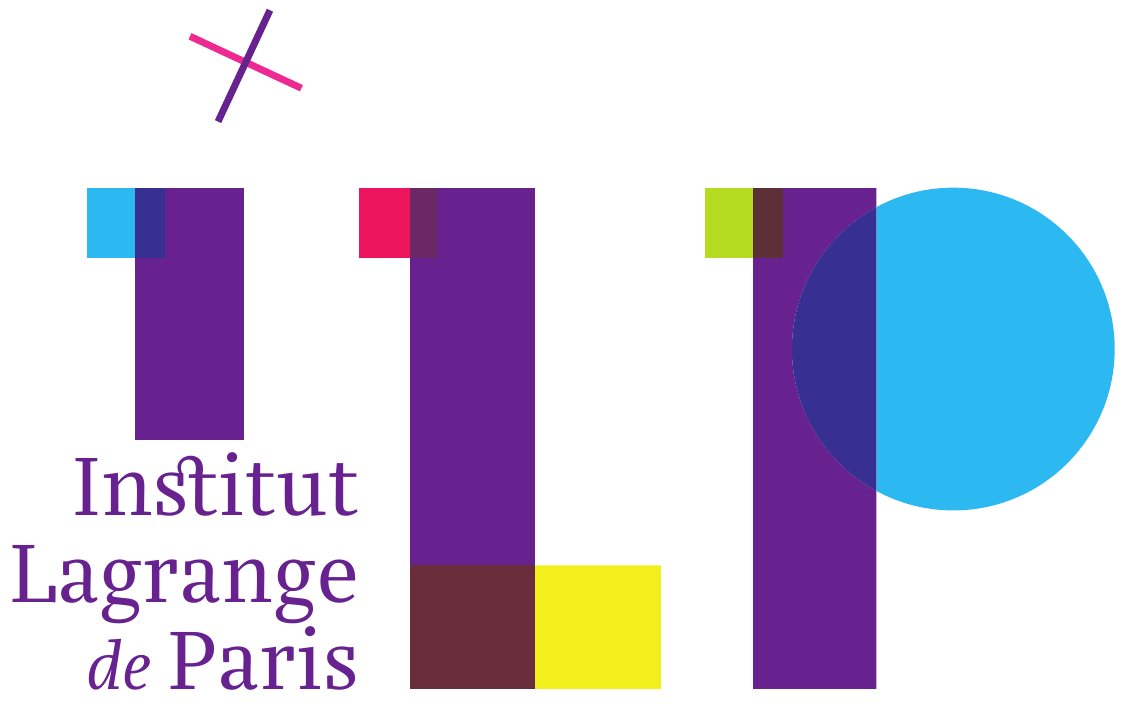}{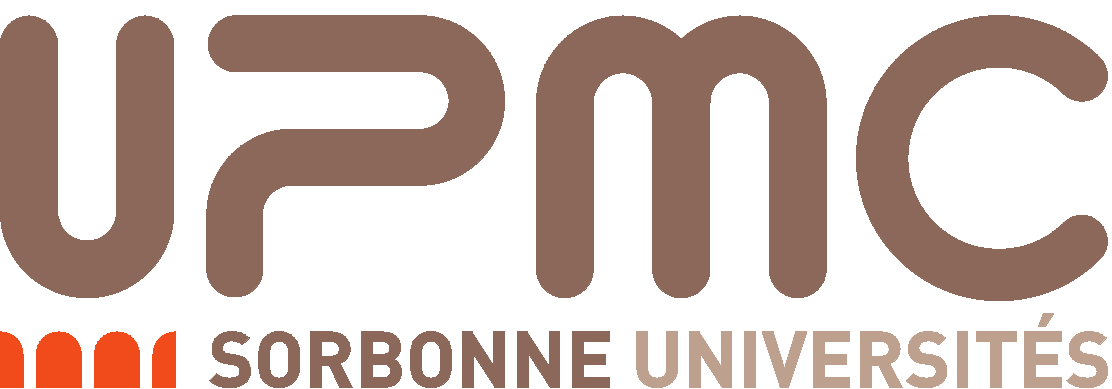}
	%
	\speciality{Physique Théorique}
	\doctoralschool[https://www.edpif.org/en/index.php]{Physique en île-de-France}
	\university[http://www.upmc.fr/]{Pierre et Marie \textsc{Curie}}
	\laboratory[http://www.lpthe.jussieu.fr/]{Laboratoire de Physique Théorique et Hautes Énergies}
	\name{Oscar}{\href{mailto:odefelice@lpthe.jussieu.fr}{de Felice}}
	\title[Solutions avec flux et Géométrie Généralisée Exceptionnelle]{Flux backgrounds and Exceptional Generalised Geometry}
	\abstract[%
			Cette thèse traite de compactifications avec flux en théorie des cordes et supergravité.\\
			D'abord, nous étudions les reductions dimensionnelles des théories de type II et de supergravité en onze dimensions, en utilisant la géométrie généralisée exceptionnelle.\\
			Nous commençons par l'introduction des techniques mathématiques necessaire à cette thèse, nous nous concentrons sur les $G$-structures et leur extension à la géométrie généralisée.\\
			Après, nous passons à discuter les compactifications à proprement parler.
			Précisément, nous nous concentrons sur type IIA, en construisant la version de la géométrie généralisée exceptionnelle décrivant cette supergravité et en trouvant les deformations de la dérivé de Lie généralisée correctes qui permettre de tenir compte et décrire correctement la mass de Romans.
			Nous présentons la méthode de Scherk-Schwarz généralisée qui nous permettre de trouver des ansatze consistants qui préservent la quantité maximale de supersymétrie.
			Aussi, nous appliquons cette méthode à des examples différents des truncations sur les spheres, nous sommes capables de reproduire l'ansatz sur la sphere six-dimensionnelle et le tensor d'imbrication, qui nous donne une supergravité jaugée $\ISO(7)$ dyoniquement  en quatre dimensions.
			Pour des spheres de dimension $d=2,3,4$, nous trouvons une obstruction à avoir des parallelisations généralisées dans les cas massifs. 
			Ceci donne une indication du fait que des reductions dimensionnelles en presence de mass de Romans peut pas exister.\\
			En outre, nous étudions les calibrations générales sur des backgrounds AdS en type IIB et M-théorie.
			Nous établissons que elles sont décrites par les structures de Sasaki-Einstein exceptionnelles, et nous focalisons notre attention sur les vectors de Reeb généralisés.
			Les inégalités pour la limite sur l'énergie peuvent être dérivées par la decomposition de la condition donnée par la symétrie $\kappa$ ou dans la même façon, par la decomposition des bilinéaires des champs spinoriels existants en literature.
			Nous expliquons comme la fermeture des formes de calibration est liée à l'integrabilité de la structure de Sasaki-Einstein exceptionnelle décrivant le background.
			En particulier, nous faisons ça pour des branes remplissants l'espace ou ponctuelles.
			En faisant ça, nous montrons que la partie de forme du vector twisté en M-théorie donne les correctes calibrations généralisées.
			Le cas au sujet des background en type IIB donne des résultats analogues.%
		]{%
					The main topic of this thesis are flux compactifications.\\
					Firstly, we study dimensional reductions of type II and eleven-dimensional supergravities using exceptional generalised geometry.\\
					We start by presenting the needed mathematical tools, focusing on $G$-structures and their extension to generalised geometry.\\
					Then, we move our focus on compactifications. 
					In particular, we mainly focus on type IIA, building the version of exceptional generalised geometry adapted to such supergravity and finding the right deformations of generalised Lie derivative to accomodate the Romans mass.
					We describe the generalised Scherk-Schwarz method to find consistent truncation ansatze preserving the maximal amount of supersymmetry.
					We apply such a method to several examples of truncations on spheres, we reproduce the truncation ansatz on $S^6$ and the embedding tensor leading to dyonically gauged $\ISO(7)$ supergravity in four dimensions.
					For spheres of dimension $d=2,3,4$, we find an obstruction to have generalised parallelisations in massive theory, giving the evidence that maximally supersymmetric reductions might not exist.\\			
					As further point, we study generalised calibrations on AdS backgrounds in type IIB and M-theory. 
					We find these are described by Exceptional Sasaki-Einstein structures and we place the focus on the generalised Reeb vectors. 
					The inequalities for the energy bound are derived by decomposing a $\kappa$-symmetry condition and equivalently, bispinors in calibration conditions from existing literature. 
					We explain how the closure of the calibration forms is related to the integrability conditions of the Exceptional Sasaki-Einstein structure, in particular for AdS space-filling or point-like branes. 
					Doing so, we show that the form parts of the twisted vector structure in M-theory provides the expected generalised calibrations. 
					The IIB case yields similar results.
				}
	\defensedate{26 Mars 2018}
	\addjurymember{M}{Henning}{\href{mailto:henning.samtleben@ens-lyon.fr}{Samtleben}}{Rapporteur}
	\addjurymember{M}{Alberto}{\href{mailto:Alberto.Zaffaroni@mib.infn.it}{Zaffaroni}}{Rapporteur}
	\addjurymember{F}{Anna}{\href{mailto:ceresole@to.infn.it}{Ceresole}}{Examinateur}
	\addjurymember{F}{Mariana}{\href{mailto:mariana.grana@cea.fr}{Gra\~na}}{Examinateur}
	\addjurymember{M}{Jan}{\href{mailto:jan.troost@lpt.ens.fr}{Troost}}{Examinateur}
	\addjurymember{F}{Michela}{\href{mailto:petrini@lpthe.jussieu.fr}{Petrini}}{Directrice de thèse}
	\titlepage
	\coverpage[Arthur \textsc{Rimbaud}]{Un soir, j'ai assis la Beauté sur mes genoux. -- Et je l'ai trouvée amère -- Et je l'ai injuriée.}
	\dedicationpage[To Martino,\\ for the hope. \\ To Gabri and Betta, \\ for the bravery.]
	\preface
	\remerciements
			Une thèse c'est une travail long et complex, et donc on peut comprendre comme soit impossible de le completer sans l'aide incommensurable des beaucoup de personnes.
			Peut-être que j'oublierai quelqu'un, mais si leur noms ne sont pas dans ma tête, ils sont dans mon coeur.
			
			Tout d'abord, je tiens à remercier vivement ma directrice de thèse, Michela pour avoir accepté de me prendre en thèse, quand je savais presque rien de théorie des cordes. 
			J'ai beaucoup appris sous sa tutelle, soit en physique, soit en general sous un point de vue humain. 
			Je la remercie pour toutes les discussions sans fin et pour le réponses à mes questions souvent bêtes.
			Je souhaite exprimer ma gratitude pour ça et pour avoir été un point de référence et un example incroyable de rigueur scientifique et personnel.
			
			L'amitié avec mes collègues thésards a été précieuse. Merci vraiment à tous.
			En particulier, sans les pauses café avec Enrico, Ruben, Sophie, Thomas, Charles, Johannes, Alessandro et Constantin, je n'aurais jamais survécu à la dernier année. 
			Sans l'aide des pauvres diables de Hugo, Matthieu et Thomas je n'aurai jamais appris le français et jamais eu l'opportunité de devenir si fort à SportsHead.
			Enfin, merci à Enrico pour sa force tranquille et à Ruben pour sa folle sagesse.
			
			Merci à tous les membres du \textsc{LPTHE} pour leur accueil chaleureux et pour avoir été une communauté de la quel je me suis senti toujours partie.
			En particulier, je suis vraiment reconnaissant envers Isabelle et Françoise pour leur aide avec la bureaucratie qui a été pour mois un obstacle parfois plus grand que la théorie des cordes.
			
			Je veux remercier aussi Dan Waldram, qui m'a introduit à la geometrie généralisée, et m'a donné confiance quand j'étais rien que un étudiant effrayé, et Charles Strickland-Constable pour les discussions précieuses sur la théorie des groups et représentations. 
			Une mention speciale à Davide Cassani, qui a été pas seulement une source de profond connaissance de la théorie des cordes, mais aussi un support pendant mes moments de difficulté.
			
			Ma vie à Paris n'aurais pas été la même sans l'amitié de Andrea, Davide, Lara, et tous les autres ``monas'', merci de coeur.
			Merci à mes amis loins, qui sont proches dans l'esprit, parce-que quand on se voit c'est comme si on avais toujours été ensemble.
			
			Je souhaite aussi souligner que je ne serais pas ici aujourd'hui sans ma famille. 
			Merci à ma mere et à mon pere pour être toujours la, sans être intrusifs, pour m'avoir enseigné à penser en façon indépendante et libre.
			Merci à ma soeur Sara, pour me donner toujours une comparaison et pour me soutenir quand je me sens perdu.
			Merci à ma soeur Camilla, pour être joyeuse et 	joviale, sans jamais être peu profonde, pour être sage et gentille avec tous.
			
			Finalement, je remercie chaleureusement les membres du jury pour avoir accepté d'être présents. Merci à Anna Ceresole, Mariana Graña, Henning Samtleben, Jan Troost and Alberto Zaffaroni.
		\abstracten
		\abstractfr
		\tableofcontents
	\content
	\chapter*{Introduction}
The major success of theoretical physics in the last half century is certainly the unification of the electromagnetic, strong and weak forces in the framework of the Standard Model.
This is is based on \emph{Quantum Field Theory} -- a framework where local excitations interact according to the laws of quantum mechanics and special relativity -- and is, 
as far as we know, the most complete, experimentally verifiable theory of fundamental interactions. 
%
	
What still remains an open question is how to include gravity in this picture, and this is due to the lack of renormalizability of the theory. 
%
Usually a quantum field theory has ultra-violet divergencies that are cured by adding to the action a finite number of terms canceling the divergencies and such that the physical quantities do not depend on them. 
This procedure does not hold for Einstein gravity and, at present, there is no 
%
satisfactory way to quantize gravity.

Among the candidates 
for a quantum theory of gravity, string theory is perhaps the most promising one since, at least in principle, it also leads to unification of gravity and the other forces.

String theory is based on the simple but revolutionary idea of replacing, at the fundamental scale, point-like particles with one-dimensional extended objects -- the strings.
		
			%
				\begin{figure}[h]
					\centering
						\scalebox{.3}{\includegraphics{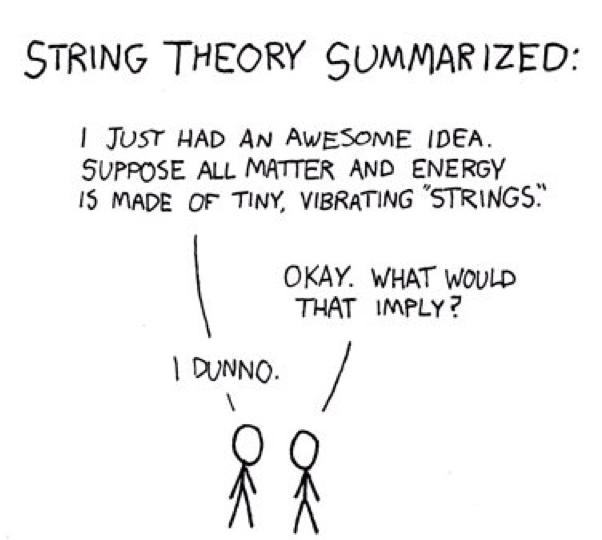}}
				\end{figure}
The fundamental constituents of the universe are extremely tiny vibrating strings moving in spacetime. Strings can have the topology of a segment -- open strings -- 
 or that of a circle -- closed strings.
%
	
The quantised string has a discrete spectrum of vibration modes, which at large distances (much larger than the characteristic string length $\ell_s = \sqrt{\alpha'}$) can be effectively
interpreted as different point particles. 
The spectrum contains a finite number of massless states and infinite tower of massive states with masses of order $1/\sqrt{\alpha'}$.
The theory was invented in another context (to describe strong interactions), but it came back to glory when it was realised that in the spectrum of the closed string there is always a massless spin 2 mode. One can identify it with the graviton, and therefore string theory automatically incorporates gravity. Since, among the massless states there can also be
gauge bosons, string theory provides a framework for the unification of all fundamental forces, that reproduces, as low energy limit, Einstein theory and gauge theories.

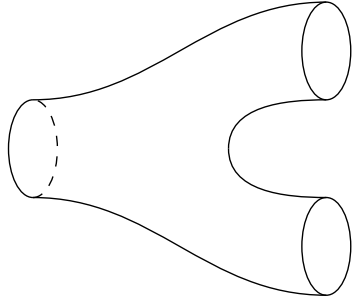
\begin{figure}[ht]
					\centering
						\scalebox{1.3}{	\begin{tikzpicture}[rotate=90]
		  \def\R{0.5}
		  \def\angEl{-30}
			   \DrawLatitudeCircle[\R]{0}
			  \draw (-1,-3) ellipse (.5 and .25);
			  \draw (1,-3) ellipse (.5 and .25);
			  \draw (-.5,0) to[out=-90,in=90] (-1.5,-3);
			  \draw (.5,0) to[out=-90,in=90] (1.5,-3);
			  \draw (-.5,-3) to[out=90,in=180] (0,-2);
			  \draw (0,-2) to[out=0,in=90] (.5,-3);
	\end{tikzpicture}}
					\caption{Feynman diagram of a string interaction vertex. 
						Imposing the finite dimension of the fundamental objects, we lose the \emph{locality} of the interactions, but we can cure the short-distance divergences.
						Now, Feynman diagrams are smooth $2$-dimensional surfaces and the interaction vertices have been ``smoothed out''.
					 }
					\label{stringdiagram}
				\end{figure}

The short distance singularities are avoided due to the extended nature of the string. A string sweeps a two-dimensional surface -- the world-sheet -- that is a smooth manifold and the interaction vertices are given by diagrams as in~\cref{stringdiagram}.
This provides an ultraviolet regularisation of the graviton scattering amplitudes, whose divergence was due to the point-wise nature of the interaction.
		
Generically the string spectrum contains 
tachyons, which can be interpreted as instabilities of the space-time.
These can be avoided by imposing that the spectrum is supersymmetry and gives rise to \emph{superstring theory}. \\
The evolution of the string is described by a two-dimensional conformal field theory defined on its world-sheet. 
To keep conformal invariance at the quantum level 
constrains the space-time where the string leaves to be ten-dimensional. \\
Taking into account all the constraints and consistency conditions, it turns out that there are only five possible superstring theories: type I, type IIA, type IIB and the two heterotic theories $\SO(32)$ and $\E_8 \times \E_8$, which have different field contents. 
In all these cases, it is possible to derive an effective quantum field theory for the massless states: these are called supergravity theories since they contain gravity and are
supersymmetric. Supergravity theories have been discovered independently from string theory in the mid seventies, 
and only later stage it was realised that they corresponded to the low energy limit of string theory.
Notice also that all supergravity theories are non renormalizable, but they make sense as effective theories of the string. \\
As we will discuss in detail later, a common feature of all ten-dimensional supergravity theories is the presence of higher-rank gauge fields -- the NS and RR fields -- 
which played a major role in all recent developments of string theory.

In mid 1990s the discovery of string dualities allowed to show that superstring theories are actually different formulations of the same theory, which are valid
in different corners of the parameter space and are related to more fundamental theory, which is conjectured to live in eleven spacetime dimensions, and has been given the name of M-theory.\footnote{%
			We know the $11$-dimensional supergravity, so the corresponding high energy fundamental theory is identified with M-theory.
			We do not have a complete formulation of M-theory yet, but we know the degrees of freedom of the theory.
			This tells us that the fundamental dynamical ingredients of this theory are not strings, but higher dimensional objects: branes.
			We are going back to branes in the following.}.	\\
The net of dualities involve other fundamental dynamical objects, besides strings, called branes. A Dp-brane is a solitonic entended\footnote{It has a (p+1)-dimensional world volume.} object that is charged under one 
of the NS and RR potentials, generalising the coupling of charged particle to the electric field. Crucially for many applications, in string theory, a 
 D-brane also has a perturbative description as dynamical hyperplanes on which open strings can end.\footnote{More precisely a Dp-brane is an open string with $p+1$ \emph{Dirichelet} boundary conditions.}
					\begin{figure}[h!]
						\centering
							\begin{tikzpicture}
		\draw[name path=braneA, smooth cycle, tension=0, fill=white, pattern color=orange, pattern=north west lines, opacity=.5] plot coordinates{(-5,2) (-6,0) (-6,-5) (-5,-3)};
		
		\draw[name path=braneB, smooth cycle, tension=0, fill=white, pattern color=blue, pattern=north west lines, opacity=.5] plot coordinates{(0,2) (-1,0) (-1,-5) (0,-3)};
		
		\draw [name path=openString, thick, red] (-5.3, 0) to[out=45, in = 210] (-1, -.5);
		\draw [name path=openString, thick, dashed, red] (-1, -.5) to[out=30, in = 190] (-.3, 0);
		
		\draw [color = green, smooth, tension=1, thick] plot coordinates{(-5.7, -1.5) (-4.7, -1) (-4.5, -2) (-4, -2.9)  (-5.7, -3.3)};

	\end{tikzpicture}
						\caption{Branes are hypersurface where open strings (or other branes) can end.}
						\label{branefig}
					\end{figure}
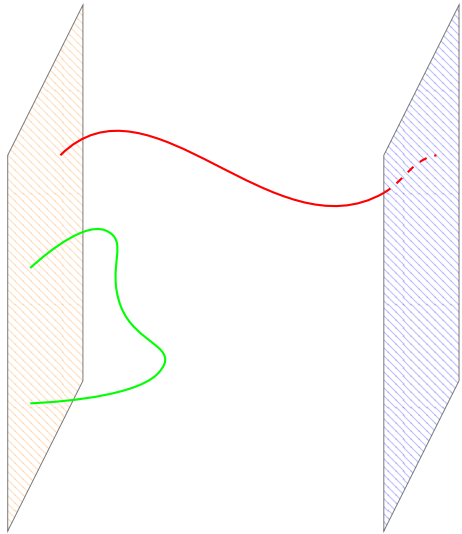
				%
			
			Because of the different spectra, each theory has its own stable branes. In M-theory there are M$2$ and M$5$ branes.
Type IIA has $\mathrm{D}_p$ branes with $p$ even, while Type IIB has $\mathrm{D}_p$ branes with $p$ odd.
Further, both type II string theories share a brane that couples electrically with the $2$-form Kalb-Ramond potential $B$, 
a $5$-brane usually called NS$5$ brane.
Branes are dynamical objects whose dynamic is governed by the fields that leaves on their world-volume. This is a very important property that lead to
the discovery of the AdS/CFT correspondence and all its developments.

		The introduction one-dimensional elementary objects seems to solve the conflict between general relativity and quantum field theory and to provide a framework for the unification of all fundamental forces.
		However, we pay a price. Space-time has extra-dimensions.
		This is one of the most striking predictions of string theory, and a model based on this framework with some hope to describe nature has to cope with this issue.
		
		One option to face this concern is called \emph{compacification}.
		It consists in assuming the original spacetime to have four large non-compact spacetime directions -- the ones we have experience of -- while the remaining ones are instead wrapped on themselves to form a very tiny compact space.
		The characteristic dimension of this space is very small, such to explain why we do not have access to it (there are actually some bounds on the maximal length these dimensions can have. These bounds are being updated constantly due to on going measures at LHC, see for example~\cite{extradim1,extradim2}).
		More formally, we are interested in string solutions with a topology like,
				\begin{equation*}
					\mathcal{M}_{10} = \mathcal{X}_4 \times M_6\, ,
				\end{equation*}
		where $\mathcal{X}$ is the non-compact \emph{external spacetime}, and $M_d$ is the compact \emph{internal} manifold.
		Because of the small size of the extra-dimensions, the motion and excitation of the strings will look to us essentially four-dimensional. 
%
However the external dimensions are not invisible: the features of the effective four-dimensional theory -- particle content, symmetries, masses, etc. -- 
depend on the geometry of the internal space.

Supersymmetry comes naturally in string theory, A major question is what amount of supersymmetry has the low-energy effective theory.
From a phenomenological point of view, the most desirable case is preserving a minimal amount of supersymmetry -- $\mathcal{N}=1$ -- since this allows for chiral fermions and it is 
compatible with the minimal extensions of the Standard Model.\footnote{%
			The scale of susy breaking can tested in present day accelators. It is highly possible that the scale of susy breaking is much higher than what expected
			until now and this must be taken into account in string compactification models.}
A residual supersymmetry is welcome also for technical purposes, since it guarantees the stability of given string solutions and also allows for simpler ways of finding
string vacua.

What is very important for this thesis is that the conditions for a given string background to be supersymmetric 
 translate into topological and differential conditions on the internal manifold and strongly constrains its geometry.
 The best known example is the case purely geometric compactifications of string theory to four-dimensions, where the internal manifold is constrained to be 
 a Calabi-Yau three-fold, namely a $6$-dimensional manifold with $\SU(3)$ holonomy~\cite{CYcomp}.
The geometry of Calbi-Yau manifolds is well known and this makes it possible to explicitly compute the low-energy four-dimensional effective actions on these compactifications.
However the theories obtained this way are not good for phenomenology since they contain many massless scalar fields -- the moduli -- that are not constrained by any potential.

 A possible way to solve the moduli problem is to find a mechanism to generate a potential for them in the lower-dimensional action.
This would have the effect of stabilise the moduli (giving them a mass and a fixed vev).
A great number of results in this direction have been reached in the last twenty years, realising that it is possible to generate a non-trivial scalar potential 
considering more general string back-grounds where some of the RR and NS field take non-trivial expectation values. \emph{fluxes}~\cite{fluxcomp1, fluxcomp2, fluxcomp3}.
One can find some nice reviews of the subject in~\cite{DuffReviewComp, MarianaFluxReview, LustReviewComp, henlect}.
 However such fields back-react on the geometry and new tools are required to analyse and study the internal compactification manifolds, which are not Calabi-Yau.
 The methods described in this thesis are promising tools in this direction.
 
 Usually, a dimensionally reduced theory has an infinite number of states with higher dimensional origin.
 We are interested in constructing effective actions with a finite number of degrees of freedom, hence we have to give a prescription to \emph{truncate} out some of the modes.
 We call \emph{truncation ansatz} this prescription.
 Among all the possible truncation ansatze, a particular class is made by the ones having the property of being \emph{consistent}.
 A \emph{consistent truncation} is a choice of a finite set of modes, where the omitted ones are not sourced by the subset chosen. 
 This is equivalent to say that the set of truncated modes has a dynamics which is not affected by the others.
 We are interested in such truncations since any solution of the lower dimensional theory can be uplifted to a solution of the higher dimensional one.

		Reasons to study compactifications are not only phenomenological.
		There are important formal motivations.
		Many supergravity theories in various dimensions are connected by compactifications.
		Historically, since the birth of supergravity, dimensional reductions have been used to build lower dimensional models from the higher dimensional ones.		
		A first milestone example is the derivation of the four-dimensional maximally supersymmetric supergravity theory from the eleven-dimensional supergravity, due to Cremmer and Julia~\cite{CremmerJulia}.
		
%

\subsection*{Outline of the thesis}

			Extension of differential geometry turned out to be powerful tools to study generic string compactifications. 
			One of these is Generalised Geometry and the main goal of this thesis is to study its applications to various contexts in supergravity, principally consistent truncations and brane calibrations.
			The thesis is organised as follows.
			
			In the first two chapters we introduce the mathematical tools which are needed in the following.
			In particular, in chapter~\ref{chap1} are given the main definitions and examples of $G$-structures. 
			The concept of torsion of a structure is analysed and we show how the torsionless conditions for some structure are equivalent to reformulate the supersymmetry conditions on the manifold.
			Finally, we expose some facts about the special holonomy of a manifold.
			
			Chapter~\ref{chapSugra} is needed to give the physical environment we are moving in. 
			We briefly review the main features of the theories of supergravity we analyse in this thesis. 
			In addition we show how the $G$-structures can be fruitfully used to describe fluxless compactifications, but how they fail to capture all the informations in the presence of fluxes.
			
			Chapter~\ref{chapEGG} is based on the exposition of the main aspect of Generalised Geometry, both complex (useful as introduction) and exceptional.
			We will face the generalisation of the $G$-structures in this context and we will build an example of how flux compactifications are elegantly encoded in this formalism.
			In particular, an appealing feature of this approach is that one can predicts the lower-dimensional supergravity independently of many explicit computations.
			We will focus on maximally supersymmetric truncations, dealing with generalised parallelisation, that we will briefly review in~\ref{secGenPar}.
			
			The core of the thesis are chapters~\ref{chapComp} and~\ref{chapbrane}.
			The~\cref{chapComp} is about flux compactifications and consistent truncations. 
			In particular, we define generalised Scherk-Schwarz reductions, and we build a concrete example of truncations of massive IIA, building also an appropriate version of Exceptional Generalised Geometry adapted to it~\cite{oscar1}.
			
			On the other hand, the~\cref{chapbrane} is centered on branes and their calibrations.
			We study brane probes in AdS backgrounds (both in M-theory and type IIB) and we look for the supersymmetric configurations of these probes. It is a notorious fact that these corrispond to probes wrapping (generalised) calibrated submanifolds.
			Using the formalism of generalised geometry and $G$-structures we show how the integrability conditions on generalised HV structures~\cite{AshmoreESE} are equivalent to have supersymmetric branes in AdS backgrounds. 
			In other words, one can state that HV structures provide generalised calibrations of branes in such backgrounds.
			
			The thesis is completed by two appendices collecting conventions and technical remarks in exceptional generalised geometry.
			
			The core of this thesis is built on the following works
					\begin{itemize}
						\item[] \cite{oscar1}~\bibentry{oscar1}
						\item[] \cite{oscar2}~\bibentry{oscar2}
					\end{itemize}
			Other projects still in progress are
					\begin{itemize}
						\item[] \cite{oscar3}~\bibentry{oscar3}
						\item[] \cite{oscar4}~\bibentry{oscar4}
						\item[] \cite{oscar5}~\bibentry{oscar5}
					\end{itemize}
\ensurepagenumbering{arabic}
	\chapter{Differential Geometry Preliminaries}
	\label{chap1}
		In this chapter we introduce the needed mathematical notions for this thesis. 
		In particular, we collect some definitions about $G$-structures, holonomy and torsion, in order to generalise them to the generalised analogous in the next chapter.
		These concepts are the mathematical way to encode the topological and differential conditions on the internal manifold coming from supersymmetry in string compactifications. 
		
		We will see in the following how precisely relate the existence and the integrability of $G$-structures to the supersymmetry of the compactified theory.
		
		\section{Introduction and Motivations}
				We are mainly interested in compactifications and dimensional reductions of 10 or 11-dimensional supergravity preserving a certain fraction of the original supersymmetry. We take the ansatz for spacetime solutions to be in the form
						\begin{equation}\label{spacetimeans}
							M = X \times M_{d}\, ,
						\end{equation}
where $X$ is a Lorentzian (\emph{external}) spacetime and $M_{d}$ is a Riemannian manifold, often called \emph{internal} space.
In order to have supersymmetry in the lower dimensional theory, the internal manifold must support spinors, namely they must be \emph{spin manifolds}.
The supersymmetry parameters in $D=10$ or $D=11$ dimensional are generically indicated by $\epsilon$ and transform in the fundamental representation of $\mathrm{Spin}(D-1,1)$.
The number of supersymmetries the lower dimensional theory is given by the decomposition ansatz for the higher dimensional supersymmetry parameters
				\begin{equation*}
				 \epsilon = \sum_{k=1}^\mathcal{N} \varepsilon^k \otimes \eta^k\, .
				\end{equation*}

				The $\varepsilon^k$ are the lower dimensional supersymmetry parameters, \emph{i.e.} $\mathrm{Spin}(D-d-1,1)$ spinors, while $\eta^k$ are commuting $\mathrm{Spin}(d)$ spinors.
				The number $\mathcal{N}$ of linearly independent spinors $\varepsilon^k$ is the number of supersymmetries of the lower dimensional theory.
				
In order to make sense of the supersymmetry in lower dimension, we assume that the internal spinors $\eta^k$ are globally defined. 
As we will see in the next section, this is a non-trivial topological condition, that imposes the reduction of the structure group of the internal manifold.
		\section{\texorpdfstring{$G$-structures}{G-structures}}
			\label{gstruc}
					Let $M$ be a manifold of real dimension $d$ and $TM$ is tangent bundle. 
					At any point $p \in M$ we can introduce a local basis, and write a generic vector $v$ as
							\begin{equation}
								v = v^a_{(\alpha)} e_a^{(\alpha)}\, .
							\end{equation}
					Its coordinates on the overlap of two patches, $U_\alpha \cap U_\beta$, are related by a local change of coordinates,
							\begin{align}
								v^a_{(\alpha)} = M_{\alpha\beta\phantom{a} b}^{\phantom{\alpha \beta}a}\ v^b_{(\beta)}\, ,
							\end{align}
					where the transformation $M_{\alpha\beta}$ is in $\GL(d, \mathbb{R})$.
					Since the construction depend on the point $p \in M$, the matrices $M_{\alpha\beta}$ can be seen as maps from the manifold to the group $\GL(d, \mathbb{R})$,
							\begin{equation*}
								\begin{tikzcd}[row sep=.3mm]
									M_{\alpha \beta} :\!\!\!\!\!\!\!\!\!\!\!\!\!\!\!\!\! & M \arrow{r} & \GL(d, \mathbb{R}) \\
 									& p \arrow[mapsto]{r} & M_{\alpha \beta}(p)
								\end{tikzcd}
							\end{equation*}
					and are called \emph{transition functions}. 
					They contain all the information about the non-trivial topology of the tangent bundle.
					The following cocycle condition is required on the triple overlap for consistency,
							\begin{equation}
								M_{\alpha\beta} M_{\beta\gamma} = M_{\alpha\gamma}\, ,
							\end{equation}
					and in addition,
							\begin{equation}
								M_{\alpha\beta}M_{\beta\alpha} = 1\, .
							\end{equation}
					Note that the last two equations are the closure and the existence of the identity axioms in a group. 
					The group $\GL(d, \mathbb{R})$ of transition functions\footnote{%
					One can define the transition functions also as the action of the group on the fibre, in order to change local frame.
						This is also mentioned above and it is related to the interpretation of active/passive transformations.%
						} %
					is called the \emph{structure group} of the tangent bundle.

\begin{figure}[h]
							\centering
								\scalebox{1.5}{\begin{tikzpicture}[>=stealth]
\draw[color=red](2,0)--(4,0)arc(0:70:4)--(90:2)arc(90:0:2)--cycle;
\fill[fill = red, opacity = .2] (2,0)--(4,0)arc(0:70:4)--(90:2)arc(90:0:2)--cycle;

\draw[->] (15:0.7)node[below,xshift=-2mm]{base manifold $M$}--(30:2);

\begin{scope}[bend right]
\foreach \i[count=\x] in {10,30,50,70}
{\node(a\x)[circle,fill,inner sep=1pt]at (\i:2.4){};
\draw[color= gray](a\x)to(a\x|-0,4);}

\foreach \i[count=\x] in {7,26,46,66}
{\node(b\x)[circle,fill,inner sep=1pt]at (\i:3){};
\draw[color= gray](b\x)to(b\x|-0,4);}

\foreach \i[count=\x] in {6,26,46,66}
{\node(c\x)[circle,fill,inner sep=1pt]at (\i:3.6){};
\draw[color= gray](c\x)to(c\x|-0,4);}

\path(c1)to coordinate[near start](d)(c1|-0,4);
\end{scope}
\draw[<-](d)--+(0.8,-0.5)node[right]{fibre};
\draw[->, color=blue](4.07,1.7) to [out=35, in= 20] +(-.2,1.77)node[above right]{$G$ action};
\node(e)[circle, fill, color=blue, inner sep=.5pt] at (4.07,1.7){};
\node(f)[circle, fill, color=blue, inner sep=.5pt] at (3.85,3.46){};

\end{tikzpicture}}
								\caption{The action of the structure group on the frame bundle is an action on the fibre only, leaving the point $p \in M$ untouched.}
								\label{gactfibre}
							\end{figure}
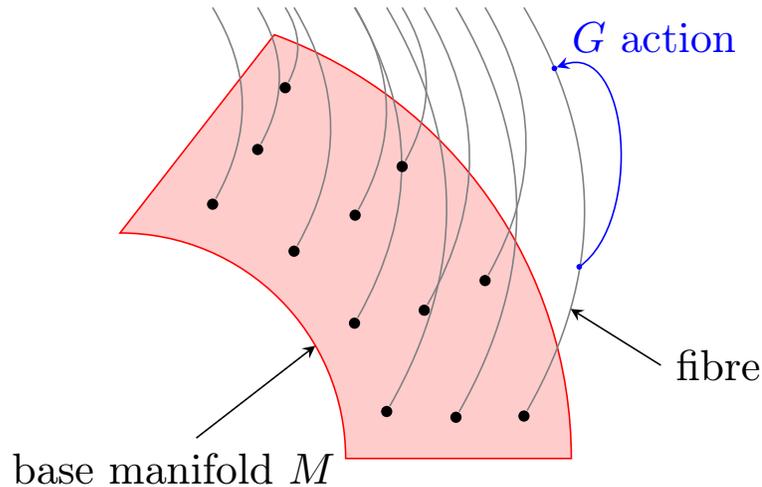

						We call \emph{Frame bundle} on $M$ the principal bundle whose fibres at any point $p \in M$ are all ordered basis -- \emph{i.e.} the frames -- of the tangent space $T_p M$, 
						\begin{equation}\label{framebund}
							F = \bigsqcup_{p \in M} F_p\, ,
						\end{equation}
				where,
						\begin{equation}
							F_p := \left\{ (p, \{ e_a \}) \mid p \in M \right\}\, .
						\end{equation}
				Generically, one can identify the fibre with $\GL(d,\mathbb{R})$.
				The group $\GL(d,\mathbb{R})$ acts freely and transitively on each fibre on the right - \emph{i.e.} $F_p$ is a principal homogeneous space for $\GL(d,\mathbb{R})$ -- to give another frame on the fibre. 
				From this point of view we can see the action of the group $\GL(d,\mathbb{R})$ as the way of changing frames keeping the point $p \in M$ fixed, so the change of frame is a transformation on the fibre only, see~\cref{gactfibre}.
				
				The set $\{e^{(\alpha)}_a\}$ defines in general a local frame on the manifold $M$ over the patch $U_{\alpha}$, namely a set of $d$ vector fields spanning $T_p M$. 
				Generally, it might not be possible to define it over the whole manifold, as it might not be possible to cover the manifold with a single chart.
				The particular case where a \emph{global frame} is defined is when the manifold is \emph{parallelisable}. 
				A noteworthy example of parallelisable manifolds are Lie groups. 
				This notion will be important for this work and we will discuss it -- and its generalisation in Generalised Geometry -- in detail in the next chapter.
				
				For simplicity we will present the various definitions using the tangent bundle, however, these and the related properties hold also for more general bundles $E$ on $M$, with a generic vector space $V$ as fibre, with a group $G$ acting on $V$.
						
					An alternative definition of the structure group is the group of the transition functions of the frame bundle $F$ in~\eqref{framebund}.
					Since a vector field $v$ is invariant across patches, we have that the frames transform with the inverse transformation with respect to the vector components,
							\begin{align}
								& & & & & & e_a^{(\alpha)} = M_{\alpha\beta\, a }^{\phantom{\alpha \beta a} b }\ e_b^{(\beta)}\, ,& & \mbox{on}\quad U_\alpha \cap U_\beta \, .& & & &
							\end{align}
					One can notice that given a reference frame, \emph{e.g.} the coordinate one $e_a \equiv \partial_a$, one can obtain a generic local frame via,
							\begin{equation}
								e^{\alpha}_a = e^{\alpha\phantom{a}b}_a \partial_b\, ,
							\end{equation}
					where $e^{\alpha\phantom{a}b}_a$ can be considered as $\GL(d,\mathbb{R})$ elements, giving us the right to identify the fibre of the frame bundle with the $\GL(d, \mathbb{R})$ group.

							\begin{figure}[h]
							\centering
									\scalebox{1}{\begin{tikzpicture}

%

	\draw[smooth cycle, tension=0.4, fill=white, pattern color=brown, pattern=north west lines, opacity=0.5] plot coordinates{(2,2) (-0.5,0) (3,-2) (5,1)} node at (0,0) {$M$};


    	\draw[smooth cycle, pattern color=orange, pattern=crosshatch dots] 
        		plot coordinates {(1,0) (1.5, 1.2) (2.5,1.3) (2.6, 0.4)} 
        		node [label={[label distance=-0.3cm, xshift=-2cm, fill=white]:$U_\alpha$}] {};
    	\draw[smooth cycle, pattern color=blue, pattern=crosshatch dots] 
        		plot coordinates {(4, 0) (3.7, 0.8) (3.0, 1.2) (2.5, 1.2) (2.2, 0.8) (2.3, 0.5) (2.6, 0.3) (3.5, 0.0)} 
        		node [label={[label distance=-0.8cm, xshift=.8cm, yshift=1cm, fill=white]:$U_\beta$}] {};
	
	\draw[dashed, thin, bend right] (1.7,1) -- (.7,2.5);
	\draw[dashed, thin] (2.5,.7) -- (2.5,2.9);
	\draw[dashed, thin] (3.5,.4) -- (4.2,2.2);
	
	\draw[orange, ->] (.7,2.5,0) -- (1.2,2.5,0);
	\draw[orange, ->] (.7,2.5,0) -- (.7,3,0);
	\draw[orange, ->] (.7,2.5,0) -- (.7,2.5,.6); 
	
	\draw[blue, ->] (4.2,2.2,0) -- (4.3,1.8,-.2);
	\draw[blue, ->] (4.2,2.2,0) -- (4.4,2.5,-.2);
	\draw[blue, ->] (4.2,2.2,0) -- (4.1,2.5,.7); 
	
	\draw[orange, ->] (2.5,2.9,0) -- (3,2.9,0);
	\draw[orange, ->] (2.5,2.9,0) -- (2.5,3.4,0);
	\draw[orange, ->] (2.5,2.9,0) -- (2.5,2.9,.6);
	\draw[blue, ->] (2.5,2.9,0) -- (2.6,2.5,-.2);
	\draw[blue, ->] (2.5,2.9,0) -- (2.7,3.1,-.2);
	\draw[blue, ->] (2.5,2.9,0) -- (2.4,3.2,.7); 
	
	\draw[-, dashed, color=gray] (3,2.9,0) to [bend left] (2.6,2.5,-.2) node[right, xshift=+2mm, color= black] {\footnotesize $O(d)$};
	\draw[-, dashed, color=gray] (2.5,3.4,0) to [bend left] (2.7,3.1,-.2);
	\draw[-, dashed, color=gray] (2.5,2.9,.6) to [bend left] (2.4,3.2,.7);

%
%
%
%
%
\end{tikzpicture}}%
								\qquad \qquad
									\scalebox{1}{\begin{tikzpicture}

	\draw[smooth cycle, tension=0.4, fill=white, pattern color=brown, pattern=north west lines, opacity=0.5] plot coordinates{(2,2) (-0.5,0) (3,-2) (5,1)} node at (0,0) {$M$};


    	\draw[smooth cycle, pattern color=orange, pattern=crosshatch dots] 
        		plot coordinates {(1,0) (1.5, 1.2) (2.5,1.3) (2.6, 0.4)} 
        		node [label={[label distance=-0.3cm, xshift=-2cm, fill=white]:$U_\alpha$}] {};
    	\draw[smooth cycle, pattern color=blue, pattern=crosshatch dots] 
        		plot coordinates {(4, 0) (3.7, 0.8) (3.0, 1.2) (2.5, 1.2) (2.2, 0.8) (2.3, 0.5) (2.6, 0.3) (3.5, 0.0)} 
        		node [label={[label distance=-0.8cm, xshift=.8cm, yshift=1cm, fill=white]:$U_\beta$}] {};
	
	\draw[dashed, thin, bend right] (1.7,1) -- (.7,2.5);
	\draw[dashed, thin] (2.5,.7) -- (2.5,2.9);
	\draw[dashed, thin] (3.5,.4) -- (4.2,2.2);
	
	\draw[orange, ->] (.7,2.5,0) -- (1.2,2.5,0);
	\draw[red, thick, ->] (.7,2.5,0) -- (.7,3,0);
	\draw[orange, ->] (.7,2.5,0) -- (.7,2.5,.6); 
	
	\draw[blue, ->] (4.2,2.2,0) -- (4.3,1.8,-.2);
	\draw[red, thick, ->] (4.2,2.2,0) -- (4.15,2.7,-.2);
	\draw[blue, ->] (4.2,2.2,0) -- (4.1,2.5,.7); 
	
	\draw[orange, ->] (2.5,2.9,0) -- (3,2.9,0);
	\draw[red, thick, ->] (2.5,2.9,0) -- (2.5,3.4,0);
	\draw[orange, ->] (2.5,2.9,0) -- (2.5,2.9,.6);
	\draw[blue, ->] (2.5,2.9,0) -- (2.6,2.5,-.2);
	\draw[blue, ->] (2.5,2.9,0) -- (2.4,3.2,.7); 
	
	\draw[-, dashed, color=gray] (3,2.9,0) to [bend left] (2.6,2.5,-.2) node[right, xshift=+1mm, yshift=+5mm, color= black] {\footnotesize $O(d-1)$};
	\draw[-, dashed, color=gray] (2.5,2.9,.6) to [bend left] (2.4,3.2,.7);

\end{tikzpicture}}
								\caption{%
									The generic structure of the frame bundle on a Riemannian manifold is $\rmO(d)$. 
									The $\rmO(d)$-action sends a generic frame to the corresponding one in an overlapping chart, preserving the metric. 
									On the right image, a simple example of reduction of the structure group, induced by the presence of a globally defined vector field.
									Frames are chosen to leave invariant the form of the vector.
									This constrains the transition functions to lie in $\rmO(d-1)$.}
								\label{figpatches}
							\end{figure}
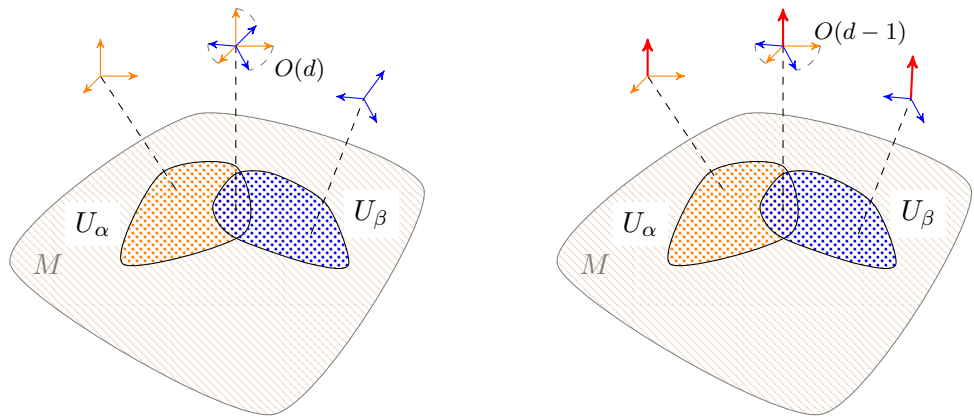
			\subsection{\texorpdfstring{$G$-structures}{G-structures}}
					A manifold $M$	 admits a $G$-structure if it is possible to reduce the structure group of $TM$ to a subgroup $G \subset \GL(d, \mathbb{R})$.
					This means that the transition functions take value in a subgroup $G$ of the general linear group.
					In other words, a $G$-structure is a principal sub-bundle of the frame bundle, $P \subset F$.
					
					An alternative definition can be given in terms of globally invariant tensors.
					We say a manifold $M$ has a $G$-structure if (and only if)\footnote{%
						Consider the inverse implication. 
						Given a non vanishing, globally defined tensor (or spinor), $\xi$, one can take the set of frames under which $\xi$ takes the same form.
						Then, it is possible to see that the structure group of the frame bundle reduces to the subgroup $G$.
						Thus, the existence of $\xi$ implies $M$ has a $G$-structure.%
						}
					there exists a globally defined $G$-invariant tensor or spinor.
					The relation between $G$-structure and globally defined invariant tensors holds also for other vector bundles, with structure groups different from $\GL(d,\mathbb{R})$. 
					In particular, it extends to spin bundles and spin structure groups.
					In these cases, the globally defined invariant objects are spinors.
					
					One can see the construction as follows. 
					The frames (\emph{i.e.} points on the fibres of the principal bundle) are elements of $\GL(d,\mathbb{R})$. 
					Once we reduce the structure group to some $G \subset \GL(d,\mathbb{R})$, they are connected by $G$ transformations.
					We can define an equivalence relation on the set of frames such that all the frames $G$-connected are equivalent.
					The coset made by modding out all the equivalent frame is $\GL(d,\mathbb{R})/G$ and the invariant tensor defining the $G$-structure is nothing else than a representative of this coset in a suitable representation.
					Concretely, all the possible (independent) choices of the invariant tensors are related to the equivalence classes of this relation. 
					Changing our choice of the invariant tensor on a manifold means to change the equivalence class, not the structure~\cite{liegroupstruc}.
					As typical example, consider a Riemannian metric $g$ on a manifold $M$.
					$g$ is an element of $\GL(d,\mathbb{R})/\rmO(d)$, thus all the metric given by an $\rmO(d)$ transformation of $g$ are equivalent.
					
					The known (and loved) structures that we are used to in differential geometry can be reinterpreted as $G$-structures on the manifold $M$, as summarised in~\cref{tabstruct}.
					Here we explore some examples in a bit more detail.

							\begin{table}[h!]
							\centering
								\begin{tabular}{l c r}
									Name					&	\begin{tabular}{@{\ }l@{}}
 																		Globally defined \\ 
																		 invariant tensor 
 																\end{tabular} 												&	$G$-structure								\\
									\midrule
									Metric					& 			$g$												&	$\rmO(d)$									\\[1.2mm]
									Orientation 				&			$\mathrm{vol}$										&	$\SL(d, \mathbb{R})$						\\[1.2mm]
									Metric Volume form 			&			$\mathrm{vol}_g$									&	$\SO(d)$									\\[1.2mm]
									Parallelisation	 			&			$\{e_a\}$											&	$\{e\}$									\\[1.2mm]
									Almost Symplectic structure	&			$\omega$	(real)	 $\Rightarrow \quad d$ even				&	$\Sp(d, \mathbb{R})$						\\[1.2mm]
									Almost Complex structure 	&			$I$ ($I^2 = - \mathrm{id}$) $\Rightarrow \quad d$ even		&	$\GL(d/2, \mathbb{C})$						\\[1.2mm]
									Almost Hermitian structure 	&			$\left.\begin{tabular}{@{\ }l@{}}
 																		$\omega$, $I$ with $I^T \omega I =\omega$ \\ 
																		$g$, $I$ with $I^T g I = g$	 \\ 
																		$\omega$, $g$ with $\omega^T g^{-1} \omega = g$
 																	 \end{tabular}\right\}$	 									&	$\U(d/2)$									\\
									\bottomrule
								\end{tabular}
								\caption{$G$-structures on a $d$-dimensional manifold $M_d$. 
									These structures are induced by globally defined tensors. The other way of thinking is also correct. 
									Given a $G$-structure, one or more invariant objects are determined.}
								\label{tabstruct}
							\end{table}
					\subsubsection{Orientation}
						A globally defined and nowhere vanishing $d$-form on a $d$-dimensional manifold is called \emph{volume form}.
						This is preserved by the group of transformations with unit determinant, \emph{i.e.} the $\SL(d,\mathbb{R})$ subgroup of $\GL(d,\mathbb{R})$.
						This structure fixes an orientation over the manifold, allowing us to define an integration operator over it.
					\subsubsection{Riemannian structures}
						A manifold is called \emph{Riemannian} when it admits a globally defined, positive-definite, symmetric covariant $2$-tensor, \emph{i.e.} a \emph{metric} $g$.
						In this case, one can choose a set of local frames $\{e_a\}$ on $M$ such that,
								\begin{equation}
									e_a^m e_b^n g_{mn} = \delta_{ab}
								\end{equation}
						and the structure group reduces from $\GL(d,\mathbb{R})$ to $\rmO(d)$.
						In the case it is possible to define a globally defined volume form associate to the metric $g$, the manifold is \emph{orientable} and the structure group further reduces to the subgroup of $\rmO(d)$ preserving this orientation,
								\begin{equation}
									\SO(d) = \SL(d,\mathbb{R}) \cap \rmO(d)\, .
								\end{equation}
					\subsubsection{Almost complex structures}
						Consider again a manifold of real dimension $d$. 
						An \emph{almost complex structure} is a globally defined tensor,
								\begin{equation}
									\begin{tikzcd}[row sep=tiny]
										I :\!\!\!\!\!\!\!\!\!\!\!\!\!\!\!\!\! & TM \arrow{r} & TM \\
 											& v^m \arrow[mapsto]{r} & I^m_{\phantom{m}n} v^n \, ,
									\end{tikzcd}
								\end{equation}
						satisfying,
								\begin{equation}
								\label{I2mid}
									I^m_{\phantom{m}n} I^n_{\phantom{n}k} = - \delta^m_{\phantom{m}k}\, .
								\end{equation}
						The condition above implies that $I$ is non-singular\footnote{%
							Note that locally one can always define a tensor with such properties, the crucial point to be an almost complex structure is that it has to be globally defined over the manifold.%
							}.
						
						Another consequence of the relation $I^2 = - \mathrm{id}$ is that the manifold dimension must be even.
						Indeed, taking the determinant of~\eqref{I2mid} we have $(\det I)^2 = (-1)^d$. 
						$I$ is a real tensor, then $\det I$ must be real, so $(-1)^d > 0$, and this implies $d$ is even.
						
						A manifold of real dimension $d=2n$ is called \emph{almost complex} if and only if it admits an almost complex structure. 
						It is easy to show that the structure group $\GL(d, \mathbb{R})$ reduces to $\GL(n, \mathbb{C})$.
						
						The structure $I$ can be used to split the tangent bundle $TM$ in two subspaces, corresponding to the eigenspaces of the map $I$. 
						They are related to the eigenvalues $\{ i, -i \}$\footnote{%
							A subtlety: in order to allow for eigenvectors related to complex eigenvalues, we have to take into account the complexification of the tangent bundle $TM \otimes \mathbb{C}$.%
							}
						and are one the complex conjugate of the other, and define a complex basis for the fibres.
						Hence, we can define projector operators,
								\begin{equation}
									P_{\pm} = \frac{1}{2}\left(\id \mp i I \right)\, ,
								\end{equation}
						onto the two eigenspaces.
						Since $I$ is globally defined, so are the projectors. This implies that the tangent bundle splits globally as,
								\begin{equation}\label{tsplit}
									TM \otimes \mathbb{C} = TM^{(1,0)} \oplus TM^{0,1}	\, .
								\end{equation}
						Since $I$ is real, $TM^{(1,0)}$ and $TM^{(0,1)}$ are sub-bundles of equal rank.
						Sub-bundles that are locally spanned by smooth vector fields are called \emph{distributions}.
						The tensor $I$, in a basis adapted to holomorphic and anti-holomorphic distributions, has the following fixed form,
								\begin{equation}
									I = \diag( i \id_n , -i \id_n )\, ,
								\end{equation}
						defined modulo $\GL(n, \mathbb{C})$ transformations.
						
						A generic $k$-tensor is decomposed into $p$ holomorphic indices and $k-p$ anti-holomorphic ones.
						Since for the complexified cotangent bundle an identical decomposition holds,
								\begin{equation}
									T^*M \otimes \mathbb{C} = T^*M^{(1,0)} \oplus T^*M^{0,1}\, ,
								\end{equation}
						then, a generic $k$-form is decomposed in holomorphic and anti-holomorphic components
								\begin{equation}
									\Lambda^k T^*M = \bigoplus_{i}^k \left( \Lambda^i T^*M^{(1,0)} \otimes \Lambda^{k-i} T^*M^{(0,1)} \right) =: \Lambda^{k,k-i} T^*M\, .
								\end{equation}
						We denote as $\Lambda^{p,q} T^*M$ the anti-symmetric bundle of rank $p+q$, whose section are the $(p,q)$-forms.\\

						The same reduction of the structure group can be seen in terms of the so-called \emph{fundamental} form, an holomorphic $n$-form $\Omega$.
						One can define a local coframe of $n$ independent $(1,0)$-forms $\phi^i \in \Gamma(\Lambda^{1,0} T^*M)$ and use it to define a local section of the bundle $\Lambda^{n,0} T^*M$, which is called \emph{canonical line bundle},
								\begin{equation}\label{Odec}
									\Omega = \phi^1 \wedge \ldots \wedge \phi^n\, .
								\end{equation}
						The fundamental form is non-degenerate,
								\begin{equation}
									\Omega \wedge \bar{\Omega} \neq 0\, .
								\end{equation}
						The form t $\Omega$ has to be \emph{simple}, that is locally decomposable into $n$ complex one-forms as in~\eqref{Odec}.
						In general, one should notice that an almost complex structure determines the forms $\phi^i$ only up to a $\GL(n,\mathbb{C})$ transformation. 
						This means that the fundamental form $\Omega$ can change between patches by an overall complex function (the determinant of the transformation).
						Thus, an almost complex structure does not need a globally defined $(n,0)$-form.
						However, if such a form exists the structure group reduces further to $\SL(n, \mathbb{C})$.
						Once we have a fundamental form on a manifold, we can extract the almost complex structure from it via the following relation,
								\begin{equation}
									I^m_{\phantom{m}n} = a\ \epsilon^{mm_1 \ldots m_{d-1}} (\RRe \Omega)_{n m_1 \ldots m_{d/2 - 1}} (\RRe \Omega)_{m_{d/2} \ldots m_d}\, ,
								\end{equation}
						where $\epsilon^{m_1 \ldots m_d}$ is the Levi-Civita symbol in $d$ dimensions and $a$ is chosen such that $I$ is suitably normalised to satisfy the~\eqref{I2mid}.
					\subsubsection{Pre-symplectic structures}
						A manifold $M$ of dimension $d = 2n$\footnote{%
							This hypothesis is not restrictive at all, in fact the existence of a non-degenerate pre-symplectic form requires the dimension to be even as for the complex structure above.%
								}
						is said to have a \emph{pre-symplectic structure} (or almost symplectic) if and only if there exists a globally defined, non-degenerate anty-symmetric real $2$-form $\omega$,
								\begin{equation}
									\omega \in \Omega^2 (M, \mathbb{R})\, .
								\end{equation}
						We can see it as an invertible linear map,
								\begin{equation}
									\begin{tikzcd}[row sep=tiny]
										\omega :\!\!\!\!\!\!\!\!\!\!\!\!\!\!\!\!\! & TM \arrow{r} & T^*M \\
 											& v \arrow[mapsto]{r} & \omega(v, \cdot) =: \iota_v \omega \, ,
									\end{tikzcd}
								\end{equation}
						and its non-degeneracy is equivalently re-written as,
								\begin{equation}
									\underbrace{\omega \wedge \ldots \wedge \omega}_{n\ \text{times}} \neq 0\, .
								\end{equation}
						In particular, an almost symplectic structure defines a volume form,
								\begin{equation}
									\mathrm{vol}_d = \frac{1}{n!} \omega^n\, ,
								\end{equation}
						and so an orientation on the manifold.
						
						The existence of such a structure reduces the structure group to $\Sp(d,\mathbb{R})$.
					\subsubsection{Almost Hermitian manifolds}\label{almHermstruct}
						A manifold $M$, endowed with a metric $g$ and an almost complex structure $I$, is \emph{almost Hermitian} if and only if the two structures are compatible, \emph{i.e.},
								\begin{equation}\label{hermetr}
									g_{pq} I^{p}_{\phantom{p}m} I^{q}_{\phantom{q}n} = g_{mn}\, .
								\end{equation}
						In this case the metric is said \emph{Hermitian}.
						
\begin{figure}
							\centering
								\begin{tikzpicture}
	\draw [blue, ultra thick] plot [smooth cycle] coordinates {(1,1) (0,-1) (-2.5,-1.5) (-1,1)};
	\draw [red, ultra thick] plot [smooth cycle] coordinates {(-1,1) (0,-1) (2.5,-1.5) (1,1)};
	\draw [yellow, ultra thick] plot [smooth cycle] coordinates {(0,3) (-1,1) (0,-1) (1,1)};
	
	\fill [green, opacity=.2] plot [smooth cycle] coordinates {(1,1) (0,-1) (-1,1)};
	
	\draw (-1.3,-.5) node [text = black] {$O(2n)$};
	\draw (1.3,-.5) node [text = black] {$Sp(2n, \mathbb{R})$};
	\draw (0,1.7) node [text = black] {$GL(n, \mathbb{C})$};
	\draw (0,.3) node [text = black] {$U(n)$};
%
\end{tikzpicture}
								\caption{%
									The intersection of two of the three groups of transformations $\GL(n, \mathbb{C})$, $\rmO(2n)$ and $\Sp(2n,\mathbb{R})$ is the same as the intersection of all the three. 
									It coincides with the group $\U(n)$ and gives an heuristic explanation of the fact that given two of the three structures, one can always build the third one.}
								\label{Uinters}
							\end{figure}
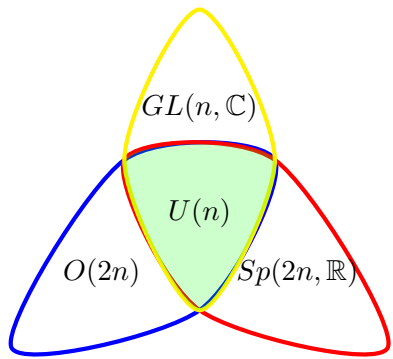

						On a manifold with two compatible invariant tensors, the structure group is reduced to the intersection of the groups leaving invariant the two compatible structures. 
						This is a general statement about $G$-structures of which we will make a large use in the rest of the thesis.
						Thus, on an almost Hermitian manifold, the structure group reduces to the intersection of $\rmO(d)$ and $\GL(d/2,\mathbb{C})$.
						This intersection group is $\U(d/2)$,
								\begin{equation}
									\U(d/2) = \rmO(d) \cap \GL(d/2, \mathbb{C}) = \Sp(d,\mathbb{R}) \cap \GL(d/2, \mathbb{C})\, .
								\end{equation}
						The last equality -- also represented in figure~\ref{Uinters} -- can be understood by the fact that given an Hermitian metric and an almost complex structure, one can alway rearrange the two invariant tensor to form a third one that has all the good properties of a pre-symplectic structure,
								\begin{equation}\label{symeco}
									\omega = \frac{1}{2} g_{mp} I^{p}_{\phantom{p}n} \dd x^m \wedge \dd x^n\, .
								\end{equation}
						The $\omega$ in~\eqref{symeco} is non degenerate by construction (from~\eqref{hermetr}).
						On the other hand, given a pre-symplectic and an almost complex structures satisfying the compatibility condition,
								\begin{equation}\label{hersym}
									\omega_{pq} I^{p}_{\phantom{p}m} I^{q}_{\phantom{q}n} = \omega_{mn}\, ,
								\end{equation}
						it is always possible to define a metric
								\begin{equation}\label{mesyco}
									g_{mn} = - \omega_{mp} I^{p}_{\phantom{p}n}\, ,
								\end{equation}
						which is automatically Hermitian, from the~\eqref{I2mid}.
						Finally, as expected, a pre-symplectic structure and a metric always define an almost complex structure.
						The compatibility is given by the following condition,
								\begin{equation}
									\omega^{T} g^{-1} \omega = g\, ,
								\end{equation}
						while, one can explicitly build the almost complex structure via,
								\begin{equation}
									I = - g^{-1} \omega\, .
								\end{equation}
					\subsubsection{Identity structure}
						Parallelisable manifolds are particularly relevant for supergravity compactifications.
						We define a parallelisable manifold in terms of vector fields and tangent spaces, follwoing~\cite{Nakahara}.
						
						Given a $d$-dimensional manifold $M$, a \emph{parallelisation} -- or \emph{absolute parallelism} -- of $M$ is a set $\{ v_1 ,\ldots, v_d\}$ of $d$ globally defined vector fields such that for all $p \in M$, the set $\{ v_1(p) ,\ldots, v_d(p)\} $ forms a basis for the tangent space in $p$, $T_p M$.
							
						If $M$ admits such a set, it is said to be \emph{parallelisable}.

						We can express the previous definition also saying that a manifold $M$ is parallelisable if admits a \emph{global frame}, since each basis vector $e_a$ is a globally defined smooth vector field.			
						From the point of view of $G$-structures, the existence of a parallelisation reduces the structure group to the identity element $\{e\}$.
						Because of this it is also said \emph{identity structure}, since transition functions on are actually identity maps, being the frame globally defined.
						This is equivalent to say that the parallelisation provides a global trivialisation of the tangent bundle $TM$.						
						Indeed, a parallelisation induces an isomorphism between tangent spaces in different points of the manifold.
						For example, aligning frames in each point, we can identify tangent spaces in that points. 
						
						An example of parallelisable manifolds are Lie groups.
						We can easily find a globally defined set of vector fields, forming a basis on $T_gG$ in each point $g \in G$. 
						These are the \emph{left(right)-invariant vector fields} $e_a$, which satisfy the Lie algebra
							\begin{equation}
							\label{liealg}
								\left[e_a , e_b \right] = f_{ab}^{\phantom{ab}c}e_c\ ,
							\end{equation}
						where $f_{ab}^{\phantom{ab}c}$ are constants called \emph{structure constants}.
						
						A further example is given by \emph{local group manifolds} \emph{i.e.} $ \mathcal{M} \cong G/\Gamma$, where $\Gamma$ is some discrete, freely-acting subgroup of the Lie group $G$.
						A possible parallelisation is given again by the left (right) invariant vector fields of $G$ if $\Gamma$ acts on the left (right). Moreover for any local group manifold, the parallelisation satisfies the~\eqref{liealg}.
						Technically, this happens because the condition $M = G / \Gamma$ implies that the left (right) invariant vector fields, which plays the role of generators of the Lie algebra $\mathfrak{g}$, must satisfy the commutation relations~\eqref{liealg}, with the structure constants that do not depend on the point on $M$.
						
						It is a well-known and remarkable result in algebraic topology -- due to Bott and Milnor et al. \cite{bott, Kervaire} -- that the only parallelisable spheres are $S^1,\ S^3,\ S^7$. This is related to the existence of the normed division algebras $\mathbb{C},\ \mathbb{H},\ \mathbb{O}$. 
						Another famous result is the non-existence of a parallelisation for a $2$-sphere. 
						This statement descends directly from the so-called \emph{hairy ball theorem}, a particular case of the \emph{Poincaré-Hopf theorem} that is considered very important since it provides a link between topological properties of manifolds and analytical ones \cite{adams, Hairy}.						
						
						Parallelisable manifolds (and their generalisation) will be very important to build maximally supersymmetric truncations of $10$- and $11$-dimensional supergravities, so we will analyse it in detail in the next chapters.
					\subsubsection{\texorpdfstring{$\SU(n)$ structures}{SU(n) structures}}
						In~\cref{almHermstruct}, we already mentioned that an almost complex structure and a compatible pre-symplectic structure reduce the structure group to $\U(d/2)$.
						In the case a globally defined $(n,0)$-form $\Omega$ (with $n = d/2$) exists associated to the almost complex structure, then we can define an orientation over the manifold and the structure group reduces further to $\SU(n)$.
						
						In terms of invariant tensors, an $\SU(n)$ structure is given by a globally defined, non-degenerate, simple $(n,0)$-form $\Omega$, a real non-degenerate two form $\omega$ and by an hermitian positive definite metric $g$.
						The forms $\omega$ and $\Omega$ satisfy the compatibility condition,
								\begin{equation}\label{compatib1}
									\omega \wedge \Omega = 0\, .
								\end{equation}
						The fundamental form $\Omega$ is usually normalised such that,
								\begin{equation}\label{compatib2}
									\Omega \wedge \overline{\Omega} = (-1)^{[n/2]}\, \frac{(2i)^{n}}{n!} \omega^n = \mathrm{vol}_{d}\, .
								\end{equation}
						The normalisation factors are chosen such that taking $n$ forms $\phi^i$ as a local basis of one-forms -- in which $\Omega$ takes the form~\eqref{Odec} -- the $\omega$ takes the canonical form,
								\begin{equation}
									\omega = - \frac{i}{2} \sum_{i} \phi^i \wedge \bar{\phi}^i\, .
								\end{equation}

						Let us assume that the manifold $M$ admits a spin structure. 
						We assume this having in mind to study supersymmetric background, which impose the internal manifold to be spin.
						Having a spin manifold means that the structure group $\SO(d)$ preserving metric and orientation can be lifted to $\mathrm{Spin}(d)$.
						Under these assumptions, we can equivalently define an $\SU(n)$ structure in terms of invariant spinors.
						
						We need some definitions.
						A spinor is said \emph{pure} if it is annihilated by half of the gamma matrices. Up to dimension $6$ this condition does not play any role, since any Weyl spinor is also pure. 
						However, in dimensions higher than $6$, this condition becomes relevant since it happens that exist spinors which are not pure.
						A concrete example is the case $d=7$, the only odd dimensional situation allowing for invariant spinors.
						The structure group reduces to $G_2$.
						
						A chiral pure spinor $\eta$, invariant under $\SU(n)$, reduces the structure group to $\SU(n)$.
						The inverse implication is also true.
						An $\SU(n)$-structure implies the existence of a globally defined, pure invariant spinor.
						This invariant spinor is related to the invariant forms we have seen above in terms of bilinears,
								\begin{equation}\label{spinorSUn}
									\begin{split}
										\omega_{mn} &= i \eta^\dagger \gamma_{mn} \eta\, , \\
										\Omega_{mnp} &= \eta^T C \gamma_{mnp} \eta\, ,
									\end{split}
								\end{equation}
						where $\gamma$ are the anti-symmetric products of gamma matrices of $\mathrm{Cliff}(d)$. 						%
				\comment{It might be useful to add a section describing $\SU(2)$ and $\SU(3)$ structures.}
				%
		\section{Holonomy and Torsion}
				So far we discussed how the existence of certain invariant, globally defined objects is equivalent of a reduction of the structure group of the frame bundle.
				This is a topological notion. There are also differential conditions one can impose, which correspond to the notion of integrability of a structure.
				In the following of the thesis we will see how these differential conditions on the $G$-structures are related in string theory to the the condition that a given
				vacuum is supersymmetric. 
					\comment[w]{%
						I don't know whether to talk about integrability for distributions and Frobenius theorem.%
						}
					\comment{%
						We have seen, for instance in the case of an almost complex structure, that the (complexified) tangent bundle is split in sub-bundles corresponding to the eigenspaces of the structure.
						These sub-bundles are called \emph{distributions}.
						The integrability of the structures (when it is possible to associate them to a distribution) is related to the integrability of the corresponding distribution.
						For this reason we give the definition of the integrability for distributions first.%
						}%
			\subsection{Integrability of structures}
					We will not give a rigorous definition of integrability.
					The curious reader can refer to many detailed references, a non-exhaustive list comprends~\cite{int1, int2, int3}.
					
					Heuristically we can say that an integrable structure (useful for our purposes) is a structure for which is possible to find a system of adapted coordinates on the manifold, such that the structure takes a particularly simple and fixed form.
					For our concerns, we will rephrase integrability in terms of intrinsic torsion.
					The structure is integrable if its compatible connection is torsion-free.
					We recover the notion of \emph{special holonomy} manifolds as the special case where all the torsions vanish.
					
					We now proceed with examples of integrability for the structures of the previous section in order to clarify the concepts.
				\subsubsection{Complex structure}
						Given an almost complex structure $I$, we say it is \emph{integrable} if and only if there exists a set of coordinates where it takes the form,
								\begin{equation}
								\label{diagI}
									I =	\begin{pmatrix}
											i \id	&		0		\\
												0		&	i \id	\\
										\end{pmatrix} \, .
								\end{equation}
						An integrable almost complex structure is called simply \emph{complex structure}.
						There are two equivalent definitions of integrability for an almost complex structure. 							%
								\begin{itemize}
									\item[-]	The first one refers to the Nijenhuis tensor
													\begin{equation}\label{nijten}
														N_I(v,w) := I \left[ I v, w \right] + I \left[ v, I w \right] - \left[ I v, I w \right] + \left[ v, w \right]\, ,
													\end{equation}
									where $v, w$ is a pair of vector fields.
											One can prove $N_I$ is actually a tensor, since it depends only on the local value of $v, w$ at each point.
											One can show that $I$ is integrable if and only if
													\begin{align*}
														& & N_I(v,w) = 0 & & \forall v, w\, .& &
													\end{align*}
									\item[-]	Alternatively, the complex structure $I$ is integrable if $TM^{(1,0)}$ part (and symmetrically the $TM^{(0,1)}$ part) are closed under the Lie bracket\footnote{
												Actually, this property of a distribution is called \emph{involutivity}. 
												A distribution is integrable if and only if it is involutive. 
												This is the content of the \emph{Frobenius theorem}.
												Because of the equivalence stated by the Frobenius theorem, often involutivity and integrability are concepts which are identified also in definitions.%
													}
													\begin{align*}
														& &	P_{\mp} \left[ P_{\pm} v, P_{\pm} w \right] = 0	& &	\forall v,w \in TM \, .	& &
													\end{align*}
												where $P_\pm$ are the projectors onto the $T^{(1,0)}$ and $T^{(0,1)}$. 
												That is, the Lie bracket of two (anti-)holomorphic vector fields must be (anti-)holomorphic.
											
												One can see that both the real and imaginary part of the relation above are proportional to $N_I$.
								\end{itemize}
						An almost complex manifold admitting a complex structure is said \emph{complex manifold}.
						We have seen that for almost complex manifolds possible to define complex coordinates in a single patch.
						For a complex manifold, the transition functions are holomorphic functions of the complex coordinates.
						Precisely, requiring holomorphic transition functions allows us to write the $I$ in the diagonal form~\eqref{diagI}.
				\subsubsection{Symplectic structure}
						The integrability for an almost symplectic structure is equivalent to the existence of a set of coordinates on $M$ such that
								\begin{equation}
									\omega = \dd x_m \wedge \dd y^m\, ,
								\end{equation}
						and such that the transition functions are symplectic with respect to the structure.
						Thanks to an important result in differential geometry, the Darboux theorem, we can translate integrability into a differential equation for the structure,
								\begin{equation}
									\dd \omega = 0\, .
								\end{equation}
						In other words, we simply demand the closure of the structure.
						
						With lack of fantasy, an integrable pre-symplectic structure is called \emph{symplectic structure} and a manifold endowed with it a \emph{symplectic manifold}.
						
					\paragraph{K\"ahler manifold} Following~\cite{KahlBook}, a \emph{K\"ahler manifold} is a complex manifold $(X, I, g)$ with an Hermitian metric $g$, whose associated symplectic structure $\omega$ is integrable.
					
					In other words, given the metric and the complex structure, the form constructed as in~\eqref{symeco} is closed,
							\begin{equation}
								\dd \omega = 0\, .
							\end{equation}
					When this happens, $\omega$ is called \emph{K\"ahler form}.
			\subsection{Intrinsic torsion}
					To any $G$-structure one can associate the notion of intrinsic torsion.
					This is because, as discussed in~\cref{gstruc} a $G$-structure can be seen as a principal bundle over the manifold $M$, with fibres given by the group $G$.
					On this bundle one can defines a connection, known as \emph{principal connection} and define its torsion.
					As argued in~\cite{joyce}, given a $G$-structure on $M$, a principal connection on the corresponding principal bundle always induces a connection $\nabla$ on the tangent bundle $TM$\footnote{%
						More precisely one can say that a general connection $D$ is compatible with a $G$-structure if and only if the corresponding connection on the principal frame bundle $F$, restricetd to the sub-bundle $P_G$ defining the $G$-structure, is still a connection on $P_G$~\cite{koba}.
						}.
					Such a connection is said to be \emph{compatible} with the structure and all the properties of the principal connection are inherited by the induced connection.
					For this reason we deal with connection on $TM$ and we are allowed to talk about torsion of the associated $G$-structure.
					
					We start the discussion about the intrinsic torsion by recalling the notion of compatibility of the connection on a manifold with a given structure.
					
					Let $(M,\Phi)$ be a manifold endowed with a $G$-structure defined by a tensor $\Phi$. A connection\footnote{%
						A connection $\nabla$ on a vector bundle $E$ is a map
								\begin{align*}
									\nabla : \Gamma\left(E\right) \rightarrow \Gamma\left(E \otimes T^*M \right)\, ,
								\end{align*}
						such that $\nabla(f v) = f \nabla(v) + v \otimes \dd f$, where $v$ is a section of $E$ and $f$ is a smooth function on $M$.%
						}	
					$\nabla$ on the tensor bundle is said to be \emph{compatible} with the structure if
							\begin{equation}
								\nabla \Phi = 0 \, .
							\end{equation}
					In general, given a structure the compatible connection is not unique.
					All the compatible connections differ by a tensor called \emph{torsion tensor},
							\begin{equation}\label{torsmap}
								T_{\nabla} (v,w) = \left[\nabla_v w , \nabla_w v \right] - \left[v , w \right]\, ,
							\end{equation}
					where $v$ and $w$ are generic sections of the tangent bundle and $[ \cdot, \cdot]$ the ordinary Lie bracket.
					
					In general given two different connections they will have a different torsion. 
					However, given a $G$-structure it is always possible to identify a part of the torsion which does not depend on the particular choice of the connection, but only on the $G$-structure.
					This part is called \emph{intrinsic torsion}.
					If the intrinsic torsion vanishes, then the associated $G$-structure is \emph{torsion-free}, \emph{i.e.} there exists at least one connection whose torsion tensor vanishes.
					The interesting fact we want to focus on is the identification of the integrability of the structure with its being torsion-free.
					
					Since we are interested in the application of these concepts in string theory compactifications, where we always deal with Riemannian manifolds, we are going to define intrinsic torsion in this case.
					For a rigorous definition of these notions for a generic manifold we refer again to~\cite{joyce}.
				\subsubsection{Metric compatible connection}
				 A Riemannian manifold has $\rmO(d)$-structure.
						Then, in this case we consider a connection $\nabla$ compatible with the metric, 
								\begin{equation}
									\nabla g = 0 \, .
								\end{equation}
						We have a torsion free connection, if the torsion of the connection vanishes.
						The interesting fact about Riemannian compatible connections is that the torsion free metric compatible connection exists and is unique, the \emph{Levi-Civita connection}.
						This result is the content of the so-called \emph{fundamental theorem of Riemannian geometry}.
						This implies that any metric on $M$ defines an $\rmO(d)$-structure which is torsion free.
						For the following, let us further restrict our attention to orientable manifolds, \emph{i.e.} to $\SO(d)$-structures.
						
						Given a generic metric compatible connection, it can be written as,
								\begin{equation}\label{contrs}
									\nabla = \nabla_{LC} + \kappa\, ,
								\end{equation}
						where we denoted with $\nabla_{LC}$ the Levi-Civita connection and with $\kappa$ a tensor called the \emph{contorsion}.
						
						The contorsion $\kappa$ is the difference between a generic torsion full connection and the torsion free connection, both metric compatible.
						From the compatibility condition, we require,
								\begin{equation}
									\kappa_{mnp} = - \kappa_{mpn}\, ,
								\end{equation}
						where indices are lowered/raised by the metric.
						One can easily show that there is a one-to-one correspondence between the torsion and contorsion tensors,
								\begin{align*}
									T_{mn}^{\phantom{mn}p} &= \kappa_{[mn]}^{\phantom{[mn]}p}\, , \\
									\kappa_{mnp} &= T_{mnp} + T_{pmn} + T_{npm} \, .
								\end{align*}
						Hence, we will use the terms torsion and contorsion equivalently.
						
						A consequence of the correspondence between torsion and contorsion is that given a torsion $T$, we can build a connection which torsion $T$ and we are guaranteed that such a connection is actually unique.
						
						Let us now take the case where a further reduction of the structure group is possible. 
						That is, a globally-defined, invariant tensor under a group $G \subset \SO(d)$ exists.
						In general a $G$-structure compatible connection will satisfy
								\begin{equation}
									\nabla g = \nabla \Phi = 0\, .
								\end{equation}
						In general, there is no reason for the torsion of this connection to be zero.
						Thus, as we have seen above in~\eqref{contrs}, we write 
								\begin{equation}
									\nabla \Phi = \nabla_{LC} \Phi + \kappa_0 \Phi = 0\, ,
								\end{equation}
						where $\kappa_0$ is called \emph{intrinsic contorsion}\footnote{%
							Heuristically speaking, the intrinsic (con)torsion is the measure of the failure of the structure to be convariantly constant with respect the Levi-Civita connection.
							}.
						It is the bit of the contorsion acting non-trivially on the structure tensor.
						
						To be more explicit, let us consider the symmetry properties of the contorsion.
						$\kappa$ is an element of $T^*M \otimes \Lambda^2 T^*M$.
						Observe that the algebra of $\SO(d)$ -- here denoted as $\mathfrak{so}(d)$ -- is isomorphic to the (linear) space of two forms $\Lambda^2 T^*M$.
						Hence, the contorsion can be interpreted as a one form taking values on the $\mathfrak{so}(d)$ algebra,
								\begin{equation}
									\kappa \in T^*M \otimes \mathfrak{so}(d)\, .
								\end{equation}
						Let us denote by $\mathfrak{g}$ the sub algebra of $\mathfrak{so}(d)$ corresponding to the algebra of the group $G \subset \SO(d)$.
						We can split the map $\kappa$ in a piece taking values in $\mathfrak{g}$ and another part taking values in $\mathfrak{g}^{\perp}$, the orthogonal complement of $\mathfrak{g}$.
						Explicitly,
								\begin{align}\label{decomk}
								&&	\kappa = \kappa_0 + \kappa_\mathfrak{g}\, , & & 	\begin{array}{l}
																				\kappa_0 			\in T^*M \otimes \mathfrak{g}^\perp \\[1mm]
																				\kappa_\mathfrak{g} \in T^*M \otimes \mathfrak{g}
																			\end{array}	&&
								\end{align}
						Since $\Phi$ is $G$-invariant, the action on it of the generators of $G$ is trivial, \emph{i.e.} $g \cdot \Phi = 0$, $\forall g \in \mathfrak{g}$.
						So,
								\begin{equation}
									\nabla \Phi = (\nabla_{LC} + \kappa_0 + \kappa_{\mathfrak{g}}) \Phi = (\nabla_{LC} + \kappa_0) \Phi = 0\, .
								\end{equation}
						Thus, the difference between two $G$-compatible connections only lies in the $\kappa_{\mathfrak{g}}$ part of the contorsion.
						All the $G$-compatible connections share the same intrinsic contorsion $\kappa_0$, which is a property of the $G$-structure itself and not of the particular choice of the connection.
						
						The \emph{intrinsic torsion} is defined from the intrinsic contorsion 
								\begin{equation}
									T^{0\phantom{mn}p}_{mn} = \kappa_{0[mn]}^{\phantom{0[mn]}p} \, .
								\end{equation}
						The intrinsic torsion is a very important tool, since it provides a classification of $G$-structures.
						The idea is that it is possible to decompose $\kappa_0$ into irreducible representations of the group $G$.
						Then a $G$-structure will be specified in terms of the representations in `$\kappa_0$. 
						When $\kappa_0$ vanishes, for instance $\nabla_{LC}\Phi = 0$, the structure is torsion-free.
				\subsubsection{Complex structure}
						An almost complex structure reduces the structure group to $\GL(n, \mathbb{C})$.
						A compatible connection $\nabla$, as we have seen, is such that $\nabla I = 0$.
						In this case, we can show that the integrability of the structure is equivalent to to the vanishing of the intrinsic torsion of the $\GL(n, \mathbb{C})$-structure.
 						One can use $\nabla I = 0$ and the definition of torsion map~\eqref{torsmap} to show that, for any $v,w$, the Nijenhuis tensor~\eqref{nijten} can be written as
								\begin{equation}
									N_{I} (v, w) = T_{\nabla} (v,w) - T_{\nabla} (Iv, Iw) + I T_{\nabla} (Iv,w) + I T_{\nabla} (v, Iw) \, .
								\end{equation}
						Thus, being the Nijenhuis tensor proportional to the torsion tensor, if the latter vanishes so does the former, and the almost complex structure is integrable.
 						Since the equation above does not depend on the choice of a particular connection, the obstruction to the integrability of the almost complex structure only comes from the intrinsic torsion.
				\subsubsection{Symplectic structure}
						As already discussed, the pre-symplectic structure corresponds to a reduction of the structure group to $\Sp(d,\mathbb{R})$.
						We can find a torsion free compatible connection if and only if $\omega$ is integrable.
						Let us consider a compatible connection $\nabla$, $\nabla \omega = 0$.
						Then, one can show,
								\begin{equation}
									\dd \omega (u, v, w) = \omega(T_\nabla (u,v), w) + \omega(T_\nabla (w,u), v) + \omega(T_\nabla (v,w), u)\, .
								\end{equation}
						Hence the vanishing of the torsion tensor implies the closure (and so the integrability) of the structure.
						Also in this case, one can prove that the result does not depend on the particular choice of the connection, but on the intrinsic torsion of the $\Sp(d, \mathbb{R})$-structure only.
			\subsection{Special Holonomy}
					We want now to analyse another differential property of a manifold: the \emph{holonomy} of a connection.
					This concept will be deeply related to supersymmetry in string compactifications.

					Let us consider a manifold $M$ and a differentiable\footnote{%
						In the rest of this thesis, unless specifically indicated, we will not distinguish between continuity and differentiability. In other words, a $\mathcal{C}^0$ function will always be also $\mathcal{C}^\infty$ and an homeomorphism will always be also a diffeomorphism.%
						}
					curve $\gamma$ on it, \emph{i.e.} $\gamma : I \rightarrow M$, where $I$ is a real open interval.
					Then, let $E$ be a vector bundle over its base manifold $M$ and $\nabla$ a connection on $E$. 
					The connection $\nabla$ provides a way of moving elements of the fibres along a curve. 
					In particular, it defines an isomorphisms between the fibres at different points along the curve, the so-called \emph{parallel transport} map,
							\begin{align}
								& & & & P_\gamma : E_{\gamma(s)} \rightarrow E_{\gamma(t)} & & \forall t,s \in I\, .
							\end{align}

					Let us now consider a \emph{loop}, \emph{i.e.} a closed curve, based at a point $x \in M$, then the parallel transport is an automorphism of the vector bundle at the point $x$,
							\begin{equation}
								P_\gamma : E_{x} \rightarrow E_{x}\, .
							\end{equation}
					As said, the map $P_\gamma$ is an automorphism, so both linear and invertible.
					It is an element of $\GL(E_x) \cong \GL(k, \mathbb{R})$, where $k$ is the dimension of the fibres.
					The set of all the possible parallel transports, for the all possible loops based at $x$ defines the \emph{holonomy group} of the connection $\nabla$ based at $x$,
							\begin{equation*}
								\mathrm{Hol}_x (\nabla) := \left\{ P_\gamma \mid \gamma \text{ based at } x \right\}\, .
							\end{equation*}
					In the case of a simply connected manifold $M$, the holonomy group depends on the base point only up to conjugation by an element of $\GL(k,\mathbb{R})$.
					More explicitly, if $\psi$ is a path connecting $x$ to $y$ in $M$, then
							\begin{equation*}
								\mathrm{Hol}_x (\nabla) = P_\psi^{\vphantom{-1}}\ \mathrm{Hol}_y (\nabla)\ P_\psi^{-1}\, .
							\end{equation*}
					Hence, given this relation one often (and we will follow this use) drops the reference to the base point, understanding that the definition holds up to group element conjugation.
					
					An important fact about holonomy is that given a a connection on a vector bundle, there is a relation between the action of the holonomy algebra $\mathfrak{hol}(\nabla)$ and the curvature of the connection (for detail on can read~\cite{holon2}).
					Recall that the curvature of a connection $\nabla$ is a two-form $R$ taking values in the Lie algebra\footnote{%
						One can write a connection as the differential operator,
								\begin{equation*}
									\nabla = \dd + A \, ,
								\end{equation*}
						where A is a one-form with values in the Lie algebra of the structure group of M, also called (often misleading) \emph{connection}.
						On a coordinate basis, the connection has the familiar form in physics literature: $\nabla = \dd x^m \otimes \nabla_m$, where and $\nabla_m$ is usually known as covariant derivative.
						
						The \emph{curvature} of the connection is defined as the two-form
							\begin{equation*}
								R := \dd A + \frac{1}{2} \left[A , A\right]\, .
							\end{equation*}
						}
					of the structure group.
					
					Consider now a tensor bundle over a manifold $M$ and suppose that this admits a covariantly constant tensor,
							\begin{equation*}
								\nabla \Phi = 0\, .
							\end{equation*}
					Then this tensor is invariant under parallel transport, and so also under the holonomy group.
					As a consequence the holonomy group cannot be the full $\GL(d, \mathbb{R})$, but it must be a subgroup, precisely the one leaving $\Phi$ invariant.
					The opposite statement holds as well, every time we have a reduced holonomy group, there exists an invariant object.
					In this case we say the manifold has a \emph{reduced holonomy}.
					
					Let us consider a connection with reduced holonomy. 
					We know we can always write it as $\nabla = \nabla_{LC} + \kappa_0$.
					This implies one can always find a $G$-structure such that the holonomy group corresponds to $G$, \emph{i.e.} $\mathrm{Hol}(\nabla) = G$.
					If the connection is Levi-Civita, then the corresponding $G$-structure is torsion-free.
							\begin{table}[h!]
							\centering
								\begin{tabular}{c | c | c | c}
									Holonomy					&			dim($M$)							&	invariant tensors		&	Manifold				\\
									\midrule
									$\SO(d)$					& 			$d$								&	$g, \mathrm{vol}_g$		&	(Riemannian) Orientable	\\[1.2mm]
									$\U(n) $	 				&			$d = 2n$							&	$g, \omega$			&	K\"ahler				\\[1.2mm]
									$\SU(n)$		 			&			$d = 2n$							&	$\omega, \Omega$		&	Calabi-Yau			\\[1.2mm]
									$\Sp(n)$		 			&			$d = 4n$							&						&	Hyperk\"ahler			\\[1.2mm]
									$\Sp(n) \Sp(1)$				&			$d = 4n$							&						&	Quaternionic K\"ahler	\\[1.2mm]
									$\mathrm{G}_2$		 	&			$d = 7$							&		$\phi_3$			&	$\mathrm{G}_2$		\\[1.2mm]
									$\mathrm{Spin}_7$		 	&			$d = 8$							&		$\Omega_4$		&	$\mathrm{Spin}_7$		\\
									\bottomrule
								\end{tabular}
								\caption{Various reduced holonomies.}
								\label{tabhol}
							\end{table}
			\section{Examples: Calabi-Yau and Sasaki-Einstein}
			\label{CYSE}
				Here we want to discuss two very important examples of manifolds with reduced holonomy: \emph{Calabi-Yau} and \emph{Sasaki-Einstein} manifolds.
				The former are even-dimensional manifolds, while the latter are odd-dimensional.
				The aim of this section is not to be complete, but to give the needed concepts for what follows in the thesis.
				A more complete source of informations (and of references) is~\cite{CalabiYauReview} for Calabi-Yau manifolds, and~\cite{SparksSE} for Sasaki-Einstein ones.
				
				We already restricted our attention to spin manifolds, since we are interested in supergravity compactifications.
				Here we want to justify this choice, saying that this is not too restrictive, since following~\cite[prop. 3.6.2]{joyce}, given a $d$-dimensional manifold $M$, with $d \geq 3$, admitting $G$-structure with $G$ simply-connected subgroup of $\SO(d)$, then $M$ is spin.
				
				For the sake of concreteness, let us restrict to the case of a six-dimensional manifold. 
				These manifolds are of particular interest since they appear in compactifications of ten-dimensional type II string theory down to four dimensions.
				The structure group $\GL(6)$ is reduced to $\SO(6) \simeq \SU(4)$ by the presence of a metric and an orientation defined by the latter.
				A globally defined spinor reduces further the structure group. The irreducible spinor representation is the $\mathbf{4}$. 
				Given a nowhere vanishing spinor $\eta$ we can always go to a frame where it takes the form $\left(0,0,0,s\right)$. 
				The $\SU(4)$ transformations leaving this invariant are precisely $\SU(3)$.
				Looking at $\SU(4)$ decomposition under $\SU(3)$, one can see that in the fundamental $\mathbf{4}$ there is a singlet under $\SU(3)$.
						\begin{equation*}
							\begin{array}{rccr}
								\SU(4) 			&	\rightarrow	&	\SU(3)					&					\\
								\mathbf{4}			&	\rightarrow	&	\mathbf{3} + \mathbf{1}		&	\eta_+			\\
								\mathbf{\bar{4}}		&	\rightarrow	&	\mathbf{\bar{3}} + \mathbf{1}	&	\eta_+^*					
							\end{array}
						\end{equation*}
				We have chosen our $6$-dimensional gamma matrices such that $\eta_+^*$ has opposite chirality.
				
				Hence, a globally defined spinor defines an $\SU(3)$-structure.
				This is also given by globally defined, nowhere-vanishing and invariant $2$-form and $3$-form satisfying compatibility conditions~\eqref{compatib1} and~\eqref{compatib2}, that is
						\begin{align*}
							\omega \wedge \Omega &= 0\, , \\
							\Omega \wedge \bar{\Omega} &= \frac{4}{3} \omega^3\, .
						\end{align*}
					%
				
					%
				
				Consider now a connection $\nabla$ compatible with the $\SU(3)$-structure,
						\begin{equation}\label{covdereta}
							\nabla \eta_+ = \nabla_{LC} \eta_+ + \kappa \cdot \eta_+ = 0\, .
						\end{equation}
				where the contorsion $\kappa$ acts on spinors as,
						\begin{equation}
							(\kappa \cdot \eta_+)_{m}^{\alpha} = \frac{1}{4} \kappa_{mnp} \gamma^{np, \alpha}_{\phantom{np,\alpha}\beta} \eta_{+}^\beta \, .
						\end{equation}
				Again, we can see the (con)torsion as the obstruction to $\eta_+$ or ($\omega$ and $\Omega$) to be covariantly constant with respect to the Levi-Civita connection.
				
				The fact that $\eta_+$ is covariantly constant with respect to the connection $\nabla$ means that $\nabla$ has holonomy $\SU(3)$.
			For $d=6$, the decomposition~\eqref{decomk} takes the following form,
						\begin{align}
							&&	\kappa = \kappa_0 + \kappa_{\mathfrak{su}(3)}\, , & & 	\begin{array}{l}
																				\kappa_0 			\in T^*M \otimes \mathfrak{su}(3)^\perp \\[1mm]
																				\kappa_{\mathfrak{su}(3)} \in T^*M \otimes \mathfrak{su}(3)
																			\end{array}	&&
						\end{align}
				where we used the fact $\mathfrak{so}(6) \cong \mathfrak{su}(4) \cong \mathfrak{su}(3) \oplus \mathfrak{su}(3)^\perp$.
				We can now apply the contorsion decomposition to the expression for $\nabla \eta_+$, the~\eqref{covdereta}.
				Recall that since $\eta_+$ is an $\SU(3)$ singlet, one has
						\begin{equation}
							\kappa_{\mathfrak{su}(3)} \cdot \eta_+ = 0\, .
						\end{equation}
				Hence, one is left with
						\begin{equation}\label{conteta}
							\nabla_{LC} \eta_+ = \kappa_0 \cdot \eta_+\, .
						\end{equation}
					%

				The intrinsic contorsion and, hence, the intrinsic torsion $T^0$ can be decomposed into irreducible representations of $\SU(3)$					%
						\begin{equation*}
							\begin{array}{rrl}
								T^0 \in T^*M \otimes \mathfrak{su}(3)^{\perp}	& \rightarrow &	\left(\mathbf{3} \oplus \mathbf{\bar{3}}\right) \otimes \left(\mathbf{1} \oplus \mathbf{3} \oplus \mathbf{\bar{3}}\right)	\\[1.5mm]
																	& \rightarrow &	\underbrace{\left(\mathbf{1} \oplus \mathbf{1} \right)}_{W_1} \oplus \underbrace{\left(\mathbf{8} \oplus \mathbf{8} \right)}_{W_2} \oplus \underbrace{\left(\mathbf{6} \oplus \mathbf{6} \right)}_{W_3} \oplus \underbrace{\left(\mathbf{3} \oplus \mathbf{\bar{3}} \right)}_{W_4} \oplus \underbrace{\left(\mathbf{3} \oplus \mathbf{\bar{3}} \right)^\prime}_{W_5} \, .
							\end{array}
						\end{equation*}
				The $W_i$ are called \emph{torsion classes}~\cite{TorsClass1, TorsClass2}. These can be seen as equivalence classes, meaning that all the structures sharing the same intrinsic torsion (or being in the same irrep of $SU(3)$) are equivalent. The representatives of the torsion classes are differential forms of different ranks, see~\cref{tabW}.
						\begin{table}[h!]
						\centering
							\begin{tabular}{| c | l |}
								\toprule
								Form						&			rank								\\
								\hline
								$W_1$					& 			complex scalar						\\	
								$W_2 $	 				&			complex primitive $(1,1)$-form			\\	
								$W_3$		 			&			real primitive $(2,1) + (1,2)$-form		\\	
								$W_4$		 			&			real one-form						\\	
								$W_5$					&			complex $(1,0)$-form				\\
								\bottomrule
							\end{tabular}
							\caption{Torsion classes as differential forms.}
							\label{tabW}
						\end{table}

				Recall that a form $\alpha$ is said \emph{primitive} if it has a zero contraction with $\omega$, \emph{i.e.} $\omega\lrcorner \alpha = 0$~\cite{joyce}.
				
				The torsions classes can be used also to express the integrability of the structure as differential conditions on the forms $\omega$ and $\Omega$				%
						\begin{subequations}
						\label{diffstruct}
							\begin{align}
								\dd \omega 	&= \frac{3}{2} \IIm \left( \overline{W}_1 \Omega \right) + W_4 \wedge \omega + W_3 \, , \\
								\dd \Omega	&= W_1 \wedge \omega \wedge \omega + W_2 \wedge \omega + \overline{W}_5 \wedge \Omega \, ,
							\end{align}
						\end{subequations}

				These formulae allow to classify the manifolds with $\SU(3)$ structures through the torsion classes.
				We collect some example in~\cref{tabTorsCl},
						\begin{table}[h!]
						\centering
							\begin{tabular}{l | l}
								Torsion Classes						&			Name							\\
								\hline
																	&											\\[-.3mm]
								$W_1 = W_2 = 0$						& 			Complex manifold					\\[1.5mm]	
								$W_1 = W_2 = W_4 = 0$					&			Symplectic manifold					\\[1.5mm]	
								$W_2 = W_3 = W_4 = W_5 = 0$ 			&			Nearly K\"ahler	manifold				\\[1.5mm]
								$W_1 = W_2 = W_3 = W_4 = 0$			&			K\"ahler manifold					\\[1.5mm]	
								$\IIm W_1 = \IIm W_2 = W_4 = W_5 = 0$		&			Half-flat manifold					\\[1.5mm]
								$W_1 = \IIm W_2 = W_3 = W_4 = W_5 = 0$ 	&			Nearly Calabi-Yau three-fold			\\[1.5mm]	
								$W_1 = W_2 = W_3 = W_4 = W_5 = 0$ 		&			Calabi-Yau three-fold					
							\end{tabular}
							\caption{Six-dimensional manifolds with $\SU(3)$ structure, classified by torsion classes.}
							\label{tabTorsCl}
						\end{table}
				\subsubsection{Calabi-Yau manifolds}
					We saw that a $d$-dimensional K\"ahler manifold can be defined as a Riemannian manifold of even dimension $d=2n$, whose 
					holonomy group is contained in $\U(n)$.
					
					In the same way we can define Calabi-Yau manifolds as K\"ahler manifolds whose holonomy group is contained in $\SU(n)$. More precisely			
									given a compact K\"ahler manifold $M$ of complex dimension $n$, with complex structure $I$, Hermitian K\"ahler metric $g$ and associated K\"ahler form $\omega$, then $(M, I, g, \omega)$ is a \emph{Calabi-Yau $n$-fold} if and only if $g$ has $\SU(n)$ holonomy.
						
						%
									Given a compact K\"ahler manifold $M$ of complex dimension $n$, with complex structure $I$, Hermitian K\"ahler metric $g$ and associated K\"ahler form $\omega$, then $(M, I, g, \omega)$ is a \emph{Calabi-Yau $n$-fold} if and only if $g$ has $\SU(n)$ holonomy.

We can also define a Calabi-Yau manifold in terms of the structures $\omega$ and $\Omega$: a Calabi-Yau structure on an $2n$-dimensional manifold $M$ is the set $(M, \omega, \Omega)$, where $\omega$ and $\Omega$ are respectively an integrable K\"ahler two-form and a complex simple $n$-form defining an integrable complex structure
							\begin{align*}
								\dd \omega 	&= 0 \, , \\
								\dd \Omega	&= 0 \, ,
							\end{align*}
that are also compatible, \emph{i.e.} $\omega \wedge \Omega = 0$, and whose metric, defined as in~\eqref{mesyco}, is Ricci-flat, \emph{i.e.} $R = 0$.

One can also show that this is equivalent to say that a Calai-Yau $n$-fold is a complex manifold, with a compatible (integrable) symplectic structure where in addition the metric associated as in~\eqref{mesyco} is Hermitian and Ricci-flat, \emph{i.e.} $R_g = 0$~\cite{joyce, KahlBook}\footnote{%
						The equivalence of these definitions is actually a consequence of the famous \emph{Calabi conjecture}~\cite{CalabiMetr}, proven by Yau~\cite{YauCalabi}, and so these manifolds were named after them.}.
						
					The standard example of trivial Calabi-Yau manifolds are the even-dimensional tori, equipped with the usual complex structure and metric.
					These are also the only Calabi-Yau compact manifolds for which an explicit Ricci-flat metric is known.
					For all the others, the Calabi theorem guarantees the existence of such a metric, but there is no explicit known construction yet.

				\subsubsection{Sasaki-Einstein manifolds}
					Sasaki-Einstein manifolds are odd-dimensional manifolds, that are both Sasakian and Einstein.
					Then, let us start with thes definition of a Sasakian manifold~\cite{SparksSE}.

					A Riemannian manifold $(S,g)$ is \emph{Sasakian} if and only if its metric cone
											\begin{equation*}
												C(S) := \mathbb{R}^+ \times S \, ,
											\end{equation*}
					equipped with the metric $\overline{g} = \dd t^2 + t^2 g$, is a K\"ahler manifold.
					
					As first consequence of this definition, we can see that $S$ has to be odd dimensional, \emph{i.e.} $d=2n-1$, where $n$ is the complex dimension of the K\"ahler cone.
					
					An Einstein manifold, on the other hand, is a manifold equipped with an Einstein metric $g$, \emph{i.e.} a metric whose Ricci curvature is proportional to the metric itself,
							\begin{equation*}
								R_g = \lambda g \, , 
							\end{equation*}
					for some real constant $\lambda$.
					For a Sasakian manifold, it turns out that $\lambda = 2(n-1)$. Moreover, one can show that a Sasakian manifold is Einstein if and only if its cone is K\"ahler-Einstein and Ricci flat, namely a Calabi-Yau~\cite{SparksSE}.
					Hence, for a Sasaki-Einstein manifold, the restricted holonomy group\footnote{%
						The restricted holonomy group is the component of the group connected to the identity element.%
						}
					of its cone is $\mathrm{Hol}^0 (\overline{g}) \subset \SU(n)$.
					
					The standard example is provided by the odd-dimensional spheres $S^{2n-1}$, equipped with the standard Einstein metric.
					In this case the K\"ahler cone is simply the space $\mathbb{C}^n / {0}$, with the standard Euclidean metric.
					
					A Sasakian manifolds inherits some properties from the K\"ahler structure of its cone.
					One of the most important one for our concerns is the \emph{contact structure}~\cite{contact1}, represented by a nowhere-vanishing vector field $\xi$, known as \emph{Reeb vector field}.
					Using the coordinate $t$ to parametrise the $\mathbb{R}^+$ direction on the cone, we can write
							\begin{equation}\label{Reeb1}
								\xi = I\left[ t \partial_t \right]\, ,
							\end{equation}
					where $I$ is the complex structure on the cone.
					We will see how the Reeb vector will play an important role in compactifications of type IIB supergravities~\cite{DavideSas1, DavideTriSas} and further in the definitions of generalised structures~\cite{AshmoreESE}.
					
					As we have seen in the last section, on Calabi-Yau manifolds the group structure defines a covariantly constant spinor.
					On a Sasaki-Einstein, we have \emph{Killing spinors}, \emph{i.e.} a spinor $\psi$ such that
							\begin{equation}
								\nabla \psi = \pm \frac{1}{2} \gamma \psi \, . 
							\end{equation}
						%
%
			A generic Sasaki-Einstein manifold supports two Killing spinors.

					
					In addition, we can characterise a Sasaki-Einstein manifold equivalently by differential forms with compatibility relations among them.
					Explicitly, a \emph{Sasaki-Einstein structure} on a manifold $M$ is the set $(M, \xi, \sigma, \omega, \Omega)$, where $\xi$ is the Reeb vector of~\eqref{Reeb1}, $\sigma$ the dual one-form defined by $\sigma\left[X \right] = g(\xi, X)$\footnote{%
						Note that the metric $g$ is defined by the two compatible structures $\omega$ and $\Omega$ as in~\eqref{mesyco}. More precisely, the complex and symplectic structure on the K\"ahler cone define a metric $\overline{g}$, whose restriction gives $g$.%
						},
					$\omega$ is a real two-form and $\Omega$ a complex simple $n$-form.
					They have to satisfy the following relations,
							\begin{equation}
								\begin{array}{c c c}
									\omega \wedge \Omega = 0 \, ,	&	\phantom{\dd}	&\iota_\xi \omega = \iota_\xi \Omega = 0 \, , \\[2mm]
									\dd \sigma = 2 \omega \, ,		&	\phantom{\dd}	&\dd \Omega = (n+1) i \sigma \wedge \Omega \, .
								\end{array}
							\end{equation}
					Note that given a Sasaki-Einstein manifold, its metric can always be re-written as,
							\begin{equation}
								\dd s^2 = \sigma \otimes \sigma + \dd s^2_{KE} \, ,
							\end{equation}
					where $\dd s^2_{KE}$ is the metric on a generic K\"ahler-Einstein base.
					From this, we can say that the Reeb vector defines a \emph{foliation}, \emph{i.e.} it splits the tangent bundle into integral sub-bundles, whose set of leaves is a K\"ahler-Einstein space.
					Such a foliation is called (not surprisingly) \emph{Reeb foliation}.

Sasaki-Einstein manifolds are very common in string compactifications.
Recently, making use of their geometry has allowed to achieve several interesting results both in flux compactification and in a wider area of research in string theory~\cite{DavideTriSas, DavideSas1, SE1, SE2, SE3, SE4, SE5}.
					
					\paragraph{Tri-Sasakian Manifolds} 
					There is a sub-class of the Sasaki-Einstein manifolds, the so-called \emph{Tri-Sasakian manifolds} which is interesting not only because it has applications in flux compactifications~\cite{DavideTriSas}, but also simply to show how reduced holonomy gives rise to structures with different properties~\cite{TriSasakiReview}.
					
					In order to define this class of manifolds, we have to give the definition of \emph{Hyperk\"ahler manifold}.
					An hyperk\"ahler manifold is a Riemannian manifold $(M,g)$ of dimension $4k$ whose holonomy group $\mathrm{Hol}(g)$ is contained in the symplectic group $\Sp(k)$.
					Note that all hyperk\"ahler manifolds are also Calabi-Yau (the metric $g$ is Ricci-flat), since $\Sp(k) \subset \SU(2k)$.
					
					We are interested in the structures defined on this kind of manifolds.
					An hyperk\"ahler manifold has a set of three complex structures $\{I, J, K\}$ with respect to which the metric is K\"ahler.
					The three complex structures respect the Hamilton algebra of quaternions~\cite{EncyclMath},
							\begin{equation}
								I^2 = J^2 = K^2 = - \id \, ,
							\end{equation}
					and they also have the property that any linear combination with real coefficients $\alpha I + \beta J + \gamma K$, and $\alpha^2 + \beta^2 + \gamma^2 = 1$ is a complex structure on $M$.
					Because of these one can say that the tangent space at any point is a quaternionic vector space.
					
					A \emph{tri-Sasakian manifold} is a Sasaki-Einstein manifold whose cone is hyperk\"ahler.
					This definition constrains the dimension of a tri-Sasakian manifold to be $4n+3$, with $n \geq 1$.
					We can also give an equivalent characterisation in terms of tensors defining structures.
					
					A tri-Sasakian manifold has three mutual orthogonal Killing vectors $\xi_i$, such that each of them is a Reeb vector, and all together they generate an $\SU(2)$ algebra,
							\begin{equation}
								\left[\xi_i , \xi_j \right] = 2 \epsilon_{ij}^{\phantom{ij}k} \xi_k \, .
							\end{equation}
					They also have their associated dual contact forms, \emph{i.e.} $\sigma^i$ such that $\sigma^i (\xi_j) = \delta^i_j$.
					Each $\sigma^i$ satisfies the differential relation,
							\begin{equation}
								\dd \sigma^i = 2 J^i - \epsilon_{\phantom{i}jk}^i \sigma^j \wedge \sigma^k \, ,
							\end{equation}
					where $\iota_{\xi_i}J^{j} = 0$.
					Taking the exterior derivative of the relation above one gets
							\begin{equation}
								\dd J^i = 2 \epsilon^{i}_{\phantom{i}jk} J^j \wedge J^k \, .
							\end{equation}

					As one may expect, the three Reeb vectors define what is known under the name of \emph{tri-Sasakian foliation}, whose space of leaves is a quaternionic K\"ahler space (whose holonomy group is contained in $\Sp(n)\Sp(1)$) $B_{QK}$, and the metric on the tri-Sasaki can be written as,
							\begin{equation}
								\dd s^2 = \dd s^2_{B_{QK}} + \sigma^i \otimes \sigma ^i \, .
							\end{equation}
					To conclude this discussion, we want to make the following remark.
					For any linear combination $\alpha^i \xi_i$ with $\alpha^i \alpha^i =1$, we can define a Sasaki-Einstein structure through,
							\begin{align*}
								\sigma &= \alpha_i \sigma^i \, , \\
								\omega &= \alpha_i \left(J^i - \frac{1}{2} \epsilon^i_{\phantom{i}jk} \sigma^j \wedge \sigma^k \right) \, , \\
								\Omega &= \left(\beta_j \sigma^j - i \gamma_j \sigma^j \right) \wedge \left(\beta_k J^k - i \gamma_k J^k \right)\, ,
							\end{align*}
					where $\beta_i$ and $\gamma_i$ are sets of coefficients such that $\beta^i \xi_i$ and $\gamma^i \xi_i$ are mutually orthogonal.
					The coefficients (up to the orthogonality condition) can be chosen arbitrarily, since the various choices gives the same Sasaki-Einstein structures up to the phase of $\Omega$.
					%
				%
			%
		%

		%
		
\ensurepagenumbering{arabic}
	\chapter{Flux Compactifications}
	\label{chapSugra}
		\section{Introduction and motivations}

			This thesis is devoted to the study of supersymmetric compactifications with non-trivial fluxes.
			We will see in the first part of this chapter how requiring some amount of supersymmetry on the lower dimensional theory constrains the geometry of the internal manifold $M$, such that it must admit geometrical structures like the ones we described in~\cref{chap1}.
			For the well-known case of fluxless compactification of a $10$-dimensional type II supergravity to a minimal supergravity in $4$ dimensions, the constraints on the internal manifold requires it to be a Calabi-Yau three-fold~\cite{CYcomp}.
			When we allow fluxes to be turned on, the supersymmetry conditions can be cast in a compact and elegant form using \emph{Generalised Geometry} and generalised structures we will introduce in~\cref{chapEGG}.
		\section{Supergravity theories}
			Supergravity theories are theories combining general relativity with supersymmetry (making this a local symmetry).
			These can be seen as low-energy effective theories of the different string theories.
			There exists also an eleven-dimensional maximally supersymmetric supergravity, which is not connected (as low-energy limit) to any string theory.
			This has been interpreted to have its higher dimensional origin in $M$-theory.
			
			The aim of this section is to describe the main feature of type II and eleven-dimensional supergravity theories, with their effective actions and the gauge symmetries of their potentials.
			\subsection{Eleven-dimensional supergravity}
				This section is devoted to the description of eleven-dimensional supergravity, \emph{i.e.} the low energy effective theory of M-theory. 
				This is not meant to be an exhaustive treatment and we refer to~\cite{polchinski, BeckerBeckerSchw} for further details.

				The bosonic degrees of freedom of eleven dimensions supergavity consist of the metric $g$, a three-form potential $A$ and its dual.
				Although it does not transport independent degrees of freedom, one often introduces the dual seven-form $\tilde G$, whose six-form potential is conventionally denoted by $\tilde A$.			
	
				The theory is invariant under diffeomorphisms and the gauge transformations			
					 	\begin{equation}
							A_3 \longrightarrow A_3 + \dd \Lambda_2 \, ,
						\end{equation}
				where $\Lambda_2$ is a two-form. 
				The gauge invariant field strength is $G_4 = \dd A_3$.
				
				The bosonic action of the eleven-dimensional supergravity is
						\begin{equation}
							\begin{split}
								S_{11} =& \frac{1}{2\kappa^2} \int \dd^{11} x \ \sqrt{g} \left[ \left(R - \frac{1}{2} \lvert G \rvert^2 \right)\right] - \\
								& - \underbrace{\frac{1}{6} G_4 \wedge G_4 \wedge A}_{S_{CS}} \, , 
							\end{split}
						\end{equation}

The equation of motion and Bianchi identity can be written (in a sourceless case) as
						\begin{equation}
							\begin{split}
								\dd &\star G + \frac{1}{2} G \wedge G = 0 \, , \\
								\dd &G = 0 \, .
							\end{split}
						\end{equation}
The theory is supersymmetric, with $\mathcal{N}=1$ supersymmetry. Notice that this is the maximal possible supersymmetry in eleven dimensions.
The fermionic degrees of freedom are completely captured by the gravitino $\Psi$.

				In addition, the equation of motion for the metric $g$, \emph{i.e.} the Einstein equation, can be written as follows,
						\begin{align}
							R_{MN} - \dfrac{1}{12}\left( G_{MPQR}G_{N}^{\phantom{N}PQR} - \dfrac{1}{12}g_{MN} G^2 \right) &= 0 \, .
						\end{align}
			\subsection{Type II theories}\label{sec:IIsugra}
Type II supergravities are the ten-dimensional effective theories for massless fields type II string theories. There are two such theories that differ in the chirality of
the fermionic fields and the rank of the form potentials. 

The bosonic sector consists of two sets of fields: the Neveu-Schwarz Neveu-Schwarz (NSNS) and the Ramond-Ramond (RR).
						\begin{table}[h!]
						\centering
							\begin{tabular}{c c c c}
									\toprule
														& 			$g$								&	metric (graviton)							\\[1.2mm]
									NSNS 				&			$B$								&	Kalb-Ramond $2$-form						\\[1.2mm]
											 			&			$\phi$							&	dilaton									\\[1.6mm]
									RR		 			&			$A_p$							&	\begin{tabular}{@{\ }l@{}}
 																											$p$ odd for type IIA \\ 
																											$p$ even for type IIB 
 																										\end{tabular}								\\[2mm]
									\midrule	
														&	$\psi_M^{\alpha, +}$, $\psi_M^{\alpha, \mp}$		&	Gravitinos									\\[1.2mm]
														&	$\lambda_\alpha^{-}$, $\lambda_\alpha^{\pm}$		&	Dilatinos									\\[1.2mm]
									\bottomrule
								\end{tabular}
							\caption{Type II supergravities spectrum in ten dimensions. 
							The different chiralities of spinors define the two theories. 
							Upper signs refer to type IIA, while lower ones to type IIB.}
							\label{tabspectr}
						\end{table}

				As one can see from~\cref{tabspectr}, the NSNS sector is the same for both type II theories.
				It contains the metric $g$, the dilaton $\phi$ and the NSNS two-form $B$.
				The latter is a $\U(1)$ gauge potential with field strength $H = \dd B$.
				
				The RR sector depends on the theory.
				Type IIA contains odd forms, while for type IIB has even ones.
				These are also $\U(1)$ gauge potentials. For applications to generalised geometry is it convenient to use the \emph{democratic formulation}~\cite{DemSugra},
of supergravity. 		This formulation considers RR potentials of all ranks $C_p$, with $p = 1, 3, \ldots, 9$ for type IIA and $p = 0, 2, \ldots, 8$ for IIB. 
				These are not all independent since their field strengths have to satisfy duality relations with respect to the Hodge dual.
				The field strength\footnote{%
					There exists another common choice for the RR potential, the so-called $A$-basis, which is related to the $C$-basis we use as $A = e^{-B} \wedge C$. 
					In this basis the field strength~\eqref{Fform} reads $F = e^B \wedge (\dd A + m)$.%
					}
				are defined by,
						\begin{equation}\label{Fform}
							F_p = \dd C_{p-1} + H \wedge C_{p-3} + e^B F_0 \, ,
						\end{equation}
				where $F_0 = m$ is the Romans mass, which can be added only in type IIA~\cite{RomansMass}, and 
				the duality relations,
						\begin{equation}\label{Fdual}
							F_p = (-1)^{\left[\frac{p+3}{2}\right]} \star F_{10-p} \, .
						\end{equation}

				 The fermionic sector of the two theories consists of two Majorana-Weyl spinors of spin $3/2$, the gravitinos $\psi_M^\alpha$, and two Majorana-Weyl spin 1/2 spinors $\lambda^\alpha$, the dilatinos.
				 Gravitinos and dilatinos have opposite chirality.
				 In type IIA the gravitinos have opposite chirality, while in type IIB they have the same chirality (chosen positive by convention).
				 As a consequence, type IIB dilatinos will both have negative chirality. 
				 This is the difference between type IIA and IIB, the former is a non-chiral theory, while the latter is chiral.
				 Nevertheless, they are both maximal supersymmetry in ten dimensions, \emph{i.e.} they are $\mathcal{N}=2$.
				 An important fact is that type IIA supergravity can be obtained by the eleven-dimensional one by a compactifiation on a circle.
				 We will analyse this reduction in a while.
				 
				 The string frame\footnote{%
				 	Einstein frame and string frame metric are related by a dilaton rescaling, \emph{i.e.}
							\begin{equation*}
								g = e^{\phi/2} g^E \, .
							\end{equation*}
					} action for the bosonic fields of type IIA is (we follow the conventions of~\cite{DemSugra}) 
						\begin{equation*}
							\begin{split}
								S_{IIA} = \frac{1}{2\kappa^2}& \int \dd^{10} x \ \underbrace{\sqrt{g} \left[ e^{-2\phi} \left(R + 4 \nabla \phi ^2 - \frac{1}{2} \lvert H \rvert^2 \right)\right]}_{S_{NS}} - \underbrace{\frac{\sqrt{g}}{2} \sum_{k=0}^2 \lvert F_{2k} \rvert^2}_{S_{R}} \\
								&- \underbrace{\frac{1}{2} B \wedge \mathcal{F}_4 \wedge \mathcal{F}_4}_{S_{CS}} \, ,
							\end{split}		
						\end{equation*}
				 where $\mathcal{F}_4 = \dd C_3$, while $F_p$ are the field strength defined above in~\eqref{Fform}.

				The bosonic action for type IIB reads 
						\begin{equation*}
							\begin{split}
							S_{IIB} = \frac{1}{2\kappa^2}& \int \dd^{10} x \ \underbrace{\sqrt{g} \left[ e^{-2\phi} \left(R + 4 \nabla \phi ^2 - \frac{1}{2} \lvert H \rvert^2 \right)\right]}_{S_{NS}} - \underbrace{\frac{\sqrt{g}}{2} \sum_{k=0}^2 \frac{1}{k!}\lvert F_{2k+1} \rvert^2}_{S_{R}} \\
															& - \underbrace{\frac{1}{2} C_4 \wedge H_3 \wedge \mathcal{F}_3}_{S_{CS}} \, .
							\end{split}							
						\end{equation*}
				Analogously to the type IIA case, we introduced $\mathcal{F}_n = \dd C_{n-1}$.
				This action has a constant shift symmetry $C_0 \rightarrow C_0 + c$, where $c$ is a constant. 
				Hence, it is referred to as an \emph{axion}~\cite{BeckerBeckerSchw, Weinberg:1977ma}.
				Furthermore, the five-form field strength $F_5$ satisfies the self-duality condition
						\begin{equation}\label{F5dual}
							F_5 = \star F_5 \, ,
						\end{equation}
				which has to be imposed as a further constraint together with the equations of motion.

				It is be useful to collect all the RR field strengths and potentials into a single polyform,
						\begin{equation*}
							\begin{array}{lr}
								C = \sum_p C_p & p\ \text{odd/even for type IIA/IIB}\, , \\[3mm]
								F = \sum_p F_p & p\ \text{even/odd for type IIA/IIB}\, .								 
							\end{array}
						\end{equation*}
				In this notation, the~\eqref{Fform} and~\eqref{Fdual} take the following form,
						\begin{align*}
							F &= \dd_H C + e^B F_0\, , \\
							F &= \star s (F) \, ,
						\end{align*}
				where we introduced the differential operator $\dd_h := \dd - H \wedge$ acting on polyforms, called \emph{$H$-twisted exterior derivative}, and the \emph{index reversal operator} $s$,
						\begin{equation}
							s(A_p) = (-1)^{\left[p/2\right]} A_p \, .
						\end{equation}
				The field strengths defined above are invariant under gauge transformations of potentials,
						\begin{equation}\label{gaugetrans}
							\begin{split}
								\delta B &= - \dd \lambda \, , \\
								\delta C &= - e^B \wedge \left(\dd \omega - m \lambda \right) \, , \\
							\end{split}
						\end{equation}
				where $\lambda$ is a one-form, $\omega$ is a polyform made of even/odd forms for type IIA/IIB and the term proportional to the Romans mass $m$ is there only in the type IIA case.			
				
				The RR field strengths have the following equations of motions and Bianchi identities (when there are no sources, like $\mathrm{D}_p$ branes).
				Bosonic fields equations for type II appear as,
						\begin{align}
							(\dd e^{-2\phi} \star H) \pm \frac{1}{2} F \wedge \star F = 0 \, , \label{eom1}
							(\dd + H) \star F &= 0 \, , 
						\end{align}
				where $\pm$ sign is referred to type IIA/B respectively, and
						\begin{equation}
							\dd F = H \wedge F \, .
						\end{equation}

				Notice that for type IIA the~\eqref{eom1},
						\begin{equation}\label{Hstar}
							\dd (e^{-2\phi} \star H) + \frac{1}{2} \left[ F \wedge \star F \right]_8 = 0 \, ,
						\end{equation}
				can be interpreted as the Bianchi identity for the dual seven-form field strength,
						\begin{equation}
							\tilde{H} = e^{-2\phi} \star H \, .
						\end{equation}
				We denoted by $[\ldots]_k$ the rank $k$ form of the polyform in the bracket.

				Making use of the self-duality relation for $F$~\eqref{Fdual}, we can rewrite the~\eqref{Hstar} as,
						\begin{equation}
							\dd \left( \tilde H + \frac{1}{2} \left[ s(F) \wedge C + m e^{-B} \wedge C \right]_7 \right) = 0\, ,
						\end{equation}
				which is solved by,
						\begin{equation}
							\tilde H = \dd \tilde{B} - \frac{1}{2} \left[ s(F) \wedge C + m e^{-B} \wedge C \right]_7 \, .
						\end{equation}
				Thus we introduce a new potential $\tilde{B}$~\cite{Bergshoeff:1997ak, Bergshoeff:2006qw}, whose (linearised) gauge transformations are fixed by requiring the invariance of its field strength,
						\begin{equation}
							\delta \tilde{B} = -(\dd \sigma + m \omega_6) - \frac{1}{2} \left[ e^{B} \wedge (\dd \omega - m \lambda ) \wedge s(C) \right]_6 \, ,
						\end{equation}
				where $\sigma$ is a five-form, while $\omega$ and $\lambda$ are the parameters of the gauge transformations~\eqref{gaugetrans}.
				
				An interesting point to notice about the massive IIA theory~\cite{RomansMass} is that one can obtain it from the non-massive one by shifting the gauge parameters as,
						\begin{equation}\label{gaugeshifts}
							\begin{split}
								\dd \omega_0 &\longrightarrow \dd \omega_0 - m \lambda \, , \\
								\dd \sigma & \longrightarrow \dd \sigma + m \omega_6 \, .
							\end{split}
						\end{equation}
				These relations will be the key of the construction of the exceptional generalised geometry for massive type IIA~\cite{oscar1}.
				
				Type IIB theory exhibits a non-compact global symmetry $\SL(2, \RR)$.
				This is not evident in the formulation we gave above, so we want to make it explicit.
				The two two-form potentials $B$ and $C_2$ can be organised into a doublet of $\SL(2, \RR)$,
						\begin{equation}
							B^i := \begin{pmatrix}
									B \\
									C_2
							\end{pmatrix}^i \, .
						\end{equation}
				Similarly, we introduce $F^i = \dd B^i$.
				Under an $\SL(2,\RR)$ transformations the $B$ fields transform linearly,
						\begin{align}
							& &	B^i \longrightarrow \Lambda_{ij} B^j \, ,	& &	\Lambda_{ij} = \begin{pmatrix}
																					a & b \\
																					c & d					
																				\end{pmatrix} \in \SL(2, \RR) \, . 
						\end{align}
				One can also define a complex scalar field $\tau$ which is the complex combination of the axion and the dilaton field, for this reason this is called \emph{axion-dilaton field}.
				This is useful since it transforms nicely under $\SL(2,\RR)$,
						\begin{equation}
							\tau \longrightarrow \frac{a \tau + b}{c \tau + d} \, .
						\end{equation}
				Then, type IIB action $S_{IIB}$ can be re-written in terms of $\SL(2,\RR)$ representations, like the symmetric matrix $h$,
						\begin{equation}
							h_{ij} = \begin{pmatrix}
								\lvert \tau \rvert^2 & -C_0 \\
								-C_0 & 1
							\end{pmatrix}_{ij} \, ,
						\end{equation} 
				transforming under $\SL(2, \RR)$ as
						\begin{equation}
							h_{ij} \longrightarrow \Lambda^{ik} h_{kl} \Lambda^{lj} \, .
						\end{equation}
				Then the action $S_{IIB}$ in terms of $\SL(2, \RR)$ covariant objects can be recast as,
						\begin{equation}
							\begin{split}
								S_{IIB} = \frac{1}{2\kappa^2}& \int \dd^{10} x \ \sqrt{g} \left[e^{-2\phi} \left(R - \frac{1}{12}F^i h_{ij} F^j + \frac{1}{4} \partial h_{ij} \partial h^{ji} \right)\right] \\
											& - \frac{1}{8 \kappa^2}\int \dd^{10} x \ \left[ \sqrt{g} \lvert F_{5} \rvert^2 - \epsilon_{ij} C_4 \wedge F^i \wedge F^j \right] \, .
							\end{split}
						\end{equation}
				The self duality condition on the $5$-form field strength~\eqref{F5dual} (which is a constraint in this formalism) is also $\SL(2,\RR)$ invariant.
				Moreover, one can re-write its definition in an $\SL(2,\RR)$ invariant form,
						\begin{equation}
							F_5 = \dd C_4 + \frac{1}{2} \epsilon_{ij} B^i \wedge H^j \, .
						\end{equation}
					%
				
					
				%
	\section{Supersymmetric backgrounds and compactifications}
				After describing the actions and the equations of motion of eleven- and ten-dimensional supergravities, we are interested in solutions.
	 Since we want to study compactifications, we look for solutions that are warped products 
		
			\begin{equation*}
					\mathcal{M}_{10} = \mathcal{X} \times M_d\, ,
				\end{equation*}
				
				of a maximally symmetric external spacetime (Minkowski, Anti-de Sitter, de Sitter) and an internal space $M_d$.
				In order to preserve Poincar\'e invariance in the external spacetime, we must set all fermionic fields to zero so the background is purely bosonic.
				
				Then, the metric ansatz reads
						\begin{equation}\label{metrsplit}
							\dd s^2_{\mathcal{M}} = e^{2A}\dd s^2_{\mathcal{X}} + \dd s^2_{M} \, ,
						\end{equation}
				where $A$ is a real function of the coordinates on $M_d$, the \emph{warp factor}.
					
				We look for supersymmetric solutions. 			
				A background is supersymmetric if all the supergravity fields (and hence the solutions) are invariant under supersymmetry transformations.
				Choosing $\epsilon$ as the quantity parameterising supersymmetry variations, one is allowed to write (schematically)
						\begin{align}\label{susyvar}
							& &	\delta(\text{boson}) = \epsilon (\text{fermion})\, ,	& &	\delta(\text{fermion}) = \epsilon (\text{boson})\, .	& &
						\end{align}
				The variations of the bosonic fields always contain a fermionic field, and since we have set these to zero the variations automatically vanish.
				On the other hand, we get non-trivial conditions from the variations of the fermionic fields. 
				Then, supersymmetry of the background is equivalent to the existence of a non-vanishing spinor $\varepsilon$ for which the supersymmetry variations vanish.
				These can be recast into differential and algebraic equations, known as \emph{Killing spinor equations}.
				The spinor $\epsilon$ solving these is then called \emph{Killing spinor}.
				A background is supersymmetric if it admits Killing spinors. The Killing spinor equation for M-theory is 
						\begin{equation}
							\nabla_M \epsilon + \frac{1}{288} \left[ \Gamma_M^{\phantom{M}NPQR} - 8 \delta_M^{\phantom{M}N} \Gamma^{PQR} \right] G_{NPQR}\ \epsilon =0 \, ,
						\end{equation}
				where $M,N, \ldots = 0,1, \dots, 10$, $\epsilon$ is a Majorana spinor and the Gamma matrices are the Clifford algebra elements in $11$ dimensions. 
				
				Type IIA Killing spinor equation can be derived by the previous one by a compactification, so we do not give it explicitly.
Finally, for type IIB the Killing spinor equations, given in terms of ten-dimensional Gamma matrices are
						\begin{subequations}
							\begin{align}
								\nabla_M \epsilon - \frac{1}{96} \left[ \Gamma_M^{\phantom{M}PQR} - 9 \Gamma^{PQ} \right] G_{MPQ}\ \epsilon^c + \frac{1}{192}\Gamma^{PQRS}F_{MPQRS} \epsilon =0 \, , \\
								i \Gamma^M P_M \epsilon^c + \frac{i}{24} \Gamma^{PQR} G_{PQR} \epsilon = 0 \, .
							\end{align}
						\end{subequations}
				where, following~\cite{Gauntlett:2005ww}, we defined $P = \tfrac{i}{2}e^\phi \dd C_0 + \tfrac{1}{2}\dd \phi$ and $G = i e^{\phi/2}(\tau \dd B - \dd C_2)$, and here $M,N = 0,1, \dots, 9$.

The Killing spinor equations are central in the study of supersymmetric string backgrounds. This is due to the fact that for backgrounds of the 	\eqref{metrsplit} one can show that
the supersymmetry variations plus the Bianchi identities for the NS and RR fields imply all other equations of motion. Thus a solution of the 			
Killing spinor equations is automatically a solution of the supergravity equations of motions. 				
	
	On backgrounds of the type~\eqref{metrsplit} the supersymmetry parameters factorise accordingly					%
						\begin{equation}
							\epsilon = \sum_{i=N} \varepsilon_i \otimes \chi \, ,
						\end{equation}
				where $\varepsilon_i$ are anticommuting spinors on the external space and $\chi$ is a generic commuting spinor on the internal manifold.
				The number $N$ of spinors $\varepsilon_i$ detemine the number of supersymmetries preserved by the background. 
				This splitting induces also a splitting of the Killing spinor equations into distinct conditions for $\varepsilon$ and for $\chi$.
				The existence of Killing spinors (and the differential conditions they have to satisfy) on the internal manifold $M$ puts several constraints on the geometry of the manifold. 
				Investigating how this happens and how this allows compactifications with fluxes is the goal of next sections.
						
			\subsection{Calabi-Yau backgrounds in type II}
				Let us start by a famous example of compactifications to four dimensions. We consider a purely geometric solutions where the only non-trivial field is the metric.
				We are going to see how supersymmetry conditions constrain the internal geometry to be Calabi-Yau.

				As discussed above, to find solutions it is enough to solve the supersymmetry variations for the spinors~\eqref{susyvar}.
				In type II theory, in absence of fluxes they reduce to 
						\begin{equation}
							\begin{array}{lcr}
								\delta \lambda_1 = \partial_M \phi \Gamma^M \epsilon_1 = 0 \, , & & \delta \lambda_2 = \partial_M \phi \Gamma^M \epsilon_2 = 0 \, ,
							\end{array}
						\end{equation}
				for the dilatino variations, while the gravitino variations reduce to the requirement that the supersymmetry parameters must be covariantly constant
						\begin{equation}\label{grav0}
							\begin{array}{lcr}
								\delta \psi^1_{M} = \nabla_M \epsilon_1 = 0 \, , & & \delta \psi^2_{M} = \nabla_M \epsilon_2 = 0 \, .
							\end{array}
						\end{equation}
				The supersymmetry parameters decompose as 
						\begin{align}\label{spinansatz}
							& \epsilon_1 = \zeta_1 \otimes \eta_1 + \text{c.c.} \, , \\
							& \epsilon_2 = \zeta_2 \otimes \eta_2 + \text{c.c.} \, .
						\end{align}
				Here $\zeta$ is a four-dimensional chiral spinor $(\gamma_5\zeta=\zeta)$ and $\eta_{1,2}$ are six-dimensional chiral spinors, of opposite chirality in IIA and same chirality in IIB
						\begin{align}
							& &	\gamma_7\eta_{1}=\eta_{1} & & \gamma_7\eta_{2}=\mp\eta_2 & & \mbox{in IIA/IIB} \, .
						\end{align}					

				Using the decomposition ansatz for the metric~\eqref{metrsplit}, and supersymmetry parameters~\eqref{spinansatz}, we obtain the six-dimensonal equations
\begin{equation}\label{intdilCY}
\slashed{\partial} \phi\,\eta_{1,2}=0\ ,
\end{equation}
where $\slashed{\partial}\phi=\gamma^m\partial_m\phi$. This implies that the dilaton must be constant, $\partial_m\phi=0$, since $||\slashed{\partial}\phi\,\eta_{1,2}||^2=(\partial\phi)^2||\eta_{1,2}||^2$.

The gravitino variations reduce to 
\begin{align}
& \nabla_\mu \rightarrow \nabla_\mu\otimes \id 
+ \frac{1}{2}e^A (\gamma_\mu \gamma_5 \otimes \slashed\partial A )\ , \\
& \nabla_m \rightarrow \id\otimes \nabla_m\ ,
\end{align}
where, on the right hand side, $\nabla_\mu$ and $\nabla_m$ are the covariant derivatives with respect to the external four-dimensional unwarped metric and the internal six-dimensional metric, respectively. Using again~\eqref{spinansatz}, we can decompose~\eqref{grav0} into
an external (four-dimensional) and an internal (six-dimensional) part as
\begin{align}
\label{extgravsplit} 
& \nabla_\mu\zeta_1 \otimes \eta_1 - \frac{1}{2}e^A (\gamma_\mu \zeta^* \otimes \slashed\partial A\, \eta_1^*)+\text{c.c.}=0 \, , \\
\label{intgravsplit}
& \zeta_1 \otimes \nabla_m \eta_1+\text{c.c.}=0 \ .
\end{align}
An identical equation holds for $\eta_2$.
The external gravitino equations~\eqref{extgravsplit} imply that $\slashed\partial A\, \eta_{1,2}^*$ should be proportional to $\eta_{1,2}$, which is impossible since $\eta_{1,2}^\dagger\gamma_m\eta_{1,2}=0$. It follows that the warping must be constant. Taking this into account,~\eqref{extgravsplit} further reduces to
\begin{equation}
\label{extgravsplit2}
\nabla_\mu\zeta_{1,2}=0 \, .
\end{equation}
The commutator of two external covariant derivatives gives
 \begin{equation}\label{extint}
[\nabla_\mu, \nabla_\nu]= \frac{1}{4\!} R_{\mu \nu \rho \sigma} \gamma^{\rho \sigma}= \frac{\Lambda}{6} \gamma_{\mu \nu} \, ,
\end{equation}
where we used the expression for the curvature tensor for a maximally symmetric (unwarped) four-dimensional metric:
$R_{\mu \nu \rho \sigma} = \frac{1}{3} \Lambda (g_{\mu \rho} g_{\nu \sigma}
- g_{\mu \sigma} g_{\nu \rho})$. 
From~\eqref{extgravsplit2} and~\eqref{extint}, it then follows $\Lambda\gamma_{\mu \nu}\zeta=0$
which implies the vanishing of the cosmological constant
\begin{equation}
\label{lapl}
\Lambda = 0 \, .
\end{equation}
Then, the external gravitino equations require that the warp factor $A$ must be constant, 
and the four-dimensional space must be Minkowski ($\Lambda =0$).

Let us now turn to the internal gravitino equations,~\eqref{intgravsplit}, which reduce to
\begin{align}
\label{intgrav2}
& & \nabla_m\eta_1=0\, , & & \nabla_m\eta_2=0\ . & &
\end{align}
Applying the same argument as below~\eqref{extint}, we see that the internal metric must be Ricci flat.

Moreover, a covariantly constant spinor implies a reduction of the holonomy group of a Riemannian manifold.
From~\eqref{intgrav2} it follows that the internal metric must have at most holonomy SU(3). A Ricci flat manifold of SU(3) holonomy is a Calabi-Yau.

If the internal metric has strict SU(3)-holonomy, then $\eta^1_+$ and $\eta^2_+$ must be proportional. Without loss of generality we can set $\eta^1_+=\eta^2_+=\eta$ with $\eta^\dagger\eta=1$. We can express 

We can construct the forms $\omega$ and $\Omega$ as spinors bilinears of the covariantly constant spinor $\eta$. Then 
we can rewrite the Calabi-Yau condition $\nabla_m\eta=0$ in the alternative form
						\begin{equation}
							\begin{array}{lcr}
								\dd \omega = 0\, , & & \dd \Omega = 0 \, .
							\end{array}
						\end{equation}

				Thus, we got an important result: in absence of fluxes, looking for a supersymmetric vacuum requires to consider a Calabi-Yau three-fold as internal manifold.
	
The properties of Calabi-Yau's manifolds are such that one can explicitely derive the four-dimensional effective action (see~\cite{CeresoleN2, Bodner:1990zm} for details).		
			This is an $\mathcal{N} = 2$ supergravity theory in four-dimensions, that is characterised by the presence of massless uncostrained scalars, the moduli. 
			In supersymmetric theories massless scalar fields are not a problem, the trouble is if some of them stay massless after SUSY breaking: massless scalar fields would provide long range interactions that are not observed in nature. 
			One solutions to the moduli problem is to find ways to generate potential terms for some or all such scalars. 
			One way is to consider compactifications admitting non-trivial fluxes.
		\subsection{Backgrounds with fluxes in type II}
			We now turn to the study of more general solutions of type II supergravity where some of the fluxes have non-zero values. 
			The presence of fluxes drastically changes the properties of the solutions. This can be seen both from the equations of motion and the
supersymmetry variations.		
Indeed, from the Einstein equation, which reads schematically
\begin{equation}
R_{MN} \sim H_{MPQ}H_N{}^{PQ}+\sum_p F_{M Q_{1} \ldots Q_{p}} F_N{}^{\, Q_{1} \ldots Q_{p}} \, ,
\end{equation}
we see that the fluxes back-react on the metric, which generically cannot be Ricci-flat (and thus Calabi-Yau) anymore. 
Another generic feature is a non trivial warp factor in the ten-dimensional metric. 

The supersymmetry variations are also modified. For example, from~\eqref{susyvar} one can see that in the presence of RR fluxes the supersymmetry conditions relate $\epsilon^{1}$ and $\epsilon^2$ so that the four-dimensional components $\zeta_{1,2}$ cannot be chosen independently anymore, as in~\eqref{spinansatz}. Therefore, in the presence of RR-fluxes one generically obtains $\mathcal{N}=1$ in four dimensions.

Repeating the strategy used in the fluxless case, we decompose the supersymmetry conditions according to the compactification
ansatz. We do not give all details here, but one can easily see that the internal gravitino variations become
\begin{equation}\label{intgravsplitagain}
\begin{split}
(\nabla_m+\frac14\slashed{H}_m)&\eta_1+\frac{1}{8} e^\phi\slashed{F}\gamma_m\gamma_7\eta_2 = 0\ , \\
(\nabla_m-\frac14\slashed{H}_m)&\eta_2-\frac{1}{8} e^\phi\slashed{F}^\dagger\gamma_m\gamma_7\eta_1 = 0\ , 
\end{split}
\end{equation}
from which we see that, generically, the internal manifold is no longer Ricci flat and hence no longer Calabi-Yau, since
\begin{equation}\label{susyintegr}
	[\nabla_m , \nabla_n ] \eta_{1,2} = \frac{1}{4} {R}_{mn}{}^{ pq} \gamma_{pq} \eta_{1,2} \neq 0 \, .
\end{equation}

It is therefore natural to wonder whether it is still possible
to say something about the geometry of the internal manifold.
Generalized Complex Geometry provides a general framework to describe flux backgrounds in string theory.

\ensurepagenumbering{arabic}
	\chapter{Generalising the geometry}
	\label{chapEGG}
		\section{Introduction and motivations}
			The aim of this chapter is to introduce Generalised Geometry, both complex (its simpler version) and Exceptional.
	
					Generalised complex geometry, as originally proposed by Hitchin~\cite{hitch1, gualtphd}, geometrises the NSNS sector of type II supergravity. 
					As described in the previous chapter, Hitchin's generalised tangent bundle is isomorphic to the sum $TM \oplus T^*M$ of the tangent and cotangent bundle to the $d$-dimensional compactification manifold $M_d$, and is patched by $\GL(d,\RR)$ transformations and gauge shifts of the NSNS two-form $B$. 
					The structure group of this extended bundle is $\rmO(d,d)$, \emph{i.e.} the T-duality group of the compactification on a $d$-dimensional torus.
					From a string theory perspective, $T$ and $T^*$ parameterise the quantum number of the string, that is momentum and winding charge.
					Extending this construction to include the RR potentials in type II supergravity~\cite{hull1, Grana:2009im, waldram4}, or adapting it to M-theory compactifications~\cite{hull1, waldram5, Coimbra:2011ky}, leads to exceptional generalised geometry. 
					In this case the structure group of the generalised tangent bundle is the U-duality group, and the bundle parameterises all the charges of the theory under study, that is momenta and winding, as well as NS- and D-brane (or M-brane) charges.
					
%
%
			%
		\section{Generalised complex geometry}
				Generalised complex geometry was introduced by Hitchin~\cite{hitch1} and Gualtieri~\cite{gualtphd} to find a structure interpolating between complex and symplectic geometries.
				
					The main idea of generalised geometry is to geometrise the gauge transformations of a two form. 
					This is done by introducing the generalised tangent bundle. 
					Given a $d$-dimensional manifiold $M$ generalised tangent bundle is the extension of the tangent space by the cotangent space
							\begin{equation}\label{gentanext}
								\begin{tikzcd}
									0 \arrow{r} &T^*M \arrow{r}{i} & E \arrow{r}{\pi}& TM \arrow[bend left=50, color = red!60]{l}{B} \arrow{r}& 0 \ ,
								\end{tikzcd}
							\end{equation}
					where $\pi$ is the so-called \emph{anchor} map, that is a projection $\pi: E \rightarrow TM$, to not be confused with the usual projection map on a bundle.
					Its action on sections is simply projecting out the form part.
					
					At any point $p\in M$, $E$ is isomorphic to the sum of the tangent and the cotangent bundle
							\begin{equation}
								E \cong \tilde{E} = TM \oplus T^*M \, .
							\end{equation}
					The sections of $E$ are called \emph{generalised vectors}. 
					On patch $U_\alpha$ they can be written as,
							\begin{equation}
								V_{(\alpha)} = v_{(\alpha)} + \mu_{(\alpha)} \in TM \oplus T^*M\, , 
							\end{equation}
					and at the intersection $U_\alpha \cap U_\beta$ the patch non-trivially 
							\begin{equation}\label{patchV1}
								V_{(\alpha)} = v_{(\alpha)} + \mu_{(\alpha)} = A_{(\alpha\beta)} v_{(\beta)} + A^{-T}_{(\alpha\beta)} \mu_{(\beta)} - \iota_{A_{(\alpha\beta)} v_{(\beta)}} \dd \Lambda_{(\alpha\beta)} \, ,
							\end{equation}
where $ A_{(\alpha\beta)}$ is a $\GL(d, \RR)$ transition function and $\Lambda$ is a one-form gauge parameter satisfying the co-cycle condition on the triple overlap 
$U_\alpha \cap U_\beta \cap U_\gamma$,
								\begin{equation}\label{cycliccond}
									\Lambda_{(\alpha\beta)} + \Lambda_{(\beta\gamma)} + \Lambda_{(\gamma\alpha)} = - i g^{-1}_{(\alpha\beta\gamma)} \dd g_{(\alpha\beta\gamma)}\, ,
								\end{equation}
						with $g$ is an element of $\U(1)$, satisfying the condition for transition functions
								\begin{align}
									& & g_{(\beta\gamma\delta)} g^{-1}_{(\alpha\gamma\delta)} g_{(\alpha\beta\delta)} g^{-1}_{(\alpha\beta\gamma)} = 1\, , & & \mbox{on}\ U_\alpha \cap U_\beta \cap U_\gamma \cap U_\delta\, .	
								\end{align}

This construction allows to introduce a two-form $B$, transforming as 	
\begin{equation}\label{Btrans}
									B_{(\alpha\beta)}:= B_{(\alpha)} - B_{(\beta)} = \dd \Lambda_{(\alpha \beta)}\, .
								\end{equation}				
which is a generalisation of the standard $\U(1)$ connection. 
					$B$ is a \emph{connection on a gerbe}~\cite{HitchinLagrangian}, that is a higher rank form generalisation of a connection over a fiber bundle~\cite{Nakahara}. 
					We can see that the map $B_{(\alpha\beta)}$ gives the non-trivial fibration of the cotangent bundle on the tangent one.
					In other words, the vector part of the generalised vector $V$ is a well-defined vector, meaning it patches correctly as a vector over the manifold.
					On the other hand, as one can observe from~\eqref{patchV1} the form part does not patch as one can expect from a one-form, but it has an extra-part parametrised by $\dd \Lambda_{(\alpha\beta)}$.
					This is what we mean when we say that the isomorphism (formally called \emph{splitting}~\cite{Hatcher}) $E \cong TM \oplus T^*M$ is not canonical, but it depends on the choice of the map $B$.
					The transition functions of the generalised tangent bundle are $\GL(d, \RR) \times \Lambda^2 T^*M$. 
					Notice that generally the $B$-field is only defined locally, however its field strength $H = \dd B$ is globally defined. 
					In applications to string theory, $B$ is identified with the NS-NS two-form potential and the patching rules~\eqref{patchV} and~\eqref{Btrans} will correctly reproduce the gauge transformations prescribed by supergravity.

					The generalised tangent bundle -- because of the duality between $TM$ and $T^*M$ as linear spaces -- is equipped with a natural $\rmO(d,d)$ symmetric bilinear form $\eta$, \emph{i.e.} a metric
							\begin{equation}
								\eta(V , W) = \eta(v+\mu , w + \lambda ) = \frac{1}{2} \left( \mu(w) + \lambda(v) \right) \, ,
							\end{equation}
					where $\mu(w) = \iota_w \mu$ denote the contraction of the vector $w$ with the one-form $\mu$. 
					One can also write the relation above as a matrix equation $\eta(V , W) = V^T \eta W$, where,
							\begin{align}
								V = \begin{pmatrix}
								 v \\
								 \mu
								\end{pmatrix}\, , & & W = \begin{pmatrix}
								 w \\
								 \lambda
								\end{pmatrix}\, , & & \eta = \frac{1}{2} \begin{pmatrix}
									0 & \id \\
									\id & 0 
								\end{pmatrix}\, .
							\end{align}
					One can diagonalise the matrix $\eta$ and make the signature $(d,d)$ explicit
							\begin{equation}
								\tilde{\eta} = \frac{1}{2} \begin{pmatrix}
									\id & 0\\
									0 & -\id 
								\end{pmatrix}\, .
							\end{equation}
					As discussed in~\cref{chap1}, defining a metric is equivalent to define a $G$-structure.
					In this case, the metric $\eta$ defines $\rmO(d,d)$ on $TM \oplus T^*M$. 
					
					
					In addition to the metric, the generalised tangent bundle $\tilde{E}$ has an orientation too~\cite{gualtphd}.
					It can be defined by the $\eta$ metric through the Levi-Civita symbol,
							\begin{equation}
								\mathrm{vol}_\eta = \frac{1}{(d!)^2} \epsilon^{m_1 \ldots m_d} \epsilon_{n_1 \ldots n_d} \partial_{m_1} \wedge \ldots \wedge \partial_{m_d} \wedge \dd x^{n_1} \wedge \ldots \wedge \dd x^{n_d} \, ,
							\end{equation}
							where $\partial_{m}$ and $ \dd x^{n}$ denote a basis on $TM$ and $T^*M$. 

					The structure group preserving both the metric and the volume form is $\SO(d,d)$.
					The Lie algebra of $\SO(TM \oplus T^{*}M) \cong \SO(d,d)$ is given by 
							\begin{equation}\label{soalg}
								\mathfrak{so}(TM \oplus T^{*}M) = \left\{ T \mid \eta( TV , W) + \eta( V , TW ) = 0 \right\},
							\end{equation}
					\textit{i.e.} generators are antisymmetric. 
					This algebra decomposes~\cite{KoerberReview} in 
							\begin{equation}\label{sodec}
								\mathrm{End}(TM) \oplus \Lambda^2 TM \oplus \Lambda^2 T^*M \, ,
							\end{equation} 
					or, equivalently, a generic element $T \in \mathfrak{so}(TM \oplus T^{*}M)$ can be written as
							\begin{equation}\label{Bbeta}
								T = \begin{pmatrix} A & \beta \\ B & -A^{T} \end{pmatrix}\ ,
							\end{equation}
					where $A \in \mathrm{End}(TM)$, $B \in \Lambda^2 T^*M$, $\beta \in \Lambda^2 TM$, and hence
							\begin{equation*}
								\begin{tikzcd}[row sep=tiny]
									A : TM \arrow{r} & TM \\
									B : TM \arrow{r} & T^*M \\
									\beta : T^*M \arrow{r} & TM\ .
								\end{tikzcd}
							\end{equation*}
					The action of the three subgroups can be done explicitly.
					For $\GL(d,\RR)$ part,
							\begin{equation}
								e^A \cdot V = A v + A^{-T} \mu \, , 
							\end{equation}
					for the so-called $B$-transformation,
							\begin{equation}\label{bact}
								e^B \cdot V = v + \mu - \iota_v B \, , 
							\end{equation}
					and, finally for the $\beta$-transformation,
							\begin{equation}
								e^\beta \cdot V = v - \beta \lrcorner \mu + \mu \, .
							\end{equation}
					Here $\cdot$ denotes the adjoint action of the $\mathfrak{so}(d,d)$ algebra.
					A noteworthy fact is that since both $B$ and $\beta$ are not invariant under $\GL(d, \RR)$, their actions do not commute with the $\GL(d, \RR)$ one.
					Another observation one can make is that, since both $B$ and $\beta$ are antisymmetric tensors, the symmetric product defined by the metric $\eta$ is invariant under $B$ and $\beta$ transformations,
							\begin{equation}
								\eta(e^{B + \beta} V, e^{B + \beta} W ) = \eta (V, W)\, .
							\end{equation}
The patching conditions can be rewritten as 												%
								\begin{equation}\label{patchV}
								\begin{array}{ccc}
								V_{(\alpha)} = e^{A_{(\alpha\beta)} + \dd \Lambda_{(\alpha\beta)}} \cdot V_{(\beta)} 
								\end{array}
							\end{equation}

The bundle $\tilde{E}$ is some-time called the untwisted generalised and its sections $\tilde{V} \in \Gamma(TM \oplus T^*M)$ are called untwisted generalised vectors.
They are related to the sections of $E$ by a $B$-transformation 							%
								\begin{equation}\label{Btwist}
									V = e^{B} \tilde{V} = e^{B} (v + \mu) = v + \mu - \iota_v B \, .
								\end{equation}
					\subsection{The generalised frame bundle}\label{genframOdd}
						The definition of generalised frame bundle is a straightforward generalisation of that of frame bundle. 
						Given a frame on $E$, that we call $\{\hat{E}_A\}$, satisfying the orthonormality condition with respect to the natural inner product
								\begin{equation}\label{eqn:frame}
									\eta\left(\hat{E}_A, \hat{E}_B\right) = \eta_{AB} = \frac{1}{2} \begin{pmatrix} 0 & \id \\ 
																					\id & 0 \end{pmatrix}_{AB} \, ,
								\end{equation}
						we can define the \emph{generalised frame bundle} as follows. 
						The \emph{frame bundle} is the bundle associated to these basis vectors. 
						Points on the fibre (frames) are connected by $\rmO(d,d)$ transformations. 
						Conversely, all frames connected by $\rmO(d,d)$ transformations to a frame that satisfies~\eqref{eqn:frame} will satisfy it too. 
						In other words, these frames form an $\rmO(d,d)$-bundle, that we call \emph{generalised frame bundle},
								\begin{equation}\label{genfr}
									F := \bigsqcup_{p\in M} \left\{\left( p,\, \hat{E}_A \right) \mid p \in M,\ \eta\left(\hat{E}_A, \hat{E}_B\right) = \eta_{AB} \right\} \, .
								\end{equation}

						Given the frame $\{\hat{e}_a\}$ for $TM$ and the coframe $\{e^a\}$ for the cotangent bundle $T^*M$, we can make a particular choice of frame, the \emph{split frame}, such that we can keep track of the vector and form part of our generalised sections~\cite{waldram1, waldram3}. 
						Explicitly we can choose,
								\begin{equation}\label{splitframe}
									\hat{E}_A := \begin{cases}
												\begin{pmatrix}\hat{e}_a\\ - i_{\hat{e}_a}B\end{pmatrix} \, , & A = a \, ,\\[3mm]
												\begin{pmatrix} e^a \\ 0 \end{pmatrix} \ , & A = a + d \, . \end{cases}
								\end{equation} 
						Note that the $B$-shift is present in our definition of the split frame. 
						This is because one can obtain the split frame by a twist of a generic basis of $TM \oplus T^*M$.
						
						We can then define a \emph{generalised $G$-structure}, as a sub-bundle of the principal generalised frame bundle associated to $E$.
						In other words, a generalised $G$-structure is a set of generalised tensors that are invariant under the action of a subgroup $G \subset \rmO(d,d)$.
					\subsection{Generalised metric}\label{genmetrOdd}
						Proceeding in with the ordinary structures, we want to describe the analogue of a Riemannian metric on the generalised tangent bundle.
						
						One can define a \emph{generalised metric} on a generalised tangent bundle $E$, as a positive definite sub-bundle of rank $d$, such that the restriction of the natural metric $\eta$ is positive definite~\cite[def. 4.1.1]{baraglia}.
						In terms of generalised $G$-structures, we say that a generalised metric is an $\rmO(d) \times \rmO(d)$-structure over $M$.
						
						The presence of such a structure splits the generalised tangent bundle $E$ into two sub-bundles,
								\begin{equation}
									E \cong C_+ \oplus C_- \, ,
								\end{equation}
						corresponding to spaces where the inner product $\eta$ has a definite sign~\cite{gualtphd, petrini3}. 	 
						This allows us to define a generalised metric~\cite{baraglia}
								\begin{equation}\label{Gmetr}
									\mathcal{G}(\cdot,\cdot) = \eta(\cdot,\cdot) \big|_{C_{+}} - \eta(\cdot,\cdot)\big|_{C_{-}} \, .
								\end{equation}
								\begin{figure}
								\centering
									\begin{tikzpicture}
	\node (o)    at ( 0,0) {};
	\node (x)    at ( 1.5,0) {};
	\node (inf)  at  ( 3,3) {}; 
	\node (inf-) at ( 3,-3) {};
	\node (-inf-) at (-3,-3) {};
	\node (+inf-) at (-3,3) {};
	\node (y) at (0,3.5) {};
	\node (x) at (3.5,0) {};
	\node (-y) at (0,-3.5) {};
	\node (-x) at (-3.5,0) {};

		\path  
		  (o) +(1.5,1.5)  coordinate  (gplus)
  		     +(1.5,-1.5) coordinate (gminus)
		     ;
       
		\draw[color=blue] (-inf-) -- (inf)
       			node[pos=0.75, above left, color= black]    {$v+ gv$}
			node[pos=1.03, above right, color= black] {$C_{g+}$};
		\draw[color= blue] (+inf-) -- (inf-)
			node[pos=0.75, below left, color= black]    {$v-gv$}
			node[pos=1.03, above right, color= black] {$C_{g-}$};
		\draw[->] (-y)--(y)
			node[pos=1.01, above] {$T^*M$};
		\draw[->] (-x) -- (x) 
			node[pos=1.04, above] {$TM$};
		\draw[dashed, gray!70] (gminus) -- (gplus)
			node[pos = 0.5, above right] {$v$};
\end{tikzpicture}
									\caption{We can represent the splitting of $E$ into the sub-bundles $C_{+} \oplus C_{-}$ by the graph of a linear map $h: TM \longrightarrow T^*M$. 
									Here is shown the particular case of a zero $B$ field transformation.}
									\label{graph}
								\end{figure}
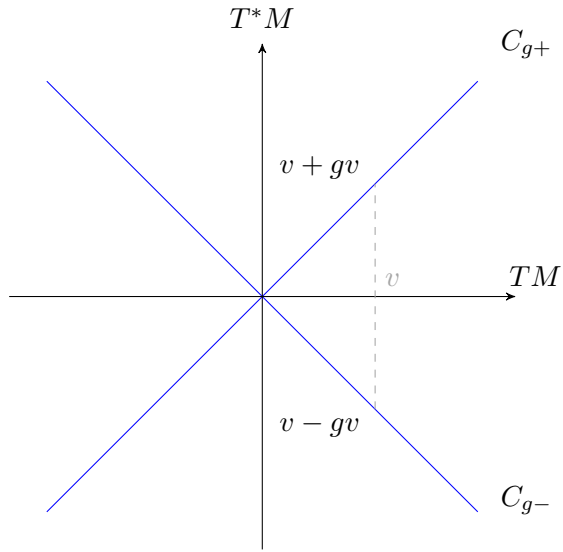

						Since any generalised section which is made only by a vector field or only a form has a zero norm with respect to the metric $\eta$, we can state for instance $T^*M \cap C_{\pm} = \{0\}$. (Analogously intersections between $TM$ and $C_{\pm}$ are just made by the zero section, as represented in~\cref{graph}.)
						Thus, we can define a map $h : TM \longrightarrow T^*M$ such that $C_{+}$ is the graph of $h$, and $C_{-}$ its orthogonal complement, and explicitly
								\begin{equation}\label{cplus}
									C_{+} = \left\{v+ hv \mid v \in \Gamma\left(TM\right)\right\}.
								\end{equation}
						The map $h$ provides an isomorphism between $TM$ and $C_{+}$.						
						One can see $h$ as an element of $T^*M \otimes T^*M$, \emph{i.e.} a $2$-tensor, and hence can be written as a sum of a symmetric and an antisymmetric part: $h=g+B$, exploiting the decomposition $T^*M \otimes T^*M \cong Sym^2 T^*M \oplus \Lambda^2 T^*M$, where $g \in Sym^2 T^*M$ and $B \in \Lambda^2 T^*M$. 
						Thus one can write a general element $V_{+} \in C_{+}$ as $V_{+} = v + \left(B + g \right)v$, where we denote with $Bv$ the contraction $\iota_v B$. 
						The orthogonality condition between $C_{+}$ and $C_{-}$ force us to write $V_{-} \in C_{-}$ as $V_{-} = v + \left(B - g \right)v$ and so
								\begin{equation}\label{cminus}
									C_{-} = \left\{v+ \left(B-g\right)v \mid v \in \Gamma\left(TM\right)\right\} \, .
								\end{equation}
						Thanks of the symmetry of the inner product induced by $\eta$ under $B$ shifts, we can identify $g$ as an ordinary Riemannian metric on $M$.
						Indeed, given any $V_{+}, W_{+} \in C_{+}$ and their inner product
								\begin{equation*}
									\begin{split}
										\eta \left( V_{+}, W_{+} \right) 	& = \eta( v + \iota_v B + g v , w + \iota_w B + g w ) \\
 																& = \eta( v + gv , w + gw ) \\
 																& = \frac{1}{2} \left( \iota_w g v + i_v g w \right) = g\left(v,w\right)\, .
									\end{split}
								\end{equation*}
						By construction, $g$ is a positive definite Riemannian metric on $M$.
						
						One can find an explicit form for the generalised metric $\mathcal{G}$~\eqref{Gmetr} by studying its action on $C_{\pm}$.
						The construction below will closely follow the one in~\cite{baraglia}, which we refer to for further details.					
						Given a generalised vector $V \in E$, one can write it as $ V= V_+ + V_{-}$, where $V_{\pm} \in C_{\pm}$.
						Thus, we have the map
								\begin{equation}
									\begin{tikzcd}[row sep=tiny]
										\mathcal{G} :\!\!\!\!\!\!\!\!\!\!\!\!\!\!\!\!\!\!\!\!& E \arrow{r} & E^* \cong E \\
 																	& V \arrow[mapsto]{r} & \mathcal{G}(V) = \mathcal{G}(V, \cdot)
									\end{tikzcd}
								\end{equation}
						where we denoted by $\mathcal{G}(V)$ the \emph{generalised one-form} $\mathcal{G}(V,\cdot)$, but since $(TM \oplus T^*M )^* \cong TM \oplus T^*M$ it can be thought as a generalised vector and then decomposed in $C_{\pm}$ components,
								\begin{equation}
									\mathcal{G}(V) = \mathcal{G}(V, \cdot) = \eta( V_{+} + V_{-}, \cdot )\big|_{C_{+}} - \eta( V_{+} + V_{-}, \cdot )\big|_{C_{-}} = V_{+} - V_{-} \, .
								\end{equation}
						From the expression above, one can state $\mathcal{G}^2 = \id$, and $C_\pm$ are the eigenspaces relative to the eigenvalues $\pm 1$ of $\mathcal{G}$. 
						Consider the usual Riemannian metric $g$ over $M$, this induces the splits of $TM \oplus T^*M$ into
								\begin{equation}
									C_{g\pm} = \left\{v \pm gv \mid v \in TM \right\}\, .
								\end{equation}
						A generic vector in $C_{g\pm}$ can be written as $V_{g\pm} = v \pm g v$. 
						In this particular case $2x= V_{g+} + V_{g-}$, thus, we are now allowed to write
								\begin{equation*}
									2\mathcal{G}(v) = V_{g+} - V_{g-} = 2 g(v),
								\end{equation*}
						and since $\mathcal{G}^2 = \id$, it holds
								\begin{equation*}
									2\mathcal{G}^2(v) = 2 \mathcal{G}(g(v)) = V_{g+} + V_{g-} = 2v \, .
								\end{equation*}
						We look for an explicit form of $\mathcal{G}$ in terms of the two quantities $g$ and $B$.
						The simplest form the matrix $\mathcal{G}$ can take (compatible with the two conditions above) is with $B=0$,
								\begin{equation*}
									\mathcal{G}_g = \begin{pmatrix} 0 & g^{-1} \\
 											g & 0 \end{pmatrix}\ .
								\end{equation*}
						Now we want to reintroduce the $B$ field we have ignored so far. 
						Recall that
								\begin{equation*}
									e^{B} V_{g\pm} = \left(v \pm gv + B v\right) = V_{\pm}
								\end{equation*}
						and also that 
								\begin{equation}\label{proj}
									\mathcal{G}\left(V_{\pm}\right) = \pm V_{\pm}.
								\end{equation}
						Using the previous relations and applying the $B$ transformation to $V_{g\pm}$, we can obtain a matrix representation for $\mathcal{G}$ as follows, 
								\begin{equation*}
									\begin{split}
										V_{\pm} = e^{B}V_{g\pm} & =\pm e^{B}\mathcal{G}_g V_{g\pm} = \\
														 & =\pm e^{B} \mathcal{G}_g e^{-B} e^{B} V_{g\pm} =\\
														 & = \pm e^{B} \mathcal{G}_g e^{-B} V_{\pm}\ .
									\end{split}
								\end{equation*}
						This is true if and only if
								\begin{equation}\label{genmet}
									\mathcal{G}= e^{-B} \mathcal{G}_g e^{B} = \begin{pmatrix} g^{-1}B & g^{-1} \\
																g - B g^{-1} B & -B g^{-1} \end{pmatrix}\, .
								\end{equation}
						The generalised metric is an element of the coset space $\rmO(d,d)/O(d) \times O(d)$ and encodes both the metric and the $B$-field~\cite{petrini3}.
						One can also introduce \emph{generalised vielbeins}, parametrising the coset, the local flat symmetry in ordinary geometry is here replaced by $O(d) \times O(d)$.
						We also require that the set of local vielbeins $\{\hat{E}_A\}$ makes the generalised metric take the following form,
								\begin{align}
									& & \eta = \hat{E}^{T} \begin{pmatrix}
										\id & 0 \\
										0 & -\id
									\end{pmatrix} \hat{E} \, , & & \mathcal{G} = \hat{E}^{T} \begin{pmatrix}
										\id & 0 \\
										0 & \id
									\end{pmatrix} \hat{E} \, ,
								\end{align}
						An explicit form is given by,
								\begin{equation}
									\hat{E} = \frac{1}{\sqrt{2}} \begin{pmatrix}
									e_+ - \hat{e}_+^T B & \hat{e}_+^T \\
									- e_{-} - \hat{e}_{-}^T B & \hat{e}_{-}^T 
									\end{pmatrix}\, ,
								\end{equation}
						where $e_{\pm}$ are two sets of vielbeins and $\hat{e}_\pm$ their inverse.
						They have to satisfy
								\begin{align}
									&& g = e_{\pm}^T e_{\pm}\, , && g^{-1} = \hat{e}_{\pm} \hat{e}_{\pm}^T \, , & &
								\end{align}
						which are the canonical conditions on ordinary vielbeins.
				\subsection{Generalised almost complex structure}
						A \emph{generalised almost complex structure} is a map
								\begin{equation}
									\mathcal{I} : E \longrightarrow E \, ,
								\end{equation}
						compatible with the bundle structure, \emph{i.e.} $\pi( \mathcal{I} V) = \pi(V)$ and with the property analogous to~\eqref{I2mid},
								\begin{equation}
									\mathcal{I}^2 = - \id \, .
								\end{equation}
						In addition, compatibility with the $\rmO(d,d)$ metric is required,
								\begin{equation}\label{hermeta}
									\mathcal{I}^T \eta \mathcal{I} = \eta \, ,
								\end{equation}
						or equivalently
								\begin{equation}\label{hermeta2}
									\eta(\mathcal{I} V, \mathcal{I} W) = \eta (V, W) \, , \qquad \forall V, W \, .
								\end{equation}

						An ordinary complex structure reduces the structure group to $\U(d/2)$. 
						Here something analogous happens, the generalised almost complex structure implies $M$ is even-dimensional, and the structure group of the exceptional tangent bundle $\rmO(d,d)$ is reduced to $\U(d/2, d/2)$. 
						
						Also in this case, one can adopt a ``matrix'' notation and describe the generalised almost complex structure as a block matrix acting on a generalised vector.
						From~\eqref{hermeta}, it follows that the generalised almost complex structure takes the form,
								\begin{equation}\label{gencompl}
									\mathcal{I} = \begin{pmatrix}
												- I	&	R \\
												L	& 	I^T
												\end{pmatrix} \, , 
								\end{equation}
						where $R$ and $L$ are an antisymmetric two-vector and a two-form respectively. 
						The requirement for $\mathcal{I}$ to square to $-\id$ imples,
								\begin{align}
									I^2 + RL &= -\id_d \, , \\
									-I R + R I^T &= 0 \, , \\
									-LI + I^T L &= 0 \, .
								\end{align}
						An important feature of the generalised almost complex structure is that it contains both the ordinary almost complex and the almost symplectic structures,
								\begin{align}
									&&	\mathcal{I}_I = \begin{pmatrix}
												- I	&	0 \\
												0	& 	I^T
												\end{pmatrix} \, ,	&&	\mathcal{I}_\omega = \begin{pmatrix}
												0		&	\omega^{-1} \\
												-\omega	& 	0
												\end{pmatrix} \, .
								\end{align}
						We can think at the general case~\eqref{gencompl} as a set of continuous intermediate structures, interpolating between the two extrema of complex and symplectic geometry, by varying the tensors $R$ and $L$.
						Indeed, the original purpose of Hitchin and Guatieri~\cite{HitchinLagrangian, gualtphd} was to find a way to unify the symplectic and complex geometry.
						
						As in the ordinary case, also the generalised almost complex structure generates two \emph{generalised distributions}, that is the two eigenbundles associated to the eigenvectors $\pm i$,
								\begin{equation}
									E \otimes \CC = L_\mathcal{I} \oplus \overline{L}_\mathcal{I} \, .
								\end{equation}
						This is somehow analogous to the split seen previously induced by the generalised metric.
						
						$L_\mathcal{I}$ and $\overline{L}_\mathcal{I}$ are \emph{maximally isotropic sub-bundles} of $E\otimes \CC$.
						Recall that a generalised sub-bundle $L$ is \emph{isotropic} if and only if it is \emph{null} with respect to the natural $\rmO(d,d)$ metric,
								\begin{equation}
									\eta(V,W) = 0 \, \qquad \forall V, W \in L \, .
								\end{equation}
						In addition, it is maximally isotropic if its rank is half of the rank of $E$.
						
						We can show that $L_\mathcal{I}$ is actually maximally isotropic.
						Given two vectors $V, W \in L_\mathcal{I}$, their inner product is,
								\begin{equation*}
									\eta(V,W) = V^T \eta W = V^T \mathcal{I}^T \eta \mathcal{I} W = (i V)^T \eta (iW) = - V^T \eta W = - \eta (V,W) = 0 \, ,
								\end{equation*}
						where we used the~\eqref{hermeta} in the second equality and the fact that both $V$ and $W$ are elements of $L_{\mathcal{I}}$ in the third one.
						Moreover, since $L_\mathcal{I}$ and $\bar{L}_\mathcal{I}$ have the same rank, we have that both have complex dimension $d$, such that $L_\mathcal{I}$ is maximally isotropic.
						%
					%
			\subsection{Dorfman derivative and Courant bracket}
					One can define a differential operator generalising the action of the Lie derivative in the ordinary case.
					We will also see how to generalised the concept of infinitesimal diffeomorphism for the generalised tangent bundle.
					
					Given two sections of $E$, for instance $V = v + \mu$ and $W = w + \lambda$, where $v, w \in \Gamma(TM)$ and $\mu, \lambda \in \Gamma(T^*M)$, we define the \emph{Dorfman derivative} or \emph{generalised Lie derivative}~\cite{hitch1, waldram5, waldram1} as 
							\begin{equation}\label{dorf}
								L_{V} W := \mathcal{L}_v w + \mathcal{L}_v \lambda - \iota_w \dd \mu \, .
							\end{equation}
					The Dorfman derivative enjoys the Leibnitz rule, \emph{i.e.}
							\begin{equation}\label{eq:Leibniz}
								L_V (L_W Z) = L_{L_V W} Z + L_W (L_V Z) \, .
							\end{equation}
					This gives the generalised tangent bundle, endowed with the generalised Lie derivative the structure of a \emph{Leibnitz algebroid}~\cite{baragliaLeib}.
					
					Note that we can give a definition that makes explicit the action of the $\rmO(d,d)$ group.
					This comes from the observation that the Lie derivative between two ordinary vectors $v$ and $v'$ of $\Gamma(TM)$ can be written in components using the $\mathfrak{gl}(d, \RR)$ action,
							\begin{equation*}
								( \mathcal{L}_v v')^m = v^n \,\partial_n v^{\prime m} - (\partial \times v)^m_{\phantom{m}n} \,v^{\prime n} \, .
							\end{equation*}
					Thus, by analogy one can use the $\rmO(d,d)$ action to write the Dorfman derivative as
							\begin{equation}\label{adjDorf}
								\left(L_V V'\right)^M = V^N \partial_N V^{\prime M} - \left(\partial \times_{\mathrm{ad}} V \right)^M_{\phantom{M}N}V^{\prime N} \, ,
							\end{equation}
					where $\times_{\mathrm{ad}}$ denotes the projection onto the adjoint bundle,
							\begin{equation}
								\times_{\mathrm{ad}} : E^* \times E \longrightarrow \mathrm{ad} \, .
							\end{equation}
					This definition is useful not only because allows to define the action of the $L_V$ operator also to representations of $\rmO(d,d)$ other than the fundamental one (like the adjoint, the second rank symmetric, etc.), but also because it is easily extendable to the exceptional case.
					
					The Dorfman derivative is not antisymmetric.
					Its antisymmetrisation defines the \emph{Courant bracket}~\cite{CourWein, waldram1},
							\begin{equation}\label{cour}
								\llbracket V, W \rrbracket := \frac{1}{2}\left( L_V W - L_W V \right) = \left[v, w\right] + \mathcal{L}_v \lambda - \mathcal{L}_w \mu - \frac{1}{2} \dd (\iota_v \lambda - \iota_w \mu) \, ,
							\end{equation}
					which makes explicit the fact that in the $\rmO(d,d)$ generalised geometry we have a \emph{Courant algebroid}~\cite{CourWein, courant}.
					A nice historical review on the subject can be found in~\cite{courhist}.
					
					The Courant bracket educes to the ordinary Lie bracket when restricted to vectors, while it vanishes on one-forms.
					It is invariant under diffeomorphisms, and under $B$-shifts parametrised by a closed $2$-form $b$,
							\begin{equation}\label{bCour}
								\llbracket e^b \cdot V, e^b \cdot W \rrbracket = e^b \llbracket V, W \rrbracket + \iota_v \iota_w \dd b \, .
							\end{equation}
					The relation above suggests the introduction of a \emph{twisted} Courant bracket by a $3$-form $H$
							\begin{equation}\label{HCour}
								 \llbracket V, W \rrbracket_H = \llbracket V, W \rrbracket + \iota_v \iota_w H \, .
							\end{equation}
					One can see that the twisted Courant bracket is the right differential operator acting on untwisted vectors, while the~\eqref{cour} is the one used in the twisted picture, where the $B$-shift is already encoded in generalised vectors.
				\subsubsection{Integrability}
						Recall that for an ordinary complex structure, integrability can be expressed in terms of the involutivity of its $\pm i$ eigenbundles 
						with respect to the Lie bracket.
						Here, we can give a definition of integrability in terms of Courant bracket and the involutivity of the generalised subbundle $L_\mathcal{I}$.
						
						Given a generalised almost complex structure $\mathcal{I}$ on $E$, we say it is \emph{integrable} if and only if its $i$-eigenspace $L_\mathcal{I}$ is closed under the Courant bracket,
								\begin{equation}
									\llbracket V, W \rrbracket \in \Gamma(L_\mathcal{I}) \, , \qquad \forall V, W \in \Gamma(L_\mathcal{I}) \, .
								\end{equation}
						A manifold admitting such a structure (called \emph{generalised complex structure}) is said \emph{generalised complex manifold}.

\subsection{Generalised geometry and compactifications} 

Generalised geometry allows to treat on the same ground diffeomorphisms and gauge transformations of the NS sector of type II supergravities. 
It is also a powerful tool to classify and study flux vacua. 
Let us consider again the case of $\mathcal{N}=1$ flux compactifications to four dimensions.

The idea is that one can define a pair of bispinors built out the 
supersymmetry parameters, $\eta_1$ and $\eta_2$ 
\begin{equation}
						\begin{split}
								\Phi^{+} & = e^{-\phi} e^{-B} \left(\eta_1^+ \otimes \overline{\eta}_2^+\right) \in \Gamma (\Lambda^{\mathrm{even}} T^*M) \, , \\
								\Phi^{-} & = e^{-\phi} e^{-B} \left(\eta_1^+ \otimes \overline{\eta}_2^-\right) \in \Gamma (\Lambda^{\mathrm{odd}} T^*M) \, ,
							\end{split}
\end{equation}
which are globally defined. The two polyforms $\Phi^{\pm}$ can be seen as sections of the positive and negative helicity $\mathrm{Spin}(6, 6)$ spinor bundles associated to $E$ through the Clifford map. And are associated to an almost generalised complex structure each. 
				Each of them is stabilised by a different $\SU(3,3)$ subrgroup of $\mathrm{Spin}(6, 6)$.
				Hence, each of them defines a different $\SU(3,3)$ generalised structure. The compatibility condition implies that the group leaving both invariant has to be the intersection of the two $\SU(3,3)$ subgroups, then $\SU(3) \times \SU(3)$.
				Thus, we see that all $\mathcal{N}=1$ flux backagrounds must have $\SU(3) \times \SU(3)$ structure~\cite{petrini2, Grana:2005sn}.

						On can show \cite{Grana:2006kf, petrini2}, that Killing spinor equations can be rewritten as differential conditions on such spinors 
							\begin{align}
									\label{susyeq1}
								& \dd_H(e^{3A}\Phi_1)= 0 \, , \\
									\label{susyeq2a}
								& \dd_H(e^{2A} \IIm \Phi_2) = 0 \, ,\\
\label{susyeq2b}
& \dd_H (e^{4A} \RRe \Phi_2) = e^{4A} \star \lambda(F) \, ,
\end{align}
where $\dd_H$ is the $H$-twisted derivative and $\Phi_1$ and $\Phi_2$ correspond to $\Phi_+$ and $\Phi_-$ in type IIA and vice-versa for IIB.
It is possible to show that such conditions correspond to the integrability of the generalised complex structure associated to $\Phi_1$. 
The supersymmetry conditions are also equivalent to the existence of a torsion-free generalised connection and structure-compatible.

%
%

%

				The ordinary Calabi-Yau case can be retrieved as a particular choice of the $\Phi^{\pm}$, 
						\begin{equation}
							\begin{array}{l c c r}
								\Phi^+ = e^{-\phi} e^{-B} e^{i\omega} \, , & & \Phi^- = i e^{-\phi} e^{-B} \Omega \, ,
							\end{array}
						\end{equation}
				with $B$ closed (eventually it can be made zero by a gauge transformation) and $\phi$ constant.
				From this particular case we can see that $\Phi^+$ is a generalisation of the symplectic structure, while $\Phi^-$ captures the generalisation of the complex structure.

		\section{Exceptional generalised geometry}
Generalised complex geometry has a natural application to string theory, since it allows to treat in a geometric way diffeomorphisms and gauge transformations of the NS sector of supergravity theories, This motivated the introduction of 
			Exceptional generalised geometry where the generalised tangent bundle admits the action of larger structure groups, $\E_{d(d)} \times \RR^+$~\cite{hull1, waldram5, waldram2}\footnote{%
				$\E_{d(p)}$ is a non-compact version of the exceptional group $\E_d$, meaning the group having as Lie algebra the exceptional one $\mathfrak{e}_d$, with a number $p$ of non-compact generators.
				$\E_{d(d)}$ is the maximal non-compact form.}, thus allowing to encode also the RR degrees of freedom of the various supergravities.
			
			The general definitions of frame bundle, generalised $G$-structures and generalised Lie derivative hold also here.
			An important difference between $\rmO(d, d)$ generalised geometry and the exceptional one is that in the $\rmO(d,d)$ case the same structure of the generalised tangent bundle allows to describe both type IIA and IIB and furthermore it does not depend on the dimension of the manifold $M_d$, as we will see in~\cref{chapComp}. 
			On the other hand, the exceptional tangent bundle takes a different form depending on whether one works in M-theory, type IIA or type IIB supergravity, and depending on the dimension of $M_d$, its fibres transform in different representations of the structure group.
			
			Thus, we are going to analyse the various cases separately, describing the suitable generalised geometry to describe the theories we will focus on.
			\subsection{M-theory}\label{sec:MthExcGeom}
				Here we review from~\cite{waldram2} the construction of the exceptional geometry for M-theory. The idea is to construct a generalised tangent bundle
				whose transition functions contain the three- and six-form potentials of M-theory.			%

						Given a $d$-dimensional manifold $M$ with $d \leq 7$. The $\rmO(d,d)$ group of generalised geometry is replaced by $\E_{d(d)}$. 
						We define the generalised tangent bundle as isomorphic to a sum of tensor bundles~\cite{hull1, waldram5},
						corresponding to the different $\GL(d,\RR)$n irreducible representations 
								\begin{equation}\label{MthExBun}
									E \cong TM \oplus \Lambda^2 T^*M \oplus \Lambda^5 T^*M \oplus (T^*M \otimes \Lambda^7 T^*M) \, .
								\end{equation}
						where for $d < 7$ some of these terms will not be present.				
						
						A generic section of $E$ is written as,
								\begin{equation}
									V = v + \omega + \sigma + \tau \, ,
								\end{equation}
						where $v \in \Gamma(TM)$ is a vector, $\omega \in \Gamma(\Lambda^2 T^*M)$, so is a two-form, etc.
						
						The bundle is defined together with patching rules. 
						These are such that, given a chart $U_\alpha$ of an atlas covering $M$, we have
								\begin{equation}
									\begin{split}
										V_{(\alpha)} = &\, v_{(\alpha)} + \omega_{(\alpha)} + \sigma_{(\alpha)} + \tau_{(\alpha)} \\[1mm]
												&\phantom{v} \in \Gamma\left( TU_{(\alpha)} \oplus \Lambda^2 T^*U_{(\alpha)} \oplus \Lambda^5 T^*U_{(\alpha)} \oplus (T^*U_{(\alpha)} \otimes \Lambda^7 T^*U_{(\alpha)}) \right) \, ,
									\end{split}
								\end{equation}
						for a local section.
						Then these sections are patched through the whole $E$ as
								\begin{align}\label{Vpatch}
									& & V_{(\alpha)} = e^{\dd \Lambda_{(\alpha \beta)} + \dd \tilde{\Lambda}_{(\alpha \beta)}} \cdot V_{(\beta)} \, , & & \mbox{on}\ U_\alpha \cap U_\beta \, . \phantom{U_\alpha \cap U_\beta}
								\end{align}
						The two quantities $\Lambda_{(\alpha \beta)}$ and $\tilde{\Lambda}_{(\alpha \beta)}$ are respectively a two- and a five-form, and $\cdot$ denotes the adjoint action of $\E_{d(d)}$. In components this reads							\comment{Maybe this in the appendix.}
								\begin{subequations}\label{expAdj}
									\begin{align}
										v_{(\alpha)} 		&= 	v_{(\beta)}									 						\, , \\
										\omega_{(\alpha)} 	&=	\omega_{(\beta)} + \iota_{v_{(\beta)}} \dd \Lambda_{(\alpha\beta)}			 	\, , \\
										\sigma_{(\alpha)} 	&=	\sigma_{(\beta)} + \dd \Lambda_{(\alpha\beta)}\wedge \omega_{(\beta)} 
																+\tfrac{1}{2}\dd \Lambda_{(\alpha\beta)} \wedge \iota_{v_{(\beta)}} \dd \Lambda_{(\alpha\beta)} 
																+ \iota_{v_{(\beta)}} \dd \tilde{\Lambda}_{(\alpha\beta)}		 			\, , \\
										\begin{split}
										\tau_{(\alpha)} 		&=	\tau_{(\beta)} +	j \dd \Lambda_{(\alpha\beta)}\wedge \sigma_{(\beta)} 
																- j \dd \tilde{\Lambda}_{(\alpha\beta)} \wedge \omega_{(\beta)} 
																+ j \dd \Lambda_{(\alpha\beta)} \wedge \iota_{v_{(\beta)}} \dd \tilde{\Lambda}_{(\alpha\beta)} \\
																& \phantom{= \tau}
																+ \tfrac{1}{2} j \dd \Lambda_{(\alpha\beta)} \wedge \dd \Lambda_{(\alpha\beta)} \wedge \omega_{(\beta)}
																+ \tfrac{1}{6} j \dd \Lambda_{(\alpha\beta)} \wedge \dd \Lambda_{(\alpha\beta)} \wedge \iota_{v_{(\beta)}} \dd \Lambda_{(\alpha\beta)}	 \, ,
										\end{split}
									\end{align}
								\end{subequations}
where $j$ denotes the operator

							 	\begin{equation}
									\left(j \lambda \wedge \mu\right)_{m,\,m_1\ldots m_d} = \frac{d!}{(p-1)!(d-p+1)!}\,\lambda_{m[m_1\ldots m_{p-1}}\mu_{m_p\ldots m_d]}\, ,
								\end{equation}
for $\lambda \in \Lambda^{p}T^*$ and $\mu \in \Lambda^{d-p+1}T^*$.

						The collection of $\Lambda_{(\alpha\beta)}$ defines this connective structure on the gerbe, satisfying the series of relations analogous to the~\eqref{cycliccond},
								\begin{equation}
									\begin{array}{lr}
										\Lambda_{(\alpha\beta)} + \Lambda_{(\beta\gamma)} +\Lambda_{(\gamma\alpha)} = \dd \Lambda_{(\alpha\beta\gamma)} & \mbox{on} \ U_\alpha \cap U_\beta \cap U_\gamma \, , \\
										\Lambda_{(\beta\gamma\delta)} - \Lambda_{(\alpha\gamma\delta)} + \Lambda_{(\alpha\beta\delta)} - \Lambda_{(\alpha\beta\gamma)} = \dd \Lambda_{(\alpha\beta\gamma\delta)} & \mbox{on} \ U_\alpha \cap U_\beta \cap U_\gamma \cap U_\delta \, .
									\end{array}
								\end{equation}
						Similar relations hold for $\tilde{\Lambda}$~\cite{waldram4}, 
								\begin{align*}
									\begin{split}
										\tilde{\Lambda}_{(\alpha\beta)} - \tilde{\Lambda}_{(\beta\gamma)} +\tilde{\Lambda}_{(\gamma\alpha)} = &\dd \tilde{\Lambda}_{(\alpha\beta\gamma)} \\
										& + \tfrac{1}{2 (3!)}\left(\Lambda_{(\alpha\beta)} \wedge \dd \Lambda_{(\beta\gamma)} + \mbox{antisymm. in}\, [\alpha\beta\gamma] \right) \, ,
									\end{split}
									\\[2mm]
									\begin{split}
										\tilde{\Lambda}_{(\alpha\beta\gamma)} - \tilde{\Lambda}_{(\alpha\beta\delta)} + \tilde{\Lambda}_{(\alpha\gamma\delta)} - \tilde{\Lambda}_{(\beta\gamma\delta)} = &\dd \tilde{\Lambda}_{(\alpha\beta\gamma\delta)} \\
										& + \tfrac{1}{2 (4!)}\left(\Lambda_{(\beta\gamma\delta)} \wedge \dd \Lambda_{(\gamma\delta)} + \mbox{antisymm. in}\, [\alpha\beta\gamma\delta] \right)\, .
									\end{split}
								\end{align*}
						In this case one can notice that the connective structure for $\tilde{\Lambda}$ depends on $\Lambda$, this further generalises the previous gerbe construction~\cite{HitchinLagrangian}, but it generates the correct patching rules that will be reinterpreted as gauge transformations of supergravity potentials in the next chapter.

						
						Technically the patching rules~\eqref{Vpatch} defines the generalised tangent bundle as a series of extensions,
								\begin{equation}
									\begin{tikzcd}[row sep=tiny]
										0 \arrow{r} &\Lambda^2 T^*M \arrow{r} &E'' \arrow{r} &TM \arrow{r} &0 	\, , \\
										0 \arrow{r} &\Lambda^5 T^*M \arrow{r} &E' \arrow{r} &E'' \arrow{r} &0 	\, , \\
										0 \arrow{r} &T^*M \otimes \Lambda^7 T^*M \arrow{r} &E \arrow{r} &E' \arrow{r} &0 \, .
									\end{tikzcd}
								\end{equation}
						Analogously to what we have seen in~\eqref{gentanext}, these extensions are splitted~\cite{Hatcher} into the isomorphism~\eqref{MthExBun} by the choice of some potentials, formally some connections on a gerbe~\cite{HitchinLagrangian}.
						
%
						As for generalised geometry, we can define a frame bundle.
						Let us define $\{\hat{E}_A\}$ as a basis of a fibre of the exceptional tangent bundle ($A$ runs over the dimension of the bundle).
						As in~\eqref{genfr}, the frame bundle is a principal bundle by construction.
						An exceptional $G$-structure is defined as a principal sub-bundle of $F$, such that its structure group is reduced to $G$.
						
						Take into account a point $p \in M$ and the exceptional fibre in that point $E_p$.
						Let $\{\hat{e}_a\}$ be a basis for $T_p M$ and $\{e^a\}$ a basis for $T^*_p M$.
						Following~\cite{waldram4}, an explicit basis of $E_p$ can be constructed as
								\begin{equation}
									\{\hat{E}_A\} = \{\hat{e}_a\} \cup \{e^{ab}\} \cup \{e^{a_1 \ldots a_5}\} \cup \{e^{a. a_1 \ldots a_6}\} \, .
								\end{equation}
						Thus, the definition for the exceptional frame bundle (analogous to~\eqref{genfr}) reads
								\begin{equation}
									F = \left\{ \left(x, \{\hat{E}_A\} \right) \mid x \in M \, , \, \, \mbox{and}\, \, \{\hat{E}_A\}\, \, \mbox{basis of}\, \, E\right\} \, .
								\end{equation}
						By construction, this is a principal bundle with fibre $\E_{d(d)} \times \RR^+$.
						It might be useful to consider the decomposition under $\GL(d,\RR)$ of the bundle transforming in the adjoint representation,
								\begin{equation}\label{eq:Ggeom-M}
									\adj \cong \RR \oplus (TM\otimes T^*M) \oplus \Lambda^3 T^*M \oplus \Lambda^6 T^*M \oplus \Lambda^3 TM \oplus \Lambda^6 TM \, ,
								\end{equation}
						then one can see it contains a scalar $l$, a $\mathfrak{gl}(d,\RR)$ element $r$, three- and a six-form $A$ and $\tilde{A}$ and a three- and a six-vector $\alpha$ and $\tilde{\alpha}$.
						Since we are interested in describing the degrees of freedom of eleven-dimensional supergravity, we will interpret $A$ and $\tilde{A}$ as the supergravity potentials.
						In exceptional generalised geometry these are gerbe connections patched on an overlap $U_\alpha \cap U_\beta$ as,
								\begin{equation}
									\begin{split}
										A_{(\alpha)} 		&= A_{(\beta)} + \dd \Lambda_{(\alpha\beta)} \, , \\
										\tilde{A}_{(\alpha)} 	&= \tilde{A}_{(\beta)} + \dd \tilde{\Lambda}_{(\alpha\beta)} - \frac{1}{2} \dd \Lambda_{(\alpha\beta)} \wedge A_{(\beta)} \, .
									\end{split}
								\end{equation}
						We will see how these reproduces the gauge transformations for potentials in eleven-dimensional supergravity~\cite{waldram4, waldram5, hull1}.
						Indeed the invariant field strengths 
								\begin{equation}
									\begin{split}
										F		&= \dd A \, , \\
										\tilde{F} 	&= \dd\tilde{A} - \frac{1}{2} A \wedge F \, ,
									\end{split}
								\end{equation}
						reproduce the supergravity ones.
						
						Also in the exceptional case we can define untwisted vectors $\tilde{V}$ as 
								\begin{equation}
									V = e^{A+ \tilde{A}} \cdot \tilde{V} \, , 
								\end{equation}
						where $\cdot$ denotes the adjoint action of the $\mathfrak{e}_7 \oplus \RR^+$ algebra (given explicitly in~\cref{app:EGG})~\cite{waldram5}.
						The sections of $E$ are called \emph{twisted} vectors, while the $\tilde{V}$ take the name of untwisted generalised sections.
							
						Also here the Dorfman derivative is constructed as a generalisation of the Lie derivative. 
						In particular, it holds also the~\eqref{adjDorf}, and similarly indicating by $V^M$ the components of $V$ in a standard coordinate basis, and embedding the standard derivative operator as a section of the dual generalised tangent bundle $E^*$, one can define the Dorfman derivative as
							\begin{equation}\label{dorfMadj}
								\left(L_V V'\right)^M = V^N \partial_N V^{\prime M} - (\partial \times_{\rm ad} V)^M_{\phantom{M}N} V^{\prime N} \, ,
							\end{equation}
						where again $ \times_{\rm ad}$ is the projection onto the adjoint bundle,
								\begin{equation}
									\times_{\mathrm{ad}} : E^* \times E \rightarrow \mathrm{ad}\, . 
								\end{equation}
						In terms of $\mathfrak{gl}(d, \RR)$ components the generalised Lie derivative can be expressed as
								\begin{equation}\label{dorfM}
									\begin{split}
									L_V V' &= \mathcal{L}_{v} v' + \left(\mathcal{L}_{v} \omega^{\prime} -\iota_{v^\prime}\dd \omega\right) + \left(\mathcal{L}_{v} \sigma' -\iota_{v^\prime}\dd\sigma - \omega^{\prime}\wedge \dd \omega\right) \\
										& \phantom{= \mathcal{L}_{v} v'}+ \left(\mathcal{L}_{v} \tau^{\prime} - j \sigma^{\prime}\wedge \dd \omega - j \omega' \wedge\dd\sigma \right) \, .
									\end{split}
								\end{equation}

						The version of Dorfman derivative being able to act on untwisted objects is called \emph{twisted Dorfman derivative} and denoted by $\mathbb{L}_{\tilde{V}}$,
								\begin{equation}
									\mathbb{L}_{\tilde{V}} \tilde{\mathcal{A}} = e^{-A-\tilde{A}}L_{e^{A+\tilde{A}} \tilde{V}} (e^{A+\tilde{A}}\tilde{\mathcal{A}}) \, ,
								\end{equation}
						where $\tilde{\mathcal{A}}$ is a generic generalised tensor.
						
						The explicit form of the twisted derivative is the same as the untwisted one, modulo the following replacements,
								\begin{equation}\label{repruleTwDorf}
									\begin{split}
										\dd \omega	&\rightarrow 	\dd \tilde{\omega} -\iota_{\tilde{v}} F \, , \\
										\dd\sigma 		&\rightarrow	\dd \tilde{\sigma} - \iota_{\tilde{v}} \tilde{F} + \tilde{\omega}\wedge F \, .
									\end{split}
								\end{equation}
						We collect in the~\cref{app:EGG} the other relevant objects and representation bundles for exceptional generalised geometry.
			\subsection{Type IIA}\label{sec:EGGIIA}
				The relevant generalised geometry for type IIA theories has been constructed in~\cite{oscar1}.
				The structure group for a $d$-dimensional manifold is $\E_{d+1(d+1)} \times \RR^+$. The generalised geometry for IIA can obtained by a reduction o
				f the M-theory one. 
				We give this construction explicitly in~\cref{app:EGG}, here we just collect the most important results and definitions.

				The generalised tangent bundle i$E$ is isomorphic to
							\begin{equation}\label{IIAtangbung}
								E \cong TM \oplus T^*M \oplus \Lambda^5 T^*M \oplus \Lambda^{\mathrm{even}} T^*M \oplus (T^*M \otimes \Lambda^6 T^*M)\, ,
							\end{equation}
and a generic section can be decomposed according to a $\GL(d,\RR)$ as 
							\begin{equation}\label{IIAgenV}
								V = v + \lambda + \sigma + \omega + \tau \, ,
							\end{equation}
where $v$ is a vector, $\lambda$ a one-form, $\sigma$ a five-form, $\omega$ a polyform in $\Gamma(\Lambda^{\mathrm{even}}T^*M)$ and $\tau \in \Gamma(T^*M \otimes \Lambda^6 T^*M)$.

				As in M-theory $E$ s defined by a series of extensions,
							\begin{equation}\label{IIAexten}
								\begin{tikzcd}[row sep=tiny]
										0 \arrow{r} & T^*M \arrow{r} &E''' \arrow{r} &TM \arrow{r} &0 	\, , \\
										0 \arrow{r} &\Lambda^{\mathrm{even}} T^*M \arrow{r} &E'' \arrow{r} &E''' \arrow{r} &0 	\, , \\
										0 \arrow{r} &\Lambda^5 T^*M \arrow{r} &E' \arrow{r} &E'' \arrow{r} &0 \, . \\
										0 \arrow{r} &T^*M \otimes \Lambda^6 T^*M \arrow{r} &E \arrow{r} &E' \arrow{r} &0 \, .
									\end{tikzcd}
							\end{equation}
In order to define the parching rules for potentials and generalised vectors, first we define the \emph{untwisted} section $\tilde{V} = \tilde{v} + \tilde{\lambda} + \tilde{\sigma} + \tilde{\omega} + \tilde{\tau}$ as
							\begin{equation}\label{eq:twistC_short}
								V = e^{\tilde{B}} e^{-B} e^C \cdot \tilde{V} \, ,
							\end{equation}
					where, as usual, $\cdot$ denotes the adjoint action of the structure group algebra.
					This twist concretely specifies the isomorphism~\eqref{IIAtangbung}. An explicit expansion of~\eqref{eq:twistC_short} can be written as,
							\begin{equation}\label{eq:twistC}
								\begin{split}
									v 		&= \tilde{v}\, , \\[1.5mm]
									\lambda 	&= \tilde{\lambda} + \iota_{\tilde{v}} B\, , \\[1.5mm]
									\sigma 	&= \tilde{\sigma} + \iota_{\tilde{v}} \tilde{B} - \left[s(C) \wedge \big(\tilde{\omega} + \tfrac{1}{2} \iota_{\tilde{v}}C + \tfrac{1}{2}\tilde{\lambda} \wedge C \big)\right]_5 \, , \\[1.5mm]
									\omega 	&= e^{-B}\wedge \big( \tilde{\omega} + \iota_{\tilde{v}}C + \tilde{\lambda} \wedge C \big) \, , \\[1.5mm]
									\tau 		&= \tilde{\tau} + j B\wedge \left[\tilde{\sigma} - s(C)\wedge \big(\tilde{\omega} + \tfrac{1}{2} \iota_{\tilde{v}}C + \tfrac{1}{2}\tilde{\lambda} \wedge C \big)\right]_5 
												+ j\tilde{B} \wedge (\tilde{\lambda} + \iota_{\tilde{v}}B) \\
											&\phantom{=\tilde{\tau}} - j s(C)\wedge\big(\tilde{\omega} + \tfrac{1}{2}\iota_{\tilde{v}}C + \tfrac{1}{2} \tilde{\lambda} \wedge C \big) \, ,
								\end{split}
							\end{equation}
					where $[\cdot]_k$ denotes the degree $k$-form of a polyform.

The patching rules for the generalised vector $V$ on the intersection $U_\alpha \cap U_\beta$ of two charts $U_\alpha$ and $U_\beta$, reads 
							\begin{equation}\label{patchingIIA}
								V_{(\alpha)} = e^{\dd \tilde{\Lambda}_{(\alpha\beta)}}e^{\dd \Omega_{(\alpha\beta)}} e^{\dd \Lambda_{(\alpha\beta)}} V_{(\beta)}\, ,
							\end{equation}
					where $\Lambda_{(\alpha \beta)}$ is a one-form, $\tilde\Lambda_{(\alpha \beta)}$ a five-form, and $\Omega_{(\alpha \beta)}$ a poly-form of even degree, all defined on $U_{\alpha} \cap U_{\beta}$.
					In the~\eqref{patchingIIA} we have dropped the transformations due to the $\GL(d,\RR)$ action.
Plugging~\eqref{eq:twistC_short} into~\eqref{patchingIIA} and reorganising the exponentials on the right hand side, one obtains the patching conditions for the adjoint fields (corresponding to gauge transformations of supergravity potentials),
							\begin{equation}\label{eq:gauge-field-patchingmassless}
								\begin{split}
									B_{(\alpha)} &= B_{(\beta)} + \dd \Lambda_{(\alpha\beta)} \, , \\
									C_{(\alpha)} &= C_{(\beta)} + e^{B_{(\beta)} +\dd \Lambda_{(\alpha\beta)}} \wedge \dd \Omega_{(\alpha\beta)} \, , \\
									\tilde{B}_{(\alpha)} &= \tilde{B}_{(\beta)} + \dd \tilde \Lambda_{(\alpha\beta)} + \tfrac{1}{2} \left[ \dd \Omega_{(\alpha\beta)} \wedge e^{B_{(\beta)} + \dd \Lambda_{(\alpha\beta)}}\wedge s(C_{(\beta)}) \right]_6 \, .
		 						\end{split}
							\end{equation}
					As we clarify in~\cref{appsec:EGGgauge}, these do indeed correspond to the finite supergravity gauge transformations between patches (here given for vanishing Romans mass, $m=0$). 
					As in the previous case, this construction generalises the standard definition of a gerbe connection.

					It is also useful to consider the decomposition under $\GL(d,\RR)$ of the adjoint bundle $\adj \subset E \otimes E^*$, 
							\begin{equation}\label{decomp_adjoint}
								\begin{split}
									\adj = 	\RR_\Delta \oplus& \RR_\phi \oplus (TM\otimes T^*M) \oplus \Lambda^2 TM \oplus \Lambda^2 T^*M \oplus \Lambda^6 TM \oplus \Lambda^6 T^*M \\
											& \oplus \Lambda^{\mathrm{odd}} TM \oplus \Lambda^{\text{odd}} T^*M \, .
								\end{split}
							\end{equation}
					Its sections $R$ can be written as 
							\begin{equation}\label{section_adjoint}
								R = l + \varphi + r + \beta + B + \tilde \beta + \tilde B + \Gamma + C \, ,
							\end{equation} 
where $r \in \mathrm{End}(T)$ will correspond to the $GL(d,\RR)$ action, while the scalars $l$ and $\varphi$ will be related to the shifts of the warp factor and dilaton, respectively. 
					The forms $B$, $\tilde B$ and $C = C_1 + C_3 + C_5$ will encode the internal components of the NSNS two-form, of its dual and of the RR potentials.
					The other elements are poly-vectors obtained by raising the indices of the forms, and do not have an immediate supergravity counterpart.

					To conclude this discussion, we want to raise an observation about the relation between the $\E_{d(d)} \times \RR^+$ generalised geometry and the $\rmO(d,d)$ one.
					Indeed, one can additionally view $E$ as an extension of Hitchin's generalised tangent space $E'$~\cite{hitch1,gualtphd} by $\rmO(d,d)\times\RR^+$ tensor bundles, as we describe in~\cref{appsec:EGGodd}.
					The Dorfman derivative for the massless type IIA can be obtained by the~\eqref{dorfM} by a dimensional reduction, or by analogy from~\eqref{dorfMadj}.
					Using an index $M$ to denote the components of a generalised vector $V$ in a standard coordinate basis, 
							\begin{equation}
								V^M = \left\{v^m , \lambda_m, \sigma_{m_1\ldots m_5}, \tau_{m,m_1\ldots m_6} , \omega, \omega_{m_1m_2}, \omega_{m_1\ldots m_4}, \omega_{m_1\ldots m_6}\right\} \, ,
							\end{equation}
					and embedding the standard derivative operator as a section of the dual generalised tangent bundle $E^*$, $\partial_M = (\partial_m, 0, \ldots,0)$, the Dorfman derivative is defined as~\cite{waldram4}
							\begin{equation}\label{eq:Liedefg}
								(L_V V')^M = V^N \partial_N V^{\prime M} - (\partial \times_{\rm ad} V)^M_{\phantom{M}N} V^{\prime N}\, , 
							\end{equation}
					where $ \times_{\mathrm{ad}}$ is again the projection onto the adjoint bundle.
					In this case, this explicitly gives 
							\begin{equation}
								\partial \times_{\ad} V = \partial \times v -\dd \lambda + \dd \sigma + \dd \omega\ .
							\end{equation}
					We recall this operator satisfies the Leibniz property~\eqref{eq:Leibniz} and is not antisymmetric.
					Hence, the generalised Lie derivative for massless type IIA in $\GL(d,\RR)$ decomposition reads
							\begin{equation}\label{dorfIIA}
								\begin{split}
									L_V V' =&\dd \mathcal{L}_v v' + \left(\mathcal{L}_v \lambda' - \iota_{v^\prime} \dd\lambda \right) + \left( \mathcal{L}_v \sigma' - \iota_{v^\prime}\dd\sigma + [s(\omega^\prime) \wedge \dd\omega]_5\right) \\
											& + \left(\mathcal{L}_v \tau' + j \sigma' \wedge \dd\lambda + \lambda^\prime \otimes \dd\sigma + j s(\omega^\prime) \wedge \dd\omega \right) \\
											& + \left(\mathcal{L}_v \omega' + \dd \lambda \wedge \omega' - (\iota_{v'}+ \lambda' \wedge)\dd\omega\right) \ .
								\end{split}
							\end{equation}
					This can also be written in terms of natural derivative operators in $\rmO(d,d)$ generalised geometry, see~\cref{appsec:EGGodd}.
					
					The action of the generalised Lie derivative on the untwisted bundle can also by defined.
					Let us denote $\mathbb{L}$ the \emph{twisted Dorfman derivative}\footnote{%
						When the generalised tangent bundle is untwisted, the Dorfman derivative is twisted, and vice-versa.%
						},
					defined as follows,
							\begin{equation}\label{eq:fromLtotildeL}
								\mathbb{L}_{\tilde{V}} \tilde{V}^\prime = e^{-C}e^{B}e^{-\tilde{B}} \cdot L_{V}V^\prime \, .
							\end{equation}
					This is completely analogous to the \emph{twisted Courant bracket} defined in~\eqref{HCour}.
					Operationally it is useful to get the twisted derivative by the untwisted one by replacing in the~\eqref{dorfIIA},
							\begin{equation}
								\begin{split}
									\dd \tilde{\lambda} 	& \longrightarrow \dd \tilde{\lambda} - \iota_{\tilde{v}} H \, , \\
									\dd \tilde{\sigma} 	& \longrightarrow \dd \tilde{\sigma} - \left[ s(\tilde{\omega}) \wedge F \right]_6 \, , \\
									\dd \tilde{\omega}	& \longrightarrow \dd_H \tilde{\omega} - (\iota_{\tilde{v}} + \tilde{\lambda}\wedge ) F \, .
								\end{split}
							\end{equation}
					where $H$ is the three-form $H = \dd B$ on $M$, $F = F_2 + F_4 + F_6$ is a polyform made out of field strengths of the potentials $C$ in~\eqref{section_adjoint}, and $\dd_H$ is the twisted exterior derivative defined by $\dd_H = \dd - H \wedge$.
					These field strength forms transform in a generalised bundle which is a the $\mathbf{912}_{-1}$ irreducible representation of $\E_{7(7)} \times \RR^+$~\cite{Grana:2009im}.
					
					
					In view of the application of this formalism to supergravity theories, it is useful to stress again that the Dorfman derivative generates the infinitesimal generalised diffeomorphisms on the internal manifold $M$. 
					Interpreting a generalised vector $V$ as a gauge parameter, the infinitesimal gauge transformation of any field is given by
							\begin{equation*}
								\delta_V = L_V \, .
							\end{equation*}
					The Leibniz property~\eqref{eq:Leibniz} then just expresses the gauge algebra $[\delta_V, \delta_{V'}] = \delta_{L_V V'}$.

					%

						We want now to see how to include in the formalism the Romans mass. 	
						In a string theory perspective this corresponds to a D$8$ brane filling the ten-dimensional spacetime.
						One of the main results of my work is the description of a generalised geometry for type IIA accommodating also the Romans mass.
						
						We have seen in the previous chapter how exceptional generalised geometry can accomodate all the fluxes of type II supergravity and M-theory by twisting the exceptional tangent bundle by their potentials.
The difficulty in incorporating the mass $m$ in the generalised geometry formalism is that the zero-form flux $m = F_0$ is not expressible as the derivative of a potential.
						This means that it is not possible to introduce the mass term as an additional twist of the generalised bundle $E$, as for the other fluxes.
						
						The key point in solving this problem is to look at the way the gauge transformations of the NSNS and RR potentials are realised in exceptional generalised geometry.
						
						We saw in~\eqref{gaugeshifts} how the mass affects the gauge transformations of type IIA supergravity.
						Since the gauge transformations of the supergravity potentials are encoded in the way the twisted generalised vectors patch, the introduction of the Romans mass requires a modification of the patching conditions~\eqref{patchingIIA} and~\eqref{eq:gauge-field-patchingmassless}.
						Following a reasoning that schematically derives the patching conditions from gauge transformations (see~\cref{appsec:EGGgauge} for a more detailed discussion), we find the new patching conditions of the form
								\begin{equation}\label{patching_m}
									V_{(\alpha)} = e^{\dd \tilde \Lambda_{(\alpha \beta)}} e^{\dd \Omega_{(\alpha \beta)} + m \Omega_{6(\alpha\beta)} } e^{-\dd \Lambda_{(\alpha \beta)} - m\,\Lambda_{(\alpha\beta)}} \cdot V_{(\beta)} \, .
								\end{equation}
						This condition reproduces the massive supergravity gauge transformations on overlapping patches $U_\alpha \cap U_\beta$.
						A first-principles derivation of this is also given in appendix~\ref{appsec:EGGgauge}.
						
						Although the structure of the exact sequences~\eqref{IIAexten} is left intact by this deformation, the precise details of the twisting~\eqref{patching_m} do change\footnote{%
							A consequence of this is the following. In massless IIA we can project a generalised vector onto its vector and zero-form parts $v+\omega_0$, giving a well-defined section of a bundle with seven-dimensional fibre. This is the dimensional reduction of the M-theory tangent bundle $TM_7$. However, with the massive IIA patching rules~\eqref{patching_m}, this projection would no longer give a section of a bundle with seven-dimensional fibre. Hence, the massive patching rules do not arise from a seven-dimensional geometry.%
							}.
 						An important feature of massive type IIA is that by virtue of the Bianchi identity we have (globally)
								\begin{equation}\label{eq:H-exact}
									H_3 = \tfrac{1}{m}\, \dd F_2 \, ,
								\end{equation}
						so that for $m\neq 0$, $H_3$ is trivial in cohomology.
						Thus, the first extension in~\eqref{IIAexten} is naturally equivalent to the trivial one.
						
						Also, a pure NSNS gauge transformation no longer acts in the $\rmO(d,d)$ subgroup of $	E_{d+1(d+1)}\times\RR^+$, simply because it also generates a $C_1$ RR potential.
						As such, there is no massive version of Hitchin's $\rmO(d,d)$ generalised geometry\footnote{%
							Though see~\cite{Hohm:2011cp} for a double field theory approach to this, where the $F_0$ flux is generated by introducing a linear dependence on the additional non-geometric coordinates dual to the winding modes of the string.%
							}.

						The modification~\eqref{patching_m} of the patching condition also requires a deformation of the Dorfman derivative. 
						Recall that the latter generates the infinitesimal gauge transformations, and that these are affected by the Romans mass via the shifts~\eqref{gaugeshifts}. 
						It follows that the massive form of the Dorfman derivative is obtained implementing the same shift in the massless expression~\eqref{dorfIIA}
								\begin{equation}\label{dorfIIAm}
									\begin{split}
										L_V V' =& \mathcal{L}_v v' + \left(\mathcal{L}_v \lambda' - \iota_{v^\prime} \dd\lambda\right) \\
												& + \left( \mathcal{L}_v \sigma' - \iota_{v^\prime}(\dd\sigma+m\omega_6) + [s(\omega^\prime) \wedge (\dd\omega-m\lambda)]_5\right) \\
												& + \left(\mathcal{L}_v \tau' + j \sigma' \wedge \dd\lambda + \lambda^\prime \otimes (\dd\sigma+m\omega_6) + j s(\omega^\prime) \wedge (\dd\omega-m\lambda) \right) \\
												& + \left(\mathcal{L}_v \omega' + \dd \lambda \wedge \omega' - (\iota_{v'}+ \lambda' \wedge)(\dd\omega-m\lambda)\right)\, ,
									\end{split}
								\end{equation}
						which contains the mass as a deformation parameter. 
						
						More formally,~\eqref{dorfIIAm} is related to the massless Dorfman derivative (here denoted by $L^{(m=0)}$) as
								\begin{equation}\label{mdefDorf}
									L_V V' = L^{(m=0)}_V V' + \underline{m}(V) \cdot V'\,,
								\end{equation}
						where, given a generalised vector $V$, we define the map $\underline{m}$ such that
								\begin{equation}
									\underline{m}(V) = m\lambda - m \omega_6 \, 
								\end{equation}
						is an object that acts in the adjoint of $\E_{7(7)}$ (see~\eqref{IIAadjvecCompact}) as
								\begin{equation}\label{madj}
									\begin{split}
										\underline{m}(V)\cdot V' = m \left( - \iota_{v'} \omega_6 - \lambda \wedge \omega_4' + \lambda' \otimes \omega_6 - \lambda \otimes \omega'_6 + \iota_{v'} \lambda + \lambda' \wedge \lambda\right)\,.
									\end{split}
								\end{equation}
						It is a tedious but straightforward computation to verify that~\eqref{dorfIIAm} satisfies the Leibniz property~\eqref{eq:Leibniz}\footnote{%
							A very subtle point is that neither of the terms on the RHS of~\eqref{mdefDorf} transforms correctly as a generalised vector under~\eqref{patching_m}, and as a consequence $\underline{m}(V)$ does not transform as a section of the adjoint bundle. However, overall $L_V V'$ defines a good section of $E$.%
							}.

						To justify further our definition, we rewrite the massive Dorfman derivative in the untwisted picture.
						Using~\eqref{eq:fromLtotildeL} we find 
								\begin{equation}\label{eq:twistedDorfmanmassive}
									\begin{split}
										\mathbb{L}_{\tilde V} {\tilde V'} = & \mathcal{L}_{\tilde v} \tilde v' + (\mathcal{L}_{\tilde v}\tilde \lambda' - \iota_{\tilde v^\prime} \dd\tilde \lambda + \iota_{\tilde v'} \iota_{\tilde v} H ) \\
																& + \mathcal{L}_{\tilde v} \tilde \sigma' - \iota_{\tilde v^\prime}\dd\tilde \sigma \\
																 & \phantom{\mathcal{L}_{\tilde v} \tilde \sigma'} + \left[ \iota_{\tilde v'}(s(\tilde\omega) \wedge F)+ s(\tilde\omega^\prime) \wedge \big(\dd \tilde\omega - H\wedge \tilde\omega - (\iota_{\tilde v}+ \tilde \lambda \wedge)F\big) \right]_5 \\
																& + \mathcal{L}_{\tilde v} \tilde\tau' + j \tilde\sigma' \wedge (\dd \tilde\lambda - \iota_{\tilde v}H) + \tilde\lambda' \otimes \big( \dd \tilde\sigma - [s(\tilde \omega)\wedge F]_6 \big) \\
																&\phantom{\mathcal{L}_{\tilde v} \tilde\tau'} + j s(\tilde\omega') \wedge \big( \dd \tilde \omega - H\wedge\tilde \omega - (\iota_{\tilde v} + \tilde \lambda\wedge)F \big) \\ 
																& + \mathcal{L}_{\tilde v}\tilde\omega' + (\dd\tilde \lambda - \iota_{\tilde v} H) \wedge\tilde\omega' \\
																& \phantom{\mathcal{L}_{\tilde v}\tilde\omega'} - (\iota_{\tilde v^\prime}+ \tilde \lambda' \wedge) \big(\dd\tilde \omega - H \wedge\tilde \omega -(\iota_{\tilde v} +\tilde\lambda\wedge)F \big)\ ,
									\end{split}
								\end{equation}
						where $F = F_0+F_2 +F_4+F_6$ is now the complete $O(6,6)$ spinor with $m\neq 0$. 
						So the twisted version of the massive Dorfman derivative produces precisely the expected gauge transformations with flux terms including the Romans mass. 
						Again, these are given by the action of~\eqref{gaugetrans}, now with $m\neq 0$. 

						Note that of all the flux terms in~\eqref{eq:twistedDorfmanmassive}, the mass term is the only one which is diffeomorphism-invariant. 
						It is also the only true deformation of the generalised Lie derivative, since it cannot be removed by twisting the generalised tangent bundle.

			\subsection{Type IIB}
				For the review of this part we will closely follow~\cite{AshmoreECY}.
For type IIB the structure group of the principal frame bundle is the same as the type IIA case, $\E_{d+1(d+1)}\times \RR^+$ for a $d$-dimensional manifold $M$~\cite{waldram5, spheres}.
On a $d$-dimensional manifold $M$, the generalised tangent bundle is
								\begin{equation}\label{IIBtanbund}
									\begin{split}
										E & \cong TM\oplus T^{*}M\oplus(T^{*}M\oplus\Lambda^{3}T^{*}M\oplus\Lambda^{5}T^{*}M)\oplus\Lambda^{5}T^{*}M \\
												& \phantom{\cong \oplus} \oplus(T^{*}M\otimes\Lambda^{6}T^{*}M) \\
										 & \cong TM\oplus(T^{*}M\otimes S)\oplus\Lambda^{3}T^{*}M\oplus(\Lambda^{5}T^{*}M\otimes S)\oplus(T^{*}M\otimes\Lambda^{6}T^{*}M) \, ,
									\end{split}
								\end{equation}
						where, as usual, $E$ is defined formally by an extension and it is isomorphic to the sum of spaces in~\eqref{IIBtanbund} by choosing the potential maps, \emph{i.e.} the connective structures on the gerbe.
						In the ~\eqref{IIBtanbund} the $S$ transforms as a doublet of $\SL(2,\RR)$. 
						We write sections of this bundle as
								\begin{equation}\label{eq:V-IIB}
									V=v+\lambda^{i}+\rho+\sigma^{i}+\tau \, ,
								\end{equation}
						where $v\in\Gamma(TM)$, $\lambda^{i}\in\Gamma(T^{*}M\otimes S)$, $\rho\in\Gamma(\Lambda^{3}T^{*}M)$, $\sigma\in\Gamma(\Lambda^{5}T^{*}M\otimes S)$ and $\tau\in\Gamma(T^{*}M\otimes\Lambda^{6}T^{*}M)$, the index $i =1, 2$ is the one labelling the fundamental representation of $\SL(2,\RR)$.
						
As before, the patching conditions reproduce the type IIB supergravity gauge transformations.
Given a cover $\{U_{\alpha}\}$ of $M$ on can define the generalised section $V_{(\alpha)}$ on $U_\alpha \cap U_\beta$ by the $V_{(\beta)}$ as follows,
								\begin{equation}
									V_{(\alpha)} = e^{\dd \Lambda^i_{(\alpha\beta)} + \dd \Omega_{(\alpha\beta)}} \cdot V_{(\beta)}\, ,
								\end{equation}
				which $\cdot$ denoting the adjoint action and $\Lambda^{(i)}$ and $\Omega$ are locally a pair of one-forms and a three-form respectively.		
By defining the untwisted vector 	
\begin{equation}\label{twistVecIIB}
									V = e^{-B^i - C} \tilde{V} \, .
								\end{equation}		
and comparing the two actions, we find 
\begin{equation}
									\begin{split}
										B^{i}_{(\alpha)} 	& = B^{i}_{(\beta)} + \dd \Lambda^{(i)}_{(\alpha \beta)} \, , \\
										C_{(\alpha)}	& = C_{(\beta)} + \dd \Omega_{(\alpha\beta)} + \frac{1}{2} \epsilon_{ij} \dd \Lambda^i_{(\alpha \beta)} \wedge B^i_{(\beta)} \, .\
									\end{split}
								\end{equation}

						The adjoint bundle is
								\begin{equation}
									\begin{split}
										\adj \tilde{F} = & \RR \oplus(TM\otimes T^{*}M) \oplus(S\otimes S^{*})_0 \oplus(S\otimes\Lambda^{2}TM) \oplus(S\otimes\Lambda^{2}T^{*}M) \\
													& \oplus\Lambda^{4}TM\oplus\Lambda^{4}T^{*}M \oplus(S \otimes\Lambda^{6}TM)\oplus(S\otimes\Lambda^{6}T^{*}M) \, ,
									\end{split}
								\end{equation}
						where the subscript on $(S\otimes S^*)_0$ denotes the traceless part. 
						We write sections of the adjoint bundle as
								\begin{equation}\label{eq:IIB-adj}
									R=l+r+a+\beta^{i}+B^{i}+\gamma+C+\tilde{\alpha}^{i}+\tilde{a}^{i} \, ,
								\end{equation}
						where $l\in\mathbb{R}$, $r\in\Gamma(\mathrm{End}(TM))$, etc. 
%

						%
					%

				The action of the Dorfman derivative on a generalised vector in type IIB is 
						
							%
								\begin{equation}\label{eq:IIB_Dorf_vector}
									\begin{split}
										L_{V}V^{\prime} =& \mathcal{L}_{v}v^{\prime}+(\mathcal{L}_{v}\lambda^{\prime i}-\imath_{v^{\prime}}\dd\lambda^{i})+(\mathcal{L}_{v}\rho^{\prime}-\imath_{v^{\prime}}\dd\rho+\epsilon_{ij}\dd\lambda^{i}\wedge\lambda^{\prime j}) \\
 														& +(\mathcal{L}_{v}\sigma^{\prime i}-\imath_{v^{\prime}}\dd\sigma^{i}+\dd\rho\wedge\lambda^{\prime i}-\dd\lambda^{i}\wedge\rho^{\prime})\\
 														& +(\mathcal{L}_{v}\tau^{\prime}-\epsilon_{ij}j\lambda^{\prime i}\wedge\dd\sigma^{j}+j\rho^{\prime}\wedge\dd\rho+\epsilon_{ij}j\sigma^{\prime i}\wedge\dd\lambda^{j})\, .
									\end{split}
								\end{equation}
						The expression for Dorfman derivative acting on a section of the adjoint bundle is given in the appendix~\ref{appsec:EGGIIB}.
					
						Also here one can give the \emph{twisted} Dorfman derivative $\mathbb{L}_{V}$ of an untwisted generalised tensor $\tilde{\mathcal{A}}$, defined by
								\begin{equation}\label{eq:IIB_twisted_dorf}
									\mathbb{L}_{\tilde{V}}\tilde{\mathcal{A}}=e^{-B^{i}-C}L_{e^{B^{i}+C}\tilde{V}}(e^{B^{i}+C}\tilde{\mathcal{A}}).
								\end{equation}
						The twisted Dorfman derivative $\mathbb{L}_{V}$ is given by the same expression as the usual Dorfman derivative but with the substitutions
								\begin{equation}
									\begin{split}
										\dd\lambda^{i}	&\rightarrow	\dd\tilde{\lambda}^{i}-\imath_{\tilde{v}}F^{i} \, , \\
										\dd\rho		&\rightarrow	\dd\tilde{\rho}-\imath_{\tilde{v}}F-\epsilon_{ij}\tilde{\lambda}^{i}\wedge F^{j} \, , \\
										\dd\sigma^{i}	&\rightarrow	\dd\tilde{\sigma}^{i}+\tilde{\lambda}^{i}\wedge F-\tilde{\rho}\wedge F^{i} \, .
									\end{split}
								\end{equation}
						We collect further details about the exceptional generalised geometry in~\cref{app:EGG}.
			\subsection{Generalised metric}\label{gen_frame_metric}
					In this section we briefly review some examples of generalised structures.
					For concreteness we restrict ourself to type IIA, however we refer to~\cite{waldram4, AshmoreECY} for a detailed discussions of the other cases.
					
					In the same way as the ordinary metric on a manifold $M$ can be seen as an $\rmO(d)$ structure on $T M$ parameterising the coset $\GL(d,\RR)/\rmO(d)$, the generalised metric can be seen as an $\SU(8) / \mathbb{Z}_2$ structure on the generalised tangent bundle, and for a six-dimensional manifold it parameterises the coset $\E_{7(7)}/(\SU(8) / \mathbb{Z}_2)$. 
					The construction of the generalised metric is a natural extension of the one we have described for the $\rmO(d,d)$ case in~\cref{genmetrOdd}.
					
					The generalised metric $\mathcal{G}$ can be defined by its action on two generalised vectors $V$ and $V'$ as
							\begin{equation}\label{GofVandV'}
								\begin{split}
									\mathcal{G}(V,V')&= \tilde{v} \lrcorner \tilde{v}' + \tilde{\lambda} \lrcorner \tilde{\lambda}' + \tilde{\sigma} \lrcorner \tilde{\sigma}' + \tilde{\tau} \lrcorner \tilde{\tau}'+ \sum_{k=0}^3\tilde{\omega}_{2k} \lrcorner \tilde{\omega}'_{2k} \\
										&= \tilde{v}^m \tilde{v}'_m + \tilde{\lambda}^m \tilde{\lambda}'_m + \tfrac{1}{5!}\tilde{\sigma}^{m_1\ldots m_5} \tilde{\sigma}'_{m_1\ldots m_5} + \tfrac{1}{6!}\tilde{\tau}^{m,m_1\ldots m_6} \tilde{\tau}'_{m,m_1\ldots m_6} \\
											& \phantom{= +} + \sum_{k=0}^3\tfrac{1}{(2k)!}\,\tilde{\omega}^{m_1 \ldots m_{2k}} \tilde{\omega}'_{m_1 \ldots m_{2k}} \, ,
								\end{split}
							\end{equation}
					where the indices are lowered/raised using the ordinary metric $g_{mn}$ and its inverse $g^{mn}$.
					
					One can also define a generalised frame $\{\hat{E}_A\}$ on $E$ and then construct the inverse generalised metric as the tensor product of two such frames
							\begin{equation}\label{genmetfor}
								\mathcal{G}^{-1} = \delta^{AB} \hat{E}_A \otimes \hat{E}_B \,. 
							\end{equation}
					We will give below a precise definition of this product. 
					To construct the generalised frame, we first consider the \emph{untwisted} generalised tangent bundle $\tilde{E}$, in an analogous way to the construction in $\rmO(d,d)$ geometry.
					Let $\hat e_a$, with $a=1,\ldots,6$, be an ordinary frame, namely a basis for the tangent space at a point of $M_6$, and let $e^a$ be the dual basis for the cotangent space\footnote{%
						We are using the hat symbol to distinguish frame vectors, $\hat{e}_a$, from co-frame one-forms, $e^a$. 
						Similarly, the hat on $\hat E_A$ indicates that this is a generalised frame vector.%
						}.
					Then we can define a frame $\tilde{\hat E}_A$ for the untwisted generalised tangent space as the collection of bases for the subspaces that compose it
							\begin{equation}\label{untframe}
								\begin{split}
									\{ \tilde{\hat{E}}_A\} = \{\hat e_a\} &\cup \{e^a\} \cup \{e^{a_1 \ldots a_5} \} \cup \{e^{a,a_1\ldots a_6}\} \cup \{1\} \\
													&\cup \{e^{a_1a_2}\} \cup \{e^{a_1\ldots a_4}\} \cup \{e^{a_1 \ldots a_6}\}\, ,
								\end{split}
							\end{equation}
					where $e^{a_1\ldots a_p}= e^{a_1}\wedge \cdots \wedge e^{a_p}$ and $e^{a,a_1\ldots a_6} = e^a\otimes e^{a_1\ldots a_6}$. 
					A frame for the \emph{twisted} generalised tangent space is obtained by twisting~\eqref{untframe} by the local $\E_{7(7)}\times \RR^+$ transformation
							\begin{equation}\label{twist_splitfr}
								\hat{E}_A = e^{\tilde{B}}e^{-B}e^{C}e^{\Delta}e^{\phi}\cdot \tilde{\hat{E}}_A \, ,
							\end{equation}
					where in addition to the twist~\eqref{eq:twistC_short} we also include a rescaling by the dilaton $\phi$ and warp factor $\Delta$, acting as specified in~\eqref{IIAadjvecCompact}. 
					Because of the rescaling by $\Delta$ the frame~\eqref{twist_splitfr} was called \emph{conformal split frame} in~\cite{Coimbra:2011ky}. 
					Note that~\eqref{twist_splitfr} is just a particular choice of frame, not the most general one. 
					Any other frame can be obtained from~\eqref{twist_splitfr} acting with an $\E_{7(7)} \times \RR^+$ transformation. 

					We denote the components of $\hat{E}_A$ carrying different flat indices as
							\begin{equation*}
								\{\hat E_A\} = \{\hat{\mathcal{E}}_a\} \cup \{\mathcal{E}^a\} \cup \{\mathcal{E}^{a_1 \ldots a_5} \} \cup \{\mathcal{E}^{a,a_1\ldots a_6}\} \cup \{\mathcal{E}\} \cup \{\mathcal{E}^{ab}\} \cup \{\mathcal{E}^{a_1\ldots a_4}\} \cup \{\mathcal{E}^{a_1 \ldots a_6}\}\, . 
							\end{equation*} 
					Explicit expressions for each of these terms are given in appendix~\ref{splitfr_MtoIIA}.

					Once we have the generalised frame, we can derive the expression for the inverse generalised metric $\mathcal{G}^{-1}$. 
					Expanded in $\GL(6)$ components, the product~\eqref{genmetfor} becomes
							\begin{equation*}
								\begin{split}
								\mathcal{G}^{-1} =& \delta^{aa'} \hat{\mathcal{E}}_a \otimes \hat{\mathcal{E}}_{a'} + \delta_{aa'} \mathcal{E}^a \otimes \mathcal{E}^{a'} + \mathcal{E} \otimes \mathcal{E} + \tfrac{1}{2}\delta_{a_1a'_1}\delta_{a_2a'_2} \mathcal{E}^{a_1a_2} \otimes \mathcal{E}^{a'_1a'_2} \\
												& + \tfrac{1}{4!}\delta_{a_1a'_1}\cdots\delta_{a_4a'_4} \mathcal{E}^{a_1\ldots a_4} \otimes \mathcal{E}^{a'_1\ldots a'_4} + \tfrac{1}{5!}\delta_{a_1a'_1}\cdots\delta_{a_5a'_5} \mathcal{E}^{a_1\ldots a_5} \otimes \mathcal{E}^{a'_1\ldots a'_5} \\
												& + \tfrac{1}{6!}\delta_{a_1a'_1}\cdots\delta_{a_6 a'_6} \mathcal{E}^{a_1\ldots a_6} \otimes \mathcal{E}^{a'_1\ldots a'_6} + \tfrac{1}{6!}\delta_{a a'}\delta_{a_1a'_1}\cdots\delta_{a_6a'_6} \mathcal{E}^{a,a_1\ldots a_6} \otimes \mathcal{E}^{a',a'_1\ldots a'_6}\, .
								\end{split}
							\end{equation*}
					which is nothing else than the inverse of~\eqref{GofVandV'} calculated on frames.
					The full expression for $\mathcal{G}^{-1}$ is long and ugly, so we only give the terms that will be relevant for the next discussion. 
					Arranging them according to their curved index structure, we have
							\begin{equation}\label{invG_comp_1}
								\begin{split}
									(\mathcal{G}^{-1})^{mn} &= e^{2\Delta}g^{mn}\, ,\\[1mm]
									(\mathcal{G}^{-1})^{m} &= e^{2\Delta}g^{mn} C_n \, ,\\[1mm]
									(\mathcal{G}^{-1})^{m}_{\phantom{m}n} &= - e^{2\Delta}g^{mp}B_{pn}\, , \\[1mm]
									(\mathcal{G}^{-1})^{m}_{\phantom{m}np} &= e^{2\Delta}g^{mq}\left(C_{qnp} - C_{q}B_{np} \right) \, , \\[1mm]
									(\mathcal{G}^{-1})^m{}_{npqr} &= {e}^{2\Delta}g^{ms}\left(C_{snpqr} -C_{s[np}B_{qr]} + \tfrac{1}{2}C_s B_{[np}B_{qr]}\right) \, , \\[1mm]
									(\mathcal{G}^{-1}) &= e^{2\Delta}\left( e^{-2\phi} + g^{mn}C_m C_n\right)\, .
								\end{split}
							\end{equation}
					These terms will be sufficient to read off all the supergravity physical fields from the generalised metric (we are omitting the formula determining $\tilde B_{m_1\ldots m_6}$).
					Some other components of $\mathcal{G}^{-1}$ are
							\begin{equation}\label{invG_comp_2}
								\begin{split}
									(\mathcal{G}^{-1})_{m} &= e^{2\Delta}g^{np}C_n B_{pm} \, , \\[1mm]
									(\mathcal{G}^{-1})_{(mn)} &= e^{2\Delta}\left( g_{mn} + g^{pq}B_{pm}B_{qn}\right)\, , \\[1mm]
									(\mathcal{G}^{-1})_{[mn]} &= - e^{2\Delta}\left( e^{-2\phi}B_{mn} - g^{pq}C_q\left( C_{pmn} - C_{p}B_{mn} \right) \right) \, , \\[1mm]
									(\mathcal{G}^{-1})_{m,np} &= -e^{2\Delta}\left(g_{m[n}C_{p]} + g^{qr}B_{qm}\left( C_{rnp} - C_{r}B_{np} \right) \right) \, .
								\end{split}
							\end{equation}

					There is also a density associated to the generalised metric which trivialises the $\RR^+$ factor of the $\E_{d+1(d+1)} \times \RR^+$ structure group. 
					In terms of the field content of type IIA it is given by
							\begin{equation}\label{gen_density}
								\Phi = (\det \mathcal{G})^{-(9-d)/ (\mathrm{dim} E)} = {g}^{1/2}\, e^{-2\phi} e^{(8-d)\Delta}\,,
							\end{equation}
					as can be seen by decomposing the corresponding M-theory density~\cite{Coimbra:2011ky}. 
					This equation provides an easy way to solve relations such as~\eqref{invG_comp_1} explicitly for the supergravity fields. For example, to solve the first, second and last of equations in~\eqref{invG_comp_1}, one can begin by setting
							\begin{equation}\label{fields_from_G_first}
								(M^{-1})^{mn} := (\mathcal{G}^{-1})^{mn} = e^{2\Delta}g^{mn}\,.
							\end{equation}
					The second of equations~\eqref{invG_comp_1} then becomes
							\begin{equation}
								C_m = M_{mn} (\mathcal{G}^{-1})^{n}\,,
							\end{equation}
					which can be substituted into the last equation in~\eqref{invG_comp_1} to give
							\begin{equation}
								e^{2\Delta} e^{-2\phi} = (\mathcal{G}^{-1}) - M_{mn} (\mathcal{G}^{-1})^{m} (\mathcal{G}^{-1})^{n} := Q\,.
							\end{equation}
					One then easily obtains the expressions for $g_{mn}, C_m, e^\Delta$ and $e^{-2\phi}$ as
							\begin{equation}\label{fields_from_G_last}
								\begin{array}{l c c r}
									e^\Delta = \bigg( \frac{\Phi}{Q \sqrt{\det M}} \bigg)^{1/6} \, ,		&	&	&	e^{-2\phi} = \bigg( \frac{Q^4 \sqrt{\det M}}{\Phi} \bigg)^{1/3} \, , \\
									g_{mn} = M_{mn} \bigg( \frac{\Phi}{Q \sqrt{\det M}} \bigg)^{1/3}\,, 	&	&	&	C_m = M_{mn} (\mathcal{G}^{-1})^n \, ,
								\end{array}
							\end{equation}
					where $M_{mn}, Q$ and $\Phi$ are given in terms of the generalised metric as above. 
					In particular, we have expressions for $e^\Delta$ and $g_{mn}$, so that solving the remaining relations in~\eqref{invG_comp_1} becomes straightforward.

					The above method to compute the warp factor from an arbitrary generalised metric involves evaluating $\det \mathcal{G}$, which is in general a slightly difficult computation. 
					A simpler way to attain the same result is to evaluate the determinant of a subset of the components of the generalised metric, denoted $\mathcal{H}$, corresponding to the degrees of freedom in the coset
							\begin{equation}
								\mathcal{H} \in \frac{\SO(d,d)\times\RR^+}{\SO(d)\times\SO(d)}\,.
							\end{equation}
					Explicitly, we construct $\mathcal{H}^{-1}$ in components via
							\begin{equation}\label{eq:GB-metric}
									\mathcal{H}^{-1} = \begin{pmatrix} (\mathcal{G}^{-1})^{mn} & (\mathcal{G}^{-1})^m{}_n \\
		 														(\mathcal{G}^{-1})_m{}^n & (\mathcal{G}^{-1})_{mn} \end{pmatrix}
		 										= e^{2\Delta} \begin{pmatrix} g^{mn} & -(g^{-1} B)^m{}_n \\
		 														(B g^{-1})_m{}^n & (g - B g^{-1} B)_{mn} \end{pmatrix}
							\end{equation}
					where in the second equality we have used~\eqref{invG_comp_1} and~\eqref{invG_comp_2}. 
					We recognise the last matrix as the components of (the inverse of) the $O(d,d)$ generalised metric~\eqref{genmet}, which has unit determinant~\cite{gualtphd}. 
					Therefore we can immediately write
							\begin{equation}
								e^{\Delta} = (\det \mathcal{H})^{-1/4d} \, .
							\end{equation}
					We comment on the appearance of the $\rmO(d,d)$ generalised metric in appendix~\ref{appsec:EGGodd}.
			\subsection{Generalised parallelisation}\label{secGenPar}
					The goal of this section is to extend the definition of identity structure given in the previous chapter to the exceptional case.
					This was firstly defined in~\cite{spheres}, and extended in~\cite{oscar1}.
					
					Namely, a generalised parallelisation $\{\hat{E}_A\}$ ($A = 1, \ldots , N$, where $N$ is the dimension of the generalised bundle) is a globally defined frame, or a set of $N$ globally defined vector fields defining a basis of $E\vert_p$ at each point $p$ of $M$.
The latter is a topological condition.
					In addition one can add a differential constraint on the frame $\{\hat{E}_A\}$, 
							\begin{equation}\label{GLP}
								L_{\hat{E}_A} \hat{E}_B = X_{AB}^{\phantom{AB}C} \hat{E}_C \, ,
							\end{equation}
					with constant coefficients $X_{AB}^{\phantom{AB}C}$.
					Following~\cite{spheres}, we call this a \emph{generalised Leibniz parallelisation}.
					The name comes from the fact that since the Dorfman derivative is not antisymmetric, the frame algebra~\eqref{GLP} is not a Lie algebra, but in general a Leibniz one.
					
					In~\cite{petrini3} using $\rmO(d,d)$ generalised geometry, it is proven that a necessary condition to admit a generalised Leibniz parallelisation is to be an homogeneous space, that is a coset space in the form $G/H$.
					This result is extendable to the full $\E_{d(d)}$ bundle.
					
%
				%
			\subsection{Generalised HV structures}\label{sec:ESE}
					Here we review briefly the results of~\cite{AshmoreESE}, in order to describe the so-called HV structures as generalised $G$-structures on the exceptional frame bundle.
					These will describe AdS backgrounds of various supergravity theories.
					
					As for ordinary $G$-structures, the existence of globally defined generalised tensors reduces the structure group of $E$ and defines generalised $G$-structures in exceptional geometry.
					
					As shown in~\cite{AshmoreESE, AshmoreECY, Grana_Ntokos}, a supergravity solution with eight supercharges is characterised by the existence of the so-called \emph{hyper}- and \emph{vector-multiplet}structures, defining the relative generalised $G$-structure.
					
					A \emph{hypermultiplet structure}~\cite{AshmoreECY}, or \emph{H structure} for short, is a triplet of sections of the weighted adjoint bundle (one can see~\cref{app:EGG}) for details) 
							\begin{equation}
								J_a \in \Gamma (\mathrm{ad} \tilde{F} \otimes (\det T^*M)^{1/2})\, , 
							\end{equation}
					such that
							\begin{equation}
								\begin{array}{ccccc}
									\com{J_a}{J_{b}} = 2 \kappa \epsilon_{abc} J_{c}& & \mbox{and} & & \tr \left( J_{a} J_{b}\right) = - \kappa^2 \delta_{ab}\, .
								\end{array}
							\end{equation}
					In the cases we are going to be interested in, \emph{i.e} $d = 6, 7$, the triplet $J_{a}$ realises a $\mathrm{Spin}^*(12) \subset \E_{7(7)} \times \RR^+$ structure and an $\SU^*(6) \subset \E_{6(6)} \times \RR^+$ respectively.
	
					A \emph{vector structure}, or \emph{V structure}, is given by a generalised vector
							\begin{equation} 
								K \in \Gamma \left( E\right) 
							\end{equation}
					that has positive norm 
							\begin{equation}
								\begin{array}{ccc}
									q(K) >0 & \mbox{or} & c(K)>0\, , 
								\end{array}
							\end{equation}
					where $q(K)$ denotes the $\E_{7(7)}$ quartic and $c(K)$ is the $\E_{6(6)}$ cubic invariant. 
					The generalised vector $K$ defines an $\E_{6(2)}$ and $\mathrm{F}_{4(4)}$ structure in $D=4$ and $D=5$, respectively.

					One can impose the following compatibility conditions on $J_a$ and $K$
							\begin{equation}\label{compHV}
								\begin{array}{ccc}
									J_{a} \cdot K=0 & \mbox{and}& \mathrm{tr}\left( J_{a}J_{b}\right) =
									\begin{cases}
										-2 \sqrt{q(K)} \delta_{ab}&\phantom{\mbox{for}} D=4\\
										-c(K) \delta_{ab}&\phantom{\mbox{for}} D=5 
									\end{cases}
								\end{array}
							\end{equation}
					The pair $(J_{a}, K)$ is then called an \emph{HV structure} and defines an $\SU(6) = \mathrm{Spin}^*(12) \cap \E_{6(2)}$ structure and a $\USp(6) = \SU^*(6) \cap F_{4(4)} $ structure in $D=4$ and $D=5$, respectively (see~\cite{AshmoreECY}). 
					The explicit form of the \emph{HV} structure depends on the theory and the dimension of the compactification manifold. 
					For instance, the generalised vector $K$ is 
							\begin{equation}\label{genvecs}
								K = \left\{ \begin{array}{lcl} 
										\xi + \omega + \sigma + \tau & \phantom{\mbox{for}} & \mbox{M theory} \\[1mm]
										\xi + \lambda^i + \rho + \sigma^i + \tau & \phantom{\mbox{for}} & \mbox{type IIB}
									\end{array} \right. \, .
							\end{equation}

					As discussed in~\cite{AshmoreECY, AshmoreESE}, and analogously to other $G$-structures, the integrability of these structures is achieved by imposing a set of integrability conditions on $(J_a, K)$,
							\begin{subequations}
								\begin{align} 
								\label{eq:moment_map}
									&& &\mu_{a} (V) = \lambda_{a} \gamma(V) && \forall \, V \in \Gamma(E)\, , && \\
								\label{eq:LK1}
									&& &L_{K}K=0\, ,\\
								\label{eq:LK2}
									&& &L_K J_a = \epsilon_{a b c} \lambda_{b} J_{c} \, , && L_{\tilde K} J_{a} =0 \, , &&
								\end{align}
							\end{subequations}
					where the second condition in~\eqref{eq:LK2} only applies for $D=4$.
					
					When $\lambda_a \neq 0$, this structure is called \emph{exceptional Sasaki-Einstein}~\cite{AshmoreESE}, while in the case of $\lambda_a = 0$ it takes the name of \emph{exceptional Calabi-Yau}~\cite{AshmoreECY}.
					
					On can show that these are equivalent to the Killing spinor equations for backgrounds preserving $\mathcal{N}=2$ supersymmetry.
	
					The functions $\mu_{a} (V)$ are a triplet of moment maps for the action of the generalised diffeomorphisms,
							\begin{equation}
								\mu_a(V) = - \frac{1}{2} \epsilon_{abc} \int_M \mathrm{tr}(J_b L_V J_c)\, .
							\end{equation}	
					We will see how the constants $\lambda_a$ depend on the theory: they are zero for Minkowski backgrounds, while for AdS are related to the inverse of the AdS radius $\lvert \lambda\rvert=2m$ for $D=4$ and $\lvert \lambda\rvert=3m$ for $D=5$, where $\lvert \lambda\rvert^2=\lambda_1^2+\lambda_2^2+\lambda_3^2$. 
					Finally, the function $\gamma$ is defined as
							\begin{equation}
								\begin{array}{llc}
									\gamma (V) = 2 \int_M q(K)^{-1/2} q(V,K,K,K) &{}& D=4\, , \\[1mm]
									\gamma (V) =\int_M c(V,K,K) & {}& D=5 \, .
								\end{array}
							\end{equation}

					The integrability conditions have important consequence, that we are going to explore further in the next chapters. 
					For instance, the generalised vector $K$ is a generalised Killing vector, that is
							\begin{equation}
								L_K \mathcal{G} = 0	\, ,
							\end{equation}
					where the generalised metric $\mathcal{G}$ in~\eqref{GofVandV'}. 
					The generalised Killing vector condition for M-theory is equivalent to
							\begin{equation}
								\begin{array}{ccccc}
									\mathcal{L}_\xi g = 0\, , &\phantom{and}& \mathcal{L}_\xi A - \dd \omega = 0\, , &\phantom{and}& \mathcal{L}_\xi \tilde{A} -\dd \sigma + \tfrac{1}{2} \dd \omega \wedge A = 0\, ,
								\end{array}
							\end{equation}
					while in type IIB one has
							\begin{equation}
								\begin{array}{lcl}
									\mathcal{L}_\xi g = 0\, , &\phantom{and}& \mathcal{L}_\xi C = \dd \rho - \tfrac{1}{2}\epsilon_{ij}\dd \lambda^i \wedge B^j \, , \\[1mm] 
									\mathcal{L}_\xi B^i - \dd \lambda^i = 0\, , &\phantom{and}& \mathcal{L}_\xi \tilde{B}^i = \dd \sigma^i + \tfrac{1}{2} \dd \lambda^i \wedge C - \tfrac{1}{2} \dd \rho \wedge B^i + \tfrac{1}{12}\epsilon_{kl} B^i \wedge B^k \wedge \dd \lambda^l \, .
								\end{array}
							\end{equation}

					The generalised Killing vector condition on $K$ means that the action of the generalised Lie derivative on the untwisted objects reduces to the usual one,
							\begin{equation}\label{dorflie}
								\hat{L}_{K} \cdot =\mathcal{L}_{\xi} \cdot \, , 
							\end{equation}
					where $\xi$ denotes the (necessarily non-vanishing) vector component of $K$. 
					By virtue of~\eqref{eq:twisted_untwisted_lie_der} this is equivalent to the vanishing of the tensor $R_{\mathbb{L}_{\tilde{V}}}$ in~\eqref{eq:tensor_r}. 
					We will refer to~\eqref{eq:moment_map} as \emph{moment map condition} while to~\eqref{eq:LK1},~\eqref{eq:LK2} and the vanishing of $R_{\mathbb{L}_{\tilde{V}}}$ in~\eqref{eq:tensor_r} as $L_K$ \emph{condition}.
					$K$ is called \emph{generalised Reeb vector} because it naturally generalises the isometry described by the usual Reeb vector in Sasakian geometry. 
	
					We will analyse further these structures in the chapters~\ref{chapComp} and~\ref{chapbrane}.

\ensurepagenumbering{arabic}
	\chapter{Consistent truncations}
	\label{chapComp}
		\section{Introduction and motivations}
			The aim of this chapter is to discuss the general approach to supergravity compactifications with fluxes.
						
			This thesis is devoted to the study of supersymmetric compactifications with some non-trivial fluxes.
			We will see in the first part of this chapter how requiring some amount of supersymmetry on the lower dimensional theory constrains the geometry of the internal manifold $M$, such that it must admit geometrical structures like the ones we described in~\cref{chap1}.
			For the well-known case of fluxless compactification of a $10$-dimensional type II supergravity to a minimal supergravity in $4$ dimensions, the constraints on the internal manifold requires it to be a Calabi-Yau three-fold~\cite{CYcomp}.
			When we allow fluxes to be turned on, the supersymmetry conditions can be cast in a compact and elegant form using \emph{Generalised Geometry} and generalised structures we introduced in~\cref{chapEGG}.
			
			We start by reviewing supersymmetric backgrounds.
			In addition, after describing the standard Calabi-Yau $\mathcal{N}=2$ four-dimensional compactification of type II theories, we summarise how to include NSNS flux, and which problems would arise if one considers the whole set of fluxes.
			Then, the discussion moves on parallelisable spaces and how this structure can be useful to define \emph{Generalised Scherk-Schwarz reductions}.
		\section{Kaluza-Klein dimensional reductions}
			%
%
			The idea of introducing extra-dimension to get some sort of unification of different phenomena is not recent.
			It dates back to the works of Kaluza and Klein~\cite{Kaluza, Klein} who studied how a spontaneous compactification of  a purely gravity theory in five dimensions i
			can give gravity and electromagnetism in four-dimensions. .
			
			We want to briefly review the Kaluza-Klein reduction in arbitrary dimension.
			We will follow and refer to these works for more details~\cite{stellereview, popeKK}.
			Firstly, we aim to reduce a $(d+1)$-dimensional purely gravitational theory down to $d$-dimensions, by a reduction on a circle $S^1_R$ of radius $R$.
			For simplicity, we suppose the $d$-th spatial coordinate is the periodic one, and we call it $y$, \emph{i.e.} for any integer $k$, $y + 2\pi k R \simeq y$.
			The $(d+1)$-dimensional vector of coordinate is $z^M \equiv (x^\mu , y)$, where $M = 0,\ldots, d$ and $\mu = 0, \ldots d-1$.
			
			Let us take into account the usual Einstein-Hilbert action for $(d+1)$-dimensional gravity,
					\begin{equation*}
						\mathcal{S} = \frac{1}{2 \mathrm{k}^2} \int \dd^{d+1}z\ \sqrt{-\mathfrak{g}}\ \mathcal{R} \, .
					\end{equation*}
			The periodicity of the $d$-th spatial dimension allows us to write the expansion in Fourier modes of the metric along that direction
					\begin{equation*}
						\mathfrak{g} (x, y) = \sum_{k} g^{(k)}(x) e^{i k y / R} \, .
					\end{equation*}
			At this point we can substitute the Fourier expansion in the action and integrate $y$ away over the $S^1$.
			However, doing this one would have a $d$-dimensional theory with an infinite tower of states, labeled by $k$.
			In order to find an effective lower dimensional theory with a finite number of degrees of freedom, we have to define a so-called \emph{truncation ansatz}, namely a prescription telling us which modes (in this case of the metric expansion) we keep and which ones are to set to zero.
			In this toy-model the criterium to truncate the spectrum is readable by the equation of motion (the Einstein equation $\mathcal{R}_{MN} = 0$).
			We linearise this for small fluctuations around the flat Minkowski solution $\mathbb{M}_{d} \times S^1$, we get
					\begin{equation*}
						\langle \mathfrak{g}_{MN} \rangle \dd z^M \dd z^N = \eta_{\mu\nu} \dd x^\mu \dd x^\nu + \dd y^2 \, , 
					\end{equation*}
			where we assumed the v.e.v. of the last component to be $\langle \mathfrak{g}_{dd} \rangle = 1$.
			Taking the circle radius $R$ small enough, we can make all the states with $n\neq 0$ very massive, thus we can neglect them in a low energy approximation of the theory.
			To conclude, we can define a truncation to lower dimensional massless modes by keeping  only the $k=0$ modes of the Fourier  expansion, which are masslees and 
			are $y$-independent.
			For instance, we can reduce the higher dimensional metric as follows,
					\begin{equation*}
						\mathfrak{g}_{MN} \dd z^M \dd z^N = e^{2\alpha \phi(x)} g_{\mu\nu} \dd x^\mu \dd x^\nu + e^{2\beta \phi(x)} (\dd y + A )^2\, ,
					\end{equation*}
			where $\alpha$ and $\beta$ are real parameters and $\phi(x)$ is a real function of the $x$ coordinates only. 
			Lastly, $A$ is a one form $A_{\mu}(x) \dd x^\mu$.
			Explicitly, this ansatz gives the prescriptions,
					\begin{align*}
						&& \mathfrak{g}_{\mu\nu} = e^{2\alpha \phi} g_{\mu\nu} + e^{2\beta \phi} A_{\mu} A_{\nu} \, , && \mathfrak{g}_{\mu d} = e^{2\beta \phi} A_{\mu} \, , & & \mathfrak{g}_{dd} = e^{2\beta\phi} \, . & &
					\end{align*}

			Substituting the quantities in the higher dimensional action functional, and integrating over $S^1$, allows us to write
					\begin{equation*}
						S = \frac{1}{2 \kappa^2} \int \dd^{d}x\ \sqrt{-g}\ \left(R - \frac{1}{2} (\partial \phi)^2 - \frac{1}{4} e^{-2(d-1) \alpha \phi} F_{\mu\nu}F^{\mu\nu} \right)\, ,
					\end{equation*}
			where $F_{\mu\nu} = 2 \partial_{[\mu} A_{\nu]}$ and $\kappa^2 = \mathrm{k}^2/ 2\pi R$.
			We imposed also a the following normalisations for $\alpha$ and $\beta$,
					\begin{align*}
						&& \alpha^2 = \frac{1}{2(d-2)(d-3)} \, , & & \beta = - (d-3) \alpha \, ,& &
					\end{align*}
			in order to get a correctly normalised kinetic term for the scalar field~\cite{popeKK}.
			
			To conclude, we obtained the Maxwell-Einstein action -- with an additional scalar $\phi$ called \emph{modulus}, on which we will come back shortly -- by reducing a purely gravitational theory on a circle.
			The gauge symmetry $A \rightarrow A + \dd \lambda$ is a consequence of the metric ansatz, when one ($x$-dependently) reparametrises the circle.
			\subsection{Consistent truncation ansatze}
				The reduction exposed above is a useful toy-model of reduction.
				However, in string and supergravity compactifications one has to cope with much more involved techniques.
				This because the action is not just gravity and also because the compactification manifold is usually more complicated than a circle.
				Despite this, there are some main points we can highlight since they are common to a large number of compactification procedures.
			
				As previously mentioned, we want a lower dimensional theory with a finite number of degrees of freedom.
				In order to achieve this, we have to give a prescription indicating which fields we keep and which ones we have to ``truncate out''.
				In other words a truncation of the higher dimensional modes on the internal manifold is necessary.
				We call this prescription \emph{truncation ansatz}.
				In our toy-example, the truncation ansatz was readable explicitly, but in general life is much harder.
			
				One way to procede is to focus on a vacuum state in low dimension and study the perturbations about this vacuum.
				The truncation prescription is given by the so-called \emph{Kaluza-Klein ansatz}.
				A nice review about this topic is given in~\cite{duffKK}.
			
				Briefly speaking, the procedure to get the truncation ansatz can be described schematically as follows.
				Firstly, one considers a vacuum of the higher dimensional theory that exhibits a spacetime solution structure as a product of spaces, the lower-dimensional spacetime and the internal space.
				Secondly, the equations of motion are linearised around vacuum.
				This produces some massive operators from the lower-dimensional perspective.
				Expanding the higher-dimensional degrees of freedom into the eigenstates of these massive operators, and keeping only the massless modes concludes the procedure.
			
				It is noteworthy that in general the ansatz gives a good way to study linear fluctuations about the chosen vacuum, but the theory coming out of this might not capture the whole information the higher-dimensional theory had originally.
				In the previous example the expansion to all-order of the ansatz was achieved by truncating out all the fields depending on the coordinate on the circle.
				However, already taking into account slightly more complicated spaces this extension will be highly non-trivial, often not possible at all.
				Therefore, Kaluza-Klein anstaze are limited to describe the physics around a chosen vacuum.
			
				One can chose to follow another approach.
				This consists of defining a truncation ansatz based on a given symmetry, that is keeping the modes which are invariant under some group of transformations.
				The typical group of transformations one takes into account is a subgroup of the isometry group of the internal manifold.
				This approach can be applied with fruitful results to the cases where the internal manifold is a group manifold or a coset space, since the isometry group is manifest~\cite{Cvetic:2003jy, schschw}.
				The simplest example to consider is the $\U(1)$ reduction.
				Notice that since $S^1 \cong \U(1)$, our toy example describes this second approach as well.
				Keeping only the higher dimensional fields independent on $y$ is equivalent to take only the invariant fields (\emph{i.e.} the singlets) under the $\U(1)$ action.
			
				At this stage, it is useful to make some remark about this second approach.
				A first point to raise, the truncated theory is not always physically relevant, since the lower dimensional physics may not be captured completely by the compactified theory.
				However, it is mathematically well-defined and independent from the choice of a specific vacuum.
				The second noteworthy aspect -- crucially important in this thesis -- is that the dimensional reduction based on the use of a symmetry is able to give ansatze that are \emph{consistent}.
				A \emph{consistent truncation} is a choice of a finite set of modes, where the omitted ones are not sourced by the subset chosen. 
				This is equivalent to say that the set of truncated modes has a dynamics which is not affected by the others.
				This fact allows us to say that a solution to equations of motion in the lower dimensional theory, which is a linear combination of only truncated modes, always lift to a solution also on the higher dimensional theory.
				It might be useful to consider a simple example to understand what we mean for consistent truncations.
				Given a theory of two scalars with Lagrangian,
						\begin{equation*}
							\mathcal{L} = \frac{1}{2}\left(\partial \varphi_1 \right)^2 + \frac{1}{2}\left(\partial \varphi_2 \right)^2 - \frac{m_1^2}{2}\varphi_1^2 - \frac{m_2^2}{2}\varphi_2^2 - g \varphi_1^2\varphi_2 \, .
						\end{equation*}
				It generates the following equations of motion,
						\begin{equation*}
							\begin{split}
								\partial_\mu \partial^\mu \varphi_1 + m_1^2 \varphi_1 &= -2g\varphi_1\varphi_2 \, , \\
								\partial_\mu \partial^\mu \varphi_2 + m_2^2 \varphi_2 &= -g\varphi_1^2 \, .
							\end{split}
						\end{equation*}
				Therefore, we can observe how $\varphi_1 = 0$ is a consistent truncation, \emph{i.e.} the evolution of $\varphi_2$ is given by a consistent (with the choice of suppressing $\varphi_1$) equation of motion, and fixed $\varphi_1 = 0$ at the initial time, it will remain fixed at all times. 
				In other words, $\varphi_1 = 0$ is both a solution of the theory reduced to the only field $\varphi_1$, and of the full theory of the two scalar fields $\varphi_1,\varphi_2$. 
				On the other hand, dropping $\varphi_2 = 0$ is not consistent, since the dynamics of $\varphi_1$ will affect $\varphi_2$, due to the fact $\varphi_1$ acts as a source term for $\varphi_2$.
			
				In this very simple example, the truncation it is easy to find by simply playing with the equations of motion.
				Not surprisingly, in the case of dimensional reductions things are not so simple and in order to have the hope of finding a consistent truncation we have to rely on symmetry.
				
				Let us reconsider our toy-model of reduction again.
				We have seen how by dimensionally reducing a theory of gravity over a circle we get a theory of gravity coupled with electromagnetism and a free massless scalar (at linear order). 
				This field is related to the radius of the compactification circle and how it varies along the $d$-dimensional spacetime.
				There is not a procedure that tells us which value $R$ has to take dynamically during the compactification and this reflects in the presence a free scalar, \emph{i.e.} with no fixed (by some potential) \emph{vacuum expectation value}, vev from now on.
				This is a characteristics that is often present in compactifications: the background exhibits a continuous degeneracy related to the variations in size and shape of the compact space.
				The fields parametrising this degeneracy are called \emph{moduli} and when the compactification does not produce any scalar potential (which constrains their vevs), one says moduli are not \emph{stabilised}.
				In our example we just have one massless scalar field, but typically Calabi-Yau compactifications produce a large number of moduli.
				This goes under the name of \emph{moduli problem}.
			
				One wants moduli to be stabilised for various reasons.
				First of all, the observables of the lower-dimensional theory depends on the moduli, so, if their vevs can be shifted arbitrarily the theory loses predictivity.
				One may be thinking that this situation is similar to handle Goldstone bosons.
				Spontaneous symmetry breaking is the reason of the origin of the Goldstone modes, indeed the physics in any vacuum connected by a Goldstone mode is the same, since all these vacua are equivalent due to symmetry.
				Moduli, however, do not need a symmetry to arise and hence in general physics will depend on their values.
				Then one finds a space of physically inequivalent vacua (the notorious \emph{moduli space}) related by varying the vev of the moduli.
				Furthermore, phenomenologically massless scalar fields should mediate long range interactions, but this contradicts the observations.
				Finally, a more formal point of view is to answer to the question ``how can the masses of particles in Standard Model come from a theory with no free parameters?''~\cite{moduliLect, fluxcomp1}.

						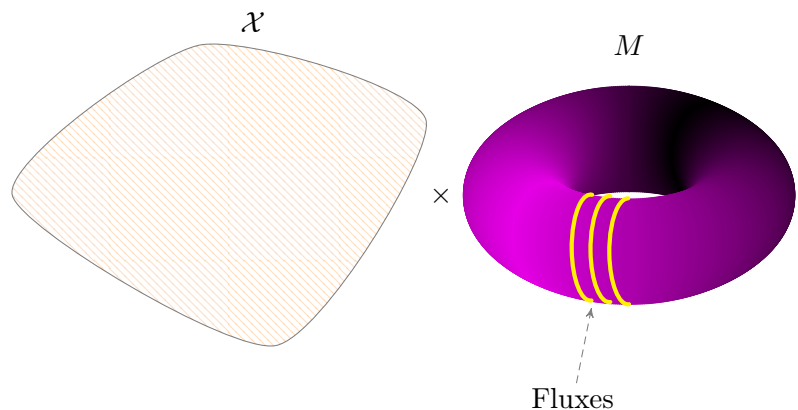
\begin{figure}[h!]
							\centering
								\begin{tikzpicture}
	    	
		\draw[smooth cycle, tension=0.4, fill=white, pattern color=orange, pattern=north west lines, opacity=.5] plot coordinates{(-5.7,2) (-8.2,0) (-4.7,-2) (-2.7,1)};
		\draw node at (-5, 2.3) {$\mathcal{X}$};
		\draw node at (0, 2) {$M$};
		\draw node at (-2.5,0) {$\times$};
	    
	    \foreach \x in {90,89,...,-90} { 
	    \pgfmathsetmacro\elrad{20*max(cos(\x),.1)}
	    \pgfmathsetmacro\ltint{.9*abs(\x-45)/180}
	    \pgfmathsetmacro\rtint{.9*(1-abs(\x+45)/180)}
	    \definecolor{currentcolor}{rgb}{\ltint, 0, \ltint}
	    \draw[color=currentcolor,fill=currentcolor] 
	        (xyz polar cs:angle=\x,y radius=.75,x radius=1.5) 
	        ellipse (\elrad pt and 20pt);
	    \definecolor{currentcolor}{rgb}{\rtint, 0, \rtint}
	    \draw[color=currentcolor,fill=currentcolor] 
	        (xyz polar cs:angle=180-\x,radius=.75,x radius=1.5) 
	        ellipse (\elrad pt and 20pt);
	    }
		\draw[yellow, ultra thick, line cap=round] (0,-1.45) arc (-93:90:-.25 and .705);
		\draw[yellow, ultra thick, line cap=round] (-.5,-1.4) arc (-93:95:-.25 and .705);
		\draw[yellow, ultra thick, line cap=round] (-.25,-1.42) arc (-93:93:-.25 and .705);
	
		\draw[densely dashed, thin, gray, <-] (-.5,-1.5) -- (-.7,-2.5)
			node at (-.75, -2.7) {\textcolor{black}{Fluxes}};

	\end{tikzpicture}
	
%
							\caption{A schematic representation of a compactification in the presence of fluxes.}
							\label{fluxcomp}
						\end{figure}
				A possible path to follow in order to solve the moduli problem is to find a mechanism to generate a scalar potential in the lower-dimensional action.
				This would have the effect of stabilise the moduli (giving them a mass and a fixed vev).
				A great number of results in this direction have been reached in the last twenty years, realising that it is possible to generate a non-trivial scalar potential in a compactification through \emph{fluxes}~\cite{fluxcomp1, fluxcomp2, fluxcomp3}.
				One can find some nice reviews of the subject in~\cite{DuffReviewComp, MarianaFluxReview, LustReviewComp, henlect}.
			
				Fluxes are higher rank objects generalisation of the electromagnetic field strength.
				They are associated with a non-zero background value of the supergravity $p$-form field strength.
				To be precise, let $F_p$ be a $p$-form field strength whose Bianchi identity is
						\begin{equation*}
							\dd F_p = 0\, ,
						\end{equation*}
				locally, one can always associate a potential $C_{p-1}$ such that $F_p = \dd C_{p-1}$. 
				When sources are present it is not possible to have a globally well-defined potential, and the integral over a a $p$-cycle $\Sigma_p$ on the internal compact manifold $M$ it is not automatically zero. 
				Then we say there is a flux of $F_p$ on $M$ supported by $\Sigma_p$,
						\begin{equation*}
							\frac{1}{(2\pi \ell_s)^{p-1}}\int_{\Sigma_p}\!\! F_p = k \neq 0\, .
						\end{equation*}
				As for the familiar Dirac's monopole, one can impose quantisation conditions on fluxes so that $k$ can take only discrete values.
				Roughly speaking, this number corresponds to how many times the extended object associated to the flux wraps around the cycle $\Sigma_p$, see~\cref{fluxcomp}.
				Requiring the presence of these quantities in the dimensional reduction we can generate a potential $V$ for the scalars in the lower dimensional theory, that will come from the kinetic term of the internal components of the fluxes.
				This can be seen from the higher dimensiona action. 
				Schematically,
						\begin{equation*}
							S= \int_{\mathcal{X}} \ldots \underbrace{\int_{M} F \wedge \star F}_{V(\phi)} \, .
						\end{equation*}

				Moduli stabilisation is not the only reason to study flux compactifications.
				For example, the presence of fluxes removes the necessity of a Ricci-flat internal space, then Calabi-Yau are no more available spaces.
				This open interesting perspectives in studying the geometry of string theory vacua and a classification of flux compactifications may be very useful in order to understand better the structure of the theory.

						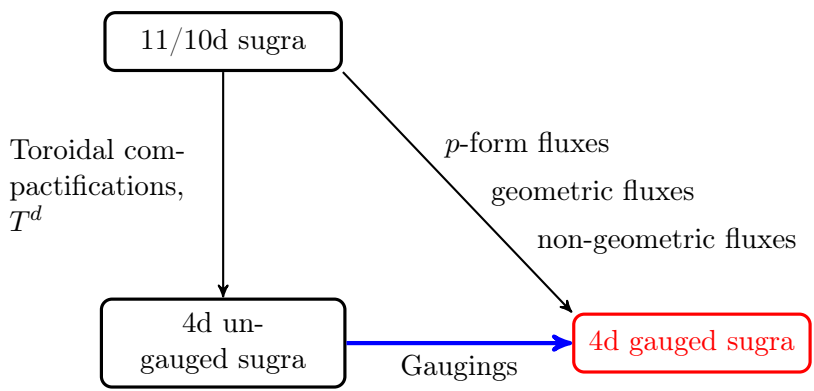
\begin{figure}[h!]
							\centering
											\begin{tikzpicture}
				%
					\node[punkt] (Msugra) {$11$/$10$d sugra};
					\node[punkt, inner sep=5pt,below=3cm of Msugra] (4sugra) {$4$d ungauged sugra};
					\node[punkt, below=3cm of Msugra, right=3cm of 4sugra, color=red] (gaugedsugra) {$4$d gauged sugra};
					\draw[->, thick] (Msugra.south) -- (4sugra.north)
						node[pos=0.5, left, text width = 2.7cm] {Toroidal compactifications, $T^d$};
					\draw[->, color=blue, ultra thick] (4sugra.east) -- (gaugedsugra.west)
						node[pos=0.5, below, color = black] {Gaugings};
					\draw[->, thick] (Msugra.south east) -- (gaugedsugra.north west)
						node[pos=0.4, above right, text width = 2.5cm] {$p$-form fluxes}
						node[pos=0.6, above right, text width = 3cm] {geometric fluxes}
						node[pos=0.8, above right, text width = 4cm] {non-geometric fluxes};
				\end{tikzpicture}
							\caption{A representation of how fluxes are a way to get gauged supergravity from higher dimensional supergravity theory.
									This gives an higher dimensional origin to gauge symmetry.}
							\label{gaugedsugra}
						\end{figure}

				Furthermore, fluxless compactifications will produce only theories in lower dimension whose gauge groups are products of $\U(1)$'s\footnote{%
					This statement holds for M- and type II theories.
					For the heterotic string theories things work a bit differently, since there gauge groups appear as a perturbative effect and they connects to type II through dualities.
					However, in this thesis we are not going to work with heterotic string theory (for more details one can look~\cite{polchinski}), so we can focus only on type II and M-theory and take the statement as true.}.
				Since we have the hope of reproducing non-Abelian gauge theories (like Standard Model), we want a compactification mechanism that also explains the higher dimensional origin of the gauge symmetry.
				Flux compactifications give us the right framework, since non-Abelian gauge groups are connected to (non-Abelian) field strength (precisely the fluxes).
				As in electrodynamics, the electromagnetic field is generated by a dynamical object, the electric charge, the $p$-form field strengths are sourced by non-perturbative dynamical objects, the \emph{D-branes} wrapping homology cycles inside the internal manifold~\cite{PolchinskiBranes}.
				
				For all these reasons a systematic study of flux compactification is useful and interesting.
				In order to achieve this, new mathematical techniques have been being introduced, and precisely the ones described in the previous chapters are an important example.
				A first inspirational remark is that through flux compactifications, we are able to explain the higher dimensional origin of the gauge symmetry.
				A subgroup of the gauge group comes from the isometries of the internal manifold -- as in Kaluza-Klein reductions-- but higher-dimensional supergravities come with form potentials, which carry their own gauge symmetry and also contribute to the gauging of the truncated theory.
				Thus, an approach treating at the same level diffeomorphisms and gauge transformation can be a promising way to better understand the structure of the theory. 
				Another interesting fact is that the supergravity theories when compactified on a torus exhibit an abelian gauge group and a large global symmetry\footnote{
					Notice that, as far as one sets the compactification scale well below the string one $\ell_s$, it is justified to work in the supergravity limit of string theory.}.
				Tori compactifications give rise to theories called \emph{ungauged supergravities}, see~\cref{gaugedsugra}.
				As seen, the global symmetry group is known under the name of $\U$-duality and it corresponds to the exceptional non-compact group $\E_{d(d)}$.
				In the mid 1990's the discrete subgroup $E_{d(d)}/\mathbb{Z}$ has been reinterpreted as part of the duality group of M-theory~\cite{hulldualities}. 
				The $\E_{d(d)}$ group relevant in our discussion depends on the dimension, we collected the various possibilities in~\cref{Udualtab}.
							\begin{table}[h!]
							\centering
								\begin{tabular}{c c c c c c}
									$d$					&		$3$			&		$4$			&		$5$		&		$6$		&		$7$		\\
										\midrule
									Global Symmetry		&	$\SL(5)$	& 	$\SO(5,5)$	&	$\E_{6(6)}$	&	$\E_{7(7)}$	&	$\E_{8(8)}$	
								\end{tabular}
								\caption{The $\U$-duality groups for the different compactifications on the tori $T^d$.}
								\label{Udualtab}
							\end{table}

				The way to get \emph{gauged supergravities}, \emph{i.e.} supergravity theories with non-abelian gauge group, is twofold.
				One can promote some subgroup of the $\U$-duality group to a local symmetry, this adds some term in the Lagrangian called \emph{gaugings}.
				It turns out that this is equivalent to consider more complicated internal manifold admitting fluxes.
				This can be inserted in a wider picture, indeed, nowadays it is a common believing that any gauged supergravity theory comes from an higher dimensional string theory compactified on some manifold supporting fluxes.
				
				The presence of these large groups of symmetry is difficult to understand from the \emph{democratic} formulation of supergravity theories~\cite{DemSugra}.
				Thus, a certain amount of effort has been made to build $\U$-duality covariant approach to supergravity: Exceptional/Double Field Theory~\cite{hull2, samt1, samt2} and (Exceptional) Generalised Geometry~\cite{Gualtieri:2003dx, hull1, waldram1, waldram2, waldram3, waldram4}.
				
				Exceptional field theory enlarges the space, such that one completes the fundamental representation of $\E_{d(d)}$ with this extra-coordinates.
				Then one gets rid of them by applying a \emph{section condition}, projecting out the unphysical degrees of freedom~\cite{samt1, samt2}.
				One can prove, once the section condition has been imposed, the two approaches are equivalent.
				
				In this thesis we have already described generalised geometry, and we are going to focus on its applications to flux compactifications.
				
				Until so far our discussion has been quite general, now in order to proceed with our analysis, we move our attention to some more technical aspects, specifying the properties our truncation ansatze must have.
		\section{Scherk-Schwarz reductions}
				As analysed in the example of Kaluza-Klein truncations, that method provides a good way to study linear fluctuations about a given vacuum, but the complete effective theory might be not completely captured.
				Further more in Kaluza-Klein reductions, there is no warranty for the truncation ansatz to be consistent.
				In order to have a construction producing a consistent truncation of the higher-dimensional theory we have to be protected by some symmetry.
				One of the first method to achieve a consistent truncation is the so-called \emph{Scherk-Schwarz reduction}~\cite{Scherk:1979zr}.
				This procedure prescribes to choose the internal manifold to be a Lie group.
				The truncation ansatz is chosen in such a way that dependence of fields on the internal coordinates is through the left-invariant objects of the Lie group.
				The consistency of this ansatz is related to this fact.
				Indeed, there is no way the singlet modes can source the truncated non-singlet ones in the equations of motion.
				
				We will present a generalisation in the context of exceptional generalised geometry to include fluxes, called \emph{Generalised Scherk-Schwarz reduction} in the next section. 
				This will be the key ingredient to find consistent truncations on spheres of massive type IIA supergravity.
				
				Before coming to the generalised Scherk-Schwarz reduction, we recall how a conventional Scherk-Schwarz reduction~\cite{Scherk:1979zr} is defined.
				For concreteness, we take the case of a type IIA supergravity, but this is not a relevant choice, since the line of reasoning is the same in any case.
				
				As said, in Scherk-Schwarz reductions, the internal manifold is chosen to be a $d$-dimensional Lie group, $M_d = G$. 
				It follows that $M_d$ is parallelisable, namely there exists a global frame $\{\hat{e}_a\}$, $a=1,\ldots,d$, trivialising the frame bundle and thus the tangent bundle $TM_d$.
				In terms of $G$-structures of~\cref{chap1}, we have an identity structure on the manifold. 
				The frame is constructed by considering a basis of vectors that are invariant under the (say) left-action of the group $G$ on itself. 
				Under the Lie derivative, the left-invariant frame satisfies the algebra
						\begin{equation}
							\mathcal{L}_{\hat{e}_a}\hat{e}_b = f_{ab}^{\phantom{ab}c} \hat{e}_c\, ,
						\end{equation}
				where $f_{ab}^{\phantom{ab}c}$ are the structure constants of $G$.
				
				The vectors $\{\hat{e}_a\}$ generate the right-isometries of the bi-invariant metric on the group manifold. 
				A truncation ansatz for the internal metric is defined by ``twisting'' the original frame on $M_d$ by a $\GL(d)$ matrix $U_a^{\phantom{a}b}$ depending on the external spacetime coordinates $x^\mu$,
						\begin{equation}
							\hat{e}'_a{}^m(x,z) = U_a{}^b(x)\, \hat{e}_b{}^m(z)\ ,
						\end{equation}
				and setting
						\begin{equation}
							g^{mn}(x,z) = \delta^{ab}\, \hat{e}'_a{}^m(x,z) \,\hat{e}'_b{}^n(x,z) = \mathcal{M}^{ab}(x) \,\hat{e}_a{}^m(z) \,\hat{e}_b{}^n(z)\ ,
						\end{equation}
				where $\mathcal{M}^{ab} = \delta^{cd} U_c{}^a U_d{}^b$. 
				As we are free to redefine the frame by $x$-dependent $\SO(d)$ transformations, the $\mathcal{M}^{ab}$ matrix parameterises the coset $\GL(d)/\SO(d)$; hence it defines $\frac{1}{2}d(d+1)$ scalars on the external spacetime. 
				It follows that $g_{mn} =\mathcal{M}_{ab} e_m{}^a e_n{}^b$, where $\mathcal{M}_{ab}$ is the inverse of $\mathcal{M}^{ab}$, and, as before, the one-forms $e^a$ are dual to the vectors $\hat{e}_a$. 
				The full ten-dimensional metric is given by
						\begin{equation}
							\dd \hat{s}^2 = g_{\mu\nu}\dd{x}^\mu\dd{x}^\nu + \mathcal{M}_{ab} (e^a - \mathcal{A}^a)(e^b - \mathcal{A}^b)\,.
						\end{equation}
				The $d$ one-forms $\mathcal{A}^a=\mathcal{A}_\mu{}^a(x)\dd{x}^\mu$ gauge the right-isometries on the group manifold, and are therefore $G$ gauge fields on the external spacetime.
				For the RR one-form one takes
						\begin{equation}
							\hat{C}_1(x) = C_\mu(x) \dd{x}^\mu + C_a(x) (e^a - \mathcal{A}^a) + \rg{C_1}\ ,
						\end{equation}
				where $\rg{C_1}$ is the potential for a background, left-invariant two-form flux. 
				This gives an additional one-form and $d$ more scalars. 
				A similar ansatz is taken for the other form potentials.
 
				The reduction defined in this way is consistent by symmetry reasons: the dependence of the type IIA fields on the internal coordinates is fully encoded in the left-invariant tensors $\hat{e}_a$ and $e^a$, and there is no way the singlet modes can source the truncated non-singlet modes in the equations of motion.
				The gauge group of the lower-dimensional, truncated theory arises from the interplay between the right-Killing symmetries generated by the left-invariant vectors $\hat{e}_a$ and the gauge transformations of the form potentials with flux, and corresponds to a semi-direct product of $G$ with a non-compact factor. 
				The full supersymmetry of the original theory is preserved in the truncation.
				
				We refer to e.g.~\cite{Kaloper:1999yr,Dall'Agata:2005ff,D'Auria:2005er,Hull:2005hk,Hull:2006tp} for a detailed account of conventional Scherk-Schwarz reductions in a context related to the one of this thesis.

		\section{Generalised Scherk-Schwarz reductions}\label{genScherkSchw}
				In~\cite{spheres}, it was observed that consistent truncations with maximal supersymmetry are related to the existence of a \emph{generalised Leibniz parallelisation}, $\{\hat{E}_A\}$ as defined in the previous chapter, by the condition~\eqref{GLP}. 
				Such a frame defines a Leibniz algebra, hence the qualification ``Leibniz'' attributed to the parallelisation. 
				Starting from a generalised Leibniz parallelisation, one can define a \emph{generalised Scherk-Schwarz reduction}. 
				As the name suggests, this is a generalisation of conventional Scherk-Schwarz reductions on local group manifolds~\cite{Scherk:1979zr} to a larger class of manifolds, which preserves the same amount of supersymmetry as the original higher-dimensional theory. 
				We will see how the constants in~\eqref{GLP}, $(X_A)_{B}{}^C$ correspond to the generators of the lower-dimensional gauge group, and are tantamount to the embedding tensor that fully determines the gauged maximal supergravity.
				For more details about the embedding tensor formalism we refer to~\cite{henlect}. 
				The truncation defined by the generalised Scherk-Schwarz procedure is conjectured to be consistent. 
				Although it has not been proved in full generality, this expectation is supported by a number of examples.
			
				A similar approach has been adopted for studying generalised Scherk-Schwarz reductions using exceptional field theory, see~e.g.~\cite{samt1,SamtExcReview,Baguet:2015sma,InversoGenSchSchw}.
			
				In particular the generalised parallelisation has been used to define, in addition, the gauge and higher-tensor fields in the truncation. 
				Formally, as mentioned earlier, under the section condition, the equations of exceptional field theory and exceptional generalised geometry are the same.
				
				We are now ready to define \emph{Generalised Scherk-Schwarz reductions}.
				
				Extensions of conventional Scherk-Schwarz reductions to reformulations (or extensions) of high-dimensional supergravity theories with larger structure groups have been considered by several authors, see~e.g.~\cite{Riccioni:2007au,Berman:2012uy,Aldazabal:2013mya,Godazgar:2013dma,spheres,Hohm:2014qga,Ciceri:2014wya,SamtExcReview,Baguet:2015sma,Malek:2015hma}. 
				Here we will follow~\cite{spheres} and define a \emph{generalised Scherk-Schwarz reduction} on a $d$-dimensional manifold $M_d$ (not necessarily a Lie group) as the direct analogue of an ordinary Scherk-Schwarz reduction, with the ordinary frame on the tangent bundle replaced by a frame on the generalised tangent bundle.
				In particular we will this will allow us to derive an explicit ansatz for the fields with one or two external legs for type IIA (in analogy to the exceptional field theory expressions for eleven-dimensional and type IIB supergravity given in~\cite{Hohm:2014qga,SamtExcReview,Baguet:2015sma}).
				
				As in any Kaluza--Klein reduction, we start by decomposing the type IIA fields according to the $\SO(1,9) \to \SO(1,9-d)\times \SO(d)$ splitting of the Lorentz group. 
				We will use coordinates $x^\mu$, $\mu = 0,\ldots, 9-d$ for the external spacetime and $z^m$, $m=1,\ldots,d$ for the internal manifold $M_d$, of dimension $d \leq 6$. Then the ten-dimensional metric can be written as
						 \begin{equation}\label{KK_decomp_metr}
							\hat{g} \,= \, e^{2\Delta}g_{\mu\nu}\dd{x}^\mu\dd{x}^\nu + g_{mn} Dz^m Dz^n\ ,
						\end{equation}
				where 
						\begin{equation}
							Dz^m = \dd{z}^m - h_\mu{}^m \dd{x}^\mu \ ,
						\end{equation} 
				and the scalar $\Delta$ is the warp factor of the external metric $g_{\mu\nu}$. 
				In this section the symbol hat denotes the original ten-dimensional fields. 
				The form fields are decomposed as
						\begin{equation}\label{expand_10dfields}
							\begin{split}
								\hat{B} &= \tfrac{1}{2} B_{m_1m_2} Dz^{m_1m_2} + \overline{B}_{\mu m} \dd{x}^\mu \wedge Dz^m + \tfrac{1}{2}\overline{B}_{\mu\nu} \dd{x}^{\mu\nu} \, , \\[1.5mm]
								\hat{\tilde{B}} &= \tfrac{1}{6!} \tilde{B}_{m_1\ldots m_6} Dz^{m_1\ldots m_6} + \tfrac{1}{5!} \overline{\tilde B}_{\mu m_1\ldots m_5} \dd{x}^\mu \!\wedge\! Dz^{m_1\ldots m_5} \\
											& \phantom{= + } + \tfrac{1}{2\cdot 4!} \overline{\tilde B}_{\mu\nu m_1\ldots m_4} \dd{x}^{\mu\nu} \!\wedge\! Dz^{m_1\ldots m_4} + \ldots \, , \\[1.5mm]
								\hat{C}_1 &= C_m Dz^m + \overline{C}_{\mu,0} \,\dd{x}^\mu \, , \\[1.5mm]
								\hat{C}_3 &= \tfrac{1}{3!} C_{m_1m_2m_3} D{z}^{m_1m_2m_3} + \tfrac{1}{2}\overline{C}_{\mu m_1m_2} \dd{x}^\mu\wedge D{z}^{m_1m_2} \\
											& \phantom{= + } + \tfrac{1}{2}\overline{C}_{\mu\nu m}\dd{x}^{\mu\nu} \wedge Dz^m + \ldots \, , \\[1.5mm]
								\hat{C}_5 &= \tfrac{1}{5!} C_{m_1\ldots m_5} D{z}^{m_1\ldots m_5} + \tfrac{1}{4!}\overline{C}_{\mu m_1\ldots m_4} \dd{x}^\mu\!\wedge\! D{z}^{m_1\ldots m_4} \\
											& \phantom{= + } +\tfrac{1}{2\cdot 3!} \overline{C}_{\mu\nu m_1m_2m_3}\dd{x}^{\mu\nu} \!\wedge\! Dz^{m_1m_2m_3}+\ldots \, , \\[1.5mm]
								\hat{C}_7 &= \tfrac{1}{6!}\overline{C}_{\mu m_1\ldots m_6} \dd{x}^\mu\wedge D{z}^{m_1\ldots m_6} +\tfrac{1}{2\cdot 5!} \overline{C}_{\mu\nu m_1\ldots m_5}\dd{x}^{\mu\nu} \wedge Dz^{m_1 \ldots m_5} + \ldots \, ,
							\end{split}
						\end{equation}
			where $\dd{x}^{\mu\nu} = \dd x^\mu\wedge \dd x^\nu$ and $Dz^{m_1\ldots m_p} = Dz^{m_1}\wedge \cdots \wedge Dz^{m_p}$.
			The ellipsis denote forms with more than two external indices, that we will not need. 
			The expansion in $Dz$ instead of $\dd z$ is standard in Kaluza--Klein reductions, and ensures that the components transform covariantly under internal diffeomorphisms. 
			We stress that at this stage the field components still depend on all the coordinates $\{x^\mu,z^m\}$: we are decomposing the various tensors according to their external or internal legs but we have not specified their dependence on the internal space yet. 
			The only exception is the external metric, which is assumed to depend just on the external coordinates: $g_{\mu\nu} = g_{\mu\nu}(x)$.
			
			The barred fields appearing in~\eqref{expand_10dfields} can also be identified by introducing the vector 
					\begin{equation}
						\partial_\mu + h_\mu = \frac{\partial}{\partial x^\mu} + h_\mu{}^m \frac{\partial}{\partial z^m}\,,
					\end{equation} 
			which satisfies $\iota_{(\partial_\mu + h_\mu)}Dz^m =0$. 
			For the the fields with one external leg we have 
					\begin{equation}
						\begin{split}
							\overline{B}_{\mu} &= \iota_{(\partial_\mu + h_\mu)} \hat B \, \big| \,, \\[1mm]
							\overline{\tilde B}_{\mu} &= \iota_{(\partial_\mu + h_\mu)} \hat{\tilde{B}} \, \big| \,, \\[1mm]
							\overline{C}_{\mu} &= \iota_{(\partial_\mu + h_\mu)} \hat C \, \big| \,,
						\end{split}
					\end{equation}
			where by the symbol ``$|$'' we mean that after having taken the contraction $\iota_{(\partial_\mu + h_\mu)}$, the forms on the right hand side are restricted to have just internal legs. 
			In other words, we set $\dd x \equiv 0$. 
			Similarly, for the fields with two external legs we find
					\begin{equation}\label{redef_one_forms_2}
						\begin{split}
							\overline{B}_{\mu\nu} &= \iota_{(\partial_\nu + h_\nu)} \iota_{(\partial_\mu + h_\mu)} \hat{B} \, , \\
							\overline{\tilde B}_{\mu\nu} &= \iota_{(\partial_\nu + h_\nu)} \iota_{(\partial_\mu + h_\mu)} \hat{\tilde B} \big|\, , \\
							\overline{C}_{\mu\nu} &= \iota_{(\partial_\nu + h_\nu)}\iota_{(\partial_\mu + h_\mu)} \hat{C} \big| \, .
						\end{split}
					\end{equation}
			Moreover, we are arrange the RR potentials in the poly-forms 
					\begin{equation}
						\begin{split}
							\overline{C}_\mu &= \overline{C}_{\mu,0} + \overline{C}_{\mu,2} + \overline{C}_{\mu,4} + \overline{C}_{\mu,6}\,, \\
							\overline{C}_{\mu\nu} &= \overline{C}_{\mu\nu,1} + \overline{C}_{\mu\nu,3} + \overline{C}_{\mu\nu,5}\,.
						\end{split}
					\end{equation}
			These barred fields need a field redefinition. 
			This can be seen by decomposing the gauge transformations of the ten-dimensional fields and imposing that they are covariant under the generalised diffeomorphisms so that they will eventually reproduce the gauge transformation of the lower-dimensional supergravity theory after the truncation is done. 
			Here we just provide the correct redefinitions, postponing their full justification to the next section. 
			We introduce the new fields
					\begin{equation}\label{redef_one_forms}
						\begin{split}
							B_{\mu} &= \overline{B}_\mu \, , \\
							C_\mu &= e^{-B}\wedge \overline{C}_\mu \, , \\
							\tilde{B}_\mu &= \overline{\tilde{B}}_\mu - \tfrac{1}{2} [ \overline{C}_\mu \wedge s(C)]_5 \, ,
						\end{split}
					\end{equation}
			where $B$, $C$ are just internal, and
					\begin{equation}\label{redef_two_forms}
						\begin{split}
							B_{\mu\nu}&= \overline{B}_{\mu\nu} + \iota_{h_{[\mu}}B_{\nu]} \,, \\
							\tilde B_{\mu\nu}&= \overline{\tilde B}_{\mu\nu} - \tfrac{1}{2} \big[ \,\overline{C}_{\mu\nu} \wedge s(C)\, \big]_4 + \iota_{h_{[\mu}} \tilde{B}_{\nu]} \,, \\
							C_{\mu\nu} &= e^{-B} \wedge \overline{C}_{\mu\nu} + \iota_{h_{[\mu}} C_{\nu]} + B_{[\mu}\wedge C_{\nu]} \,.
						\end{split}
					\end{equation}
			Note that we are using a notation where the various tensors are treated as differential forms on the internal manifold, while we explicitly display their external indices.

			Having decomposed the higher-dimensional fields in a suitable way, we are now ready to construct our truncation ansatz. 
			As a first thing we rearrange the type IIA fields with zero, one or two external indices in terms of generalised geometry objects. The fields with purely internal legs, i.e. 
					\begin{equation}
						\left\{g_{mn},\, B_{m_1m_2},\, \tilde B_{m_1\ldots m_6} ,\, C_m,\, C_{m_1m_2m_3},\, C_{m_1 \ldots m_5} \right\} \, ,
					\end{equation}
			together with the warp factor $\Delta$ and the dilaton $\phi$, parameterise a generalised metric $\mathcal{G}^{MN}$.
			The (redefined) fields with one external index are collected in the generalised vector $\mathcal{A}_\mu{}^M$,
					\begin{equation}\label{def_calA_mu^M}
						 \{ h_\mu{}^m ,\, B_{\mu m} ,\, \tilde B_{\mu m_1\ldots m_5},\, \tilde{g}_{\mu m_1\ldots m_6,m} ,\, C_{\mu,0},\, C_{\mu m_1m_2},\, C_{\mu m_1\ldots m_4},\, C_{\mu m_1\ldots m_6} \} \, ,
					\end{equation}
			Here, $\tilde{g}$ is a tensor belonging to $ \Lambda^7T^*M_{10}\otimes T^*M_{10}$, related to the dual graviton. 
			This is not part of type IIA supergravity in its standard form and we will thus ignore it by projecting $\mathcal{A}_\mu$ on the $E'''$ bundle introduced in~\eqref{IIAexten},
					\begin{equation}
						\mathcal{A}_\mu{}^M \eqs \{ h_\mu{}^m ,\, B_{\mu m} ,\, \tilde B_{\mu m_1\ldots m_5},\, C_{\mu,0},\, C_{\mu m_1m_2},\, C_{\mu m_1\ldots m_4},\, C_{\mu m_1\ldots m_6} \} \, .
					\end{equation}
			Here and below, the $\eqs$ symbol in an equation involving generalised vectors means that the equality holds after projecting on the bundle $E'''$ using the natural mappings~\eqref{IIAexten}, namely after dropping the $T^*\otimes \Lambda^{6}T^*$ component.

			The fields with $\mu\nu$ indices defined in~\eqref{redef_two_forms} are components of a generalised tensor $\mathcal{B}_{\mu\nu}{}^{MN}$, which is a two-form in the external spacetime and a section of the bundle $N$ on $M_6$ defined in~\eqref{Nbundle}. 
			They actually correspond to the components of this object living on the bundle $N'$ given in~\eqref{NprimeBundle}, that is
					\begin{equation}
						\mathcal{B}_{\mu\nu}{}^{MN} \,\eqs\, \{ B_{\mu\nu},\, \tilde{B}_{\mu\nu m_1\ldots m_4} ,\, C_{\mu\nu m},\, C_{\mu\nu m_1m_2m_3} ,\, C_{\mu\nu m_1\ldots m_5}\}\,.
					\end{equation}
			For the equations involving sections of the bundle $N$, by the $\eqs$ symbol we mean that the equality holds after having projected on the bundle $N'$, see~\cref{app:EGG} for details.

			Suppressing the internal indices, the objects introduced above read
					\begin{equation}
						\begin{split}
							\mathcal{A}_\mu \,&\eqs\, h_\mu + B_{\mu} + \tilde B_{\mu} + C_{\mu,0}+ C_{\mu,2}+ C_{\mu ,4} + C_{\mu,6} \, , \\
							\mathcal{B}_{\mu\nu} \,&\eqs\, B_{\mu\nu} + \tilde{B}_{\mu\nu} + C_{\mu\nu,1}+ C_{\mu\nu ,3} + C_{\mu\nu,5} \, .
						\end{split}
					\end{equation}

			The construction of a (bosonic) truncation ansatz leading to a $(10-d)$-dimensional theory preserving maximal supersymmetry is then specified by the following steps:
					\begin{itemize}
					\item[\textit{1.}] 	One should find a generalised parallelisation $\{\hat E_A\}$, namely a globally-defined frame for the $\E_{d+1(d+1)}\times \RR^+$ generalised tangent bundle on $M_d$.
									This means that the frame $\{\hat E_A\}$ must be an $E_{d+1(d+1)}$ frame, namely that it is given by an $E_{d+1(d+1)}$ transformation of the coordinate frame\footnote{%
										By coordinate frame we mean
												\begin{equation*}
													\{ \tilde{\hat{E}}_A\} = \{\partial_m\} \cup \{\dd x^m\} \cup \{\dd x^{m_1 \ldots m_5} \} \cup \{\dd x^{m,m_1\ldots m_6}\} \cup \{1\} \cup \{\dd x^{m_1m_2}\} \cup \{\dd x^{m_1\ldots m_4}\} \cup \{\dd x^{m_1 \ldots ,_6}\} \, . 
 												\end{equation*}
										}.
									We will see how this condition applies in the examples below. 
									In addition, the frame must satisfy the algebra~\eqref{GLP},
											\begin{equation}\label{LeibnizParall}
												L_{\hat E_A}\hat E_B = X_{AB}{}^C \hat E_C\ ,
											\end{equation}
									with constant coefficients $X_{AB}{}^C$. 
									It is then a \emph{generalised Leibniz parallelisation}, as seen in~\cref{secGenPar}. 
									The constants $X_{AB}{}^C$ correspond to the generators of the gauge group: in gauged supergravity they are defined by contracting the \emph{embedding tensor} $\Theta_A{}^\alpha$ encoding the gauging of the theory with the generators $(t_\alpha)_B{}^C$ of the U-duality group, $X_{AB}{}^C = \Theta_A{}^\alpha (t_\alpha)_B{}^C$ (we refer to e.g.~\cite{henlect} for a review of the embedding tensor formalism). 
									Using the Leibniz property of the Dorfman derivative together with~\eqref{LeibnizParall}, we see that indeed the constants $X_{AB}{}^C$ realise the gauge algebra
											\begin{equation}\label{eq:gauge-alg-X}
												[X_A,X_B] \ = \ -X_{AB}{}^C X_C\ .
											\end{equation}
									We emphasise that, provided the dimensional reduction goes through consistently, the knowledge of $X_{AB}{}^C$ alone is sufficient to completely determine the resulting gauged maximal supergravity.
					\item[\textit{2.}] One twists the parallelising frame by an $\E_{d+1(d+1)}$ matrix $U_A{}^B$ depending on the external spacetime coordinates $x^\mu$:
											\begin{equation}
												\hat E'_A{}^M(x,z) = U_A{}^B(x)\hat E_B{}^M(z)\ ,
											\end{equation}
									and use this to construct a generalised inverse metric:
											\begin{equation}\label{invG_from_parall}
												\mathcal{G}^{MN}(x,z) = \delta^{AB}\hat E'_A{}^M (x,z) \hat E'_B{}^N (x,z) = \mathcal{M}^{AB}(x) \hat E_A{}^M(z) \hat E_B{}^N(z)\ .
											\end{equation}
									The matrix 
											\begin{equation}
												\mathcal{M}^{AB} = \delta^{CD}U_C{}^A U_D{}^B
											\end{equation}
									parameterises the coset $\E_{d+1(d+1)}/K$, where $K$ is the maximal compact subgroup of $\E_{d+1(d+1)}$ (indeed, we are free to redefine the generalised frame by $x$-dependent $K$ transformations). 
									Hence it accommodates all the scalars of the lower-dimensional theory.

									Now one equates~\eqref{invG_from_parall} to the generic form of the generalised inverse metric $\mathcal{G}^{-1}$ introduced in section~\ref{gen_frame_metric}, whose relevant components are given in~\eqref{invG_comp_1} and~\eqref{invG_comp_2}. 
									In this way we obtain the truncation ansatz for the full set of higher-dimensional degrees of freedom with purely internal components, which gives the scalar fields in the lower-dimensional theory. 
									This also provides the expression for the warp factor $\Delta$. 
									Concretely, these can be extracted following eqs.~\eqref{fields_from_G_first}--\eqref{fields_from_G_last}. 
									Note that, since the generalised density $\Phi$ appearing in~\eqref{fields_from_G_last} is independent of the twist matrix $U_A{}^B$, it can be advantageously computed at the origin of the scalar manifold, where $\mathcal{M}^{AB}=\delta^{AB}$. 
									So at any point on the scalar manifold the density is given by
											\begin{equation}\label{gen_density_background}
												\Phi = \rg{g}{}^{1/2} e^{ -2 \rg\phi} \ e^{(8-d)\rg\Delta} \, ,
											\end{equation}
									where the {\large${\rg{\,}}$} symbol denotes the ``reference'' values of the corresponding fields, namely the values for trivial twist matrix.
					\item[\textit{3.}]	Finally, the full set of vector fields in the lower-dimensional theory is specified by taking the following ansatz for the generalised vector $\mathcal{A}_\mu{}^M$ introduced in~\eqref{def_calA_mu^M},
											\begin{equation}\label{trunc_ansatz_vec}
												\mathcal{A}_\mu{}^M(x,z) = \mathcal{A}_\mu{}^A(x) \hat{E}_A{}^M(z) \, .
											\end{equation}
									The ansatz for the two-forms is
											\begin{equation}\label{ansatz_two-forms}
												\mathcal{B}_{\mu\nu}{}^{MN}(x,z) \eqs \tfrac{1}{2}\,\mathcal{B}_{\mu\nu}{}^{AB}(x) (\hat{E}_A \otimes_{N'} \!\hat{E}_B)^{MN}(z)\,, 
											\end{equation}
									where $\mathcal{B}_{\mu\nu}{}^{AB} = \mathcal{B}_{\mu\nu}{}^{(AB)}$, and the product $\otimes_{N'}$ is defined in~\eqref{N'prod_IIA}.
					
					\end{itemize}
			A few comments in order. 
			Although the conditions in \emph{Step~1} above are definitely non-trivial to satisfy, they are not as constraining as requiring that $M_d$ is a Lie group as needed in ordinary Scherk-Schwarz reductions. 
			Indeed, one can see that a necessary condition for the existence of a generalised parallelisation satisfying~\eqref{LeibnizParall} is that $M_d$ is a coset manifold, $M_d = G/H$ for some $G$ and $H\subset G$~\cite{spheres, petrini3}.
			In the particular case that $M_d$ is a Lie group, a generalised Scherk-Schwarz reduction coincides with an ordinary Scherk-Schwarz reduction if the chosen generalised parallelisation uses just left-invariant tensors\footnote{%
				See~\cite[app. C]{spheres} for a discussion.
				In this case, adopting a generalised geometry approach still has some advantage in that~\eqref{LeibnizParall} directly provides the full embedding tensor.}.
			However, when reducing the NSNS sector, it is possible to obtain a generalised parallelisation which realises a $G\times G$ gauge group rather than just $G$~\cite{Baguet:2015iou}. 
			In the next section we will provide a frame for the full type IIA generalised geometry on $S^3$ which gives rise to an $\SU(2)\times\SU(2)$ gauging (this has also appeared in~\cite{Malek:2015hma}).
			
			The spheres $S^d = \SO(d+1)/SO(d)$ provide examples of generalised parallelisations that are not based on Lie groups.
			In~\cite{spheres}, the ideas above were applied to give evidence that the sphere consistent truncations based on eleven-dimensional supergravity on $S^7$~\cite{deWit:1986oxb}, eleven-dimensional supergravity on $S^4$~\cite{Nastase:1999kf}, type IIB supergravity on $S^5$ and the NSNS sector of type II supergravity on $S^3$, can be interpreted as generalised Scherk-Schwarz reductions. 
			In section~\ref{sec:examples} we will provide additional examples.
			\subsection{Consistent reduction of gauge transformations}
					In this section we provide a partial proof of the consistency of our generalised Scherk-Schwarz truncation ansatz by showing that the internal diffeomorphisms together with the NSNS and RR gauge transformations consistently reduce to the appropriate gauge variations in lower-dimensional maximal supergravity\footnote{%
						A more thorough proof would require studying the reduction of the supersymmetry variations or the equations of motion.}.
					This will also justify the field redefinitions performed in~\eqref{redef_one_forms} and~\eqref{redef_two_forms}. 
					The reader not interested in the details of this computation, which is rather technical, can safely skip to the next section.
					
					The gauge transformations of the ten-dimensional fields were given in section~\ref{sec:IIsugra}. Including also the diffeomorphisms, they read
							\begin{equation}\label{gauge_var_full}
								\begin{split}
									\delta \hat{g} &= \mathcal{L}_{\hat{v}} \hat{g} \,, \\
									\delta \hat{B} &= \mathcal{L}_{\hat{v}} \hat{B} - \dd \hat{\lambda} \ , \\
									\delta \hat{C} &= \mathcal{L}_{\hat{v}} \hat{C} -e^{\hat{B}}\wedge(\dd \hat{\omega} - m \hat{\lambda}) \ , \\
									\delta \hat{\tilde{B}} &= \mathcal{L}_{\hat{v}} \hat{\tilde{B}} - (\dd \hat{\sigma} + m \hat{\omega}_6) - \tfrac{1}{2} [e^{\hat{B}} \wedge (\dd\hat{\omega} - m \hat{\lambda}) \wedge s(\hat{C})]_6 \ .
								\end{split}
							\end{equation}

					We can immediately see why the redefinition of the RR potentials in~\eqref{redef_one_forms} is needed: for the gauge transformation of $C_\mu$ to start with $\partial_\mu \omega$ (as required for a gauge field in supergravity), we need to remove the $B$-terms with internal legs appearing in front of $\dd \omega$. 
					The same argument determines the redefinition of the six-form NSNS potential in~\eqref{redef_one_forms}.

					In order to decompose the gauge transformations, we express the gauge parameters as
							\begin{equation}\label{10d_gauge_param}
								\begin{split}
									{\hat{v}} &= v = v^m \frac{\partial}{\partial z^m}\,, \\[1mm]
									\hat{\lambda} &= \lambda + \overline{\lambda}_\mu = \lambda_m \dd{z}^m + \overline{\lambda}_\mu \dd{x}^\mu \, , \\
\hat\sigma &= \sigma + \overline{\sigma}_\mu + \overline{\sigma}_{\mu\nu} \, =\, \tfrac{1}{5!}\sigma_{m_1\ldots m_5}\dd z^{m_1\ldots m_5} + \tfrac{1}{4!}\,\overline{\sigma}_{\mu m_1\ldots m_4}\dd x^\mu\wedge \dd z^{m_1\ldots m_4}
 \\
\,&\qquad \qquad \qquad \qquad\,\ + \tfrac{1}{2\cdot3!}\,\overline{\sigma}_{\mu\nu m_1\ldots m_3}\dd x^{\mu\nu}\wedge \dd z^{m_1\ldots m_3} + \ldots \, ,
								\end{split}
							\end{equation}
					where the ellipsis denote terms with more than two external indices, that we will ignore. 
					Note that the vector $\hat v$ is purely internal, that is the diffeomorphisms we consider are just the internal ones. Similarly for the RR poly-form gauge parameter we find
							\begin{equation}
								\begin{split}
									\hat\omega = \omega + \overline{\omega}_\mu + \overline\omega_{\mu\nu} & = (\omega_0+\omega_2+\omega_4+\omega_6) + (\overline{\omega}_{\mu,1}+ \overline{\omega}_{\mu,3}+\overline{\omega}_{\mu,5}) \\
															& \phantom{=} + (\overline{\omega}_{\mu\nu,0}+\overline{\omega}_{\mu\nu,2}+\overline{\omega}_{\mu\nu,4}+\overline{\omega}_{\mu\nu,6}) + \ldots \, .
								\end{split}
							\end{equation}
					As in~\eqref{expand_10dfields}, initially we impose no restriction on the dependence of the components of the gauge parameters on the coordinates $\{x^\mu,z^m\}$. 
					However, differently from~\eqref{expand_10dfields}, note that the expansion of the gauge parameters is made in $\dd{z}^m$ and not in $Dz^m=\dd z^m - h_\mu{}^m \dd x^\mu$. 
					The fields marked with a bar require a redefinition, which will be introduced below.

					The gauge transformations of the fields with purely internal legs maintain precisely the same form as in~\eqref{gauge_var_full}. 
					As for the fields with one external leg, redefined as in~\eqref{redef_one_forms}, after some computation we find that their variations are
							\begin{equation}\label{var_one_ext_index}
								\begin{split}
									\delta h_\mu &= -\partial_\mu v + \mathcal{L}_v h_\mu \, , \\
									\delta B_\mu &= -\partial_\mu \lambda + \din \overline{\lambda}_\mu + \mathcal{L}_v B_\mu - \iota_{h_\mu} \din\lambda\, , \\
									\delta \tilde B_\mu &= -\partial_\mu \sigma + \din{\overline{\sigma}_{\mu}} - m\,\overline{\omega}_{\mu,5} + \mathcal{L}_v \tilde{B}_{\mu} \\
												& \phantom{=} - \iota_{h_\mu}(\din\sigma+m\omega_6) + \left[ C_\mu \wedge s(\din \omega - m\lambda) \right]_5 \, , \\
									\delta C_\mu &= - \partial_\mu \omega + \din \overline{\omega}_\mu + m \overline{\lambda}_\mu + \mathcal{L}_v C_\mu + C_{\mu} \wedge \din \lambda \\
												& \phantom{=} - (\iota_{h_\mu}+ B_\mu \wedge)(\din\omega -m\lambda)\,,
								\end{split}
							\end{equation}
					where the exterior derivative $\din := \dd z^m \partial_m$ and $\mathcal{L}$ act on the internal coordinates only.
					The fields with two external legs have the following gauge variations
							\begin{equation}\label{transf_2form_BtildeB}
								\begin{split}
									\delta B_{\mu\nu} &= -2 \partial_{[\mu} \overline{\lambda}_{\nu]} +\iota_{h_{[\mu}}\partial_{\nu]} \lambda - \iota_{h_{[\mu}}\din \overline{\lambda}_{\nu]} + \mathcal{L}_v B_{\mu\nu} - \iota_{\partial_{[\mu}v} B_{\nu]}\, , \\[2mm]
									\delta \tilde B_{\mu\nu} &= - 2 \partial_{[\mu}\overline{\sigma}_{\nu]} -\din \overline{\sigma}_{\mu\nu} -m\, \overline{\omega}_{\mu\nu,4} + \iota_{h_{[\mu}}\big(\partial_{\nu]}\sigma - \din \overline{\sigma}_{\nu]} + m \,\overline{\omega}_{\nu],5} \big) \\
														&\phantom{= \, } + \mathcal{L}_v \tilde B_{\mu\nu}- \iota_{\partial_{[\mu}v}\tilde{B}_{\nu]} \\
														&\phantom{= \, } + \big[C_{\mu\nu} \wedge s(\din\omega-m\lambda) + ( -\partial_{[\mu}\omega + \din \overline{\omega}_{[\mu} + m\overline{\lambda}_{[\mu} ) \wedge s(C_{\nu]}) \big]_4 \, ,
								\end{split}
							\end{equation}
					and (we give the transformations for the barred fields, as those of the unbarred field $C_{\mu\nu}$ are more cumbersome)
							\begin{equation}\label{Cbar_transf}
								\begin{split}
									\delta (e^{-B} \wedge \overline{C}_{\mu\nu}) &= - 2\partial_{[\mu} \overline{\omega}_{\nu]} - 2\iota_{h_{[\mu}}\din \overline{\omega}_{\nu]} +2\iota_{h_{[\mu}}\partial_{\nu]}\omega -\iota_{h_{\nu}} \iota_{h_\mu} \din \omega - \din \overline{\omega}_{\mu\nu} \\
																		&\phantom{= \,} + \mathcal{L}_v (e^{-B}\wedge\overline{C}_{\mu\nu}) +\din\lambda \wedge (e^{-B}\wedge\overline{C}_{\mu\nu}) - \overline{B}_{\mu\nu} (\din \omega -m\lambda) \\
																		&\phantom{= \, } + 2 B_{[\mu} \wedge \iota_{h_{\nu]}}(\din \omega -m\lambda) + 2 B_{[\mu}\wedge ( \partial_{\nu]} \omega - \din\overline{\omega}_{\nu]} -m\overline{\lambda}_{\nu]}) \\
																		&\phantom{= \, } + B_{\mu} \wedge B_\nu \wedge \left(\din\omega - m \lambda\right) \,.
								\end{split}
							\end{equation}
					The gauge parameters with purely internal indices can be arranged into a generalised vector with the $T^*\otimes \Lambda^6T^*$ component projected out,
							\begin{equation}
								\Lambda^M \eqs \{ v^m,\, \lambda_m,\, \sigma_{m_1\ldots m_5},\, \omega_0,\, \omega_{m_1m_2} ,\,\omega_{m_1\ldots m_4},\,\omega_{m_1\ldots m_6} \} \, ,
							\end{equation}
					while the gauge parameters with one external leg form a section of the bundle $N'$,
							\begin{equation}
								\overline{\Xi}_\mu^{\phantom{\mu}(MN)} \eqs \{ \overline{\lambda}_\mu ,\, \overline{\sigma}_{\mu n_1\ldots n_4} ,\, \overline{\omega}_{\mu n},\, \overline{\omega}_{\mu n_1n_2n_3} ,\, \overline{\omega}_{\mu n_1\ldots n_5} \} \, .
							\end{equation}
					The transformations for the fields with two external legs will be discussed below.
 
					The gauge transformation of the fields with purely internal indices is given by the compact expression
							\begin{equation}\label{gauge_var_G}
								\delta_\Lambda \mathcal{G}^{-1} = L_\Lambda \mathcal{G}^{-1}\,,
							\end{equation}
					where $L_\Lambda$ is the massive Dorfman derivative~\eqref{dorfIIAm}.
					The gauge variation~\eqref{var_one_ext_index} of fields with one external leg can be repackaged into
							\begin{equation}\label{Avaria0}
								\delta \mathcal{A}_\mu \, \eqs\, -\partial_\mu \Lambda + L_\Lambda \mathcal{A}_\mu + \dd_{\mathrm{m}} \overline{\Xi}_\mu \, ,
							\end{equation}
					where it is understood that the differentials in the generalised Lie derivative act on the internal coordinates only. The operator $\dd_{\mathrm{m}}$ is defined on any element $W = W_0 + W_4 + W_{\rm odd}$ of the bundle $N'$ as
							\begin{equation}\label{m_twist_ext_der}
								\dd_{\mathrm{m}} W = \dd W + m(W_0 - W_5)\,,
							\end{equation}
					and can be seen as an exterior derivative twisted by the Romans mass. Then in the present case we have
							\begin{equation}
								\dd_{\mathrm{m}} \overline{\Xi}_\mu =\din \overline{\Xi}_\mu + m (\overline{\lambda}_\mu - \overline{\omega}_{\mu,5})\,.
							\end{equation}
					It is easy to verify that for $W = V \otimes_{N'} V' $,
							\begin{equation}\label{eq:symmetricL}
								\dd_{\mathrm{m}} W \,\eqs\, L_V V' + L_{V'}V \,\,.
							\end{equation}
					If we now redefine the gauge parameter with one external leg as
							\begin{equation}\label{redefXi}
								\overline{\Xi}_\mu\, =\, \Xi_\mu - \mathcal{A}_\mu \xN \Lambda\,,
							\end{equation}
					and use the property~\eqref{eq:symmetricL}, we obtain 
							\begin{equation}
								\dd_{\mathrm{m}} \overline{\Xi}_\mu \,\eqs\, \dd_{\mathrm{m}} \Xi_\mu - L_{\mathcal{A}_\mu} \Lambda - L_\Lambda \mathcal{A}_\mu\, . 
							\end{equation}
					This redefinition allows to cast~\eqref{Avaria0} in the form
							\begin{equation} \label{Avaria}
								\delta \mathcal{A}_\mu \, \eqs\, -\partial_\mu \Lambda - L_{\mathcal{A}_\mu} \Lambda + \dd_{\mathrm{m}} \Xi_\mu \,,
							\end{equation}
					where one recognise the derivative $(\partial_\mu + L_{\mathcal A_\mu})\Lambda$, covariant under generalised diffeomorphisms. 
					This is the appropriate form for matching the gauged supergravity covariant derivative after Scherk-Schwarz reduction.

					We need to express the gauge transformations~\eqref{transf_2form_BtildeB} and~\eqref{Cbar_transf} of the external two-form fields in generalised geometry terms. 
					This requires a rather complicated redefinition of the gauge parameters $\overline{\omega}_{\mu\nu}= \overline{\omega}_{\mu\nu,0}+\overline{\omega}_{\mu\nu,2}+\overline{\omega}_{\mu\nu,4}$ and $\overline{\sigma}_{\mu\nu}$:
							\begin{equation}\label{expr_2form_gauge_par}
								\begin{split}
									\overline{\omega}_{\mu\nu} &= \omega_{\mu\nu} + (\iota_v+\lambda\wedge) C_{\mu\nu} - \omega\,B_{\mu\nu} + (2\lambda_{[\mu} + \iota_{h_{[\mu}}\lambda + \iota_v B_{[\mu})C_{\nu]} \\
														&\phantom{=} + (\iota_{h_{[\mu}} + B_{[\mu}\wedge\, )( 2\omega_{\nu]}+\iota_v C_{\nu]} + \lambda\wedge C_{\nu]} + \iota_{h_{\nu]}}\omega + B_{\nu]}\wedge \omega) \\[2mm]
									\overline{\sigma}_{\mu\nu} &= \sigma_{\mu\nu} + 2\iota_{h_{[\mu}}\sigma_{\nu]} +\iota_{h_\mu}\iota_{h_\nu}\sigma + \iota_v(\tilde{B}_{\mu\nu} - \iota_{h_{[\mu}}\tilde{B}_{\nu]}) + \iota_v C_{[\mu,4}C_{\nu],0} \\
														&\phantom{=} -\iota_v C_{[\mu,2}\wedge C_{\nu],2} + 2\lambda\wedge (C_{[\mu,2}C_{\nu],0}) \\
														&\phantom{=} - \big[(C_{\mu\nu} - \iota_{h_{[\mu}}C_{\nu]} + B_{[\mu}\wedge C_{\nu]})\wedge s(\omega) - 2C_{[\mu}\wedge s(\omega_{\nu]})\big]_3\,.
								\end{split}
							\end{equation}
					We repackage the new parameters $\sigma_{\mu\nu}$ and $\omega_{\mu\nu} = \omega_{\mu\nu,0} +\omega_{\mu\nu,2}+\omega_{\mu\nu,4}$ into
							\begin{equation}
								\Phi_{\mu\nu}= \sigma_{\mu\nu} + \omega_{\mu\nu}\,.
							\end{equation}
					This object lives in a sub-bundle of a bundle transforming in the $\mathbf{912}$ representation of $\E_{7(7)}$ (see~\cref{tab:EddRep}), and collects the gauge parameters of the potentials that are three-forms in the external spacetime. 
					One can then show that, with the identifications~\eqref{expr_2form_gauge_par}, the gauge transformations for $B_{\mu, \nu}$, $\tilde B_{\mu \nu}$,~\eqref{transf_2form_BtildeB}, and $C_{\mu\nu}$ (these follow from~\eqref{Cbar_transf} and the last in~\eqref{redef_two_forms}) can be expressed as
							\begin{equation}\label{delta_calB_munu}
								\begin{split}
									\delta \mathcal{B}_{\mu\nu} =& -2\partial_{[\mu} \overline{\Xi}_{\nu]} - 2 L_{\mathcal A_{[\mu}} \overline{\Xi}_{\nu]} - \dd_{\mathrm{m}} \overline{\Xi}_{[\mu} \xN \mathcal{A}_{\nu]} - \partial_{[\mu} \Lambda \xN \mathcal{A}_{\nu]} \\
															& + \dd_{\mathrm{m}} \mathcal{B}_{\mu\nu}\xN \Lambda - Y_{\mu\nu} - \dd_{\mathrm{m}}\Phi_{\mu\nu}\,,
								\end{split}
							\end{equation}
					where the action of $\dd_{\mathrm{m}}$ on an element of $N'$ is given in~\eqref{m_twist_ext_der}, and we define 
							\begin{equation}\label{dmOnPhimunu}
								\dd_{\mathrm{m}}\Phi_{\mu\nu} = \din (\sigma_{\mu\nu} + \omega_{\mu\nu}) + m \omega_{\mu\nu,4} \,.
							\end{equation}
					The tensor $Y_{\mu\nu}$ is given in terms of $W_\nu \equiv \mathcal{A}_\nu \xN \Lambda$ by
							\begin{equation}
								\begin{split}
									Y_{\mu\nu} =	& \dd \big( \iota_{h_{[\mu}} W_{\nu]} + B_{[\mu}\wedge W_{\nu],{\rm odd}} - C_{[\mu}W_{\nu],0} - C_{[\mu,0} W_{\nu],3} + C_{[\mu,2}W_{\nu],1} \big) \\
												& + m\big(\iota_{h_{[\mu}} W_{\nu],5} + B_{[\mu}\wedge W_{\nu],3} -C_{\mu,4}W_{\nu,0} \big) \, .
								\end{split}
							\end{equation}
					After some manipulations, this can be re-expressed as
							\begin{equation}
								Y_{\mu\nu}= L_{\mathcal{A_{[\mu}}}\mathcal{A}_{\nu]} \xN \Lambda + 2 L_{\mathcal{A_{[\mu}}}\Lambda \xN \mathcal{A}_{\nu]} + L_\Lambda \mathcal{A}_{[\mu} \xN \mathcal{A}_{\nu]}\,,
							\end{equation}
					which in turn allows to rewrite~\eqref{delta_calB_munu} as
							\begin{equation}
								\begin{split}
									\delta \mathcal{B}_{\mu\nu} =	& -2\partial_{[\mu} \overline{\Xi}_{\nu]} - 2 L_{\mathcal A_{[\mu}} \overline{\Xi}_{\nu]} - \dd_{\mathrm{m}} \overline{\Xi}_{[\mu} \xN \mathcal{A}_{\nu]} - \partial_{[\mu} \Lambda \xN \mathcal{A}_{\nu]} \\
															&+ \dd_{\mathrm{m}} \mathcal{B}_{\mu\nu}\xN \Lambda - L_{\mathcal{A_{[\mu}}}\mathcal{A}_{\nu]} \xN \Lambda - 2 L_{\mathcal{A_{[\mu}}}\Lambda \xN \mathcal{A}_{\nu]} \\
															& - L_\Lambda \mathcal{A}_{[\mu} \xN \mathcal{A}_{\nu]} - \dd_{\mathrm{m}}\Phi_{\mu\nu} \, .
								\end{split}
							\end{equation}
					Introducing the gauge field strength
							\begin{equation}\label{eq:defHmunu}
								\mathcal{H}_{\mu\nu} = 2 \partial_{[\mu}\mathcal{A}_{\nu]} + L_{\mathcal{A}_{[\mu}}\mathcal A_{\nu]} + \dd_{\mathrm{m}} \mathcal{B}_{\mu\nu}\,,
							\end{equation}
					and recalling the expression for $\delta{\mathcal A}_\mu$ given in~\eqref{Avaria} and the redefinition of the gauge parameter $\overline \Xi_\mu$ in~\eqref{redefXi}, the variation of $\mathcal{B}_{\mu\nu}$ eventually takes the compact form
							\begin{equation}\label{var_Bmunu}
								\delta \mathcal{B}_{\mu\nu} = -2\partial_{[\mu} \Xi_{\nu]} - 2 L_{\mathcal A_{[\mu}} \Xi_{\nu]} + \Lambda\xN \mathcal{H}_{\mu\nu} + \mathcal{A}_{[\mu} \xN \delta\mathcal{A}_{\nu]} - \dd_{\mathrm{m}}\Phi_{\mu\nu} \, .
							\end{equation}

					We can now plug in our truncation ansatz and show that it reproduces the correct lower-dimensional gauge-transformations. 
					For the gauge parameters we take an ansatz similar to the one for the physical fields, that is
							\begin{equation}
								\begin{split}
									\Lambda^M(x,z) &= -\Lambda^A(x) \hat{E}_A{}^M(z)\ , \\
									\widetilde{\Xi}_\mu{}^{MN}(x,z) &= -\tfrac{1}{2}\,\widetilde{\Xi}_\mu{}^{AB}(x)\, (\hat{E}_A{} \otimes_N \!\hat{E}_B)^{MN}(z)\,.
								\end{split}
							\end{equation}

					Plugging the ansatz into the variation~\eqref{gauge_var_G} of the generalised metric, and using the action~\eqref{LeibnizParall} of the generalised Lie derivative on the parallelisation, we obtain
							\begin{equation}
								\delta_{\Lambda} \mathcal{M}^{AB} = -\Lambda^C (X_{CD}{}^A \mathcal{M}^{DB} + X_{CD}{}^B\mathcal{M}^{AD}) \, ,
							\end{equation}
					which is the correct variation of the scalar fields in gauged maximal supergravity, see e.g.~\cite{henlect}.

					In order to write the variation of $\mathcal A_\mu$, let us first observe that the ansatz together with the property~\eqref{eq:symmetricL} implies
							\begin{equation}
								\dd_{\mathrm{m}} \Xi_\mu \eqs -\tfrac{1}{2}(L_{\hat{E}_B}\hat{E}_C + L_{\hat{E}_C}\hat{E}_B)\, \Xi_\mu{}^{BC} = - Z^A{}_{BC}\, \Xi_\mu{}^{BC}\hat{E}_A \, ,
							\end{equation}
					where we introduced the symmetrised structure constants $Z^A{}_{BC}=X_{(BC)}{}^A$.
					Then, interpreting the variation of $\mathcal A_\mu$ in the~\eqref{Avaria} as $(\delta \mathcal{A}_\mu{}^A) \hat{E}_A $ and plugging the ansatz in, we get
							\begin{equation}
								\delta \mathcal{A}_\mu{}^A = \partial_\mu \Lambda^A + \mathcal{A}_\mu^B X_{BC}{}^A \Lambda^C - Z^A{}_{BC}\, \Xi_\mu{}^{BC} \, .
							\end{equation}
					This is the correct gauge variation of the gauge fields in maximal supergravity (see again~\cite{henlect}).

					Finally, we need to consider the transformation of $\mathcal{B}_{\mu\nu}$.
					Equation~\eqref{eq:defHmunu} yields
							\begin{equation}
								\mathcal{H}_{\mu\nu} = \mathcal{H}_{\mu\nu}^A \hat{E}_A \,,
							\end{equation}
					with
							\begin{equation}\label{eq:defHmunu_comp}
								\mathcal{H}_{\mu\nu}^{A} = 2 \partial_{[\mu} \mathcal{A}_{\nu]}{}^A + X_{BC}{}^A \mathcal{A}_{[\mu}{}^B\mathcal{A}_{\nu]}{}^C + Z^{A}{}_{BC} \, \mathcal{B}_{\mu\nu}{}^{BC}\,.
							\end{equation}
					This is the expression for the covariant field strengths used in gauged supergravity.
					We also obtain
							\begin{equation}
								\begin{split}
									L_{\mathcal{A}_\mu} \Xi_\nu &= -\tfrac{1}{2}\mathcal{A}_\mu{}^C\, \Xi_\nu{}^{AB} L_{\hat{E}_C}( \hat{E}_A\otimes_{N'}\!\hat{E}_B) \\
														&= - \mathcal{A}_\mu{}^C \,\Xi_\nu{}^{(DA)} X_{CD}{}^B \hat{E}_A\otimes_{N'}\!\hat{E}_B\, ,
								\end{split}
							\end{equation}
					where to pass from the first to the second line we distributed the Lie derivative on the factors of the $\otimes_{N'}$ product and used the Leibniz property of the generalised frame. Therefore:
							\begin{equation}
								-2\partial_{[\mu} \Xi_{\nu]} - 2 L_{\mathcal A_{[\mu}} \Xi_{\nu]} = D_{[\mu} \Xi_{\nu]}{}^{AB} \hat{E}_A \xN \hat{E}_B\,,
							\end{equation}
					where
							\begin{equation}
								D_{[\mu} \Xi_{\nu]}{}^{AB} = \partial_{[\mu} \Xi_{\nu]}{}^{AB} + 2\mathcal{A}_{[\mu}{}^C \,\Xi_{\nu]}{}^{(DA)} X_{CD}{}^B \,.
							\end{equation}

					Putting everything together,~\eqref{var_Bmunu} eventually takes the appropriate form to describe the two-form gauge transformations in gauged supergravity:
							\begin{equation}
								\delta \mathcal{B}_{\mu\nu}{}^{AB} = 2 \,D_{[\mu} \Xi_{\nu]}{}^{AB} - 2\, \Lambda^{(A} \mathcal{H}_{\mu\nu}{}^{B)} + 2\,\mathcal{A}_{[\mu}{}^{(A} \delta\mathcal{A}_{\nu]}{}^{B)} + \ldots\,,
							\end{equation}
					where $\mathcal{H}_{\mu\nu}^{A}$ was given in~\eqref{eq:defHmunu_comp}. 
					The ellipsis denote a term coming from expressing $\dd_{\mathrm{m}}\Phi_{\mu\nu}$ in~\eqref{var_Bmunu} by means of the parallelisation that we will not discuss in detail. 
					This eventually gives the two-form gauge parameters in the lower-dimensional supergravity theory, contracted with the gauge group generators $X$. 
					In four-dimensional supergravity, this term drops from all relevant equations, because the two-forms $\mathcal{B}_{\mu\nu}{}^{AB}$ always appear contracted with the embedding tensor, namely as $Z^A{}_{BC}\mathcal{B}_{\mu\nu}{}^{BC}$~\cite{deWit:2007kvg}, which implies that the term in the ellipsis is projected out due to the quadratic constraint. 
					From a generalised geometry perspective, the corresponding statement is that in a reduction to four dimensions~\eqref{var_Bmunu} always appears under the action of the exterior derivative twisted by the Romans mass, $\dd_{\mathrm{m}}$; given the definitions~\eqref{dmOnPhimunu} and~\eqref{m_twist_ext_der}, it is immediate to check that $\dd_{\mathrm{m}}(\dd_{\mathrm{m}}\Phi_{\mu\nu})=0$, hence the gauge parameters with two external indices drop from all relevant equations. 
					This is no longer the case in reductions to supergravities in dimension six or higher, where the tensor hierarchy stops at one form degree higher, so that the three-form gauge potentials, as well as their two-form gauge parameters, also play a role. 

					In conclusion, we have shown that under the generalised Scherk--Schwarz ansatz, the (massive) type IIA gauge transformations consistently reduce to the correct gauge transformation in lower-dimensional supergravity.	
			\subsection{Examples of consistent sphere truncations}\label{sec:examples}
					In this section, we apply the generalised Scherk-Schwarz procedure to study consistent reductions of massless and massive type IIA supergravity on the spheres $S^6$, $S^4$, $S^3$ and $S^2$, as well as on six-dimensional hyperboloids. 
					While for the massless case it is always possible to find generalised parallelisations that reproduce the known reductions to maximal gauged supergravities in lower-dimensions, for the massive theory we could only find a suitable generalised parallelisation on $S^6$. 
					We propose a general argument of why this is the case.
					The analysis closely follows the one given in~\cite{oscar1}.

				\subsubsection{\texorpdfstring{$S^6$ parallelisation and $D=4$, $\ISO(7)_m$ supergravity}{S6 parallelisation and D=4, ISO(7)m supergravity}}
						We start our series of examples by revisiting the consistent reduction of type IIA supergravity on the six-sphere $S^6$ down to $D=4$ maximal supergravity with $\ISO(7)$ gauge group that was recently studied in detail in~\cite{Guarino:2015jca,Guarino:2015qaa,Guarino:2015vca}. 
						For vanishing Romans mass, this reduction can be understood as a limit of the consistent truncation of eleven-dimensional supergravity on $S^7$ (or on a seven-dimensional hyperboloid), where the seven-dimensional manifold degenerates into the cylinder $S^6 \times \mathbb {R}$~\cite{Hull:1988jw,Boonstra:1998mp}. 
						In that case the group $\ISO(7)$ is gauged purely electrically. 
						This means that only the $28$ electric vector fields participate in the gauging, while the $28$ magnetic duals do not appear in the Lagrangian. 
						When the Romans mass $m$ is switched on, the truncation ansatz remains consistent with no modifications required. 
						However one finds that the magnetic vectors now also enter in the gauge covariant derivatives~\cite{Guarino:2015vca}, thus providing a \emph{dyonic} gauging. 
						The resulting four-dimensional supergravity is not equivalent to the theory with purely electric $\ISO(7)$ gauging~\cite{Dall'Agata:2014ita}; for this reason, we will denote it as the $\ISO(7)_m$ theory.
						This is an example of \emph{symplectic deformation} of maximal supergravity of the type first discovered for the $D=4$, $\SO(8)$ theory in~\cite{Dall'Agata:2012bb}. 
						The $\ISO(7)_m$ theory admits several supersymmetric and non-supersymmetric AdS$_4$ solutions~\cite{DallAgata:2011aa,Gallerati:2014xra,Guarino:2015qaa}, which all disappear when the parameter $m$ is sent to zero\footnote{%
							Specific formulae uplifting these AdS$_4$ vacua to massive type IIA supergravity were given in~\cite{Guarino:2015jca,Guarino:2015vca,Varela:2015uca}. 
							Three of them are $G_2$-invariant and also included in the truncation of massive IIA supergravity on $S^6\simeq G_2/\SU(3)$ of~\cite{Cassani:2009ck}.%
							}.
						The structure of the $\ISO(7)_m$ theory was analysed in detail in~\cite{Guarino:2015qaa}.
						
						In the following, we introduce a parallelisation of the $\E_{7(7)}\times \RR^+$ tangent bundle on $S^6$. 
						Then, evaluating our massive generalised Lie derivative on the frame we obtain precisely the embedding tensor characterising the dyonic $\ISO(7)_m$ gauging. 
						We also re-derive the truncation ansatz for the four-dimensional bosonic fields from generalised geometry.
						
						A generalised parallelisation on $S^6$ is defined as follows. 
						Let $y^i, i = 1, \dots , 7$, with $\delta_{ij} y^i y^j = 1$, be the constrained coordinates on $S^6$, describing its embedding in $\RR^7$ (see appendix~\ref{app:constrcoo} for some useful details about spheres in constrained coordinates). 
						Let $v_{ij}$ be the $\SO(7)$ Killing vectors and define the following forms
							\begin{equation}\label{deflof}
								\begin{array}{l c r}
									\omega^{ij} = R^2 \dd y^i \wedge \dd y^j					& \phantom{T^*\otimes \Lambda^6T^*} 	& \in \Lambda^2 T^*\, , \\
									\rho_{ij} = \rg{*}(R^2 \dd y_i \wedge \dd y_j)				& \phantom{T^*\otimes \Lambda^6T^*} 	& \in \Lambda^4 T^*\, , \\
									\kappa_i = -\rg{*}(R \dd y_i) 							& \phantom{T^*\otimes \Lambda^6T^*} 	& \in \Lambda^5 T^*\, , \\
									\tau^{ij} = R (y^i \dd y^j - y^j \dd y^i)\otimes \rg{\rm vol}_6 & \phantom{T^*\otimes \Lambda^6T^*} 		& \in T^*\otimes \Lambda^6T^*\, .
								\end{array}
							\end{equation}

					Here and in the rest of this section, the symbol $\rg{}$ means that the corresponding quantity is computed using the reference round metric of radius $R$. 
					The index on the coordinates $y^i$ is lowered with the $\RR^7$ metric $\delta_{ij}$.
					We also twist the generalised tangent bundle with a five-form RR potential $\rg{C_5}$ such that 
							\begin{equation}\label{F6onS6}
								\rg{F_6} =\dd \rg{C_5} = \frac{5}{R} \rg{\rm vol}_6 \, ,
							\end{equation} 
					with all other $p$-form potentials vanishing; the reason for this choice will become clear soon.

					The generalised frame can be split according to the decomposition 
							\begin{equation}\label{decomp_E7}
								\begin{split}
									\E_{7(7)} &\supset\; \SL(8,\mathbb R) \supset \SL(7,\mathbb R) \\
											\mathbf{56}\, &\to\, \mathbf{28} + \mathbf{28'} \mapsto \mathbf{21} + \mathbf{7} + \mathbf{21'}+ \mathbf{7'} \, .
								\end{split}
							\end{equation}
					as
							\begin{equation}
								\{\hat{E}_A\} = \{\hat{E}_{IJ}, \hat{E}^{IJ}\} = \{\hat{E}_{ij},\hat{E}_{i8},\hat{E}^{ij},\hat{E}^{i8}\}\ .
							\end{equation}
					We will call ``electric'' the $\hat{E}_{IJ}$ frame elements, transforming in the $\mathbf{28}$ of $\SL(8)$, and ``magnetic'' the $\hat{E}^{IJ}$, transforming in the $\mathbf{28'}$.

					A generalised parallelisation is given by
							\begin{equation}\label{s6frame}
								\hat{E}_{A} = \begin{cases}
											\hat{E}_{ij} = v_{ij} + \rho_{ij} + \iota_{v_{ij}}\!\rg{C_5}\ , \\
											\hat{E}_{i8} = y_i + \kappa_i - y_i \rg{C_5}\ , \\
											\hat{E}^{ij} = -\omega^{ij} - \tau^{ij} + j \!\rg{C_5} \wedge \omega^{ij}\ , \\
											\hat{E}^{i8} = R\,\dd y^i - y^i \rg{\rm vol}_6 + R\,\dd y^i \wedge \rg{C_5} \ .
										\end{cases}
							\end{equation}
					It is not hard to see that this is globally defined. 
					For instance, $\hat{E}_{ij}$ is nowhere vanishing as the Killing vectors $v_{ij}$ vanish at $y_i = y_j =0$, while the four-forms $\rho_{ij}$ vanish at $y_i^2+y_j^2=1$. 
					Moreover, $\hat{E}_{i8}$ never vanishes as the locus $\kappa_i = 0$ does not overlap with $y_i =0$; similar considerations hold for the magnetic part of the frame.
					The frame is also orthonormal with respect to the generalised metric~\eqref{GofVandV'}. 
					Indeed, invoking the contraction formulae in~\eqref{contractions_sphere}, we have
							\begin{equation}
								\begin{split}
									\mathcal{G}(\hat{E}_{ij} , \hat{E}_{kl}) &= v_{ij}\lrcorner v_{kl} + \rho_{ij} \lrcorner \rho_{kl} = \delta_{ik}\delta_{jl} - \delta_{il}\delta_{jk} \, , \\
									\mathcal{G}(\hat{E}_{i8} , \hat{E}_{k8}) &= y_i \, y_k + \kappa_i\lrcorner \kappa_k = \delta_{ik}\ , \\
									\mathcal{G}(\hat{E}^{ij} , \hat{E}^{kl}) &= \omega^{ij}\lrcorner \omega^{kl} + \tau^{ij} \lrcorner \tau^{kl} = \delta^{ik}\delta^{jl} - \delta^{il}\delta^{jk} \, , \\
									\mathcal{G}(\hat{E}^{i8} , \hat{E}^{k8}) &= R^2 \dd y^i \lrcorner\dd y^k + y^iy^k \rg{\vol}_6\lrcorner \rg{\vol}_6 = \delta^{ik} \ ,
								\end{split}
							\end{equation}
					with all other pairings vanishing.

					We now evaluate the massive Dorfman derivative~\eqref{dorfIIAm} between two arbitrary frame elements, making use of various properties of the round spheres given in appendix~\ref{app:constrcoo}.
					In particular, we need identity~\eqref{eq:contrvol}, which together with our choice~\eqref{F6onS6} for $\rg{C_5}$ implies
							\begin{equation}
								\iota_{v_{ij}}\rg{F_6}= \dd \rho_{ij} \, .
							\end{equation}
					We find that the electric-electric pairings give
							\begin{equation}
								\begin{split}
									L_{\hat{E}_{ij}} \hat{E}_{kl} &= \tfrac{2}{R}\big( \delta_{i[k}\hat{E}_{l]j} - \delta_{j[k}\hat{E}_{l]i} \big) \, ,\\
									L_{\hat{E}_{ij}}\hat{E}_{k8} &= -\tfrac{2}{R} \delta_{k[i}\hat{E}_{j]8} \, ,\\
									L_{\hat{E}_{i8}} \hat{E}_{kl} &= \tfrac{2}{R} \delta_{i[k}\hat{E}_{l]8} \, ,\\
									L_{\hat{E}_{i8}}\hat{E}_{k8} &= 0\ ,
								\end{split}
							\end{equation}
					while for the electric-magnetic ones we have
							\begin{equation}
								\begin{split}
									L_{\hat{E}_{ij}}\hat{E}^{kl} &= \tfrac{4}{R} \delta_{[i}^{[k} \delta_{j]j'} \hat{E}^{l]j'}\, , \\
									L_{\hat{E}_{ij}}\hat{E}^{k8} &= - \tfrac{2}{R} \delta^k_{[i} \delta_{j]j'} \hat{E}^{j'8} \, ,\\
									L_{\hat{E}_{i8}}\hat{E}^{kl} &= 0 \, ,\\
									L_{\hat{E}_{i8}}\hat{E}^{k8} &= -\tfrac{1}{R}\delta_{ij} \hat{E}^{jk}\, ,
								\end{split}
							\end{equation}
					for the magnetic-electric 
							\begin{equation}
								\begin{split}
									L_{\hat{E}^{ij}}\hat{E}_{kl} &= L_{\hat{E}^{ij}}\hat{E}_{k8} = L_{\hat{E}^{i8}}\hat{E}_{k8} = 0\, , \\
									L_{\hat{E}^{i8}}\hat{E}_{kl} &= -2\, m \,\delta^i_{[k} \hat{E}_{l]8}\, ,
								\end{split}
							\end{equation}
					and for the magnetic-magnetic
							\begin{equation}
								\begin{split}
									L_{\hat{E}^{ij}}\hat{E}^{kl} &= L_{\hat{E}^{ij}}\hat{E}^{k8} \, =\, L_{\hat{E}^{i8}}\hat{E}^{kl} = 0 \ ,\\
									L_{\hat{E}^{i8}}\hat{E}^{k8} &= m\, \hat{E}^{ik}\ .
								\end{split}
							\end{equation}
					We thus obtain that condition~\eqref{LeibnizParall} is satisfied, namely the frame defines a Leibniz algebra under the massive Dorfman derivative. 
					The non-vanishing constants $X_{AB}{}^C$ read in $\SL(8)$ indices
							\begin{equation}\label{ISO7m_emb_tens}
								\begin{split}
									X_{[II'][JJ']}{}^{[KK']} 		&= - X_{[II']}{}^{[KK']}{}_{[JJ']} = 8 \, \delta_{[I}^{[K} \, \theta_{I'][J} \ \delta_{J']}^{K']}\ , \\[2mm]
									X^{[II']}{}_{[JJ']}{}^{[KK']} 	&= - X^{[II'][KK']}{}_{[JJ']} = 8\, \delta^{[I}_{[J} \, \xi^{I'][K} \ \delta^{K']}_{J']}\ ,
								\end{split}
							\end{equation}
					with
							\begin{align}
								& &\theta_{IJ} = \frac{1}{2R} \mathrm{diag}\big(\underbrace{1,\ldots,1}_{7},0\big)\, ,& & \xi^{IJ} = \frac{m}{2} \,{\rm diag}\big(\underbrace{0,\ldots,0}_{7},1\big)\, . & & 
							\end{align}
					These match precisely the embedding tensor given in~\cite{Guarino:2015qaa} (modulo renormalising the generators by a $-1/2$ factor, see appendix C therein). 
					The latter determines a dyonic $ISO(7)_m$ gauging of maximal $D=4$ supergravity, where the $\SO(7)$ rotations are gauged electrically while the seven translations are gauged dyonically. 
					When $m=0$, we have $\xi^{IJ} =0$ and the $\ISO(7)$ gauging becomes purely electric.
					
					Following the procedure for a generalised Scherk-Schwarz reduction described in the previous section, we can use our generalised parallelisation to deduce the truncation ansatz for the bosonic supergravity fields. 
					We start from the scalar ansatz. 
					In four-dimensional maximal supergravity, the scalar matrix $\mathcal{M}^{AB}$ parameterises the coset $\E_{7(7)}/\SU(8)$. 
					Under the decomposition~\eqref{decomp_E7}, this splits as 
							\begin{equation}
								\begin{split}
									\mathcal{M}^{AB} &= \{ \mathcal{M}^{II',JJ'}, \, \mathcal{M}^{II'}{}_{JJ'} ,\,\mathcal{M}_{II'}{}^{JJ'} ,\,\mathcal{M}_{II',JJ'}\} \\
												& = \{ \mathcal{M}^{ii',jj'}, \, \mathcal{M}^{ii',j8},\, \ldots ,\,\mathcal{M}_{i8,j8}\}\ .
								\end{split}
							\end{equation}
					Equating the components~\eqref{invG_comp_1} of the inverse generalised metric to those constructed from the parallelisation as in~\eqref{invG_from_parall}, we obtain
							\begin{equation}\label{scalar_ansatz_S6}
								\begin{aligned}{l}
									e^{2\Delta} g^{mn} &= \tfrac{1}{4}\mathcal{M}^{ii',jj'} v^m_{ii'}\, v^n_{jj'}\ , \\
									e^{2\Delta} g^{mn}C_n &= \tfrac{1}{2}\mathcal{M}^{ii',j8}\, v^m_{ii'}\,y_{j}\ , \\
									-e^{2\Delta} g^{mp}B_{pn} &= \tfrac{1}{2} \mathcal{M}^{ii'}{}_{j8}\, v^m_{ii'}\,R\, \partial_n y^{j}\ , \\
									e^{2\Delta}g^{mq}\left(C_{qnp} - C_{q}B_{np} \right) &= - \tfrac{1}{4}\mathcal{M}^{ii'}{}_{jj'}\, v^{m}_{ii'}\, \omega^{jj'}_{np}\ , \\
									e^{2\Delta}\left( e^{-2\phi} + g^{mn}C_m C_n\right) &= \mathcal{M}^{i8,j8}\, y_i\, y_j \ ,\\
									{e}^{2\Delta}g^{ms}\big(C_{snpqr} - \rg{C}_{snpqr} -C_{s[np}B_{qr]} + \tfrac{1}{2}C_s B_{[np}B_{qr]}\big) &=\tfrac{1}{4} \mathcal{M}^{ii',jj'} v^{m}_{ii'}\, (\rho_{jj'})_{npqr} \ ,
								\end{aligned}
							\end{equation}
					where we recall that the indices $i,i',j,j' =1\ldots, 7$ label the constrained coordinates while $m,n,\ldots=1,\ldots,6$ are curved indices on $S^6$.
					The scalar ansatz obtained in this way agrees with the formulae given in~\cite{Guarino:2015vca} (cf. eqs.~(3.14)--(3.18) therein). 
					The additional relations appearing in eqs.~(3.19)--(3.22) of~\cite{Guarino:2015vca} can also be retrieved in the same way. 
					The last equation in~\eqref{scalar_ansatz_S6} does not appear in~\cite{Guarino:2015vca}, and determines how the four-dimensional scalars enter in $C_{m_1\ldots m_5}$. 
					Dualising its field strength $F_{m_1\ldots m_6}$ it should be possible to derive the expression of the Freund-Rubin term.

					One can disentangle the different supergravity fields in~\eqref{scalar_ansatz_S6} by following the procedure in eqs.~\eqref{fields_from_G_first}--\eqref{fields_from_G_last}. 
					We recall that the generalised density $\Phi$ appearing in~\eqref{fields_from_G_last} can be computed at the origin of the scalar manifold, where $\mathcal{M}^{AB}=\delta^{AB}$, and is given by eq.~\eqref{gen_density_background}. 
					Evaluating the first, second and second-last line of~\eqref{scalar_ansatz_S6} with $\mathcal{M}^{ii',jj'}=\delta^{i[j}\delta^{j']i'}$, $\mathcal{M}^{ii',j8}=0$, $\mathcal{M}^{i8,j8}=\delta^{ij}$, we find that $\rg{\Delta}=\rg{\phi} = 0$. Hence for the present truncation the generalised density is simply $\Phi = \rg{g}{}^{1/2}$.

					We can also provide the ansatz for the vector fields as explained in section~\ref{genScherkSchw}. Separating the components of eq.~\eqref{trunc_ansatz_vec}, we obtain
							\begin{align*}
								h_\mu &= \tfrac{1}{2}\mathcal{A}_\mu{}^{ii'} v_{ii'} \ ,\\
								B_{\mu} &= \mathcal{A}_{\mu\,i8} \,R \,\dd y^i\ ,\\
								C_{\mu,0} &= \mathcal{A}_\mu{}^{i8} \,y_i \ ,\\
								C_{\mu,2} &= -\tfrac{1}{2}\mathcal{A}_{\mu\,ii'}\,R^2 \dd y^i \wedge \dd y^{i'} \, ,
							\end{align*}
					which again agrees with~\cite{Guarino:2015vca}. 
					Here, $\mathcal{A}^{IJ} = \{\mathcal{A}^{ij}, \mathcal{A}^{i8} \}$ are the electric one-form fields in the four-dimensional theory while $\mathcal{A}_{IJ} = \{\mathcal{A}_{ij}, \mathcal{A}_{i8} \}$ are their magnetic duals.
					We can also provide an ansatz for the type IIA dual fields with one external leg
							\begin{equation}
								\begin{split}
									C_{\mu ,4} &= \tfrac{1}{2}\mathcal{A}_\mu{}^{ii'} \big(\rho_{ii'} + \iota_{v_{ii'}}\! \rg{C_5} \!\big)\ ,\\
									C_{\mu ,6} &= \mathcal{A}_{\mu\,i8} \big(-y^i \rg{\rm vol}_6 + R\, \dd y^i\wedge \rg{C_5}\big) \ ,\\
									\tilde{B}_{\mu} &= \mathcal{A}_\mu{}^{i8} \, \big(\kappa_i -y_i \!\rg{C_5}\big) \, .
								\end{split}
							\end{equation}

					Finally, the ansatz for the fields with two external legs follows from the general formula~\eqref{ansatz_two-forms} 
							\begin{equation}
								\begin{split}
									B_{\mu\nu}&= \mathcal{B}_{\mu\nu}{}^{ij}{}_{j8}\, y_i \,,\\
									\tilde{B}_{\mu\nu} &= \tfrac{1}{8}\big(\tfrac 12 \mathcal{B}_{\mu\nu}{}^{i_1i_2,i_38} y^j y_{[i_1}\epsilon_{i_2i_3]j k_1\ldots k_4} - \mathcal{B}_{\mu\nu\,k_1k_2,k_3k_4}\big) R^4 \dd y^{k_1} \wedge \dd y^{k_2} \wedge\dd y^{k_3} \wedge\dd y^{k_4} \,,\\
									C_{\mu\nu ,1}&= \big(\mathcal{B}_{\mu\nu\,ij}{}^{kj}+ \mathcal{B}_{\mu\nu\,i8}{}^{k8} \big) y_k R\, \dd y^i \,,\\
									C_{\mu\nu ,3}&= \big(\tfrac{1}{12} \mathcal{B}_{\mu\nu}{}^{ii',jj'} y_{[i}\epsilon_{i']jj' k_1\ldots k_4} y^{k_4} - \tfrac12 \mathcal{B}_{\mu\nu\,k_1k_2,k_38} \big)R^3 \dd y^{k_1}\wedge \dd y^{k_2}\wedge \dd y^{k_3} \,,\\
									C_{\mu\nu,5}&= \mathcal{B}_{\mu\nu}{}^{ij}{}_{j8}\big(-\kappa_i + y_i\! \rg{C_5}\big) \,.
								\end{split}
							\end{equation}
			\subsubsection{\texorpdfstring{Hyperboloids and $D=4$, $\ISO(p,7-p)_m$ supergravity}{Hyperboloids and D=4, ISO(p,7-p)m supergravity}}\label{Hpq}
					The generalised Leibniz parallelisation on $S^6$ presented above can be adapted to construct a similar one on the six-dimensional hyperboloids $H^{p,7-p}$. 
					This leads to a consistent truncation of massive type IIA supergravity to four-dimensional $\ISO(p,7-p)_m$ maximal supergravity.

					The hyperboloid $H^{p,q}$ is the homogeneous space
							\begin{equation}
								H^{p,q} = \frac{\SO(p,q)}{\SO(p-1,q)}\ ,
							\end{equation}
					and can be seen as the hypersurface in the Euclidean space $\RR^{p+q}$ defined by the equation
							\begin{equation}\label{embH}
									\eta_{ij}\, y^i y^j = 1\, ,
							\end{equation}
					where $i,j=1,\ldots,p+q$ and
							\begin{equation}
									\eta_{ij} = \mathrm{diag}\big( \underbrace{+1,\ldots, +1}_{p}, \underbrace{-1,\ldots, -1}_{q}\big)\, .
							\end{equation}
					Clearly, taking $q=0$ yields the sphere $S^{p-1}$.

					Let us focus on the six-dimensional hyperboloids $H^{p,7-p}$, with $1\leq p < 7$.
					A parallelisation on these manifolds can be introduced following the same path as for $S^6$, replacing the Kronecker $\delta_{ij}$ by $\eta_{ij}$ where appropriate.
					In particular, the Killing vectors $v_{ij}$, that for the six-sphere satisfy the $\mathfrak{so}(7)$ algebra~\eqref{algebra_Killing_v}, now respect the $\mathfrak{so}(p,7-p)$ algebra,
							\begin{equation}
								\mathcal{L}_{v_{ij}} v_{kl} = 2R^{-1}\left(\eta_{i[k}v_{l]j} - \eta_{j[k}v_{l]i}\right)\ .
							\end{equation}
					The equations~\eqref{Lie_on_y}-\eqref{Lie_on_omega} also need to be modified by replacing $\delta_{ij}$ with $\eta_{ij}$ everywhere. 
					We can keep the definitions~\eqref{deflof}, noting however that they now transform in representations of $\SO(p,7-p)$ instead of $\SO(7)$.
					Then~\eqref{s6frame} defines a generalised parallelisation on $H^{p,7-p}$. 
					The Dorfman derivative between two frame elements satisfies~\eqref{LeibnizParall}, with the non-vanishing embedding tensor components being still given by~\eqref{ISO7m_emb_tens}, where however now
							\begin{equation}
								\theta_{IJ} = \frac{1}{2R} \,{\rm diag}\big(\underbrace{1,\ldots,1}_{p },\underbrace{-1,\ldots,-1}_{7-p },0\big)\ ,
							\end{equation}
					while $\xi^{IJ}$ remains unchanged.
					
					 This corresponds to an $\ISO(p,7-p)\simeq \CSO(p,7-p,1)$ frame algebra, where the seven translational symmetries are gauged dyonically.

					The truncation ansatz remains formally the same as for the reduction on $S^6$. 
					We thus infer that there exists a consistent truncation of massive IIA supergravity on the six-dimensional hyperboloids $H^{p,7-p}$, down to $\ISO(p,7-p)_m$ gauged supergravity. 
					As above, the subscript $m$ emphasises that the translational isometries are gauged dyonically. Setting $m=0$, one recovers a truncation of massless type IIA supergravity on $H^{p,7-p}$ down to the $\ISO(p,7-p)$ theory with purely electric gauging (see also~\cite{Hohm:2014qga}).

					It was found in~\cite{Dall'Agata:2012bb} that the only gaugings of four-dimensional maximal supergravity in the $\CSO(p,q,r)$ class (with $r >0$) admitting a symplectic deformation are $\CSO(p,7-p,1)\simeq \ISO(p,7-p)$\footnote{%
						For these gaugings, the symplectic deformation is of on/off type: all non-zero values of the parameter controlling the magnetic gauging are equivalent.%
						}.
					Here we have established that all these symplectic deformations arise as consistent truncations of massive type IIA supergravity: while for $p=7$ the internal manifold is $S^6$, for $1\leq p < 7 $ the internal manifold is the hyperboloid $H^{p,7-p}$.

					The same ideas could be applied to products of hyperboloids and tori, $H^{p,q}\times T^r$, with $p+q+r=7$. 
					In this case, the parallelisation would satisfy the $\CSO(p,q,r+1)$ algebra.
			\subsubsection{\texorpdfstring{$S^4$ parallelisation with $m=0$ and $D=6$, $\SO(5)$ supergravity}{S4 parallelisation with m=0 and D=6, SO(5) supergravity}}
					The U-duality group for type IIA on a four-dimensional manifold $M_4$ is $E_{5(5)}\simeq \SO(5,5)$ and the generalised tangent bundle is
							\begin{equation}
								E \simeq T \oplus T^* \oplus \RR \oplus \Lambda^2 T^*\oplus \Lambda^4 T^*\ .
							\end{equation}
				A section of $E$
							\begin{equation}
								V = v + \lambda + \omega_0 + \omega_2 + \omega_4 
							\end{equation}
				transforms in the spinorial $\mathbf{16^+}$ representation of $\SO(5,5)$.

				We are interested in the case where $M_4$ is the four sphere $S^4$ and we describe it using constrained coordinates $y^i$ in $\RR^5$.
				It is then convenient to consider the decomposition of the generalised frame $\hat{E}_A$, $A = 1\ldots,16$ under $\SL(5, \RR)$ 
							\begin{equation}
								\begin{split}
									\SO(5,5) &\supset \SL(5,\RR) \\
									\mathbf{16^+} &\mapsto \mathbf{10} + \mathbf{5} + \mathbf{1} \, ,
								\end{split}
							\end{equation}
				so that $\{\hat{E}_{A}\} = \{\hat{E}_{ij}\} \cup \{\hat{E}_{i}\} \cup \{\hat{E}\}$, with $i,j=1,\ldots,5$.

				For \emph{massless} type IIA supergravity on $S^4$, we take the frame
						\begin{equation}\label{parall_S4_meq0}
							\hat{E}_{A} = \begin{cases} \hat{E}_{ij} = v_{ij} + \rho_{ij} + \iota_{v_{ij}} \rg{C_3} \ ,\\
							\hat{E}_{i} = R\,\dd y_i + y_i \! \rg{\rm vol}_4 + R\, \dd y_i\, \wedge\! \rg{C_3}\ ,\\
							 \hat{E} = 1 \ ,
									\end{cases} 
						\end{equation}
				where $v_{ij}$ are the $\SO(4)$ Killing vectors and
						\begin{equation}
							\rho_{ij} = \rg{*} (R^2 \dd y_i \wedge \dd y_j) = \frac{R^2}{2}\, \epsilon_{ijk_1k_2k_3} y^{k_1}\dd y^{k_2}\wedge \dd y^{k_3} \ .							
						\end{equation}
				Note that we have twisted the frame by a background RR potential $\rg{C_3}$, that is the supergravity potential whose field strength threads the whole $S^4$\footnote{%
					The twist by $C_3$ acts on a vector $\tilde V$ of the untwisted generalised tangent bundle $\tilde E$ on $M_4$ as (cf.~eq.~\eqref{eq:twistC}):
							\begin{equation*}
								V = e^{C_3}\cdot \tilde V = \tilde v + \tilde \lambda + \tilde \omega_0 + (\tilde \omega_2 + \iota_{\tilde v}C_3) + (\tilde \omega_4 + \tilde \lambda \wedge C_3)\ .
							\end{equation*}
					}.
				This is chosen such that
						\begin{equation}
							\rg{F_4} \ = \dd \rg{C_3} = \frac{3}{R} \rg{\vol}_4\ , 
						\end{equation}
				which, recalling~\eqref{eq:contrvol}, implies
						\begin{equation}\label{Fcn}
							\iota_{v_{ij}} \!\rg{F_4} = \dd\rho_{ij}\ .
						\end{equation}
				We will not twist by $C_1$ or $B$ instead as there are no two- or three-cycles on $S^4$.
				Following similar reasoning as for $S^6$, it is easy to see that the frame above is globally defined and orthonormal with respect to the generalised metric~\eqref{GofVandV'}, thus it specifies a generalised parallelisation.

				In four dimensions (or lower), the massive generalised Lie derivative simplifies considerably and reads
						\begin{equation}\label{dorf4m}
							\begin{split}
								L_V V' =& \mathcal{L}_v v' + \left(\mathcal{L}_v \lambda' - \iota_{v^\prime} \dd\lambda\right) + \left(\iota_v \dd\omega_0' - \iota_{v^\prime} (\dd\omega_0 - m\lambda) \right) \\
										&\phantom{=} + \left(\mathcal{L}_v \omega_2^\prime - \iota_{v^\prime}\dd\omega_2 - \lambda' \wedge (\dd\omega_0 - m\lambda) + \omega_0' \dd\lambda \right) \\
										&\phantom{=} + \left( \mathcal{L}_v \omega_4^\prime - \iota_{v^\prime}\dd\omega_4 - \lambda' \wedge \dd\omega_2 + \omega_2^\prime \wedge \dd\lambda\right) \ .
							\end{split}
						\end{equation}
				Using the relations in appendix~\ref{app:constrcoo}, we compute the \emph{massless} Dorfman derivative (that is expression~\eqref{dorf4m} with $m=0$) between the frame elements. 
				We find that the only non-vanishing pairings are
						\begin{equation}\label{Dorfman_par_S4}
							\begin{split}
								L_{\hat{E}_{ij}}\hat{E}_{kl} &= 2R^{-1}\big(\delta_{i[k}\hat{E}_{l]j} - \delta_{j[k}\hat{E}_{l]i} \big)\, , \\
								L_{\hat{E}_{ij}}\hat{E}_{k} &= -2R^{-1}\delta_{k[i}\hat{E}_{j]} \, .
							\end{split}
						\end{equation}
				This defines a Leibniz algebra since $L_{\hat{E}_{ij}}\hat{E}_{k} \neq -L_{\hat{E}_{k}}\hat{E}_{ij}=0$; the associated gauge algebra, following from~\eqref{eq:gauge-alg-X}, is the $\SO(5)$ algebra.

				A consistent truncation of massless type IIA supergravity on $S^4$ preserving maximal supersymmetry has been constructed in~\cite{Cowdall:1998rs,Cvetic:2000ah} by simply reducing on a circle the seven-dimensional theory defined by eleven-dimensional supergravity on $S^4$. 
				The gauge group of the resulting $\mathcal{N} =(2,2)$ six-dimensional theory is indeed~$\SO(5)$ (see also~\cite{Bergshoeff:2007ef} for a discussion of the gauging in six dimensions). 
				This theory does not admit AdS$_6$ vacua: the most symmetric solution is a half-BPS domain-wall, originating from a circle reduction of the AdS$_7\times S^4$ vacuum of eleven-dimensional supergravity, and describing the near-horizon geometry of D4-branes.

				Following the example of $S^6$, one might expect that the same frame~\eqref{parall_S4_meq0} would lead to a generalised parallelisation for $m\neq0$ with a modified gauge group in six-dimensions. 
				However, it is easy to check by direct computation that with the massive Dorfman derivative the frame~\eqref{parall_S4_meq0} does not satisfy a Leibniz algebra. 
				We will further comment on this in section~\ref{massive_algebras}.
		\subsubsection{$S^3$ parallelisation with $m=0$ and $D=7$, $\ISO(4)$ supergravity}\label{S3and7dsugra}
				The U-duality group of type IIA supergravity on a three-dimensional manifold $M_3$ is $E_{4(4)} \simeq \SL(5,\RR)$, and the corresponding generalised tangent bundle is
							\begin{equation}
								E \cong T \oplus T^* \oplus \RR \oplus \Lambda^2 T^*\, , 
							\end{equation}
				with sections
							\begin{equation}
								V = v + \lambda + \omega_0 + \omega_2 
							\end{equation}
				transforming in the $\mathbf{10}$ of $\SL(5,\RR)$.
				A generalised frame $\{\hat{E}_A\}$, $A = 1\ldots,10$, can equivalently be denoted as $\{\hat{E}_{IJ}= \hat{E}_{[IJ]}\}$, with $I,J =1,\ldots, 5$.
				We consider again $M_3 = S^3$ in constrained coordinates $y^i$ in $\RR^4$, and we decompose the frame under $\SL(4, \RR)$ as
							\begin{equation}
								\begin{split}
									\SL(5,\RR) \; &\supset \SL(4,\RR)  \\[1mm]
									\mathbf{10}\, &\mapsto \mathbf{6} + \mathbf{4} \, .
								\end{split}
							\end{equation}
				so that $\{\hat{E}_{IJ}\} = \{\hat{E}_{ij},\,\hat{E}_{i5}\}$, with $i,j=1,\ldots,4$.

				For vanishing Romans mass, $m=0$, we can easily construct a generalised parallelisation that realises the $\mathfrak{iso}(4)$ algebra.
				We choose the frame
						\begin{equation}\label{parall_S3_meq0}
									\hat{E}_{IJ} = \begin{cases} \hat{E}_{ij} = v_{ij} + \rho_{ij} + \iota_{v_{ij}}\!\rg{B} \, ,\\[1mm]
										\hat{E}_{i5} = y_i + \kappa_i - y_i \!\rg{B}\, , \end{cases} 
						\end{equation}
				where $v_{ij}$ are the $\SO(4)$ Killing vectors and
							\begin{equation}
								\begin{split}
									\rho_{ij} &= \rg{*}(R^2 \dd y_i\wedge \dd y_j) = R\, \epsilon_{ijkl}\, y^k\dd y^l \ , \\
									\kappa_{i} &= \rg{*}(R\, \dd y_i) = \frac{R^2}{2}\,\epsilon_{ijkl}\, y^j \dd y^k \wedge \dd y^l\ .
								\end{split}
							\end{equation}
				Here, we have twisted the frame by the $B$ field\footnote{%
					The twist by $B$ acts on a vector $\tilde V$ of the untwisted generalised tangent bundle $\tilde E$ on $M_3$ as
							\begin{equation*}
								V = e^{-B}\cdot \tilde V = \tilde v + (\tilde \lambda + \iota_{\tilde v} B) + \tilde \omega_0 + (\tilde \omega_2 -\tilde \omega_0 B)\ .
							\end{equation*}
						},
				chosen in such a way that
						\begin{equation}\label{HproptoVol_3}
								\rg{H} = \dd \!\rg{B_{}}  = \frac{2}{R} \rg{\vol}_3\ , 
							\end{equation}
				which, again recalling~\eqref{eq:contrvol}, implies
							\begin{equation}\label{hcn}
								\iota_{v_{ij}} \rg{H} = \dd\rho_{ij}\ .
							\end{equation}
				This frame is globally defined and orthonormal; hence it defines a generalised parallelisation.
				Recalling appendix~\ref{app:constrcoo} and relation~\eqref{hcn}, one can check that the Dorfman derivative with $m=0$ yields
							\begin{equation}\label{ISO4-alg}
								\begin{split}
									L_{\hat{E}_{ij}}\hat{E}_{kl} &= 2R^{-1}\big(\delta_{i[k}\hat{E}_{l]j} - \delta_{j[k}\hat{E}_{l]i} \big)\ ,\\
									L_{\hat{E}_{ij}}\hat{E}_{k5} &= -2R^{-1}\delta_{k[i}\hat{E}_{j]5} \ ,\\
									L_{\hat{E}_{i5}}\hat{E}_{kl} &= 2R^{-1}\delta_{i[k}\hat{E}_{l]5}\ , \\
									L_{\hat{E}_{i5}}\hat{E}_{k5} &= 0\ , 
								\end{split}
							\end{equation}
				and the relation~\eqref{LeibnizParall} is satisfied, with structure constants
						\begin{equation}\label{XofISO4}
							X_{[II'][JJ']}{}^{[KK']} = 2\, \delta_{[I}^{[K}Y_{I'][J}\delta_{J']}^{K']}\ , \qquad Y_{II'} = \frac{2}{R} \mathrm{diag}(1,1,1,1,0)\ .
							\end{equation}
				Note that, as the Dorfman derivative is antisymmetric on this frame, it realises a Lie algebra (rather than just a Leibniz algebra), which in this case is the $\ISO(4)\simeq\CSO(4,0,1)$ algebra. 

				A consistent truncation of massless type IIA supergravity to maximal $D=7$ supergravity with gauge group $\ISO(4)$ has been known for some time. 
				This can be obtained starting from the well-known reduction of eleven-dimensional supergravity on $S^4$, which yields maximal $D=7$, $\SO(5)$ supergravity~\cite{Nastase:1999kf}, and implementing the limiting procedure of~\cite{Hull:1988jw}. 
				In the limit, $S^4$ degenerates into $\RR \times S^3$; correspondingly, the $\SO(5)$ gauge group of the seven-dimensional theory is contracted to $\ISO(4)$\footnote{%
					This is analogous to the way the $\ISO(7)$ reduction of massless IIA supergravity on $S^6$ is obtained from the $\SO(8)$ reduction of eleven-dimensional supergravity on $S^7$.}.
				The bosonic part of this $S^3$ reduction was worked out in detail in~\cite{Cvetic:2000ah} (where the $\SO(4)$ subgroup of the gauge group was emphasised). 
				A discussion of the resulting maximal supergravity can be found in~\cite{Samtleben:2005bp}. 
				In seven dimensions the embedding tensor determining the gauging transforms in the $\mathbf{15} + \mathbf{40'}$ representation of the global symmetry group $\SL(5)$~\cite{Samtleben:2005bp}. 
				For the $\ISO(4)$ gauging, its non-vanishing components are solely in the $\mathbf{15}$, and match those in~\eqref{XofISO4} obtained from the parallelisation. 
				In addition to the metric, the fourteen $\SL(5)/\SO(5)$ scalars and the ten $\ISO(4)$ gauge vectors, the bosonic field content of the seven-dimensional theory is made of a massless two-form and four massive self-dual three-forms.
				The scalar potential does not admit stationary points, and the most symmetric ground state solution is a domain wall, describing the near-horizon geometry of NS5-branes.

				We would now like to see whether the frame~\eqref{parall_S3_meq0} gives a generalised parallelisation also for $m\neq0$. 
				In this case the problems appear even before considering the action of the massive Dorfman derivative. 
				Indeed the frame~\eqref{parall_S3_meq0} requires the existence of a non trivial field strength $H$, while we know from~\eqref{eq:H-exact} that for $m\neq 0$ $H$ is exact. 
		\subsubsection{Massive algebras on $S^3$ and $S^4$}\label{massive_algebras}
				In the previous sections we saw that, contrary to the case of $S^6$, the massless frames for $S^3$ and $S^4$ do not lead to good parallelisations when the Romans mass is turned on. 
				In this section, we provide some understanding of why the frame on $S^6$ is the only one that satisfies a good algebra also in the massive Dorfman derivative. 
				We also explore the possibility of finding other parallelisations that do satisfy an algebra of the desired type.
				For $S^3$ we derive a no-go theorem showing that, under mild assumptions, one cannot find a frame which gives rise to a maximally supersymmetric consistent truncation.

				Given a $d$-dimensional sphere $S^d$ with a non-zero flux for a $d$-form field-strength, one can build a $\GL(d+1)$ generalised tangent bundle, which is isomorphic to $T\oplus \Lambda^{d-2}T^*$. 
				Since this admits a global generalised frame, the sphere is generalised parallelisable~\cite{spheres}. 
				This generalised frame is a $\GL(d+1)$ rotation of the coordinate frame. 
				For spheres, the $\GL(d+1)$ generalised tangent bundle is always a sub-bundle of the full $\E_{d+1(d+1)}\times\RR^+$ bundle and, in fact, it is possible to decompose the whole generalised tangent bundle into representations of the $\GL(d+1)$ subgroup. 
				Moreover, all the parts of the parallelisations of the bundle $E$ are related to the corresponding coordinate frames by the same $\GL(d+1)$ transformation.

				In the previous sections we constructed the frame $\hat{E}_A$ and the respective Leibniz algebra for type IIA on $S^d$, $d=3,4,6$. 
				We consider now the effect of adding the Romans mass to the massless Dorfman derivative. 
				As the given frame on $S^d$ already satisfies a Leibniz algebra for the massless Dorfman derivative with constant structure constants $X_{AB}{}^C$, the structure constants of the same frame with the massive Dorfman derivative will be $X_{AB}{}^C + Y_{AB}{}^C$, where
							\begin{equation}
								Y_{AB}{}^C = \hat{E}_A{}^M \hat{E}_B{}^N E^C{}_P\, \underline{m}_{MN}{}^P \, , 
							\end{equation}
				are the frame components of the Romans mass map $\underline{m}_{MN}{}^P$ (see section~\ref{sec:EGGIIA}).
				The frame $\hat{E}_A$ will thus give a generalised Leibniz parallelisation in the massive Dorfman derivative if the additional structure constants $Y_{AB}{}^C$ are constant.
 
				A natural way for this to happen would be if the components $Y_{AB}{}^C$ are equal to the components $\underline{m}_{MN}{}^P$, which are constant by definition.
				This would mean that the frame $\hat{E}_A{}^M$ must lie in the stabiliser group of the Romans mass. 
				The stabiliser is the subgroup of $E_{d+1(d+1)} \times\RR^+$ that leaves $\underline{m}_{MN}{}^P$ invariant. It can be determined by combining~\eqref{IIAadjvecCompact} and~\eqref{eq:commAdjIIA} with %
							\begin{equation}
								(R\cdot \underline{m}) (V) = [ R, \underline{m}(V) ] - \underline{m} (R\cdot V) \, ,
							\end{equation}
				where $R$ is an element of the adjoint of $\E_{d+1(d+1)}\times\RR^+$, see~\eqref{section_adjoint}. 
				For instance, in six dimensions we find that $R \cdot \underline{m} = 0$ for $R$ of the form\footnote{For lower-dimensional spheres it is enough to truncate to the relevant potentials.}
							\begin{equation}\label{eq:stabiliser}
								R = l + \varphi + r + \beta + \tilde B + \Gamma_5 + C \, ,
							\end{equation}
				where $l = -\varphi$ and $\Gamma_5$ is a five-vector, while $C=C_1+ C_3 + C_5$. 
				The stabiliser group is the semi-direct product of a Lie group $G$ with a nilpotent group $G'$. 
				The Lie algebra $\mathfrak{g}$ of $G$ is generated by $r$, $\Gamma_5$, $C_5$ and $l = -\varphi$ in~\eqref{eq:stabiliser}. 
				The Lie algebra of $G'$ is $\mathfrak{g}' = \mathfrak{g}'_1 \oplus \mathfrak{g}'_2$ where $\mathfrak{g}_1$ and $\mathfrak{g}_2$ are generated by $\beta$ and $C_3$, and $C_1$ and $\tilde{B}$, respectively.
				The algebra $\mathfrak{g}'$ is graded so that the commutator of two $\mathfrak{g}_1$ elements is in $\mathfrak{g}_2$ and all other commutators vanish.
				The stabiliser groups of $\underline{m}$ for the dimensions of interest in this work are summarised in table~\ref{tab:stab}. 
				In the table, $\mathbf{R}_1$ and $\mathbf{R}_2$ denote the representations of $G$ in which $\mathfrak{g}'_1$ and $\mathfrak{g}'_2$ transform.
						\begin{table}[h]
						\centering
							\begin{tabular}{rccc}
									$d$ & $G$ & $\mathbf{R}_1$ & $\mathbf{R}_2$ \\ 
									\hline
 										$6$ & $\GL(7)$ & $\mathbf{35}$ & $\mathbf{7'}$ \\
		 								$5$ & $\SL(5)\times\SL(2)\times\RR^+$ & $(\mathbf{10},\mathbf{2})_{+1}$ & $(\mathbf{5},\mathbf{1})_{+2}$ \\
										$\leq 4$ & $\GL(d) \times \RR^+$ & $(\Lambda^2 T)_{+1} \oplus (\Lambda^3 T^*)_{+1}$ & $(T^*)_{+2}$
							\end{tabular}
						\caption{Constituents of the stabiliser group of $\underline{m}_{MN}{}^P$.}
						\label{tab:stab}
						\end{table}

				It is noteworthy that only for $d=6$ the group $G$ coincides with $\GL(d+1)$. 
				Since the frame $\hat{E}_A{}^M$ is an element of $\GL(d+1)$, we see that for $S^6$ the frame does lie in the relevant stabiliser group\footnote{%
					Note that for $d=6$ the full stabiliser group is isomorphic to the geometric subgroup of $E_{7(7)}\times\RR^+$ for M-theory.}.
				Hence the massless frame remains a good Leibniz parallelisation when the Romans mass is switched on. 
				However, for $d \leq 5$ it does not, and this provides a partial explanation for why these frames do not give Leibniz parallelisations in massive IIA. 
				By this reasoning, one is not surprised that $S^6$ is the only case which works in massive IIA without modifying the frame.

				However, the above argument does not rule out the possibility that there are alternative Leibniz generalised parallelisations of the lower-dimensional spheres in the massive IIA. 
				In what follows, we explore this possibility focusing on the case of $S^3$, for simplicity. 
				As noted before, in massive type IIA $H_3$ must be trivial in cohomology. 
				As $S^3$ has only a non-trivial 3-cycle, this means that there can be no cohomologically non-trivial field strengths. 
				We thus assume that the background field configuration has non-zero Romans mass and all other fields are zero. 
				This implies that the generalised tangent space has no twisting and is given by the direct sum 
							\begin{equation}
								E = \tilde{E} = T \oplus T^* \oplus \Lambda^0 T^* \oplus \Lambda^2 T^* \, .
							\end{equation}
				Suppose now that there exists a generalised Leibniz parallelisation $\hat{E}_A$ that gives an $SO(4)$ algebra
							\begin{equation}\label{massive_algebra_S3}
								\begin{split}
									L_{\hat{E}_{ij}}\hat{E}_{kl} &= 2R^{-1}\big(\delta_{i[k}\hat{E}_{l]j} - \delta_{j[k}\hat{E}_{l]i} \big)\ ,\\
									L_{\hat{E}_{ij}}\hat{E}_{k5} &= -2R^{-1}\delta_{k[i}\hat{E}_{j]5} \ , \\
									L_{\hat{E}_{i5}}\hat{E}_{kl} &=0\ , \\
									L_{\hat{E}_{i5}}\hat{E}_{k5} &= 0\ ,
								\end{split}
							\end{equation}
				where $L$ is the massive Dorfman derivative. 
				This implies that the generalised metric $\mathcal{G}^{-1} = \delta^{AB} \hat{E}_A \otimes \hat{E}_B$ is preserved by the Dorfman derivative so that the $\hat{E}_A$ are generalised Killing vectors~\cite{Grana:2008yw, Lee:2015xga}. 
				Thus the gauge transformations of the background fields generated by the $\hat{E}_A$ all vanish. 
				As we have no gauge fields, this leads to the conditions
							\begin{align}\label{eq:gen-Killing}
								& & \mathcal{L}_{v_A} g = 0 \, , & &  \dd \lambda_A = 0 \, ,& & \dd \omega_A - m \lambda_A = 0 \, ,& & 
							\end{align}
				which imply that the Dorfman derivative reduces to the Lie derivative term only
							\begin{equation}
								L_{\hat{E}_A} \equiv \mathcal{L}_{v_A} \, . 
							\end{equation}
				As the vector parts of the $\hat{E}_{ij}$ satisfy the $\SO(4)$ algebra, these must be the $S^3$ Killing vectors (up to an overall constant automorphism), and we have that
							\begin{equation}
								L_{\hat{E}_{ij}} \equiv \mathcal{L}_{v_{ij}} 
							\end{equation}
				is the action of the $\SO(4)$ isometry group. 
				The second of~\eqref{massive_algebra_S3} then says that the $\hat{E}_{k 5}$ components of the frame transform in the vector representation. 
				This implies that
							\begin{equation}
								\hat{E}_{i5} = a_1 k_i + a_2 y_i + a_3 \dd y_i + a_4 * \dd y_i
							\end{equation}
				for some real coefficients $a_n$, where $y^i$, with $i=1, \ldots, 4$ are the constrained coordinates on $\RR^4$. 
				Here, $k_i$ are the standard conformal Killing vectors on the sphere (cf.~appendix~\ref{app:constrcoo}). 
				As $L_{\hat{E}_{i5}} \hat{E}_{j5} = 0$ we have $a_1=0$ and~\eqref{eq:gen-Killing} gives us $a_2 = ma_3$ and $a_4=0$. 
 				One can then see that
						\begin{equation}\label{parall_S3_m}
							\hat{E}_A = \hat{E}_{IJ} = \begin{cases} \hat{E}_{ij} = v_{ij} + R^2 \,\dd y_i \wedge \dd y_j \\[1mm]
														\hat{E}_{i5} = R\,(m\,y_i + \dd y_i)\ , \end{cases} 
							\end{equation}
				where $R$ is the radius of $S^3$, is the unique frame giving a parallelisation of the generalised tangent bundle on $S^3$ which satisfies the $SO(4)$ algebra~\eqref{massive_algebra_S3}\footnote{%
					In appendix~\ref{IIBonS3} we show that in type IIB it is possible to find a parallelisation for the generalised tangent bundle on $S^3$ that satisfies the same Leibniz algebra~\eqref{massive_algebra_S3}.}.
				If $mR=1$, the frame is also orthonormal in the generalised metric.
				However, the frame~\eqref{parall_S3_m} fails to be in the $\SL(5,\RR)\times\RR^+$ generalised frame bundle. 
				We recall from~\cite{Coimbra:2011ky} that the generalised frame bundle is defined to be those frames which are related to the coordinate frame by an $\E_{d+1(d+1)}\times\RR^+$ transformation. 
				In the $\SL(5,\RR)\times\RR^+$ case, this means that there must also be a parallelisation $\hat{E}_I$ of the bundle $W \simeq (\det T)^{-1/2} \otimes (T + \det T)$, discussed in~\cite{spheres}, such that
							\begin{equation}
								\hat{E}_{IJ} = \hat{E}_I \wedge \hat{E}_J \,.
							\end{equation}
				It is simple to show that our frame~\eqref{parall_S3_m} is not of this form, and is thus outside of the generalised frame bundle. 
				This means that one cannot use it to describe a consistent truncation of supergravity. 
				For example, the Scherk-Schwarz twist of this frame does not define a generalised metric which can be parameterised in terms of supergravity fields, and as such it does not provide a scalar ansatz for such a reduction.
				
				Having ruled out the possibility of the algebra~\eqref{massive_algebra_S3}, one could still wonder if there are other frame algebras containing $\SO(4)$ which could fare better. 
				The obvious alternative would be the $ISO(4)$ algebra~\eqref{ISO4-alg}. 
				However, we will now see that just requiring this algebra already leads to a contradiction.

				For the $\hat{E}_{ij}$ parts of the frame, we can use the same generalised Killing vector arguments as above to deduce that $L_{\hat{E}_{ij}} \equiv \mathcal{L}_{v_{ij}}$, so we can again decompose the frames into $\SO(4)$ representations. 
				This decomposition implies that the one-form part of $\hat{E}_i$ is closed, and, together with the generalised Killing vector condition, that the one-form part of $\hat{E}_{ij}$ vanishes. 
				The constraint $L_{\hat{E}_{i5}} \hat{E}_{j5} = 0$ then gives that $L_{\hat{E}_{i5}} \equiv - (\dd \omega_{2,i}) \cdot$ is the adjoint action of $\dd \omega_{2,i} \in \Lambda^3 T^* \subset \adj$, where $\omega_{2,i}$ is the two-form part of $\hat E_{i5}$.
				However, this contradicts another of the hypothesised algebra relations $L_{\hat{E}_{i5}}\hat{E}_{kl} = 2R^{-1}\delta_{i[k}\hat{E}_{l]5}$ as the image of $\dd \omega_{2,i} \in \adj$ is contained in $\Lambda^2 T^* \subset E$, while $\hat{E}_{i5}$ must feature one-form parts in order for $\hat{E}_{IJ}$ to give a parallelisation.

				We have thus shown that the two most likely frame algebras featuring $\SO(4)$ in the gauge group cannot be realised in massive type IIA parallelisations. While these arguments do not systematically rule out all possibilities, they are highly suggestive that there is no maximally supersymmetric consistent truncation of massive type IIA on $S^3$ with gauge group $\SO(4)$ (or larger).
				It seems that a similar conclusion can be reached for the $S^4$ case. 
				We note that~\eqref{parall_S3_m}, augmented by an additional piece $\hat{E} = \vol_4$, also yields a Leibniz parallelisation of the type IIA generalised tangent bundle on $S^4$, satisfying the $\SO(5)$ algebra. 					However, again one can prove this is not an $\SO(5,5)\times {\mathbb R}^+$ frame. 

				One can construct an $\SO(5,5)\times\RR^+$ covariant projection acting on four generalised vectors $E^4 \rightarrow \Lambda^4 T^*$. 
				This is done by taking the projections to the bundle $N$ of the two pairs of generalised vectors and then contracting the resulting sections of $N$, which transform in the vector representation of $\SO(5,5)$, using the $\SO(5,5)$ invariant metric. Due to the $\RR^+$ weights, the inner product is in fact a volume form and transforms under $\RR^+$, but it is $\SO(5,5)$ invariant. 
				By explicit computation, one can check that the components of this quartic $\SO(5,5)$ invariant on $E$ are not preserved, or rescaled, when one moves to the frame~\eqref{parall_S3_m} combined with $\hat{E} = \vol_4$, showing that this frame is not an $\SO(5,5)\times\RR^+$ frame.
		\subsubsection{\texorpdfstring{$S^2$ parallelisation and $D=8$, $\SO(3)$ supergravity}{S2 parallelisation and D=8, SO(3) supergravity}}
				We conclude our set of examples by considering type IIA supergravity on the two-sphere $S^2$. 
				Again, we will see that while it is easy to define a generalised Leibniz parallelisation for $m=0$, in the massive case the most likely frames do not work.

				On a two-dimensional manifold, the U-duality group is $\SL(3)\times \SL(2)$, and the generalised tangent bundle reads
						\begin{equation}
							E \cong T \oplus T^* \oplus \RR \oplus \Lambda^2 T^*\, ,
						\end{equation}
				which factorises as
						\begin{equation}\label{2dEfactorised}
	E\cong (\RR \oplus \det T^* ) \otimes ( T \oplus \RR)\, =\, U \otimes W \,,
						\end{equation}
				where $U$ transforms as an $\SL(2)$ doublet and $W$ as an $\SL(3)$ triplet.

				An $\SL(3)\times \SL(2)$ frame is specified by $\{\hat E_{i\alpha}\}$, where $i=1,2,3$ is an $\SL(3)$ index while $\alpha =\pm$ is an $\SL(2)$ index.
				According to the factorisation~\eqref{2dEfactorised}, it can be written as 
						\begin{equation}\label{factorSL3SL2}
							\hat{E}_{i \alpha} = \hat{E}_\alpha \otimes \hat{E}_i\,,
						\end{equation}
				where $\hat{E}_\alpha$ is a frame for $V$ and $\hat{E}_i$ is a frame for $W$. 
				This guarantees that the scalar matrix $\mathcal{M}^{i\alpha,j\beta}$ defined by the generalised Scherk-Schwarz ansatz parameterises the seven-dimensional coset $\frac{\SL(3)}{\SO(3)}\times \frac{\SL(2)}{\SO(2)}$, as expected for maximal supergravity in eight dimensions.

				For vanishing Romans mass, a generalised Leibniz parallelisation on $S^2$ is given by
						\begin{equation}\label{S2parallel}
							\begin{cases}
								\hat E_{i+} = v_i + y_i + \iota_{v_i} \rg{C_1}\ , \\[1mm]
								\hat E_{i-} = \dd y_i + y_i \rg{\rm vol}_2 -\, \dd y_i\, \wedge \rg{C_1}\ ,
							\end{cases}
						\end{equation}
				where $v_i = \frac 12\epsilon_{i}{}^{jk}v_{jk}$ are the $\SO(3)$ Killing vectors and $\rg{\rm vol}_2$ is the volume on the round $S^2$ of unitary radius. 
				Notice that (before twisting by $\rg{C_1}$) the factorisation condition~\eqref{factorSL3SL2} is satisfied by taking
						\begin{equation}
							\begin{split}
								\hat{E}_i &= v_i + y_i\,, \\
								\hat{E}_\alpha &= { 1 \choose {\rm vol}_2 }_\alpha\,.
							\end{split}
						\end{equation}
				Moreover, choosing the two-form flux as
						\begin{equation}
							\rg{F_2} = \dd\! \rg{C_1}\ =\, \frac{1}{R} \rg{\vol}_2\ , 
						\end{equation}
				so that $\iota_{v_i}\dd \!\rg{C_1} = c\,R\,\dd y_i$, the massless Dorfman derivative yields
						\begin{equation}\label{S2Leibniz}
							\begin{array}{lcc}
								L_{\hat E_{i+}}\hat E_{j+} = - \tfrac{1}{R}\epsilon_{ij}{}^k \hat E_{k+} \, , & \phantom{L_{\hat{E}_{i+}} \hat{E}_{j-} = - \tfrac{1}{R}} & L_{\hat{E}_{i+}} \hat{E}_{j-} = - \tfrac{1}{R}\epsilon_{ij}{}^k \hat{E}_{k+}\, , \\ 
								L_{\hat{E}_{i-}}\hat{E}_{j+} = 0 \ , & \phantom{L_{\hat{E}_{i+}} \hat{E}_{j-} = - \tfrac{1}{R}} & L_{\hat{E}_{i-}} \hat{E}_{j-} \ =\ 0 \, ,
							\end{array}
						\end{equation}
				which is a Leibniz algebra leading to an $\SO(3)$ gauge algebra.

				Hence we have an $\SL(3)\times \SL(2)$ Leibniz parallelisation with associated $\SO(3)$ gauge algebra. 
				This can be used to define a generalised Scherk--Schwarz reduction of massless type IIA supergravity on $S^2$, down to to maximal supergravity in eight dimensions with gauge group $\SO(3)$. 
				As pointed out in~\cite{Boonstra:1998mp}, this consistent reduction on $S^2$ is the same as the conventional Scherk--Schwarz reduction of eleven-dimensional supergravity on the group manifold $\SU(2)\simeq S^3$, presented long ago in~\cite{Salam:1984ft}.
				The explicit truncation ansatz for the metric, dilaton and RR two-form on $S^2$ can be found in~\cite[sect.$\:$6]{Cvetic:2000dm}, and its relation with the $S^3$ reduction of eleven-dimensional supergravity is explained in~\cite{Cvetic:2003jy}.

 				When the Romans mass is switched on, the frame~\eqref{S2parallel} fails to satisfy an algebra under the Dorfman derivative with $m\neq 0$. One could consider the alternative generalised frame
						\begin{equation}\label{alternateS2frame}
							\begin{cases}
								\hat E_{i+} = v_i + y_i \!\rg{\rm vol}_2\,, \\[1mm]
								\hat E_{i-} = \dd y_i + y_i \,,
							\end{cases}
						\end{equation}
				which compared to~\eqref{S2parallel} has the role of the $\RR$ and $\Lambda^2 T^*$ terms exchanged, and is not twisted by $\rg{C_1}$.
				This frame is still globally defined, orthonormal and can easily be checked to satisfy the $\SO(3)$ algebra under the massive Dorfman derivative for $mR=1$. 
				However, it cannot be put in the form~\eqref{factorSL3SL2}, so it is not an acceptable $\SL(3)\times \SL(2)$ frame. 
				This means that a Scherk-Schwarz reduction based on~\eqref{alternateS2frame} would not define a generalised metric of the type given by the supergravity degrees of freedom~\eqref{genmetfor}, so it would not make sense to define an ansatz like~\eqref{invG_from_parall}.
				The $S^2$ case is thus on the same footing as $S^3$ and $S^4$, that is it does not seem to allow for a consistent truncation of massive type IIA supergravity preserving maximal supersymmetry.
\ensurepagenumbering{arabic}
	\chapter{Generalised Calibrations in \texorpdfstring{$\mathrm{AdS}$}{AdS} backgrounds}
	\label{chapbrane}
			
			\section{Introduction and Motivations}
			In this chapter, based mainly on~\cite{oscar2}, we are interested in investigating the relation between the Exceptional Sasaki-Einstein structures~\cite{AshmoreESE} presented in~\cref{sec:ESE} and generalised brane calibrations, in $\mathrm{AdS}_5 \times M_5$ backgrounds in type IIB and in $\mathrm{AdS}_5 \times M_6$ and $\mathrm{AdS}_4 \times M_7$ compactifications in M-theory.
			
			Also in this case it is useful to analyse the problem through the lens of $G$-structures.
			We have mentioned how requiring the AdS background to be supersymmetric is equivalent to put integrability conditions on HV structures, that in that case take the name of \emph{Exceptional Sasaki-Einstein structures}~\cite{AshmoreESE}, or of \emph{Exceptional Calabi-Yau spaces (ECY)} for compactifications to Minkowski spacetimes~\cite{AshmoreECY}.
			
			G-structures also appear naturally in defining calibration forms on the compactification manifolds. 
			
			A $p$-form $\phi$ on a $d$-dimensional manifold $M$ ($d> p$) is a \emph{calibration} if and only if it is closed, \emph{i.e.} $\dd  \phi =0$, and its pull-back to any tangent $p$-plane $\mathcal{S}$ satisfies the inequality
					\begin{equation}\label{eq:def_cal}
						P_{\mathcal{S}}[\phi] \leq \vol_{\mathcal{S}}\, ,
					\end{equation}
			where $\vol_{\mathcal{S}}$ is the volume form on the plane $\mathcal{S}$ induced from the metric on $M$~\cite{Cal_Geo}.
			
			The ordering relation in~\eqref{eq:def_cal} has to be read as $P_{\mathcal{S}}[\phi] = \alpha  \vol_{\mathcal{S}}$, with $\alpha  \in \mathbb{R}^+$ and $\alpha  \leq 1$.
			For the supersymmetric backgrounds relevant in string and M-theory the calibration forms can be written as bilinears in the supersymmetry Killing spinors. 
			For instance, on Calabi-Yau manifolds there are two types of calibration forms, which corresponds to products of the K\"ahler form and to the real part of the holomorphic form on the Calabi-Yau. 

			A $p$-dimensional submanifold $\Sigma_p$ is called \emph{calibrated} if it saturates the condition~\eqref{eq:def_cal} at each point: $P_{\Sigma_p} [\phi] = \vol_{\Sigma_p}$. 
			One can show that a calibrated submanifold minimises the volume in its homology class. 
			Indeed, given another submanifold $\Sigma'$, such that $\Sigma - \Sigma' =\partial B$ is the boundary of a $p+1$-dimensional manifold $B$, one has (see e.g.~\cite{Cal_Geo})
	\begin{equation*}
		\mathrm{Vol}\left( \Sigma' \right) = \int_{\Sigma'} \vol_{\Sigma'} \geq \int_{\Sigma} P_{\Sigma}[ \phi] + \int_{\partial B} P_{\partial B}[\phi] 
			= \int_{\Sigma} \vol_{\Sigma} + \int_{B} \dd  P_B[ \phi] = \mathrm{Vol} \left( \Sigma\right)\, , 
	\end{equation*}
where we used the definition of calibration form, Stokes theorem and the fact that $\Sigma$ saturates the inequality~\eqref{eq:def_cal}.
For a nice review on these arguments, one can refer, for instance, to~\cite{Joyce:2001nm}. 

Calibrations are useful tools in string theory because they provide a classification of supersymmetric branes in a given background. 
In a purely geometric background (no fluxes) supersymmetric branes wrap calibrated submanifolds, so that they minimise their volume~\cite{Becker:1995kb, Becker:1996ay, Gibbons:1998hm, Gauntlett:1998vk}. 
The calibration form is constructed as a bilinear in the Killing spinors of the background geometry, and its closure follows from the Killing spinor equations of the background. 

In the more general case of a background with non-trivial fluxes supersymmetric branes are associated with \emph{generalised calibrations}. 
Since the branes couple with the background fluxes, they do not correspond to minimal volume submanifolds but to configurations that minimise the energy. 
As in the fluxless case the generalised calibration form is related to the Killing spinors of the background~\cite{Gutowski:1999iu, Gutowski:1999tu, Townsend:1999nf, Gauntlett:2001ur, Gauntlett:2002sc, Gauntlett:2003cy, MS03, Cascales:2004qp, HPS03, HPS04}. 
Also in these cases the calibration forms can be written as bilinears in the Killing spinors and the closure of the generalised calibration form can then be deduced from the Killing spinor equations of the supersymmetric background~\cite{Cascales:2004qp, Gutowski:1999tu, Martucci:2005ht, MS03}.

In~\cref{sec:ESE} we have seen how the exceptional HV structure contains a generalised vector $K$ that generalises the Reeb vector field and the contact structure of usual Sasakian geometry.
For this reason, it is believed to encode information on brane configurations and conformal dimensions of chiral operators, as the contact structure does in~\cite{Martelli:2006yb}. 
In particular, in~\cite{AshmoreESE} the form parts of the generalised vector $K$ were conjectured to describe generalised calibrations for brane configurations dual to barionic operators in the dual gauge theory.
The aim of this chapter is to prove this conjecture and to show that the vector structure is indeed associated to generalised calibrations.

In the following, we focus on the calibrations forms associated to branes wrapping cycles in the internal manifolds and that are point-like in the AdS space. 
We show that for these configurations the general expression for the calibration forms that can be constructed using $\kappa$-symmetry can be expressed in terms of the generalised Killing vector $K$ defining the Exceptional Sasaki-Einstein structure and that the closure of the calibration forms is given by the integrability (more precisely the $L_K$ condition) of the ESE structure. 
Our results proves the conjecture appeared in~\cite{AshmoreESE}, that the generalised Killing vector is a generalised calibration.
We also partially discuss other brane configurations that are calibrated by the vector $K$.

The analysis in this chapter is far from being complete. 
For instance we did not fully study the calibration forms for branes with world-volumes spanning different directions in the AdS space. 
These should be related to components of the hypermultiplet structures and their closure to the moment map conditions. 
It would also be interesting to perform a similar analysis for compactifications to Minkowski space where the relevant structures are Generalised Calabi-Yau's~\cite{AshmoreECY}. 
We leave this analysis for future work.

Conventions for Clifford algebras and bilinears of spinor notations are relegated to~\cref{app:notation}.

\section{Generalised calibrations in M-theory}
\label{sec:Mcal}

The aim of this section is to study calibrations for supersymmetric brane configurations of M-theory on AdS backgrounds of the form
		\begin{equation*}
			\dd s^2 = e^{2 \Delta} \dd s^2(X_D) + \dd s^2(M_d)\, ,
		\end{equation*}		
 in terms of exceptional geometry.
AdS calibrations have been thoroughly discussed in the literature~\cite{Gutowski:1999iu, Gutowski:1999tu, MS03, Koerber:2007jb} and led to the notion of generalised calibration. 
In this section we will interpret these calibrations in terms of the Exceptional Sasaki-Einstein structures describing the AdS background.

Supersymmetry static M-theory backgrounds have been studied in~\cite{GauntlettGeomKill}. 
As we have seen in~\cref{chapComp}, a supersymmetric background admits a Majorana Killing spinor $\varepsilon$ satisfying,
		\begin{equation}
			\nabla_M \varepsilon + \frac{1}{288} \left[ \Gamma_M^{\phantom{M}NPQR} - 8 \delta_M^{\phantom{M}N} \Gamma^{PQR} \right] G_{NPQR}\ \varepsilon =0 \, ,
		\end{equation}
where $M,N, \ldots = 0,1, \dots, 10$, $G= \dd  A $ is the four-form field strength and the Gamma matrices are the Clifford algebra elements in $11$ dimensions. 
The four-form $G$ and the metric $g$ satisfy the relative equations of motion
		\begin{align}
			R_{MN} - \dfrac{1}{12}\left( G_{MPQR}G_{N}^{\phantom{N}PQR} - \dfrac{1}{12}g_{MN} G^2 \right) &= 0 \, ,\\
			\dd  \star G + \dfrac{1}{2} G \wedge G &= 0\, .
		\end{align}

	The Killing spinor can then be used to build one-, two- and five-forms
	\begin{subequations}
		\label{11df}
			\begin{align}
				\label{11d1f}
				\mathcal{K}_M &= \bar{\varepsilon} \Gamma_M \varepsilon \, , \\
				\label{11d2f}
				\omega_{M N} &= \bar{\varepsilon} \Gamma_{M N} \varepsilon \, , \\
				\label{11d5f}
				\Sigma_{MNPQR} & = \bar{\varepsilon} \Gamma_{M N PQ R} \varepsilon \, , 
			\end{align}
	\end{subequations}
and the supersymmetry conditions imply that
	\begin{eqnarray}
		& \dd  \mathcal{K} = \frac{2}{3} \iota_\omega G + \frac{1}{3} \iota_\Sigma \star G \, , \\
		& \dd  \omega = \iota_\mathcal{K} G \, , \\
		& \dd  \Sigma = \iota_\mathcal{K} \star G - \omega \wedge G \, . 
	\end{eqnarray}

Supersymmetry also implies that the vector $\hat{\mathcal{K}}^M$ dual to the one-form~\eqref{11d1f} is a Killing vector, \emph{i.e.}
		\begin{equation}
			\begin{array}{ccc}
				\mathcal{L}_{\hat{\mathcal{K}}} g = 0\, , & \phantom{and} & \mathcal{L}_{\hat{\mathcal{K}}} G = 0 \, . 
			\end{array}
		\end{equation}	
	The vector $\hat{\mathcal{K}}^M$ can be either null or time-like, and for the backgrounds of interest here it is time-like%
		\footnote{%
			In this case the forms $\mathcal{K}$, $\omega$ and $\Sigma$ define an $SU(5)$ structure in 11 dimensions.%
		}.%

	Now let us focus on AdS backgrounds,
		\begin{equation}
		\label{eq:metric_mtheory_ads}
			\dd  s^2 = e^{2\Delta}\dd  s^2(\mathrm{AdS}) + \dd  s^2(M)\, ,
		\end{equation}
	where $\Delta$ is a real function on $M$, the warp factor.
	
	As usual, to construct the generalised calibrations for M-branes we can make use of $\kappa$-symmetry. 
	A supersymmetric brane satisfies the bound
		\begin{equation}
		\label{kprojector}
			\hat{\Gamma} \varepsilon = \varepsilon \, ,
		\end{equation}
	where $\varepsilon$ is the background Killing spinor and the $\kappa$-symmetry operator $\hat \Gamma$ depends on the type of brane. For an $\mathrm{M}5$-brane this is defined as~\cite{MS03,Gabella:2012rc},
		\begin{equation}
		\label{KopMth}
			\hat{\Gamma} = \frac{1}{L_{DBI}}\Gamma_0 \left[\frac{1}{4}\Gamma^\alpha  (\tilde{H}\lrcorner H)_\alpha  + \frac{1}{2} \Gamma^{\alpha \beta} \tilde{H}_{\alpha \beta}+ \frac{1}{5!}\Gamma^{\alpha _1 \ldots \alpha _5} \epsilon_{\alpha _1 \ldots \alpha _5} \right]\, , 
		\end{equation} 
	where $H = d B + P[A]$ is the world-volume three-form, $\tilde{H}$ is its world-space dual~\cite{Pasti:1997gx,Pasti:1995tn,Bergshoeff:1998vx}%
		\footnote{%
			The field $\tilde{H}$ is defined in terms of an auxiliary scalar field $a$, which is needed to ensure the Lorentz covariance of the world-volume Lagrangian~\cite{Pasti:1997gx,Pasti:1995tn},
				\begin{equation}
					\tilde{H}_{\mu\nu} = \frac{1}{\sqrt{\vert\partial a \vert^2}} \left(\star H\right)_{\mu\nu\alpha }\partial^{\alpha } a (\sigma)\, .
				\end{equation}
			The scalar $a$ is subject to a gauge transformation and one usually fixes it by going to the \emph{temporal gauge}, \emph{i.e.} $a= \sigma^0 = t $. 
			This gauge fixing procedure breaks the Lorentz invariance $SO(1,5)$ down to $SO(5)$ and sets $\tilde{H}$ equal to the world-space dual of $H$.%
			}%
		and $L_{DBI}$ is the Dirac-Born-Infeld Lagrangian for the $\mathrm{M}5$ brane,
		\begin{equation}
		\label{DBIlagr}
			L_{DBI} = - \sqrt{-\det(P[g] + \tilde{H})}\, .
		\end{equation} 
	Per usual, $P[\bullet]$ denotes the pull-back on the $\mathrm{M}5$ world-volume and we defined
		\begin{equation}
			\Gamma_{\alpha _1 \ldots \alpha _s} = \Gamma_{M_1 \ldots M_s} \partial_{\alpha _1}X^{M_1} \ldots \partial_{\alpha _s}X^{M_s}\, .
		\end{equation}

	As discussed in~\cite{MS03, Gabella:2012rc}, the $\kappa$-symmetry condition~\eqref{kprojector} can be used to derive the following bound~\cite{Barwald:1999hx},
		\begin{equation}
		\label{M5ksym}
			\lVert \varepsilon \rVert^2 L_{DBI}\ \vol_5 \geq \left[ \frac{1}{2} P[ \iota_{\hat{\mathcal{K}}} H ] \wedge H + P[\omega] \wedge H + P[ \Sigma] \right]\, ,
		\end{equation}
	where $K$, $\Sigma$ and $\omega$ are defined in~\eqref{11df}.
	To satisfy the bound one has to take into account that the space is Anti-de Sitter. 
	As discussed in~\cite{Gabella:2012rc}, the norm $\varepsilon^\dagger \varepsilon$ depends on the AdS coordinates and the bound is saturated when the $\mathrm{M}5$ brane sits at the center of AdS.
	Explicitly, the metric of $\mathrm{AdS}_n$ in global coordinates can be written as,
		\begin{equation}
			\dd  s^2 = R^2\left(-\cosh^2 \varrho\, \dd  t^2 + \dd  \varrho^2 + \sinh^2 \varrho\, \dd  \Omega_{n-2} \right)\, ,
		\end{equation}
	and $\varepsilon^\dagger \varepsilon \propto \cosh \varrho$, thus, the~\eqref{M5ksym} can be saturated only for $\varrho = 0$, \emph{i.e.} in the center of AdS.

	Further, the bound~\eqref{M5ksym} can be used to derive a bound on the energy of the $\mathrm{M}5$ brane \cite{MS03,Gabella:2012rc}. The energy of the an $\mathrm{M}5$-brane is given by 
		\begin{equation}
		\label{energyM5}
			E_{\mathrm{M}5} =- \int_{\mathcal{S}} \dd ^5 \sigma\ g(\hat{P},\hat{\mathcal{K}}) \, , 
		\end{equation}
	where $\mathcal{S}$ denotes the $5$-dimensional world-space of the brane, $\hat{P}_M$ is the conjugate momentum%
		\footnote{%
		To write this expression, we have chosen again the static gauge $X^M = ( t, \sigma^\alpha )$.%
		}~\cite{Bergshoeff:1998vx},
		\begin{equation}
			\begin{split}
				\hat{P}_M =& \frac{\partial L_{M 5}}{\partial (\partial_\tau X^M)} = P_M + \dfrac{1}{4} \dfrac{1}{5!} \epsilon^{\tau \alpha _1\ldots\alpha _5} H_{\alpha _1 \alpha _2\alpha _3}H_{\alpha _4\alpha _5\alpha _6} \partial^{\alpha _1} X_M \\
						&\phantom{= P_M + + } - \dfrac{\tau_5}{5!} \epsilon^{\tau \alpha _1\ldots\alpha _5}\left[ \iota_M \tilde{A} - \tfrac{1}{2} \iota_M A \wedge (A - 2 H)\right]_{ \alpha _1\ldots\alpha _5} \, ,
			\end{split}
		\end{equation}
	where $X^M$ are the embedding coordinates of the brane. 
	The quantity $g(\hat{P},\hat{\mathcal{K}}) = \hat{P}^M \hat{\mathcal{K}}^N g_{MN}$ can be interpreted as a Noether charge density of the symmetry generated by $\hat{\mathcal{K}}$~\cite{Martucci:2011dn}. 
	Then the inequality~\eqref{M5ksym} gives a bound on the energy of the brane,
		\begin{equation}
		\label{BPScon}
			E_{\mathrm{M}5} \geq E^{BPS}_{\mathrm{M}5}\, ,
		\end{equation}
	where
		\begin{equation}
			E^{BPS}_{\mathrm{M}5} = \int_{\mathcal{S}} P[\Sigma] + P[\iota_{\hat{\mathcal{K}}} \tilde{A}] + P[\omega] \wedge H + \tfrac{1}{2} P[\iota_{\hat{\mathcal{K}}} H] \wedge (A - 2H)\, .
		\end{equation}
	As shown in~\cite{Gabella:2012rc}, the form
		\begin{equation}
		\label{genkcal}
			\Phi_{\mathrm{M}5} = \Sigma + \iota_{\hat{\mathcal{K}}} \tilde{A}+\omega \wedge H + \tfrac{1}{2} \iota_{\hat{\mathcal{K}}} H \wedge (A - 2H)\, ,
		\end{equation}
	is a generalised calibration, namely is closed by supersymmetry and it minimise the energy in its homology class being a topological quantity~\cite{MS03}.

	The discussion for $\mathrm{M}2$ works analogously, and the calibration form is 
		\begin{equation}
		\label{genkcalM2}
			\Phi_{\mathrm{M}2} = \omega + \iota_{\hat{\mathcal{K}}} H\, .
		\end{equation}

As final comment, we want just to point out that the construction above can also be derived by the supersymmetry algebra. The same calibration forms emerge in the supersymmetry algebra with the central extensions due to the presence of BPS extended objects, and one can prove their closure by using the Killing spinor equations~\cite{HPS03,Gutowski:1999tu, Cascales:2004qp}.

\subsection{\texorpdfstring{Calibrations on $\mathrm{AdS}_5 \times M_6$}{Calibration on AdS5 x M6}}
Even if the formalism described in the previous section is completely general, in what follows we will focus on static M-branes in backgrounds of the type~\eqref{eq:metric_mtheory_ads} and we will show how the calibration forms~\eqref{genkcal} and~\eqref{genkcalM2} are naturally encoded in the generalised Sasaki-Einstein structure.

We consider first the case of compactifications to $5$-dimensional AdS. 
The supersymmetry conditions for backgrounds of this type are give in \cite{Gauntlett:2004zh} while the corresponding exceptional generalised geometry is given in~\cite{AshmoreESE,Grana_Ntokos}, and we briefly review it below. 
We refer to these works also for notation and conventions, and we collect, for convenience, again the relevant conventions used here in~\cref{app:mtheoryconv}.

The metric takes the form
\begin{equation}
\label{eq:metric_mtheory_ads5}
	\dd  s^2 = e^{2\Delta}\dd  s^2_{\mathrm{AdS}_5} + \dd  s^2_{M_6}\, ,
\end{equation}
where we denote the inverse AdS radius as $m$. As shown in~\cite{Gauntlett:2004zh}, supersymmetry constrains the geometry of the six-dimensional internal manifold: $M_6$ has a local $SU(2)$ structure and topologically is a two-sphere bundle over a four-dimensional base%
		\footnote{%
			The four dimensional base can be either a K\"ahler-Einstein manifold with positive curvature or a product of two constant curvature Riemannian surfaces. The latter case in non-Einstein~\cite{Gauntlett:2004zh}.%
			}. 

There is a non-trivial four-form field strength $\mathcal{F}$ with non-zero components along the internal manifold $M_6$, 
	\begin{equation}
		F_{m_1 \ldots m_4} = \left(\mathcal{F}\right)_{m_1 \ldots m_4}\, ,
	\end{equation}
while the external components are set to zero, $\mathcal{F}_{\mu_1 \ldots \mu_4} = 0$.

The internal flux $F$ satisfies the equations of motion and the Bianchi identity
	\begin{equation}
		\begin{array}{ccccc}
			\dd  F = 0\, , & & & & \dd  (e^\Delta \star_6 F) = 0\, , 
		\end{array}
	\end{equation}
with $\star_6$ the Hodge star on $M_6$, while the dual form $\tilde{F}_{m_1 \ldots m_7} = \left(\star_{11}\mathcal{F}\right)_{m_1 \ldots m_7}$ identically vanishes on the six-dimensional internal manifold $M_6$.

The Clifford algebra $\mathrm{Cliff}(1,10)$ decomposes in $\mathrm{Cliff}(1,4)$ and $\mathrm{Cliff}(0,6)$:
	\begin{equation}
	\label{eq:dec_gammas_m6}
		\begin{array}{ccc}
			\hat{\Gamma}^\mu = e^{\Delta}\,\rho^\mu \otimes \gamma_7\, , & \phantom{and} &\hat{\Gamma}^{m+4} = \mathds{1}_4 \otimes \gamma^m\, ,
		\end{array}
	\end{equation}
with $\gamma_7 = - i \gamma^1 \ldots \gamma^6$ the chiral operator in $6$ dimensions, and $\rho$ and $\gamma$ matrices satisfying
	\begin{equation}
	\label{eq:cliff}
		\begin{array}{ccc}
			\left\{\rho_\alpha , \rho_\beta \right\} = 2 \eta_{\alpha \beta} \mathds{1}\, , & \phantom{and} &\left\{\gamma_{a} , \gamma_b \right\} = 2 \delta_{ab} \mathds{1}\, ,
		\end{array}
	\end{equation}
in terms of the frame indices $\alpha ,\beta = 0,\ldots,4$ on $\mathrm{AdS}_5$ and $a,b = 1, \ldots, 6$ on $M_6$. We collect further conventions about spinors and Clifford algebras in~\cref{app:mtheoryconv}.

To have an $\mathcal{N}=2$ supersymmetric background we decompose the $11$-dimensional spinor as
	\begin{equation}
	\label{eq:decomp_mtheory_spinors}
		\varepsilon = \psi \otimes \chi + \psi^c \otimes \chi^c\, ,
	\end{equation}
where $\psi$ is an element of $\mathrm{Cliff}(1,4)$. 
Notice that, in order to have an AdS backgrounds, the internal spinor $\chi$ cannot be a chirality eigenstate~\cite{Gauntlett:2004zh}. 
Hence, it can be written as,
	\begin{equation}
	\label{chiredef}
		\chi = \sqrt{2} \left(\cos \alpha  \chi_1 + \sin \alpha  \chi_2^* \right)\, ,
	\end{equation}
where $\alpha $ is a parameter chosen to get the unit norm for the spinor, as in~\cite{Gauntlett:2004zh}.\\

The exceptional geometry for these backgrounds is given in~\cite{AshmoreESE,Grana_Ntokos}.
The exceptional bundles to consider are again~\eqref{MthExBun} and~\eqref{eq:Ggeom-M}. 
The vector structure $K\in E$ and the hypermultiplet structure $J_a \in \mathrm{ad}F$ can be expressed in terms of the $SU(2)$ structure of~\cite{Gauntlett:2004zh}. 

In this discussion we are mostly concerned with the generalised Killing vector $K$. 
Its untwisted version is given by
	\begin{equation}
	\label{vecK}
		\begin{array}{cc}
			\tilde{K}=\xi -e^{\Delta} Y' + e^{\Delta} Z \equiv \xi + \tilde{\omega} + \tilde{\sigma} & \in \tilde{E} 
		\end{array}
	\end{equation}
with the vector $\xi$, the two-form $Y'$ and five-form $Z$ defined in terms of spinor bilinears as in~\cite{AshmoreESE}, 
	\begin{align}
		\label{vecbilM6}
		\xi &= -i \left(\bar{\chi}_1+\chi_2^T\right)\gamma^{(1)}\left(\chi_1 - \chi_2^* \right)\, ,\\[1mm]
		\label{YbilM6}
		Y' &= -i \left(\bar{\chi}_1+\chi_2^T\right)\gamma_{(2)}\left(\chi_1 - \chi_2^* \right)\, ,\\[1mm]
		\label{ZbilM6}
		Z &= -i \left(\bar{\chi}_1+\chi_2^T\right) \gamma_{(5)} \left(\chi_1 - \chi_2^* \right)\, .
	\end{align}
The twisted version of the V structure is obtained by the (exponentiated) adjoint action
\begin{equation}
	K = e^{A+\tilde{A}} \tilde{K}\, ,
\end{equation}
where $A$ is the three-form and $\tilde{A}$ the six-form potential of $M$-theory. Using the expressions for the commutator and the adjoint action from~\cite{AshmoreECY}, one obtains~\cite{AshmoreESE} 
	\begin{equation}
	\label{eq:twisted_K_M}
		K= \xi + \left( \iota_{\xi}A - e^{\Delta}Y'\right) + \left( e^{\Delta} Z - e^{\Delta} A \wedge Y' + \tfrac{1}{2} \iota_{\xi} A \wedge A\right)\, .
	\end{equation}
As discussed above, the tensor $\tilde{R}$ must vanish for the generalised Lie derivative to reduce to the usual one, and this is equivalent to~\cite{AshmoreESE}
	\begin{align} \label{isom5}
		& &\dd  \tilde{\omega} = \iota_{\xi} F\, , & & \dd  \tilde{\sigma} = \iota_{\xi} \tilde{F} - \tilde{\omega} \wedge F\, . & &
	\end{align}
On $M_6$, this yields the differential conditions
	\begin{subequations}
	\label{eq:m-theory_structure_eqs}
		\begin{align}
			\dd  \left( e^{\Delta} Y'\right) &= - \iota_{\xi} F\, , \\[1mm]
			\dd  \left( e^{\Delta}Z\right) &= e^{\Delta} Y' \wedge F\, ,
		\end{align}
	\end{subequations}
which we refer to as $L_K$ conditions in the language of exceptional generalised geometry.

Supersymmetry gives also the Killing vector condition
	\begin{equation}
		\mathcal{L}_\xi F = \mathcal{L}_\xi \Delta = \mathcal{L}_\xi g = 0\, . 
	\end{equation}

We can now discuss how the generalised Killing vector $K$ is related to the calibration forms for supersymmetric branes.
The general calibrations for $\mathrm{M}5$ and $\mathrm{M}2$ branes are given by the~\eqref{genkcal} and~\eqref{genkcalM2}. 
With an appropriate choice of the $\mathrm{AdS}_5$ gamma matrices (see~\cref{app:mtheoryconv}), the 11-dimensional Killing vector $\mathcal{K}_M$ in~\eqref{11d1f} has only two non-zero components,
\begin{subequations}
\begin{align}
& \mathcal{K}_0 = \bar{\psi} \rho_0 \psi \\
& \mathcal{K}_m = -i \left(\bar{\chi}_1+\chi_2^T\right)\gamma_{m}\left(\chi_1 - \chi_2^* \right) = \xi_m \, ,
\end{align}
\end{subequations}
where we fixed the norms of the spinors to $\bar \chi \chi = 1$ and $(\bar{\psi}\psi) = i/2$ and $\xi_m$ is the Reeb vector on $M_6$. Consistently, we also fixed the value of the angular parameter to $\alpha  = \pi/4$ in~\eqref{chiredef}.

The specific expression of the calibration forms $\Phi_{\mathrm{M}5}$ and $\Phi_{\mathrm{M}2}$ in~\eqref{genkcal} and~\eqref{genkcalM2} depends on how many AdS directions are spanned by the world-volume of the branes.

Consider an $\mathrm{M}5$ wrapping a $5$-cycle in $M_6$. 
We choose again the static gauge for the brane embedding and we set to zero the world-volume gauge field (so $H = A$). 
The the relevant components of the $\Sigma$ and $\omega$ in~\eqref{11d2f} and~\eqref{11d5f} are the internal ones,
	\begin{subequations}
		\begin{align}
			\omega_{m_1 m_2} &= e^\Delta \, \bar{\chi}\gamma_7 \gamma_{m_1 m_2} \chi = e^\Delta Y^\prime \, , \\
			\Sigma_{m_1 \ldots m_5} &= e^\Delta \, \bar{\chi}\gamma_7 \gamma_{m_1 \ldots m_5} \chi = e^\Delta Z \, , 
		\end{align}
	\end{subequations}
and the calibration form in~\eqref{genkcal} reads (recall that the pull-back of $\tilde A$ is zero),
	\begin{equation}
		\Phi_{\mathrm{M}5} = e^{\Delta} Z -e^{\Delta} A \wedge Y' + \tfrac{1}{2}\iota_{\xi} A\wedge A \, . 
	\end{equation}
Note that this is exactly the pull-back on the brane of the twisted generalised vector $K$ in~\eqref{eq:twisted_K_M}. 
A similar computation for an $\mathrm{M}2$-brane wrapping a $2$-cycle in $M_6$ gives
	\begin{equation}
		\Phi_{\mathrm{M}2} = e^{\Delta} Y' - \iota_{\xi} A \, ,
	\end{equation}
which is again the pull-back on the $\mathrm{M}2$-brane of the twisted generalised vector $K$.
Using the $L_K$ conditions~\eqref{eq:m-theory_structure_eqs} and choosing the a gauge for $A$ such that 
$\mathcal{L}_\xi A =0$, it is straightforward to check that $\Phi_{\mathrm{M}5}$ and $\Phi_{\mathrm{M}2}$ are closed.
Explicitly, for instance for $\mathrm{M}5$, one has,
	\begin{equation}
		\begin{split}
			\dd  \Phi_{\mathrm{M}5} &= \dd  (e^{\Delta} Z ) - \dd  (e^{\Delta} A \wedge Y') + \tfrac{1}{2}\dd (\iota_{\xi} A\wedge A) = \\
					&= e^{\Delta} Y' \wedge F - F \wedge e^\Delta Y' + \iota_\xi F \wedge A + \tfrac{1}{2}\dd (\iota_\xi A)\wedge A + \tfrac{1}{2}\iota_\xi A \wedge F = \\
					&= \iota_\xi F \wedge A + \tfrac{1}{2} \mathcal{L}_\xi A \wedge A - \tfrac{1}{2}\iota_\xi F \wedge A - \tfrac{1}{2} A \wedge \iota_\xi F = 0\, .
		\end{split}
	\end{equation}
Analogously, one can verify that $\Phi_{\mathrm{M}2}$ is also closed, showing that the purely internal configuration of the membrane is supersymmetric.

The generalised vector $K$ is also related to the calibration forms for other types of brane probes. 
Here we focus on branes with one one leg in the external space-time, that is a string moving in AdS. 
We leave the study of other membrane configurations to future work.
The calibration forms for $\mathrm{M}2$ and $\mathrm{M}5$-branes of this kind are given by~\eqref{genkcal} and~\eqref{genkcalM2} in this case, take the following form
	\begin{align}
		& \Phi_{\mathrm{M}2} = e^{2\Delta} \tilde{\zeta}_1 \\
		& \Phi_{\mathrm{M}5} = e^{2\Delta} \star Y' + e^{2\Delta} \tilde{\zeta}_1 \wedge A 
	\end{align}
where $Z = \star \tilde{\zeta}_1$. 
The two calibrations are components of the (poly)-form 
	\begin{equation}
	\label{eq:cal_M_q1}
		\Phi= e^{2\Delta}\tilde{\zeta}_1 +e^{2\Delta}\star_6 Y'\, + e^{2\Delta} \tilde{\zeta}_1 \wedge A\, .
	\end{equation}
which is the Hodge dual of the vector structure ~\eqref{eq:twisted_K_M} . 
We want now to study it's closure and its relation to the integrability conditions.
In this case, the closure follows from the moment map condition $\mu_3 \equiv 0$, rather than from the $L_{K}$ condition. 
In~\cite{AshmoreESE}, it is shown that this moment map condition requires
\begin{equation}
	\dd  \bigl( e^{2\Delta} \tilde{\zeta}_1\bigr) = 0\, ,
\end{equation}
so that this form calibrates a $\mathrm{M}2$-brane. 
Again, combining the two conditions, we get
\begin{equation}
	\begin{array}{lcr}
		\dd  \left( e^{3\Delta} \sin\Theta \right) = 2 m e^{2\Delta} \tilde{\zeta}_1 & \mbox{and} & \dd  \left( e^{3\Delta} V\right) = e^{3\Delta} \sin \Theta F + 2m e^{2\Delta} \star Y'\, .
	 \end{array}
\end{equation}
From the vanishing of $\mu_3$ in~\cite{AshmoreESE}, it is easy to verify that the form $e^{2\Delta} \star_6 Y' + e^{2\Delta}\tilde{\zeta}_1 \wedge A$ is closed (for non-vanishing $m$).
%
%
%
\subsection{\texorpdfstring{Calibrations in $\mathrm{AdS}_4 \times M_7$}{Calibrations on AdS4 x M7}}
\label{M-thAdS4}
In this section, we discuss M-theory calibrations on $\mathrm{AdS}_4$ backgrounds. 
Again, we first review the exceptional generalised geometry~\cite{AshmoreESE} and then relate it to generalised calibrations.
Conventions for the spinor bilinears and the supersymmetry equations for the internal forms can be found in~\cite{Gabella:2012rc} and the relevant ones for this work are collected in appendix~\ref{app:conv}. 

The background metric has the following form 
	\begin{equation}
	\label{eq:metric_mtheory_ads4}
		\dd  s^2 = e^{2\Delta} \dd  s^2_{\mathrm{AdS}_4} + \dd  s^2_{M_7} \, .
	\end{equation}
We set the inverse $\mathrm{AdS}_4$ radius to $m=2$. In addition, there is a non trivial four-form flux 
	\begin{equation}
		G = m \vol_4 + F \, ,
	\end{equation}
where $F = \dd  A$ is the flux component on $M_7$ and it satisfies the following Bianchi identity and equations of motion 
	\begin{equation}
		\begin{array}{ccccc}
			\dd  F = 0\, , & & & &\dd ( e^{2\Delta} \star_7 F ) = - m F \, , 
	\end{array}
	\end{equation}
with $\star_7$ the Hodge star on $M_7$. 
We will also need its dual $\tilde{F} = \dd  \tilde{A} - \tfrac{1}{2} A \wedge F$.

The $11$-dimensional gamma matrices split as 
	\begin{equation}
	\label{eq:dec_gammas_m7}
		\begin{array}{ccc}
			\Gamma_{\mu} = e^{\Delta} \rho_{\mu} \otimes \id & \mbox{and} & \Gamma_m = e^{\Delta} \rho_5 \otimes \gamma_m\, ,
		\end{array}
	\end{equation}
with $\{\rho_{\mu}, \rho_{\nu}\}=2 g_{\mu \nu}$ and $\{\gamma_m, \gamma_n\}= g_{mn}$. 
The matrix $\rho_5 = i \rho_{0123}$ is the chirality operator in four dimensions, and $\gamma_{1 \ldots 7}=i \id$.
For further details about Clifford algebra conventions we refer to the~\cref{app:conv}.

The spinor ansatz preserving eight supercharges reads~\cite{Gabella:2012rc, Gabella:2011sg}
	\begin{equation}
	\label{eq:spinor_m7}
		\begin{split}
			\varepsilon &= \sum_{i=1,2} \psi_i \otimes e^{\Delta/2} \chi_i + \psi^c_i \otimes e^{\Delta/2} \chi_i^c\\
					&= e^{\Delta/2} \psi_+ \otimes \chi_- + e^{\Delta/2} \psi_- \otimes \chi_+ + \mathrm{c.c.} 
		\end{split}
	\end{equation}
where $\chi_{\pm} \coloneqq \tfrac{1}{\sqrt{2}} \left( \chi_1 \pm i \chi_2\right)$ and $\psi_{\pm} \coloneqq \tfrac{1}{\sqrt{2}} \left( \psi_1 \pm \psi_2\right)$. 
In addition, we take the $\mathrm{AdS}_4$ spinors $\psi_i$ to have positive chirality, \emph{i.e.} $\rho_5 \psi_i = \psi_i$.\\

Combining the supersymmetry conditions and equations of motion for the fluxes one can express the internal fluxes in terms of spinor bilinears by~\cite{Gabella:2012rc},
	\begin{align}
		F &= \dfrac{3 m}{\tilde{f}}\ \dd  (e^{6 \Delta} i( \bar{\chi}_+^c \gamma_{(3)} \chi_-))\, , \\
		\tilde{F} &= - \tilde{f}\ \vol_7 \, . 
	\end{align}

The features of the solutions depend on the \emph{electric} charge $m$. 
When $m=0$ the solutions correspond to near horizon geometries of $\mathrm{M}5$-branes wrapped on internal cycles (no $\mathrm{M}2$ charge). 
The geometries with $m\neq0$ correspond to the presence of a non-vanishing $\mathrm{M}2$ charge. 
For $m \neq 0$ the internal manifolds always admit a canonical contact structure, as shown in~\cite{Gabella:2012rc}. \\

%
%
	The generalised geometry relevant for backgrounds of this kind is discussed in~\cite{AshmoreESE}. 
	The $HV$ structure is given by a generalised vector $X$ in the fundamental of $E_{7(7)}$ and a triplet $J_a$ in the adjoint representation. 
	The untwisted vector reads
		\begin{equation}
			\tilde{X}= \xi + e^{3\Delta} Y + e^{6\Delta} Z - i e^{9\Delta} \tau\, ,
		\end{equation}	
where the forms are bilinears in the internal background spinors 
		\begin{equation}
		\label{bilrel}
			\begin{array}{lcccccr}
				\sigma = i \bar{\chi}_+^c \gamma_{(1)} \chi_- \, , &\phantom{and}& Y = i \bar{\chi}_+^c \gamma_{(2)} \chi_- \, ,&\phantom{and}& Z=\star_7 Y \, ,& \phantom{and}& \tau = \sigma \otimes \vol_7\, ,
			\end{array}
		\end{equation}
and $\xi$ is the vector dual to the one-form $\sigma$. 
Notice that the vector structure has the same form in both cases of a Sasaki-Einsten internal manifold and of a generic flux background~\cite{AshmoreESE}. 
Indeed the seven-dimensional manifolds giving $\mathcal{N}=2$ supersymmetry always admit a local $SU(2)$ structure. 
Moreover, the Killing vector constructed by spinor bilinears in~\eqref{bilrel} (or equivalently its dual one-form $\sigma$) defines a contact structure.
This allows us to write the $M_7$ metric as a Reeb foliation, analogously to the case of a Sasaki-Einstein manifold~\cite{Gabella:2012rc}. 
As a consequence, the volume form can be written making use of the contact structure,
		\begin{equation}
			\dfrac{1}{3!} \sigma \wedge \dd  \sigma \wedge \dd  \sigma \wedge \dd  \sigma = \left(\dfrac{3m^2}{\tilde{f}}\right)^3 e^{9\Delta} \vol_7 = 2\left(\dfrac{3m^2}{\tilde{f}}\right)^3 \sqrt{q(K)}\, ,
		\end{equation}
where $q(K)$ is the $E_{7(7)}$ invariant and $K$ is the real part of the twisted vector structure $X$
		\begin{equation}
		\label{Kads4}
			K = \xi - \frac{1}{2} \sigma \wedge \omega \wedge \omega + \iota_\xi \tilde{A}\, .
		\end{equation}
As already mentioned in~\cref{sec:ESE}, supersymmetry implies that $X$ is a generalised vector and its vector part, $\xi$, is a Killing vector. 
Through AdS/CFT, $\xi$ is the dual of the R-symmetry of the conformal $\mathcal{N}=2$ gauge theory in three dimensions. 
Then, as discussed in~\cref{sec:ESE}, the generalised Lie derivative along $X$ must reduce to $\mathcal{L}_\xi$, which implies the vanishing of the tensor $\tilde{R}$, or more explicitly
		\begin{equation}
		\label{M7Rvan}
			\begin{array}{l}
				\dd  (e^{3\Delta}Y)=\iota_{\xi} F\, , \\
				\dd  (e^{6\Delta}Z)= \iota_{\xi} \tilde{F} -e^{3\Delta} Y \wedge F\, .
			\end{array}
		\end{equation}
As expected, these reproduce part of the supersymmetry equations in~\cite{Gabella:2012rc}.

One can choose the gamma matrices and spinors in such a way that the Killing vector $\mathcal{K}$ has components~\cite{Gabella:2012rc} 
	\begin{equation}
		\begin{split}
			\mathcal{K}_0 & = \sum_i \bar{\psi}_i\rho_0 \psi_i\, , \\
			\mathcal{K}_m & = - \tfrac{i}{2} e^{2\Delta}\, \bar{\chi}^c_+ \gamma_m \chi_- \, .
		\end{split}
	\end{equation}

As in the previous section, the form of the generalised calibrations for $\mathrm{M}5$ and $\mathrm{M}2$ branes,~\eqref{genkcal} and~\eqref{genkcalM2}, depends on the direction spanned by the branes. 
Again, we considered first an $\mathrm{M}5$ wrapping a 5-cycle in $M_7$, with a zero world-volume gauge field ($H = A$) and in the static gauge. 

In this case, the relevant components of the forms~\eqref{11d2f} and~\eqref{11d5f} are 
	\begin{equation*}
		\begin{aligned}
			& \omega = \tfrac{i}{2} e^{3\Delta} \bar{\chi}_{+}^c \gamma_{(2)} \chi_{-} = e^{3\Delta} Y \, ,\\[1mm]
			& \Sigma = -e^{6\Delta}(\bar{\chi}_+ \gamma_{(5)} \chi_+^c + \bar{\chi}_-^c \gamma_{(5)} \chi_- ) = e^{6\Delta} Z\, ,
		\end{aligned}
	\end{equation*}
and the calibration $\Phi_{\mathrm{M}5}$ gives
	\begin{equation}
		\Phi_{\mathrm{M}5} = (e^{6\Delta} Z + A \wedge e^{3\Delta} Y + \tfrac{1}{2}\iota_\xi A \wedge A + \iota_{\xi}\tilde{A})\, .
	\end{equation}	
One can also add an $\mathrm{M}2$ completely arranged along the internal directions. 
The corresponding calibration form is given by,
	\begin{equation}
		\Phi_{\mathrm{M}2} = (e^{3\Delta} Y + \iota_{\xi} A )\, ,
	\end{equation}
which together with $\Phi_{\mathrm{M}5}$ gives rise to a poly-form,
	\begin{equation}
		\Phi = (e^{3\Delta} Y + \iota_{\xi} A ) + (e^{6\Delta} Z + A \wedge e^{3\Delta} Y + \tfrac{1}{2}\iota_\xi A \wedge A + \iota_{\xi}\tilde{A})\, ,
	\end{equation}
and this, again, corresponds to the vector structure. 
The closure of $\Phi$, follows from supersymmetry%
		\footnote{%
			Note that $L_K$ conditions imply that $\Phi_{\mathrm{M}5}$ and $\Phi_{\mathrm{M}2}$ are separately closed.%
			}%
			, more precisely, from the $L_K$ conditions~\eqref{M7Rvan},
		\begin{equation}
			\begin{split}
				\dd  \Phi &= \dd  (e^{3\Delta} Y + \iota_{\xi} A ) + \dd  (e^{6\Delta} Z + A \wedge e^{3\Delta} Y + \tfrac{1}{2}\iota_\xi A \wedge A + \iota_{\xi}\tilde{A}) \\[1mm]
						& = \iota_\xi F + \dd  (\iota_\xi A) + \iota_\xi \tilde{F} - e^{3\Delta} Y \wedge F + F \wedge e^{3\Delta} Y - A \wedge \iota_\xi F \\
						& \phantom{=} + \tfrac{1}{2} \dd  (\iota_\xi A) \wedge A + \tfrac{1}{2} \iota_\xi A \wedge F + \dd  (\iota_\xi \tilde{A}) \\[1mm]
						&= \mathcal{L}_\xi A + \mathcal{L}_\xi \tilde{A} + \tfrac{1}{2} \mathcal{L}_\xi A \wedge F + \tfrac{1}{2} A \wedge \iota_\xi F - A \wedge \iota_\xi F + \tfrac{1}{2}\iota_\xi A \wedge F = 0\, ,
			\end{split}	
		\end{equation}
where in the last line we made a gauge choice, such that,
	\begin{align}
		& & & \mathcal{L}_\xi A= 0\, , & & \mathcal{L}_\xi \tilde{A}= 0\, . & 
	\end{align}

One can consider not only branes wrapping internal cycles. 
For instance, for $\mathrm{M}5$-brane spanning two spatial directions, we can show that -- also in this case -- the related calibration form comes from the vector $K$. 
The relevant form is given by
		\begin{equation}
			\Phi = (e^{3\Delta} Y + \iota_{\xi} A) \, .
		\end{equation}
The closure of this form comes directly from~\eqref{M7Rvan}.

It is also easy to see that, in this case, branes with one leg aligned with an external space direction 
are not supersymmetric, as indeed already discussed in~\cite{SanchezLoureda:2005ap}.

Indeed, for instance in the case of an $\mathrm{M}2$ wrapping an internal cycle and with one external leg, the candidate calibration form is proportional to $\sigma$ in~\eqref{bilrel}, which is not closed -- \emph{i.e.} $\dd  \sigma \sim \omega$. 
This condition, in the language of Exceptional generalised geometry, is a part of $ L_K J_a = \epsilon_{abc}  \lambda_b J_c $. 
It is interesting to note that, on the other hand, this configuration is supersymmetric in the case of a Minkowski background, since the analogous condition reads $L_K J_a = 0$,~\cite{AshmoreECY}.

To conclude the analysis, let us focus on space filling brane configurations. 
This case corresponds to the $J_a$ components of the Exceptional Sasaki-Einstein structure. 
For instance, for an $\mathrm{M}5$-brane we find
		\begin{equation}
		\label{q3cal}
			\Phi = -e^{4\Delta}{V}_-\, ,
		\end{equation}
where $V_{-}$ is the two-form defined from spinor bilinears as follows,
		\begin{equation}
		\label{Vbil}
			V_{\pm} \coloneqq \dfrac{1}{2i} \left( \bar{\chi}_+ \gamma_{(2)}\chi_+ \pm \bar{\chi}_- \gamma_{(2)}\chi_-\right)\, ,
		\end{equation}
	which gives the $TM \otimes T^*M$-component of $J_a$ by raising one index. 
	In particular, in the limit of a Sasaki-Einstein manifold (the only one for which the expression of $J_a$ is given explicitly in~\cite{AshmoreESE}), the calibration form~\eqref{q3cal} corresponds to $J_3$, and we have good reasons to trust this result also for the cases where generic fluxes are turned on. 
	We leave the complete discussion of these cases for future work.

\section{Supersymmetric branes in type IIB}
\label{sec:IIBcal}
In this section, we want to discuss the analogous conditions to have supersymmetric extended objects in a type IIB supergravity AdS background and their connections to Exceptional Sasaki-Einstein structures defining such backgrounds.

We are interested in backgrounds with non trivial fluxes. The NS three-form is $H = \dd  B$ and the RR fields are 
	\begin{equation} 
	 F_1=\dd  C_0 \qquad 
	 F_3=\dd  C_2 \qquad 
	 F_5=\dd  C_4 - \frac{1}{2} H \wedge C_2 + \frac{1}{2} F_3 \wedge B
	\end{equation} 
The field strengths $F$ satisfy the duality condition
	\begin{equation} 
	\label{fdual}
		F_p = (-1)^{\left[\tfrac{p}{2}\right]} \star F_{10-p}\, , 
	\end{equation} 
while the S-duality of type IIB is reflected in the fact that $B$ with $C_2$ form an $SL(2)$ doublet. It could be useful to define a complexified version of the three-form flux~\cite{Schwarz:1983qr},
	\begin{equation} 
		G = F_3 + i H_3\, .
	\end{equation} 
The Bianchi identities are written as
	\begin{align}
		& &	\dd  F_5= \frac{1}{8}\, \IIm\, G\wedge G^*\, ,
			& & 
		\dd  G= 0\, . & & 
	\end{align}

The generalised calibrations for backgrounds with non-trivial NS-NS flux have been constructed in~\cite{Martucci:2011dn} (see also~\cite{Cascales:2004qp, HPS04} for an equivalent derivation in terms of the supersymmetry algebra). 
The two Majorana-Weyl supersymmetry parameters $\varepsilon_1$ and $\varepsilon_2$ can be used to construct the following bilinears~\cite{BJRT99},
	\begin{align}
		\label{K10}
		\mathcal{K} &= \dfrac{1}{2} ( \bar{\varepsilon}_1 \Gamma_M \varepsilon_1 + \bar{\varepsilon}_2 \Gamma_M \varepsilon_2)\ \dd  x^M\, , \\
		\label{omega10}
		\omega &= \dfrac{1}{2} ( \bar{\varepsilon}_1 \Gamma_M \varepsilon_1 - \bar{\varepsilon}_2 \Gamma_M \varepsilon_2)\ \dd  x^M\, , \\
		\label{psiIIB}
		\Psi &= \sum_{k=0}^2 \dfrac{1}{(2 k+ 1)!} \bar{\varepsilon}_1 \Gamma_{M_1 \ldots M_{2k +1} } \varepsilon_2 \ \dd  x^{M_1} \wedge \ldots \wedge \dd  x^{M_{2k +1}} \, . 
	\end{align}
Using the Killing spinor equations for type IIB, one can show that the vector $\hat{\mathcal{K}}$ dual to the one form $\mathcal{K}$ is a Killing vector~\cite{Tomasiello:2011eb},
	\begin{equation} 
		\begin{array}{ccc}
			\mathcal{L}_{\hat{\mathcal{K}}} g = 0\, ,& \phantom{and}& \mathcal{L}_{\hat{\mathcal{K}}} F = 0 \, . 
		\end{array}
	\end{equation} 
Notice that also the spinor bilinears~\eqref{K10}--\eqref{psiIIB} are invariant under the transformation generated by~$\hat{K}$
	\begin{equation} 
		\begin{array}{ccc}
			\mathcal{L}_{\hat{\mathcal{K}}} \omega = 0\, , & \phantom{and} & \mathcal{L}_{\hat{\mathcal{K}}} \Psi = 0 \, .
		\end{array}
	\end{equation} 

As discussed in~\cite{Martucci:2011dn}, we may write the $\kappa$-symmetry condition to have a supersymmetric $\mathrm{D}p$-brane 
	\begin{equation} 
		\hat{\Gamma}_{\mathrm{D}p}\ \varepsilon_2 =\varepsilon_1\, ,
	\end{equation} 
where, the $\kappa$-symmetry operator is defined as~\cite{BT97, Marolf:2003vf} 
	\begin{equation} 
	\label{eq:brane_kappa}
		\hat{\Gamma}_{\mathrm{D}p}=\frac{1}{\sqrt{-\det\left( P[G]+\mathcal{F} \right)}} \sum_{2l+s=p+1}\frac{\epsilon^{\alpha _1 \ldots \alpha _{2l}\beta_1 \ldots \beta_s}}{l!s!2^l} \mathcal{F}_{\alpha _1\alpha _2} \ldots \mathcal{F}_{\alpha _{2l-1}\alpha _{2l}} \Gamma_{\beta_1 \ldots \beta_s}\, ,
	\end{equation} 
and $P[\bullet]$ denotes the pullback to the $(p+1)$-dimensional brane world-volume and $\mathcal{F}= F + P[B]$, with $B$ the NS two-form and $F$ the world-volume gauge field-strength.

The energy of the brane (the charge associated to the transformation generated by $\hat{\mathcal{K}}$) is
	\begin{equation} 
		E = - \int_{\mathcal{S}}\ \dd ^p \sigma\ \hat{P}_M \hat{\mathcal{K}}^M \, ,
	\end{equation} 
where $\mathcal{S}$ is the brane world-space and $\hat{P} = \tfrac{\partial L_{\mathrm{D}p}}{\partial (\partial_\tau X^M)}$ takes the form
	\begin{equation} 
		\hat{P}_M = - \mu_{\mathrm{D}p} e^{- \phi} \sqrt{- \det \mathcal{M}} (\mathcal{M}^{-1})^{( \alpha  \tau)} B_{MN} \partial_\alpha  X^N 
			+ \frac{\mu_{\mathrm{D}p}}{p!} \epsilon^{\tau \alpha _1 \ldots \alpha _p} \left[ \iota_M (C \wedge e^\mathcal{F}) \right]_{\alpha _1 \ldots \alpha_p} \, ,
	\end{equation} 
where we denoted $\mathcal{M} = P[g] + \mathcal{F}$. 
Note that again we are in the temporal gauge in adapted coordinates, such that the world-volume of the brane is $\mathbb{R}\times \mathcal{S}$.
One has the usual BPS bound,
	\begin{equation} 
		E \geq E_{BPS}\, ,
	\end{equation} 
with
	\begin{equation} 
		\label{BPScal}
			\begin{split}
				E_{BPS} =\mu_{\mathrm{D}p} \int_\mathcal{S} \dd ^p \sigma\, &P\left[e^{- \phi} \Psi - \iota_{\hat{\mathcal{K}}} C - \omega \wedge C \right] \wedge e^{\mathcal{F}} \\
					&+ \mu_{\mathrm{D}p}\int_{\mathcal{S}}\dd ^p \sigma\, P\left[\omega-\iota_{\hat{\mathcal{K}}}B\right] \wedge \left.\left(C \wedge e^{\mathcal{F}}\right)\right\vert_{p-1} \, .
			\end{split}
	\end{equation} 
Thus, one can read the generalised calibration form from the last expression,
	\begin{equation} 
	\label{BPScal2}
		\Phi_{\mathrm{D}p} = e^{- \phi} \Psi - \iota_{\hat{\mathcal{K}}} C - \omega \wedge C \wedge e^{\mathcal{F}} + \omega-\iota_{\hat{\mathcal{K}}}B \wedge \left.\left(C \wedge e^{\mathcal{F}}\right)\right\vert_{p-1}\, .
	\end{equation} 
One can show~\cite{Martucci:2011dn, Evslin:2007ti} that this is a topological quantity.
In addition, one can also show that this form is closed, making use of potential configurations preserving the symmetry generated by $\hat{\mathcal{K}}$, \emph{i.e.}
	\begin{align}
	\label{vanpot}
		&&\mathcal{L}_{\hat{\mathcal{K}}} B = 0\, , & &\mathcal{L}_{\hat{\mathcal{K}}} C = 0\, ,&&
	\end{align}
analogously to what has been done in the previous section for M-theory. 
As a final observation, we would like to point out that the same conclusions about calibration forms can be obtained by supertranslation algebra, as done for example in~\cite{Cascales:2004qp, Gutowski:1999tu,HPS04}. \\

Let us now focus on type IIB compactifications to $\mathrm{AdS}_5$-backgrounds. 
As for the discussion of supersymmetric extended objects in M-theory above, we now apply the supersymmetry conditions and the aforementioned approach to branes in type IIB string theory on 
	\begin{equation} 
	\label{eq:metric_IIB_ads5}
		\dd  s^2= e^{2\Delta} \dd  s^2_{\mathrm{AdS}_5} + \dd  s^2_{M_5}\, ,
	\end{equation} 
and relate the calibration forms to the geometric description by the vector and hypermultiplet structures. 
The exceptional geometry of this setup is discussed in~\cite{AshmoreESE,Grana_Ntokos}, based on the geometric description in~\cite{Gauntlett:2005ww}. 
For $\mathcal{N}=2$ backgrounds of the form~\eqref{eq:metric_IIB_ads5} with generic fluxes, the internal manifold $m_5$ admits a (local) identity structure~\cite{Gauntlett:2005ww,GGPSW09, GGPSW09_02}. 

%
%
%
%
The two ten-dimensional Majorana-Weyl spinors of the same chirality which describe a IIB background of the form~\eqref{eq:metric_IIB_ads5} can be decomposed as in~\cite{Grana_Ntokos}%
		\footnote{%
		We follow the conventions given in the appendix of~\cite{Grana_Ntokos}.
		We collect them in~\cref{sect:IIB_notation}.
		},
	\begin{equation} 
	\label{eq:splitting_IIB_ads5}
		\varepsilon_i= \psi \otimes \chi_i \otimes u + \psi^c \otimes \chi_i^c \otimes u\, . 
	\end{equation} 
Here $\psi$ denotes the external $\mathrm{Spin(4,1)}$ spinor, $\chi_i$ are the internal $\mathrm{Spin}(5)$ spinors and $u$ a two-component spinor. 
It might be convenient to define the complex spinors $\zeta_1 = \chi_1 + i \chi_2$ and $\zeta_2^c = \chi_1^c + i \chi_2^c$.
As for the previous cases, one can construct the relevant bilinears defining a local identity structure on the internal manifold~\cite{Gauntlett:2005ww}. 
One introduces the vectors
\begin{equation} 
\label{IIBbil}
	\begin{aligned}
 		K_0^m &:= \bar{\zeta}_1^c\gamma^m\zeta_2\, , \\
 		K^m_3 &:= \bar{\zeta}_2\gamma^m\zeta_1 \, ,\\
 		K^m_4 &:= \tfrac{1}{2}\left(\bar{\zeta}_1\gamma^m\zeta_1 - \bar{\zeta}_2\gamma^m\zeta_2\right)\, , \\
 		K^m_5 &:= \tfrac{1}{2}\left(\bar{\zeta}_1\gamma^m\zeta_1 + \bar{\zeta}_2\gamma^m\zeta_2\right)\, ,
\end{aligned}
\end{equation} 
that are not all linear independent. 
Relations between these forms comes from supersymmetry, as shown in~\cite{Gauntlett:2005ww}. 
Then, one can use the following scalars to parametrise the norms of the spinors
\begin{equation} 
\label{scalarbilIIB}
	\begin{aligned}
 		A &:= \tfrac{1}{2}\left(\bar{\zeta}_1\zeta_1 + \bar{\zeta}_2\zeta_2\right)\, , \\
		A\sin\Theta &:= \tfrac{1}{2}\left(\bar{\zeta}_1\zeta_1 - \bar{\zeta}_2\zeta_2\right)\, , \\
		S &:= \bar{\zeta}^c_2\zeta_1\, , \\
		Z &:= \bar{\zeta}_2\zeta_1\, .
	\end{aligned}
\end{equation} 
Finally, one considers the two-forms
\begin{equation} 
\label{IIB2formsbil}
	\begin{aligned}
		U_{mn} &:= -\tfrac{i}{2}\left(\bar{\zeta}_1\gamma_{mn}\zeta_1 + \bar{\zeta}_2\gamma_{mn}\zeta_2\right)\, , \\
		V_{mn} &:= -\tfrac{i}{2}\left(\bar{\zeta}_1\gamma_{mn}\zeta_1 - \bar{\zeta}_2\gamma_{mn}\zeta_2\right)\, , \\
		W_{mn} &:= -\bar{\zeta}_2\gamma_{mn}\zeta_1\, .
	\end{aligned}
\end{equation} 

The HV structure for these backgrounds can be found in \cite{AshmoreESE, Grana_Ntokos}. 
The untwisted generalised vector structure 
$K \in \Gamma(\tilde{E})$ in \eqref{genvecs} is given in terms of the identity structure above by 
	\begin{equation} 
	\label{eq:K_tilde_IIB}
		\begin{aligned}
			\tilde{K}= \tilde{\xi} + \tilde{\lambda}^i +\tilde{\rho} + \tilde{\sigma}^i = K_5^{\sharp} +e^{2\Delta - \frac{\phi}{2} } \begin{pmatrix} \RRe K_3 \\ \IIm K_3 \end{pmatrix} - e^{4\Delta - \phi} \star V \, ,
		\end{aligned}
	\end{equation} 
where $\phi$ is the dilaton and $\Delta$ the warp factor.
Notice that for these backgrounds the five-forms vanish $\sigma^i =0$. 
The twisted vector structure is obtained by acting on $\tilde K$ with the adjoint element as in appendix E of~\cite{AshmoreECY}
	\begin{equation} 
	\label{eq:IIB_twisted_explicit}
	K = \xi + \lambda^i + \rho \, ,
	\end{equation} 
where the twisted quantities are~\cite{AshmoreESE,AshmoreECY}
	\begin{subequations}
	\label{IIbtwist}
		\begin{align}
			\xi &= \tilde{\xi}\, , \label{IIbtwistv} \\[1mm]
			\lambda^i & = \tilde{\lambda}^i + \iota_{\xi}B^i\, , \label{IIbtwist1} \\[1mm]
			\rho & = \tilde{\rho} + \iota_\xi C + \epsilon_{ij}\tilde{\lambda}^i \wedge B^j + \tfrac{1}{2}\epsilon_{ij}\left( \iota_\xi B^i \right)\wedge B^j\, . \label{IIbtwist3}
		\end{align}
	\end{subequations}
and we defined $B^1 = B$, $B^2 = C_2$, $C = C_4$, $F^1= H$, $F^2 = F_3$ and $F=F_5$.

As already discussed the condition that the generalised Lie derivative $L$ along the Reeb vector $K$ has to reduce to the conventional one, $\mathcal{L}_{\xi}$ implies some differential equations on the elements of the vector structure that reproduce some of the supersymmetry conditions on the identity structure derived in~\cite{Gauntlett:2005ww},
	\begin{equation} 
	\label{eq:IIB_ESE_structure_eqs}
		\begin{split}
			\dd  \tilde{\lambda}^i&= \iota_{\xi} F^i\, ,\\
			\dd  \tilde{\rho} &= \iota_{\xi} F +\epsilon_{ij} \tilde{\lambda}^i \wedge F^j\, .
		\end{split}
	\end{equation} 
%
%

Analogously to the M-theory case, we want now to express the calibration conditions for a D$p$ probe in these backgrounds in terms of the generalised structure and check that their closure in implied by differential conditions on Exceptional Sasaki-Einstein structures. To this purpose we have to specialise the calibrations \eqref{K10}, \eqref{omega10} and \eqref{psiIIB} to the various brane configurations.
The $\mathrm{AdS}_5$ geometry and, in particular the products of external spinors, is the same as in the previous section. Thus, in our conventions, the Killing vector $\mathcal{K}$ has the following components,
	\begin{equation} 
		\begin{split}
			\mathcal{K}_0 &= e^{-2\Delta} \bar{\psi} \rho_0 \psi \otimes A \, , \\
			\mathcal{K}_m & = \tfrac{1}{2}\left(\bar{\zeta}_1\gamma^m\zeta_1 + \bar{\zeta}_2\gamma^m\zeta_2\right) = \xi_m \, ,
		\end{split}
	\end{equation} 
where $\xi$ is the Reeb vector. 
We also fix the norm of the internal spinors such that $A=1$.

As in the previous sections, we focus on the cases of point-like AdS particles and space-filling branes where the calibrations are related to the generalised vector $K$. 
Consider first a D$1$ wrapping an internal one-cycle. 
The relevant terms is \eqref{BPScal2} are 
	\begin{equation} 
		\begin{aligned}
			&\omega =- e^{2\Delta-\phi/2} \tfrac{1}{2} \left(\bar{\zeta}_2 \gamma_m \zeta_1 + \zeta_2^T\gamma_m \zeta_1^* \right) = - e^{2\Delta-\phi/2} \RRe K_3\, , \\
			&\Psi =- e^{2\Delta+\phi/2}\tfrac{1}{2i} \left(\bar{\zeta}_2 \gamma_m \zeta_1 - \zeta_2^T\gamma_m \zeta_1^* \right) = - e^{2\Delta+\phi/2} \IIm K_3\, , \\
			&\iota_{\hat{\mathcal{K}}} B = \iota_\xi B^1 =: \iota_\xi B\, , \\
			&\iota_{\hat{\mathcal{K}}} C = \iota_\xi B^2\, .
		\end{aligned}
	\end{equation} 
and it is immediate to see that the calibration form is given by the generalised vector $K$
	\begin{equation} 
		\Phi_{\mathrm{D}1} = -\tilde{\lambda}^2 -\iota_\xi B^2 = -e^{2\Delta-\phi/2} \IIm K_3 -\iota_\xi B^2\, .
	\end{equation} 
Using equation~\eqref{eq:IIB_ESE_structure_eqs} one can show that $\Phi_{\mathrm{D}1}$ is closed. 
Using again the properties of the $\mathrm{AdS}_5$ spinors, it is easy to show that a space-filling D$5$-brane is also calibrated by the same form,
	\begin{equation} 
		\Phi_{\mathrm{D}5} = ( -\tilde{\lambda}^2 -\iota_\xi B^2) \otimes \star 1 =( -\tilde{\lambda}^2 -\iota_\xi B^2) \otimes \vol_{\mathrm{AdS}_5} \, .
	\end{equation} 

Similarly, one can find the calibration form for a D$3$-brane wrapping purely internal cycles,
	\begin{equation} 
	\label{D3cal}
		\Phi_{\mathrm{D}3}= \tilde{\rho} +\iota_\xi C +\epsilon_{ij}\tilde{\lambda}^i \wedge B^j + \tfrac{1}{2}\epsilon_{ij}\iota_\xi B^i \wedge B^j\, .
	\end{equation} 
Its closure follows from the $L_K$ conditions under the gauge choice%
		\footnote{%
		We also need to use $\tilde{\rho} \propto \star V$.%
		} that the potential are invariant under $K$.
Again, this form provides also the calibration for a space-filling D$7$-brane.

In the particular case where the only non-trivial background flux is the five-form, the generalised Sasaki-Einstein structure reduces to the standard one and the internal manifold is Sasaki-Einstein.
In this case the spinor ansatz~\eqref{eq:splitting_IIB_ads5} simplifies since the two internal spinors are proportional to each other, \emph{i.e.} $\chi_2=i \chi_1$, and, consequently, the one-form part vanishes.
The twisteed vector~\eqref{eq:IIB_twisted_explicit} simplifies to
	\begin{equation} 
		K = \xi - \sigma \wedge \omega + \iota_{\xi}C\, , 
	\end{equation}  
and the (untwisted) hypermultiplet structure is~\cite{AshmoreESE}
	\begin{align}
	\label{Jstruct}
		\tilde{J}_+ &= \tfrac{1}{2}\kappa u^i \Omega - \tfrac{i}{2} \kappa u^i \Omega^{\sharp}\, ,\\[1mm]
		\tilde{J}_3 &= \tfrac{1}{2}\kappa I + \tfrac{1}{2} \kappa \hat{\tau}^{i}_{\phantom{i}j} + \tfrac{1}{8}\kappa \Omega^{\sharp}\wedge \bar{\Omega}^{\sharp} - \tfrac{1}{8}\kappa \Omega \wedge \bar{\Omega} \, , 
	\end{align}
where $u^i = (-i, 1)^i$ and $I$, $\omega$ and $\Omega$ are the complex structure, the symplectic and the holomorphic two-forms on the K\"ahler-Einstein basis of $M_5$. 

Note that in the vector structure, the three-form is $\sigma \wedge \omega$ and by the structure equation~\eqref{eq:IIB_ESE_structure_eqs} one immediately sees that adding the potential part~$\iota_{\xi}C$ yields to a closed form (up to a gauge choice),
	\begin{equation} 
		\dd  \left( \tilde{\rho}+\iota_{\xi}C \right)= \iota_{\xi} \dd  C + \mathcal{L}_{\xi} C - \iota_{\xi} \dd  C = \mathcal{L}_{\xi} C = 0\, .
	\end{equation} 

Since in this case, the form of the hypermultiplet structure is simple, we can also study calibrations that are not associated to the vector $K$. 
We will do it in the simplest Sasaki-Einstein background, namely $\mathrm{AdS}_5 \times S^5$, where $S^5$ is the five-dimensional sphere. 
The background is given by
	\begin{equation} 
	\label{ads5bkgrd}
		\begin{split}
			\dd  s^2 &= \dfrac{R^2}{r^2} \dd  r^2 + \dfrac{r^2}{R^2} \eta_{\mu\nu} \dd  x^\mu \dd  x^\nu + \dd  s^2(S^5)\, , \\[1mm]
			C_4 &= \left(\dfrac{r^4}{R^4} - 1\right) \dd  x^0 \wedge \ldots \wedge \dd  x^3\, .
		\end{split}
	\end{equation} 
while all other fluxes, dilaton and warp factors vanish.
The $S^5$ can be written as a $U(1)$ fibration over $\mathbb{CP}^2$,
	\begin{equation} 
		\dd  s^2(S^5) = \dd  \Sigma_{4}^2 + \sigma \otimes \sigma\, ,
	\end{equation} 
where $\dd  \Sigma_{4}^2$ is the \emph{Fubini-Study metric} over $\mathbb{CP}^2$. The form $\sigma$ is given by $\sigma = \dd  \psi + A$, where $A$ is a connection such that $\mathcal{F} = \dd  A = 2 \omega$, and $\psi$ is the periodic coordinate on the circle $U(1)$ with period $6\pi$.

Explicitly, the sphere $S^5$ takes the form~\cite{Gauntlett:2004yd}
	\begin{equation} 
		\begin{split}
			\dd  s^2(S^5) &= \dd  \alpha ^2 + \dfrac{1}{4} \sin^2 \alpha  (\dd  \theta^2 + \sin^2 \theta \dd  \phi^2) + \dfrac{1}{4} \cos^2 \alpha  \sin^2\alpha  (\dd  \beta + \cos\theta \dd  \phi)^2 \\
				 & \phantom{=} + \dfrac{1}{9}\left[\dd  \psi - \dfrac{3}{2}\sin^2\alpha  (\dd  \beta + \cos\theta \dd  \phi)\right]^2\, ,
		\end{split}
	\end{equation} 
with $\psi \in [0,6\pi]$, $\beta \in [0,4\pi]$, $\alpha  \in [0,\pi/2]$, $\theta \in [0,\pi]$ and $\phi \in [0,2\pi]$.
In these coordinates the holomorphic form has the following expression,
	\begin{equation} 
		\Omega= - \frac{1+i}{\cos \theta}\ \dd  \beta \wedge \dd  \theta + \frac{1+i}{8} \cos \sigma \cos \theta \sin \theta \sin^3 \sigma\ \dd  \sigma \wedge \dd  \phi + (1+i)\ \dd  \theta \wedge \dd  \phi\, .
	\end{equation} 

First, consider a D$5$-brane spanning the directions $0,1,2, r$ in $\mathrm{AdS}_5$. 
The world-volume of the brane is $\mathrm{AdS}_4 \times S^2$, where $S^2$ is the sphere parametrized by the angles $(\theta, \phi)$. 
Then the expression~\eqref{BPScal2} reduces to 
	\begin{equation} 
		\begin{split}
			\Phi_{\mathrm{D}5} &= -\ \dd  x^0 \wedge \dd  x^1 \wedge \dd  x^2 \wedge \dd  x^4 \wedge \tfrac{(1-i)}{2} e^{4\Delta}\Omega \\
				& = -\ \dd  x^0 \wedge \dd  x^1 \wedge \dd  x^2 \wedge \dd  r \wedge \vol_{S^2}\, ,
		\end{split}
	\end{equation} 
and we see that it corresponds to the two-form part of $\tilde{J}_+$ in~\eqref{Jstruct}.
Modulo choice of coordinates,%
		\footnote{%
		Here $\dd  x^4 \propto \sin \theta\ \dd  r + \dd  \sigma + \dd  \beta$.%
		} it agrees with the analogous form in~\cite{Cascales:2004qp}.

We can also consider a D$3$-brane probe spanning the directions $0,1,r $ of $AdS_5$. 
The world-volume is now $\mathrm{AdS}_3 \times S^1$ and the calibration is given by the Hodge dual of the $4$-form part of $\tilde{J_3}$,
	\begin{equation} 
		\Phi_{\mathrm{D}3} = \dd  x^0 \wedge \dd  x^1 \wedge \dd  r \wedge \tfrac{1}{8} e^{4\Delta} \star( \Omega \wedge \bar{\Omega})\, .
	\end{equation} 
One can prove the closure of this form by the $L_K J$ relations. In particular,
	\begin{equation} 
		\dd  (e^{4\Delta} \star ( \Omega \wedge \bar{\Omega})) = - m\ \iota_\xi \vol_5 = 0\, ,
	\end{equation} 
where the first equality comes from the conditions to have a vanishing $\tilde{R}$-tensor~\cite{AshmoreECY,AshmoreESE}. 
In other words, it is the rewriting of the~\eqref{eq:IIB_ESE_structure_eqs} in the Sasaki-Einstein case.

\ensurepagenumbering{arabic}
	\chapter*{Conclusions}
	
		The major aim of this thesis was to study flux backgrounds.
		In the first part we focused on the problem of seeking for a systematic way to find consistent truncations in the presence of fluxes.
		In the second part, we were more concerned in reformulating the supersymmetry conditions for brane probes in AdS backgrounds, in terms of integrability of exceptional structures in exceptional geometry.
	
		This thesis contains the construction of a generalised geometric description of type IIA flux backgrounds.
		
		We made a large use of the formalism of $G$-structures and generalised geometry.
		As in conventional geometry, integrability is defined as the existence of a generalised torsion-free connection that is compatible with the structure, or equivalently as the vanishing of the generalised intrinsic torsion.
		In the case analysed, \emph{i.e.} truncations of massive type IIA, the integrability conditions correspond to the Leibnitz algebra for parallelisations.
		These enquires the trucation preserves maximal supersymmetry in lower dimension~\cite{spheres}.
		Moreover, the notion of \emph{generalised Leibniz parallelisation} is the key ingredient to construct a consistent truncation ansatz.
		We showed in chapter~\ref{chapComp} how to build consistent truncation ansatze, by the so-called~\emph{Generalised Scherk-Scwharz reductions}, making use of generalised Leibniz parallelisations.
		We constructed various examples of massive type IIA spheres truncations.
		In the case of the truncation on a six-dimensional sphere, we obtained a generalised parallelisation on $S^6$ satisfying the $\ISO(7)$ algebra, and spelled out the corresponding truncation ansatz as obtained from the generalised Scherk--Schwarz prescription. 
		As recently described in \cite{Guarino:2015jca,Guarino:2015vca}, the Romans mass introduces a magnetic gauging of the $\ISO(7)$ translations in the truncated four-dimensional theory, yielding a symplectic deformation~\cite{Dall'Agata:2014ita} of the type first found in~\cite{Dall'Agata:2012bb} for the $\SO(8)$ gauging.
		We found the same phenomenon for type IIA supergravity on the six-dimensional hyperboloids $H^{p,7-p}$: on these spaces one can define a consistent truncation down to $\ISO(p,7-p)$ supergravity in four-dimensions; switching the Romans mass on leads to the symplectically-deformed $\ISO(p,7-p)$ gauging described in~\cite{Dall'Agata:2014ita}.
		We also obtained generalised Leibniz parallelisations on $S^4$, $S^3$ and $S^2$ for vanishing Romans mass, reproducing the Leibniz algebra of known consistent truncations of massless type IIA supergravity on these manifolds. 
		When the Romans mass is switched on, these parallelisations no more satisfy a Leibniz algebra. 
		We offered an explanation of why this is the case by showing that the frame lies in the stabiliser group of the Romans mass only for the parallelisation on $S^6$. 
		For massive type IIA on $S^3$ we presented a no-go result indicating that a consistent truncation including the $\SO(4)$ algebra does not exist. 
		It would be interesting to see whether similar no-go theorems can be proved for the $S^4$ and $S^2$ cases.
		
		As said, in order to construct apply the generalised Scherk-Schwarz prescription, we built the adapted version of generalised geometry for massive type IIA. 
		The principal issue in this construction is accommodating the flux due to the Romans mass. 
		This is achieved by deforming the generalised Lie derivative such that the deformed one generates the gauge transformations of type IIA supergravity with a non-zero $m$-flux.

		An interesting fact is the existence of an alternative massive type IIA~\cite{Howe:1997qt}. 
		This can be obtained from eleven-dimensional supergravity by gauging a combination of the $\GL(1)$ global symmetry and the trombone symmetry of the equations of motion.
		It is not known a description of such a theory through a Lagrangian.
		One may wonder whether other massive extensions of type IIA can exist.
		However, in~\cite{Tsimpis:2005vu}, making use of superspace arguments, it was discussed how this and the Romans mass are the only possible extensions.
		It is natural to ask how this deformation appears in our formalism.
		We want deformation parameters to be diffeomorphism invariant then we require them to appear as $\GL(6)$ singlets with zero $\RR^+$ weight. 
		There are precisely two such singlets in the $\mathbf{912}_{-1}$ representation of $\E_{7(7)}\times\RR^+$, one of which we have already identified as the Romans mass deformation.
		There is also a singlet in the $\mathbf{56}_{-1}$ representation, which is another part of the generalised torsion~\cite{Coimbra:2011ky}, and which could also be used to deform the Dorfman derivative. 
		When performing generalised Scherk-Schwarz reductions, this additional $\mathbf{56}_{-1}$ part of the embedding tensor is generated by gauging the trombone symmetry~\cite{LeDiffon:2008sh}, and the resulting theory does not have an action. 
		It is natural to conjecture that deforming the Dorfman derivative by switching on a combination of the second singlet in $\mathbf{912}_{-1}$ and the singlet in $\mathbf{56}_{-1}$ would give the relevant gauge algebra for the theory described in~\cite{Howe:1997qt}.
		The result of~\cite{Tsimpis:2005vu} can be verified in this case, since, by considering the closure of the gauge algebra, one can argue that there are no others deformations, as singlets of $\GL(6)$ in the torsion representation bundle.
		
		In the last chapter of the thesis, we focused on brane calibrations.
		The tools used are again generalised $G$-structures.
		In particular, we concentrated our attention on AdS backgrounds with eight supercharges.
		These have an elegant description in generalised geometry in terms of exceptional Sasaki-Einstein structures~\cite{AshmoreESE, Grana_Ntokos}.
		In~\cref{chapbrane}, we studied the relation between the Exceptional Sasaki-Einstein structures and generalised brane calibrations in $\mathrm{AdS}_5 \times M_5$ backgrounds in type IIB and in $\mathrm{AdS}_5 \times M_6$ and $\mathrm{AdS}_4 \times M_7$ compactifications in M-theory.
		We focussed on the calibrations forms associated to branes wrapping cycles in the internal manifolds and that are point-like in the AdS space. 
		We showed that for these configurations the general expression for the calibration forms that can be constructed using $\kappa$-symmetry can be expressed in terms of the generalised Killing vector $K$ defining the Exceptional Sasaki-Einstein structure and that the closure of the calibration forms is given by the integrability (more precisely the $L_K$ condition) of the ESE structure. 
		The results of the chapters prove the conjecture appeared in~\cite{AshmoreESE} that the (form part of the) generalised Killing vector is a generalised calibration.
		The motivation of this conjecture can be found in holography~\cite{Maldacena:1997re}.
		One can observe that the generalised killing vector $K$ generates the global $R$-symmetry of the filed theory dual to the AdS background in supergravity.
		It is made by a combination of the vector part (generating diffeomorphism) and $p$-forms (parametrising gauge transformations), under which the generalised metric is invariant.
		Thus there is a sign that $K$ is an object related to $R$-symmetry in a non-trivial way, since it encodes informations about the gauge transformations of flux potentials.
		In AdS/CFT correspondence, BPS branes have volume associated to the conformal dimension of chiral operator in the dual SCFT, thus finding a calibration (that is a supersymmetric circle on which branes wrap) is equivalent to find the conformal dimensions of the dual operators.
		
		We have seen how to construct these calibrations in generalised geometry and how BPS conditions correspond to the integrability of the structures.
		
		We also partially discussed other brane configurations that are calibrated by the vector $K$.
		However we did not perform a complete analysis, leaving the discussion for a future work.

	\section*{Future works}
		There are many other directions for future study.
		
		For intance, so far, there is not a description of supergravity flux backgrounds in terms of generalised geometry for any amount of supersymmetry.
		As stated, supersymmetric background preserving $\mathcal{N}$ supersymmetries are given by integrable $G$-structures, where $G$ is the stabiliser group of the $\mathcal{N}$ Killing spinors~\cite{AndCharSpecHol}.
		Maximally supersymmetric backgrounds, as seen are described by parallelisations~\cite{spheres, Baguet:2015sma}.
		Half-maximal supergravity truncations have been recently described in exceptional field theory~\cite{Malek:2017cjn}, but a generalised geometry formulation is not known at the moment.
		The $\mathcal{N}=2$ backgrounds, as seen in~\cite{AshmoreESE, AshmoreECY, Grana:2009im, Malek:2016bpu}, is a rich field for generalised $G$-structure applications.
		An $\mathcal{N}=1$ formalism to describe vacua is known in $\rmO(d,d)$ generalised geometry~\cite{petrini2, Grana:2005sn, Grana:2006kf} and recently an exceptional picture has been found in~\cite{CoimbraN1}. 
		However, the analysis is far to be complete.
		Hence, obvious extension is to consider backgrounds with different amounts of supersymmetry, which will be described by new geometric structures within generalised geometry.
		In~\cite{oscar5}, there is a work in progress development of the structures to describe supersymmetric backgrounds with sixteen supercharges.
		Furthermore, the hope is not just to find a new descriptions of such backgrounds, but that this would lead to discover new examples of truncations.
		In addition, an ambitious project would be a complete classification of such structures analysing the mathematical constraints on the internal geometries and finding a coherent structure describing them.
		
		Moreover, all these techniques are may also be applied in different contexts from consistent truncations, like for example, studies of holography effects, marginal deformations of the dual field theories, etc.
		
		More specifically, the formalism developed in the first part of this thesis about massive type IIA may also be applied to investigate marginal deformations of conformal field theories holographically dual to (massive) type IIA AdS backgrounds.
		Recently this theories have been identified with a class of Chern-Simons-matter theories~\cite{Guarino:2015jca}, and in~\cite{oscar4}, we aim to describe exactly marginal deformations of such conformal theories by constructing (and studying their deformations) the exceptional Sasaki-Einstein structures for massive type IIA, describing the AdS background in~\cite{Varela:2015uca} preserving eight supercharges.
		The work of~\cite{AshmoreDef} is the first example in this sense.
		There, the authors study the exceptional Sasaki-Einstein structures describing AdS$_5$ background in both type IIB and eleven-dimensional supergravity to analyse marginal deformations of the dual $\mathcal{N}=1$ CFT in four dimensions.
		A famous result in gauge theory~\cite{Green:2010da} states that marginal deformations are determined by imposing $F$-term conditions on operators of conformal dimension three and then quotienting by the complexified global symmetry group.
		In~\cite{AshmoreDef}, it was shown that this result has a geometrical interpretation: the marginal deformations are obtained as solutions of moment maps for the generalised diffeomorphism group that have the correct charge under the Reeb vector.
		Indeed, the Reeb generates the $U(1)_R$ symmetry group.
		In the case this is the only symmetry of the background, then all the marginal deformations are exactly marginal.
		When there are other global symmetries, the field theory result predicts one has to quotient out these symmetries.
		On the supergravity side, this can be read as fixed points of the moment maps, being an obstruction for marginal deformations to be exactly marginal.
		
		This analysis holds for any kind of internal geometry in an AdS background, however, so far the examples to which this has been applied are all Sasaki-Einstein geometries.
		So, it would be interesting to apply it to one of the few examples of non-Sasaki–Einstein backgrounds, such as the Pilch–Warner solution~\cite{PilchWarner}.
		This would give the marginal deformations of the dual SCFT, the so-calle Leigh-Strassler theory~\cite{LSexact}.
		We aim to study this example in~\cite{oscar3}.
		
		To conclude, generalised geometry gives tools to better analyse several aspects of supergravity and string theory.
		It has also applications in holography and field theory.
		Furthermore, it is interesting also in pure mathematics, since it is related to various areas beyond differential geometry, like algebraic topology, algebraic geometry and group theory.
		Thus it provides an example of a topic that lies at the frontier between mathematics and physics, on the one hand, receiving deep insights from both fields, but on the other hand it could also give some useful tools to understand and answer questions in both areas.
		For these reasons, generalised geometry seems to be worth of further efforts and studies, since its U-duality covariant approach may reveal something hidden so far and perhaps help us to understand the geometrical nature of dualities in string theory.

	\appendix
	\chapter{Notations and Conventions}
	\label{app:notation}
			The indices used in this thesis -- if not differently indicated -- are:
					\begin{align*}
						\mu, \nu  \ &: \ \text{external spacetime indices} \, ,\\
						m,n \ &:\ \text{curved indices on the internal manifold $M_d$} \, ,\\
						a,b \ &:\ \text{frame indices on $M_d$} \, ,\\
						i,j \ &:\ \text{indices for the embedding coordinates of $S^d$ in $\mathbb R^{d+1}$ (or $H^{p,q}$ in $\RR^{p+q}$)} \, ,\\
						I,J \ &:\ \text{$\SL(d+2,\RR)$ indices}\, , \\
						M,N \ &: \ \text{curved indices for the $E_{d+1(d+1)}\times \RR^+$ generalised tangent space on $M_d$} \, ,\\
						A,B \ &: \ \text{frame indices for the $E_{d+1(d+1)}\times \RR^+$ generalised tangent space on $M_d$} \, .
					\end{align*}

			Our tensor conventions are the same as in~\cite{waldram2}. 
			We collect here the ones relevant for our computations. 
			On a $d$-dimensional manifold $M_d$, given a form $\lambda \in \Lambda^p T^*$ and a poly-vector $w \in \Lambda^q T$, 
					\begin{equation}
						\lambda = \frac{1}{p!} \lambda_{m_1\ldots m_p} \dd x^{m_1} \wedge \cdots \wedge \dd x^{m_p}\ ,\qquad w = \frac{1}{q!} w^{m_1 \ldots m_q} \frac{\partial}{\partial x^{m_1}}\wedge \cdots \wedge\frac{\partial}{\partial x^{m_q}} \,,
					\end{equation}
			we define the contraction
					\begin{equation}\label{deflrcorner}
						\begin{array}{l r}
							(w \,\lrcorner\,\lambda)_{m_1\ldots m_{p-q}} = \frac{1}{q!}w^{n_1 \ldots n_q}\lambda_{n_1\ldots n_q m_1 \ldots m_{p-q}} &\mbox{if}\ q \leq p \, , \\[1mm]
							(w \,\lrcorner\,\lambda)^{m_1\ldots m_{q-p}} = \frac{1}{p!}w^{m_1 \ldots m_{q-p}n_1 \ldots n_p}\lambda_{n_1\ldots n_p} &\mbox{if}\ p< q\, .
						\end{array}
					\end{equation}
			The contraction of a vector $v\in T$ with a form $\lambda$ is also denoted by $\iota_v \lambda \equiv v\,\lrcorner \,\lambda$.

			The contraction of a poly-vector $w$ with a tensor $\tau \in T^*\otimes \Lambda^d T^*$ is defined as
					\begin{equation}
						(w\,\lrcorner\, \tau)_{m_1\ldots m_{d-q+1}} = \frac{1}{(q-1)!} w^{n_1\ldots n_q} \tau_{n_1,\,n_2\ldots n_q m_1\ldots m_{d-q+1}} \,.
					\end{equation}

			Moreover, for $\lambda \in \Lambda^{p}T^*$ and $\mu \in \Lambda^{d-p+1}T^*$, we define the \emph{$j$-operator} giving  $j \lambda \wedge \mu \in T^* \otimes \Lambda^d T^*$ as: 
					\begin{equation}\label{joper}
						\left(j \lambda \wedge \mu\right)_{m,\,m_1\ldots m_d} = \frac{d!}{(p-1)!(d-p+1)!}\,\lambda_{m[m_1\ldots m_{p-1}}\mu_{m_p\ldots m_d]}\ .
					\end{equation} 
			This is the same as $j\lambda \wedge \mu = \dd x^m \otimes (\iota_m \lambda \wedge \mu)$. Upon exchanging $\lambda$ and $\mu$ one has
					\begin{equation} 
						j \lambda \wedge \mu = (-1)^{p(d-p+1)+1}\, j\mu \wedge \lambda\ .
					\end{equation}

			For the Hodge star we take
					\begin{equation}
						(*\lambda)_{m_1\cdots m_{d-p}} = \frac{1}{p!}\sqrt{g}\,\epsilon_{m_1\cdots m_{d-p}}^{\phantom{m_1\cdots m_{d-p}} n_1\ldots n_p}\lambda_{n_1\ldots n_p} \ ,
					\end{equation}
			with $\epsilon_{1\ldots d} = +1$.

			The action of a $\mathfrak{gl}(d)$ element $r \in T \otimes T^*$ on a vector $v\in T $ and on a $p$-form is defined as
					\begin{equation}
						(r \cdot v)^m = r^{m}_{\phantom{m}n} v^n\,, \qquad (r\cdot \lambda)_{m_1\ldots m_p} = -p\, r^{n}_{\phantom{n}[m_1} \lambda_{|n| m_2\ldots m_p]}  \,. 
					\end{equation}
			\section{Constrained coordinates on the spheres}\label{app:constrcoo}
				In the following we provide some useful formulae for the embedding coordinate description of the round sphere $S^d$, mostly taken from~\cite{spheres}. 
				These are needed to study the parallelisations of the exceptional tangent bundle presented in the main text.

				We parameterise $\RR^{d+1}$ in Cartesian coordinates as $x^i = r\, y^i$, $i=1,\ldots d+1$, with 
					\begin{equation}
						\delta_{ij}\,y^iy^j \,=\, 1\ .
					\end{equation}
				Then the $d$-dimensional sphere $S^d$ of radius $R$ is obtained by fixing $r=R$. 
				The standard metric and volume form on $\RR^{d+1}$ induce the following round metric and \hbox{volume form on $S^d$:}
					\begin{equation}\label{roundSdmetric}
						\rg{g} = R^2 \, \delta_{ij}\dd y^i \dd y^j \ ,
					\end{equation}
					\begin{equation}
						\rg{\rm vol}_{d} \, = \, \frac{R^d}{d!} \,\epsilon_{i_1 \ldots i_{d+1}} y^{i_1} \dd y^{i_2}\wedge \cdots \wedge \dd y^{i_{d+1}}\ .
					\end{equation}
				The Killing vector fields generating the $SO(n+1)$ isometries can be written as 
						\begin{equation}
							v_{ij} \,=\, R^{-1}\left(y_i k_j - y_j k_i \right)\ ,
						\end{equation}
				where $k_i$ are conformal Killing vectors, satisfying 
						\begin{align}
							\mathcal{L}_{k_i}\! \rg{g} &= - 2 y^i \rg{g}\ , \\
										k_i (y_j) &:= \iota_{k_i}\dd y_j \, = \, \delta_{ij} - y_i y_j\ .
						\end{align}
				The index on the coordinates $y^i$ is lowered using the $\RR^{d+1}$ metric $\delta_{ij}$.
				The Killing vectors $v_{ij}$ generate the $\mathfrak{so}(n+1)$ algebra,
						\begin{equation}\label{algebra_Killing_v}
							\mathcal{L}_{ v_{ij}} v_{kl} \, =\, R^{-1}\left(\delta_{ik}v_{lj} - \delta_{il}v_{kj} - \delta_{jk}v_{li} + \delta_{jl} v_{ki} \right)\ ,
						\end{equation}
				while the constrained coordinates $y_k$ and their differentials $\dd y_k$ transform in the fundamental representation of $SO(n+1)$ under the Lie derivative, 
						\begin{equation}\label{Lie_on_y}
						\begin{split}
							\mathcal{L}_{v_{ij}}y_k \,&\equiv\, \iota_{v_{ij}}\dd y_k \,=\, R^{-1}\left( y_i \delta_{jk} -  y_j\delta_{ik}\right)\ , \\[1mm]
							\mathcal{L}_{v_{ij}} \mathrm{d} y_k \, &=\,  R^{-1}\left(\mathrm{d}y_i \delta_{jk} - \mathrm{d}y_j\delta_{ik}\right) \ .
						\end{split}
						\end{equation}
				The $(d-1)$-form
						\begin{equation}
							\kappa_i = - \rg{*}(R\,\dd y_i) = \frac{R^{d-1}}{(d-1)!}\,\epsilon_{ij_1\ldots j_d}\, y^{j_1}\dd y^{j_2}\wedge \cdots \wedge \dd y^{j_d}
						\end{equation}
				transforms under $\mathcal{L}_{v_{ij}}$ exactly as $\dd y_k$ (since $\mathcal{L}_{v_{ij}}$ preserves the round metric~\eqref{roundSdmetric}, it commutes with the Hodge star):
						\begin{equation}
							\mathcal{L}_{v_{ij}} \kappa_k \,=\, R^{-1}\left(\kappa_i \delta_{jk} - \kappa_j\delta_{ik}\right) \ .
						\end{equation}

				We also introduce the forms
						\begin{equation}
							\begin{split}
								\omega_{ij} &= R^2\, \dd y_i \wedge \dd y_j \ , \\[1mm]
								\rho_{ij} &= \rg{*}\omega_{ij} = \frac{R^{d-2}}{(d-2)!} \,\epsilon_{ijk_1\ldots k_{d-1}} y^{k_1} \dd y^{k_2} \wedge \cdots \wedge \dd y^{k_{d-1}}\ , \\[1mm]
								\tau_{ij} &= R\left(y_i\mathrm{d}y_j  - y_j\mathrm{d}y_i\right) \otimes \rg{\rm vol}_d\ ,
							\end{split}
						\end{equation}
				which transform in the adjoint representation of $\SO(d+1)$ under $\mathcal{L}_{v_{ij}}$. Namely,
						\begin{equation}\label{Lie_on_omega}
							\mathcal{L}_{v_{ij}} \omega_{kl} \, = \, R^{-1}\left(\delta_{ik}\omega_{lj} - \delta_{il}\omega_{kj} - \delta_{jk}\omega_{li} + \delta_{jl} \omega_{ki} \right) \ ,
						\end{equation}
				and similarly for the others, with the same overall factor $R^{-1}$.

				Furthermore, one can show the relations
						\begin{align}
								\iota_{v_{ij}} \rg{\vol}_d &= \frac{R}{d-1} \mathrm{d}\rho_{ij}\, , \label{eq:contrvol}\\[1mm]
								\dd\kappa_i &= \frac{d}{R}\, y_i \rg{\rm vol}_d\ , \\[1mm]
								\dd \,\iota_{v_{ij}}\kappa_k &= - \dd\,(y_k \rho_{ij}) \ ,
						\end{align}
				which are proven by making use of the trivial identity $y_{[i_1}\epsilon_{i_2\ldots i_{d+2}]}=0$.

				When computing the norm of our generalised frames, we will need the following ``squares'' of the forms defined above:
						\begin{equation}\label{contractions_sphere}
							\begin{split}
								v_{ij} \,\lrcorner\, v_{kl} &= y_iy_k \delta_{jl} - y_jy_k \delta_{il} - y_iy_l \delta_{jk} +y_jy_l \delta_{ik}\ , \\[1mm]
								\omega_{ij} \,\lrcorner\, \omega_{kl} \,=\, \rho_{ij}\,\lrcorner\,\rho_{kl} &= \delta_{ik}\delta_{jl}- \delta_{il}\delta_{jk} - (y_iy_k \delta_{jl} - y_jy_k \delta_{il} - y_iy_l \delta_{jk} +y_jy_l \delta_{ik})\ ,  \\[1mm]
								\tau_{ij}\,\lrcorner\,\tau_{kl} &= y_iy_k \delta_{jl} - y_jy_k \delta_{il} - y_iy_l \delta_{jk} +y_jy_l \delta_{ik}\ ,  \\[1mm]
								\kappa_i \,\lrcorner\, \kappa_j \,=\, R^2\, \dd y_i \,\lrcorner\, \dd y_j &=  \delta_{ij} - y_iy_j \ .
							\end{split}
						\end{equation}
Here, the round metric $\rg{g}$ and its inverse are used to lower/raise the indices; for instance, $\omega_{ij} \,\lrcorner\, \omega_{kl} \equiv \frac{1}{2} \rg{g}{}^{\!mp}\rg{g}{}^{\!nq}(\omega_{ij})_{mn}(\omega_{kl})_{pq}$, and so on.

		\section{Conventions for spinors and gamma matrices}
\label{app:conv}
In this appendix we collect the conventions for spinors and gamma matrices 
that are relevant for the thesis, in particular for~\cref{chapbrane}. 

\subsection{\texorpdfstring{Type IIB on $\mathrm{AdS}_5 \times M_5$}{Type IIB on AdS5 x M5}}
\label{sect:IIB_notation}

We follow the conventions in~\cite{Grana_Ntokos}. The ten-dimensional metric is
\begin{equation}
	\dd  s^2 = e^{2A}\dd  s^2_{\mathrm{AdS}_5} + \dd  s^2_{M_5} \, , 
\end{equation}
and the ten-dimensional gamma matrices $\Gamma^M$ are chosen as 
\begin{equation}
	\begin{array}{lcc}
		\Gamma^{\mu} = e^{-A} \rho^{\mu} \otimes \id_4 \otimes \sigma^3\, ,& \phantom{\mbox{with}} & \mu=0, \ldots 4\, , \\
		\Gamma^{m+ 4} = \id_4 \otimes \gamma^m \otimes \sigma^1\, , & \phantom{\mbox{with}} & m=1, \ldots 5 \, ,
	\end{array}
\end{equation}
where $\rho^\mu$ and $\gamma^m$ generate $\mathrm{Cliff}(1,4)$ and $\mathrm{Cliff}(5)$ respectively, satisfying 
\begin{equation}
	\begin{array}{ccc}
		\{\rho^{\mu},\rho^{\nu}\}= 2 g^{\mu\nu}\, , & \phantom{\mbox{with}}& \{\gamma^{m},\gamma^{n}\}= 2 g^{mn} \, , 
	\end{array}
\end{equation}
and $\eta^{\mu \nu} = \mathrm{diag}(-1, 1,1,1,1)$. We also have 
\begin{equation}
	\begin{array}{ccc}
		\rho^{01 \ldots 4}=-\IIm \id\, , & \phantom{\mbox{with}}& \gamma^{1 \ldots 5}=\id\, .
	\end{array}
\end{equation}
We choose the $\mathrm{Cliff}(1,4)$ and $\mathrm{Cliff}(5)$ intertwiners as
\begin{equation}
\label{Achoice}
	\begin{array}{lcl}
		A_{1,4} = \rho_0 & \quad & C_{1,4} = D_{1,4} A_{1,4}\, , \\
		A_5 = 1 & \quad & C_5 = D_5 \, ,
	\end{array}
\end{equation}
where $C = - C^T$ in any dimension, so that 
\begin{equation}
\label{IIbintertw}
	\begin{array}{lcl}
		\rho^{\mu \, \dagger} = - A_{1,4}\rho^{\mu} A_{1,4}^{-1} & \quad \qquad \qquad & \gamma^{m \, \dagger} = \gamma^m\, , \\ 
		\rho^{\mu \, T}= C_{1,4}\rho^{\mu}C_{1,4}^{-1} & \quad \qquad \qquad & \gamma^{m \, T}= C_5 \gamma^m C_5^{-1}\, , \\ 
		\rho^{\mu *}= - D_{1,4} \rho^{\mu} D_{1,4}^{-1} & \quad \qquad \qquad & \gamma^{m *} = C_5 \gamma^m C_5^{-1} \, .
	\end{array}
\end{equation}

An explicit choice of a basis for the $\mathrm{Cliff}(1,4)$ gamma matrices is
\begin{equation}
\label{ads5gamma}
	\begin{array}{ccccc}
		\rho^0=\IIm \sigma^2 \otimes \sigma^0\, , & \rho^i = \sigma^1 \otimes \sigma^i\, , & \rho^4=-\sigma^3 \otimes \sigma^0\, , &\phantom{\mbox{and}} & i=1,2,3 \, ,
	\end{array}
\end{equation}
while for the $\mathrm{Cliff}(5)$ gamma matrices we take 
\begin{equation}
	\begin{array}{ccccc}
		\gamma^1=\sigma^1 \otimes \sigma^0\, , & \gamma^2=\sigma^2 \otimes \sigma^0\, , & \gamma^3=\sigma^3 \otimes \sigma^1\, , & \gamma^4=\sigma^3 \otimes \sigma^2\, , & \gamma^5=-\sigma^3 \otimes \sigma^3\, , 
	\end{array}
\end{equation}
with intertwiners 
\begin{equation*}
	\begin{array}{ccc}
		A_{1,4}=\rho^0\, , & C_{1,4}=\rho^0 \rho^2\, ,& C_5=\sigma^1 \otimes \sigma^2\, .
	\end{array}
\end{equation*}

With these choices for the gamma matrices the ten-dimensional chiral gamma decomposes as 
\begin{equation}
	\Gamma_{11} = \Gamma_{0, \ldots 9} = \id_4 \otimes \id_4 \otimes \sigma^2 \, . 
\end{equation}
The ten-dimensional supersymmetry parameters are Majorana-Weyl spinors of negative chirality $\Gamma_{11} \varepsilon_i = - \varepsilon_i$ ($i=1,2$) and decompose as
\begin{equation}
	\varepsilon_i = \psi \otimes \chi_i \otimes u + \psi^c \otimes \chi_i^c \otimes u\, ,
\end{equation}
where $\psi$ is an external $\mathrm{Spin}(4,1)$ spinor, $\chi_i$ are internal $\mathrm{Spin}(5)$ spinors and $u$ a two-component spinor satisfying
\begin{equation}
	\sigma^2 u = - u \qquad \qquad u^* = \sigma^1 u \, . 
\end{equation}

Charge conjugation of the external and internal spinors is defined as
\begin{align}
	\psi^c= D_{1,4} \psi^* && \chi^c= C_5 \chi^* \, . 
\end{align}
One can easily check that, with the above choices, 
\begin{equation}
\label{eq:IIB_spinor_cc_properties}
	\begin{array}{lcl}
		\psi^{cc}=-\psi\, , & \quad \quad & \left( \rho^{\mu_1} \ldots \rho^{\mu_k}\psi\right)^c= \left( -1\right)^k \rho^{\mu_1} \ldots \rho^{\mu_k} \psi^c \, , \\
		\chi^{cc}=-\chi\, , & \quad \quad & \left( \gamma^{m_1} \ldots \gamma^{m_k} \chi\right)^c= \gamma^{m_1} \ldots \gamma^{m_k} \chi^c\, .
	\end{array}
\end{equation}

For $5$-dimensional internal spinors, from the properties listed above, one can derive
\begin{equation}
	\left( \overline{\chi^c} \gamma_{m_1 \ldots m_r} \phi^c\right) = \left( \overline{\chi}\gamma_{m_1 \ldots m_r} \phi\right)^*\, ,
\end{equation}
and
\begin{equation}
	\left( \overline{\chi^c}\gamma_{m_1 \ldots m_r}\phi\right) = -\left( \overline{\chi} \gamma_{m_1 \ldots m_r} \phi^c\right)^*\, .
\end{equation}

Similarly we can derive some useful identities for the internal spinors.
Let us consider the expression,
\begin{equation}
	\left( \overline{\psi^c}\rho_{\mu_1 \ldots \mu_q} \psi^c\right) = \left( -1\right)^{q+1} \left( \overline{\psi} \rho_{\mu_1} \ldots \rho_{\mu_q} \psi\right)^*\, .
\end{equation}
where, as always, $\overline{\psi}=\psi^{\dagger}\rho_0$. Next, we obtain
\begin{equation}
	\left( \overline{\psi}\rho_{\mu_1} \ldots \rho_{\mu_q}\psi\right)^* = - \left( -1\right)^{\frac{q(q+1)}{2}} \left( \overline{\psi} \rho_{\mu_1} \ldots \rho_{\mu_q}\psi\right)\, .
\end{equation}
Combining these two equations yields
\begin{equation}
	\left( \overline{\psi^c}\rho_{\mu_1 \ldots \mu_q} \psi^c\right) = -\left( -1\right)^{\frac{(q+1)(q+2)}{2}} \left( \overline{\psi} \rho_{\mu_1} \ldots \rho_{\mu_q}\psi\right)\, .
\end{equation}
For the other combinations, one obtains
\begin{equation}
	\left( \overline{\psi^c}\rho_{\mu_1 \ldots \mu_q}\psi\right)^* = - \left( -1\right)^{q+1} \left( \overline{\psi} \rho_{\mu_1} \ldots \rho_{\mu_q} \psi^c\right)\, ,
\end{equation}
and 
\begin{equation}
	\left( \overline{\psi^c} \rho_{\mu_1 \ldots \mu_q}\psi\right)^* = -\left( -1\right)^{\frac{q(q+1)}{2}} \left( \overline{\psi} \rho_{\mu_1 \ldots \mu_q} \psi^c\right)\, . 
\end{equation}
Again combining the last two relations gives
\begin{equation}
	 \left( \overline{\psi}\rho_{\mu_1 \ldots \mu_q} \psi^c\right) = \left( -1\right)^{\frac{(q+1)(q+2)}{2}}\left( \overline{\psi}\rho_{\mu_1 \ldots \mu_q} \psi^c\right)\, ,
\end{equation}
so that these terms vanish for $q=0,1,4$.

\subsection{M-theory}
\label{app:mtheoryconv}
We follow again the conventions in~\cite{Grana_Ntokos} for the metric ansatz
\begin{equation}
	\dd  s^2 = e^{2\Delta}\dd  s^2_{\mathrm{AdS}} + \dd  s^2_{M} \, .
\end{equation}
We consider two M-theory setups: $\mathrm{AdS}_4$ compactifications with a $7$-dimensional internal manifold $M_7$
and $\mathrm{AdS}_5$ ones on a $6$-dimensional internal manifold $M_6$.
The eleven-dimensional gamma matrices are $\hat{\Gamma}^M$, $M=0,\ldots, 10$, satisfying the Clifford algebra $\mathrm{Cliff}(1,10)$ relations, 
\begin{equation}
	\{\hat{\Gamma}^A , \hat{\Gamma}^B \} = 2 \eta^{AB}\, .
\end{equation}
They will decompose as in~\eqref{eq:dec_gammas_m6} and~\eqref{eq:dec_gammas_m7}, and for convenience, we report them here,
\begin{equation}
\label{11gammadec}
	\begin{array}{lccr}
		\hat{\Gamma}^\mu = e^{-\Delta}\,\rho^\mu \otimes \Gamma_7\, , & \hat{\Gamma}^{m+4} = \id_4 \otimes \Gamma^m\, & \phantom{\mbox{for}} &\mbox{for}~\mathrm{AdS}_5 \times M_6\, , \\[2mm]
		\hat{\Gamma}^\mu = e^{-\Delta}\,\rho^\mu \otimes \id_8\, , & \hat{\Gamma}^{m+3} =e^{-\Delta} \rho_5 \otimes \Gamma^m\, &\phantom{\mbox{for}} &\mbox{for}~\mathrm{AdS}_4 \times M_7\, .
	\end{array}
\end{equation} 
In the expressions above we denoted the internal gamma matrices with the same symbol for both cases, with $\Gamma_7$ the chiral operator in $6$ dimensions, defined below, and with $\rho_5$ the external $\mathrm{Cliff}(1,4)$ chiral operator.
\subsubsection{\texorpdfstring{M-theory on $\mathrm{AdS}_5 \times M_6$}{M-theory on AdS5 x M6}}
Here we give the conventions for M-theory solutions of the form $\mathrm{AdS}_5 \times M_6$. The external part is the same as the type IIB compactification.
So we refer to~\cref{sect:IIB_notation}. For the internal part we take as reference~\cite{VanProeyen:1999ni} and we write all the gamma matrices as tensor products of Pauli matrices
\begin{equation}
\label{gamma6}
	\begin{aligned}
		\Gamma^1 &= \sigma_1 \otimes \id \otimes \sigma_{1}\, , \\[1mm]
		\Gamma^2 &= \sigma_1 \otimes \id \otimes \sigma_{2}\, , \\[1mm]
		\Gamma^3 &= \sigma_1 \otimes \sigma_{1} \otimes \sigma_{3}\, , \\[1mm]
		\Gamma^4 &= \sigma_2 \otimes \sigma_{1} \otimes \sigma_{3}\, , \\[1mm]
		\Gamma^5 &= -\sigma_1 \otimes \sigma_{3} \otimes \sigma_{3}\, , \\[1mm]
		\Gamma^6 &= \sigma_3 \otimes \id \otimes \id\, . 
	\end{aligned}
\end{equation}
In even dimensions we can define a chiral operator
\begin{equation}
\label{chiralgamma}
	\Gamma_7 = - \IIm \Gamma^1 \ldots \Gamma^6\, .
\end{equation}
The chiral operator $\Gamma_7$ squares to the identity and satisfies the following relations with the other gamma matrices
\begin{align*}
	\left\{\Gamma_7\, , \Gamma_{m} \right\} &= 0\, ,\\
	\left[\Gamma_7\, , \Gamma_{mn} \right] &= 0\, .
\end{align*}
By induction, these properties can be extended to any odd/even rank element of the Clifford algebra.

The intertwiners of $\mathrm{Cliff(6)}$ can be written as follows
\begin{align*}
\Gamma_m^T & = C_6^{-1} \Gamma_m C_6 \, , \\
\Gamma_m^* & = D_6^{-1} \Gamma_m D_6 \, , \\
\Gamma_m^{\dagger} & = A_6 \Gamma_m A^{-1}_6 \, .
\end{align*}
and, for our conventions, 
\begin{equation*}
	\begin{array}{lcr}
		A_6 = 1 \, ,& & D_6 = C_6\, .
	\end{array}
\end{equation*}

\subsubsection{\texorpdfstring{M-theory on $\mathrm{AdS}_4 \times M_7$}{M-theory on on AdS4 x M7}}
Here we give the conventions which are relevant for the $\mathrm{AdS}_4$ solutions of M-theory. We made the choice of having compatible conventions with the previous section, such that one can embed all the relations above in the following ones.

The Clifford algebra $\mathrm{Cliff}(7)$ and its generators are constructed by the same set of gamma matrices~\eqref{gamma6} of $\mathrm{Cliff}(6)$ 
plus the chiral gamma $\Gamma_7$ in~\eqref{chiralgamma}.

The $\mathrm{Cliff}(7)$ intertwiners are written as
\begin{align*}
	\Gamma_m^T &= C_7^{-1} \Gamma_m C_{7}\, , \\[1mm]
	\Gamma_m^\dagger &= A_7 \Gamma_m A_7^{-1}\, , \\[1mm]
	\Gamma_m^* &= D_7^{-1} \Gamma_m D_7\, .
\end{align*}
Numerically the matrices $A_7$, $C_7$, $D_7$ are the same as $A_6$, $C_6$, $D_6$.

The four-dimensional gamma matrices on $\mathrm{AdS}_4$ satisfy 
\begin{equation}
	\{\rho_a , \rho_b \} = 2 \eta_{ab}\id \, ,
\end{equation}
where $a, b$ are frame indices. Hence it holds $\eta^{ab}e^\mu_a \otimes e^\nu_b = g^{\mu\nu}$, where $g^{\mu\nu}$ is the $\mathrm{AdS}_4$ inverse metric. 
In terms of flat frame indices, we choose a basis for explicit calculations for $\mathrm{Cliff}(1,3)$,
\begin{equation}
\begin{array}{lrcc}
	\rho^0=\IIm \sigma^2 \otimes \sigma^0\, , & \rho^i = \sigma^1 \otimes \sigma^i\, , &\phantom{\mbox{and}} & i=1,2,3 \, . 
\end{array}
\end{equation}
As for the internal part we have chosen the basis above such that we can embed it into~\eqref{ads5gamma}.
The intertwiners can be written as in~\eqref{Achoice},
\begin{equation}
	\begin{array}{ccc}
		\rho^{\mu\dagger} = -A_{1,3} \rho^\mu A_{1,3}^{-1}\, , &\phantom{\mbox{for}} &A_{1,3} = \rho_0\, , \\[1mm]
		\rho^{\mu T} = C_{1,3} \rho^\mu C_{1,3}^{-1}\, , &\phantom{\mbox{for}} & C_{1,3} = D_{1,3} A_{1,3} \, , \\[1mm]
		\rho^{\mu *} = -D_{1,3} \rho^\mu D_{1,3}^{-1}\, . & \phantom{\mbox{for}} & 
	\end{array}
\end{equation}

	\chapter{Exceptional Generalised Geometry}
	\label{app:EGG}
		%
		\section{Generalised geometry for M-theory}\label{appsec:EGGMth}
			We review the construction of $\E_{d(d)} \times \RR^+$ given in~\cite{hull1, waldram2, waldram4}.
			This will also be useful since by a dimensional reduction we will build the appropriate geometry for type IIA.
			We use the same notation and conventions as~\cite{waldram2, waldram4}. 
			
			In M-theory compactified on a seven-dimensional manifold $M_7$, the fibres of the generalised tangent bundle $E$ transform in the fundamental $\mathbf{56}_1$ representation of the $E_{7(7)}\times\mathbb R^+$ structure group. 
			Under $\GL(7)$, $E$ decomposes as
					\begin{equation}\label{gense7}
 						E \cong TM_7 \oplus \Lambda^2T^*M_7 \oplus \Lambda^5T^*M_7 \oplus (T^*M_7\otimes\Lambda^7T^*M_7)\, .
					\end{equation} 
			A section can be written as 
					\begin{equation}
 						V = v + \omega + \sigma + \tau\, ,
					\end{equation}
			where at each point on $M_7$, $v \in \Gamma(TM_7)$ is an ordinary vector field, $\omega\in \Gamma(T^*M_7)$, $\sigma \in \Gamma(\Lambda^5T^*M_7)$ and $\tau \in \Gamma((T^*\otimes \Lambda^7 T^*)M_7)$.
			
			The adjoint bundle $\ad$ decomposes under $\GL(7)$ as
					\begin{equation}
						\adj = \RR \oplus (TM_7\otimes T^*M_7) \oplus \Lambda^3 T^*M_7 \oplus \Lambda^6 T^*M_7 \oplus \Lambda^3 TM_7 \oplus \Lambda^6 TM_7 \, ,
					\end{equation}
			with sections transforming in the $\mathbf{133}_0+ \mathbf{1}_0$ representation of $\E_{7(7)}\times \RR^+$ given by
					\begin{equation}\label{eq:Radj}
						R = l + r + a + \tilde a + \alpha + \tilde\alpha \ ,
					\end{equation}
			where $l \in \RR$ gives the shift of the warp factor, $r \in \mathrm{End}(T M_7)$, $a \in \Lambda^3 T^*M_7$ is related to the three-form potential of M-theory, $\tilde a \in \Lambda^6T^*M_7$ to its dual, while $\alpha \in \Lambda^3 TM_7$ and $ \tilde\alpha \in \Lambda^6 TM_7$ are a three- and a six-vector.
				
			The adjoint action of the $\mathfrak{e}_{7(7)}\times \RR^+$ algebra on a generalised vector is denoted as $V' = R \cdot V$ and reads:
					\begin{equation}\label{Mth_adjoint_act}
						\begin{split}
							v' &= l\ v + r \cdot v + \alpha \lrcorner \omega - \tilde{\alpha} \lrcorner \sigma \, ,\\
							\omega' &= l \, \omega + r \cdot \omega +v \lrcorner a+ \alpha \lrcorner \sigma + \tilde{\alpha} \lrcorner \tau \, \\
							\sigma' &= l \,\sigma + r \cdot \sigma +v \lrcorner \tilde a + a \wedge \omega+ \alpha \lrcorner \tau \, , \\
							\tau'&= l\, \tau+ r \cdot \tau - j \tilde a \wedge \omega+ j a \wedge \sigma \, .
						\end{split}
					\end{equation}
			The $\mathfrak{e}_{7(7)}$ subalgebra is given by imposing $ \frac{1}{2} \tr(r) = l$.
			
			The adjoint commutator $R''= [R , R']$ is
					\begin{equation}\label{comm_Mth_adj}
						\begin{split}
							l'' &= \tfrac{1}{3} (\alpha \lrcorner a' - \alpha' \lrcorner a) + \tfrac{2}{3} (\tilde \alpha' \lrcorner \tilde a - \tilde \alpha \lrcorner \tilde a') \, , \\
							r'' &= [ r, r'] + j \alpha \lrcorner j a' - j \alpha' \lrcorner j a - \tfrac{1}{3} (\alpha \lrcorner a' - \alpha' \lrcorner a) \id \\
								& \phantom{=} + j \tilde \alpha' \lrcorner j \tilde a - j \tilde \alpha \lrcorner j \tilde a' - \tfrac{2}{3} (\tilde \alpha' \lrcorner \tilde a - \tilde \alpha \lrcorner \tilde a')\id \, ,\\
							a'' &= r \cdot a' - r' \cdot a + \alpha' \lrcorner \tilde a - \alpha \lrcorner \tilde a' \, , \\
							\tilde a'' &= r \cdot \tilde a' - r' \cdot \tilde a - a \wedge a' \, , \\
							\alpha'' &= r \cdot \alpha' - r' \cdot \alpha +\tilde \alpha' \lrcorner a - \tilde \alpha \lrcorner a' \, , \\
							\tilde \alpha'' &= r \cdot \tilde \alpha' - r' \cdot \tilde \alpha - \alpha \wedge \alpha' \, .
						\end{split}
					\end{equation}

			 As seen in the main text, the generalised tangent bundle $E$ is actually twisted to take into account the non-trivial gauge potentials of M-theory, and this is why it is only isomorphic to the sum of bundles in~\eqref{gense7}.  The twist is implemented by an action by adjoint elements.
			 Given a section $\tilde{V}$ of the untwisted tangent bundle $\tilde{E}$, a section $V$ of $E$ is defined as
					\begin{equation}\label{twist_Mth}
						V = e^{A + \tilde A}  \cdot \tilde{V} \, , 
					\end{equation}
			where $A + \tilde A$ is an element of the adjoint bundle.
			The patching condition on the overlaps $U_{\alpha} \cap U_{\beta}$ is 
					\begin{equation}
						V_{(\alpha)} = e^{\dd \Lambda_{(\alpha \beta)} + \dd \tilde \Lambda_ {(\alpha \beta)}} \cdot V_{(\beta)} \, , 
					\end{equation}
			where $\Lambda_{(\alpha \beta)}$ and $ \tilde\Lambda_ {(\alpha \beta)}$ are a two- and five-form, respectively. 
			This corresponds to gauge-transforming the three- and six-form potentials in~\eqref{twist_Mth} as 
					\begin{equation}
						\begin{split}
							A_{(\alpha)} &= A_{(\beta)} + \dd \Lambda_{(\alpha \beta)} \, ,  \\
							\tilde A_{(\alpha)} &= \tilde A_{(\beta)} + \dd \tilde \Lambda_{(\alpha \beta)}  -\frac{1}{2}   \dd \Lambda_{(\alpha \beta)}  \wedge A_{(\beta)} \, .
						\end{split}
					\end{equation}
			Then, the respective gauge-invariant field-strengths reproduce the supergravity ones are
					\begin{equation}
						\begin{split}
								F &= \dd A \, ,  \\
							\tilde F &= \dd \tilde A - \frac{1}{2} A \wedge F \, .
						\end{split}
					\end{equation}

			Finally, we discuss the relevant representations of $\E_{d(d)}\times \RR^+$, defining generalised tensors.
			Further than the vector one, there are other tensor bundles which will be of particular importance in the construction of consistent truncations, we collect some of them in~\cref{tab:EddRep}.
			
			Amongst the interesting bundle representations, an object we will need is the bundle $N$ first introduced in~\cite{Coimbra:2011ky}. 
			This is a sub-bundle of the symmetric product of two generalised tangent bundles, $N \subset S^2 E$, and can be expressed as
				\begin{equation}\label{MthNbundle}
					\begin{split}
						N \cong & T^*M_7 \oplus \Lambda^4 T^*M_7 \oplus (T^*M_7 \otimes \Lambda^6T^*M_7) \\[1mm]
						& \oplus (\Lambda^3T^*M_7\otimes \Lambda^7T^*M)\oplus (\Lambda^6T^*M_7 \otimes \Lambda^7T^*M_7) \,.
					\end{split}
				\end{equation}
Formally, $N$ can be described via a series of exact sequences
\begin{equation}
\label{eq:N-sequences}
\begin{tikzcd}[row sep=tiny]
										0 \arrow{r} &\Lambda^4 T^*M \arrow{r} &N' \arrow{r} &T^*M \arrow{r} &0 	\, , \\
										0 \arrow{r} & T^*M\otimes \Lambda^6T^*M  \arrow{r} &N'' \arrow{r} &N' \arrow{r} &0 	\, , \\
										0 \arrow{r} &\Lambda^7 T^*M \otimes \Lambda^3 T^*M \arrow{r} &N''' \arrow{r} &N'' \arrow{r} &0 \, . \\
										0 \arrow{r} &\Lambda^7 T^*M \otimes \Lambda^6 T^*M \arrow{r} &N \arrow{r} &N''' \arrow{r} &0 \, .
									\end{tikzcd}
\end{equation}
Under $E_{7(7)}\times \mathbb{R}^+$, sections of $N$ transform in the $\mathbf{133}_{2}$ representation. 
Their expression in terms of the symmetric product of generalised vectors can be found in~\cite{Coimbra:2011ky}. 

The simplest of the intermediate bundles appearing in~\eqref{eq:N-sequences} is $N'$, whose type IIA counterpart will be relevant for the scopes of this paper. This can be expressed as
\begin{equation}\label{MthN'}
N' \simeq  T^*M_7 \oplus \Lambda^4 T^*M_7\,.
\end{equation}
Given a basis $\{\hat E_A\}$, $A = 1,\ldots, 56$, for the generalised tangent bundle $E$, a section $S$ of $N'$ has the form
\begin{equation}
S \,=\, S^{AB}\hat E_A \otimes_{N'} \hat E_B \,,
\end{equation}
where $S^{AB}$ are functions on the manifold and the map $\otimes_{N'}: E \otimes E \to N'$ is defined by
\begin{equation}\label{N'prod_Mth}
V \otimes_{N'} V' \,=\, (v \,\lrcorner\, \omega' + v'\,\lrcorner\,\omega) + (v \,\lrcorner\, \sigma' + v' \,\lrcorner\, \sigma - \omega \wedge \omega')\,.
\end{equation}
We make this definition as it is the result of taking the $E_{7(7)}\times\RR^+$ covariant projection of $V\otimes V'$ onto $N$ (from~\cite{Coimbra:2011ky}) and then projecting onto $N'$ using the natural mappings in~\eqref{eq:N-sequences}. We stress that the sections of $N'$ themselves do not transform in a definite representation of $\E_{7(7)}\times \mathbb{R}^+$.

			We refer to~\cite{waldram4} for a detailed discussion.
					\begin{table}[h!]
						\centering
						\begin{tabular}{l l l l l }
							$d$	&	$E^*$					&	$\adj F$													& 	$N \subset S^2 E$			&$K \subset E^* \otimes \adj F$	\\
							\midrule
							7	& 	$\mathbf{56}_{-1}$			&	$\mathbf{133}_0 + \mathbf{1}_0$								&	$\mathbf{133}_2$			&$\mathbf{912}_{-1}$		\\[1.2mm]
							6 	&	$\mathbf{27}_{-1}$			&	$\mathbf{78}_0 + \mathbf{1}_0$								&	$\mathbf{27'}_2$			&$\mathbf{351'}_{-1}$		\\[1.2mm]
							5  	&	$\mathbf{16}^c_{-1}$			&	$\mathbf{45}_0 + \mathbf{1}_0$								&	$\mathbf{10}_2$			&$\mathbf{144}^c_{-1}$		\\[1.2mm]
							4	&	$\mathbf{10}_{-1}$			&	$\mathbf{24}_0 + \mathbf{1}_0$								&	$\mathbf{5'}_2$				&$\mathbf{40}_{-1} + \mathbf{15}'_{-1}$		\\[1.2mm]
							3	&	$(\mathbf{3},\mathbf{2})_{-1}$	&	$(\mathbf{8},\mathbf{1})_0 + (\mathbf{1}, \mathbf{3})_0 + \mathbf{1}_0$	&	$(\mathbf{3'},\mathbf{1})_{2}$	&$(\mathbf{3'},\mathbf{2})_{-1} + (\mathbf{6},\mathbf{2})_{-1}$		\\[1.2mm]
							\bottomrule
						\end{tabular}
						\caption{Some generalised tensor bundles.}
						\label{tab:EddRep}
					\end{table}
			Note that these bundle representations also appear in the tensor hierarchy formulation of gauged supergravity~\cite{Ciceri:2014wya, TensorHier1} and in $E_{11}$ dimensional reduction~\cite{E11PetW, E11Berg}.
			\subsection{Generalised Lie derivative}
				The Dorfman derivative is constructed as a generalisation of the Lie derivative,
						\begin{equation}
							(L_V V')^M =  V^N \partial_N  V^{\prime M} - (\partial \times_{\ad} V)^M_{\phantom{M}N} V^{\prime N} \, , 
						\end{equation}
				Its expression for two generalised vectors is given in~\eqref{dorfM}.
				
				The action of the Dorfman derivative acting on an element of the adjoint can also be constructed~\cite{waldram4} and reads,
						\begin{equation}\label{eq:M_Dorf_adjoint}
							\begin{split}
								L_{V} R & =(\mathcal{L}_{v}r+ j\alpha \lrcorner j\dd\omega-\tfrac{1}{3}\id\alpha\lrcorner\dd\omega-j\tilde{\alpha}\lrcorner j\dd\sigma+\tfrac{2}{3}\id\tilde{\alpha}\lrcorner\dd\sigma) +(\mathcal{L}_{v}\tilde{\alpha}) \\
 & \phantom{=} + (\mathcal{L}_{v}a+r\cdot\dd\omega-\alpha\lrcorner\dd\sigma) +(\mathcal{L}_{v}\tilde{a}+r\cdot\dd\sigma+\dd\omega\wedge a)+(\mathcal{L}_{v}\alpha-\tilde{\alpha}\lrcorner\dd\omega)\, .
							\end{split}
						\end{equation}

				The expression for the twisted Lie derivative is
						\begin{equation}
							\mathbb{L}_{\tilde{V}} \tilde{\mathcal{A}} = \mathcal{L}_{\tilde{v}} \tilde{\mathcal{A}} + R_{ \mathbb{L}_{\tilde{V}} } \cdot \tilde{\mathcal{A}} \, ,
						\end{equation}
				where $R_{ \mathbb{L}_{\tilde{V}} }$ is a element in the adjoint representation of $\mathfrak{e}_{7(7)}\times \RR^+$ explicitly given by
						\begin{equation}
							R_{ \mathbb{L}_{\tilde{V}} } = \dd \tilde{\omega} - \iota_{v} F + \dd \tilde{\sigma} - \iota_{\tilde{v}} \tilde{F} +\tilde{\omega} \wedge F \, .
						\end{equation}
				This coincides with the replacements rules given in~\eqref{repruleTwDorf}.
		\section{Generalised geometry for type IIA from M-theory reduction}\label{appsec:EGGIIA}
			We can now proceed and reduce the structures above to type IIA supergravity (in string frame) on a six-dimensional manifold $M_6$.
			Consider M-theory exceptional generalised geometry on a seven dimensional manifold $M_7$, then decomposing the $E_{7(7)}\times \RR^+$ generalised tangent bundle $E$ under the $\GL(6, \RR)$ structure group of $M_6$, we get 
					\begin{equation}\label{app:gentb}
						E \simeq T \oplus T^* \oplus \Lambda^5 T^*\oplus (T^* \otimes \Lambda^6 T^*) \oplus \Lambda^\mathrm{even}T^* \, ,
					\end{equation}
			where $\Lambda^{\mathrm{even}}T^*=\mathbb{R} \oplus \Lambda^2 T^* \oplus \Lambda^4 T^* \oplus \Lambda^6 T^*$ and each term in the direct sum is now on $M_6$. 
			A section, transforming again in the fundamental of $E_{7(7)}\times \RR^+$, can be written as
					\begin{equation}\label{app:genvec}	
						V = v + \lambda + \sigma + \tau + \omega\ ,
					\end{equation}
			where $\omega=\omega_0 + \omega_2 + \omega_4 + \omega_6$ is a poly-form in $ \Lambda^{\mathrm{even}} T^*$.

			The $\GL(6)$ decomposition of the adjoint bundle is
					\begin{equation}
					\adj F = \RR_\Delta \oplus \RR_\phi \oplus (T\otimes T^*) \oplus \Lambda^2 T \oplus \Lambda^2 T^*	\oplus \Lambda^6 T \oplus \Lambda^6 T^* \oplus \Lambda^{\mathrm{odd}} T \oplus 												\Lambda^{\mathrm{odd}} T^*\ ,
					\end{equation}
			with generic section
					\begin{equation}\label{Edd_adjoint}
						R = l + \varphi + r + \beta + b + \tilde \beta + \tilde b + \alpha + a \, ,
					\end{equation}
			where $\alpha = \alpha_1 + \alpha_3 + \alpha_5 \in \Lambda^{\mathrm{odd}} T$ and $a= a_1 + a_3 + a_5 \in \Lambda^{\mathrm{odd}} T^*$ are antisymmetric poly-vectors and poly-forms, respectively. 

			Let us now derive the action of the adjoint of $E_{7(7)}\times \RR^+$ on a generalised vector and the commutators of two adjoints in type IIA language. 
			Denoting by $z$ the coordinate along the seventh direction, a type IIA generalised vector is related to an M-theory one as
					\begin{equation}\label{redrules}
						\begin{split}
							v_{\mathrm{M}} &= v + \omega_0 \partial_z\, , \\
							\omega_{\mathrm{M}} &= \omega_2 - \lambda \wedge \dd z \, , \\
							\sigma_{\mathrm{M}} &= \sigma + \omega_4 \wedge \dd z\, , \\
							\tau_{\mathrm{M}} &= \tau \wedge \dd z + \dd z \otimes (\omega_6 \wedge \dd z)\, ,
						\end{split}
					\end{equation}
			where, as in the main text, $\tau = \tau_1 \otimes \tau_6$ and the subscript M denotes the M-theory quantities defined in section~\ref{sec:MthExcGeom}. 
			Similarly, the M-theory adjoint~\eqref{eq:Ggeom-M} decomposes as
					\begin{equation}\label{redrulesad}
						\begin{array}{l}
							l_{\mathrm{M}} \ =\ l - \tfrac13 \varphi\\
							a_{\mathrm{M}} \ =\ a_3 + b \wedge \dd z \\
							\tilde a_{\mathrm{M}} \ = \ \tilde b + a_5 \wedge \dd z \\
							\alpha_{\mathrm{M}} \ =\ \alpha_3 + \beta \wedge \partial_z \\ 
							\tilde \alpha_{\mathrm{M}} \ =\ \tilde \beta + \alpha_5 \wedge \partial_z 
						\end{array}
							\qquad \qquad 
						r_{\mathrm{M}} = 	\begin{pmatrix} 
										r + \tfrac13 \varphi \,\id \,&\, - \alpha_1 \\ 
										a_1 \,&\, -\tfrac23 \varphi
									\end{pmatrix} \, ,
					\end{equation}
			where the identification $l_{\mathrm{M}} = l - \tfrac13 \varphi$ follows from the relation between the M-theory and IIA warp factors $\Delta_{\mathrm{M}} = \Delta_{\mathrm{IIA}} - \tfrac13 \phi$.

			Decomposing the M-theory adjoint action given in~\eqref{Mth_adjoint_act} yields the IIA adjoint action on a generalised vector. Denoting this by $V' = R\cdot V$, we have
					\begin{align}\label{IIAadjvecCompact}
						v' &= l v + r \cdot v - [ \alpha \lrcorner s(\omega)]_{-1} - \beta \lrcorner \lambda - \tilde \beta \lrcorner \sigma \, , \\
						\lambda' &= l \lambda + r \cdot \lambda - v \lrcorner b - [\alpha\lrcorner s(\omega)]_1 - \tilde \beta \lrcorner \tau \, , \\
						\sigma' &= (l-2\varphi) \sigma + r \cdot \sigma + v \lrcorner \tilde b - [\omega\wedge s(a)]_5 - \beta \lrcorner \tau \, , \\
						\tau' &= (l-2\varphi) \tau + r \cdot \tau + j a \wedge s(\omega) + j \tilde b \wedge \lambda - j b \wedge \sigma \, , \\
						\omega' &= (l-\varphi) \omega + r \cdot \omega + b\wedge \omega + v \lrcorner a + \lambda \wedge a + \beta \lrcorner \omega + \alpha \lrcorner \sigma +\alpha\lrcorner \tau \, ,
					\end{align}
			where $s$ is the sign operator $s(\omega_n) = (-1)^{[n/2]} \omega_n$ for $\omega_n \in \Lambda^n T^*$, and $[\ldots]_p$ denotes the form of degree $p$ in the formal sum inside the parenthesis (by $-1$ we mean we pick the vector component). 
			The $E_{7(7)}$ subalgebra is specified by $\frac 12{\rm tr}(r) = l - \varphi$. 
			In particular, the $\rmO(6,6) \subset \E_{7}$ action is generated by $r, b$ and~$\beta$, also setting $\varphi = -\tfrac{1}{2}\mathrm{tr}(r)$ and all other generators to zero. 

			Reducing the M-theory commutator~\eqref{comm_Mth_adj} with the decomposition~\eqref{redrulesad} we find that the IIA adjoint commutator $R'' = [R,R']$ reads 
				\begin{subequations}\label{eq:commAdjIIA}
					\begin{align}
								\begin{split}
									l'' &= -\tfrac{1}{2} (\alpha_1 \lrcorner a_1' - \alpha_1' \lrcorner a_1)+\tfrac{1}{2} (\alpha_3 \lrcorner a_3' - \alpha_3' \lrcorner a_3) \\
									&\phantom{=} - \tfrac{1}{2} (\alpha_5 \lrcorner a_5' - \alpha_5' \lrcorner a_5) + (\tilde \beta' \lrcorner \tilde b - \tilde \beta \lrcorner \tilde b') \, ,
								\end{split}
								\\
								\begin{split}
									\phi'' &= \tfrac{3}{2} (\alpha_1' \lrcorner a_1 - \alpha_1 \lrcorner a_1') + \tfrac{1}{2} (\alpha_3 \lrcorner a_3' - \alpha_3' \lrcorner a_3) - \tfrac{1}{2}(\alpha_5' \lrcorner a_5 - \alpha_5 \lrcorner a_5') \\
										&\phantom{=} - ( \beta \lrcorner b' - \beta' \lrcorner b) + (\tilde \beta' \lrcorner \tilde b - \tilde \beta \lrcorner \tilde b') \, ,
								\end{split}
							\\
								\begin{split}
									r'' 	&= [ r, r'] + j \alpha_1' \lrcorner j a_1 - j \alpha_1\lrcorner j a_1' + j\alpha_3 \lrcorner j a_3' - j\alpha_3' \lrcorner ja_3 - j\alpha_5 \lrcorner j a_5' +j \alpha_5' \lrcorner j a_5 \\ 
										&\phantom{=} + j \beta \lrcorner j b' - j \beta' \lrcorner j b - j \tilde \beta \lrcorner j \tilde b' + j \tilde \beta' \lrcorner j \tilde b + \tfrac{1}{2}\id (\alpha_1' \lrcorner a_1 - \alpha_1 \lrcorner a_1') \\
										&\phantom{=} + \tfrac{1}{2} \id (\alpha_3' \lrcorner a_3 - \alpha_3 \lrcorner a_3')+ \tfrac{1}{2} \id (\alpha_5' \lrcorner a_5 - \alpha_5 \lrcorner a_5') + \id(\tilde \beta \lrcorner \tilde b' - \tilde\beta' \lrcorner \tilde b) \, ,
								\end{split}
							\\
							b'' &= r \cdot b'- r' \cdot b +\alpha_1 \lrcorner a_3' - \alpha_1' \lrcorner a_3 - \alpha_3 \lrcorner a_5' + \alpha_3' \lrcorner a_5 \, , \\
							\tilde b'' &= r \cdot \tilde b' - r' \cdot \tilde b -2\varphi \tilde b' + 2\varphi'\tilde b + a_1 \wedge a_5' - a_1' \wedge a_5 - a_3 \wedge a_3' \, ,\\
							a'' &= r \cdot a' - r' \cdot a - \varphi a' + \varphi' a + b\wedge a' - b'\wedge a + \beta \lrcorner a' - \beta' \lrcorner a -\alpha \lrcorner \tilde b' + \alpha' \lrcorner \tilde b \, ,\\
							\beta'' &= r \cdot \beta' - r' \cdot \beta + \alpha_3' \lrcorner a_1 - \alpha_3 \lrcorner a_1' - \alpha_5' \lrcorner a_3 + \alpha_5 \lrcorner a_3' \, , \\
							\tilde \beta'' &= r \cdot \tilde \beta' - r' \cdot \tilde \beta + 2\varphi \tilde \beta' - 2\varphi' \tilde \beta + \alpha_1 \wedge \alpha_5' - \alpha_3 \wedge \alpha_3' + \alpha_5\wedge \alpha_1' \, ,\\
								\begin{split}
									\alpha'' &= r \cdot \alpha' - r' \cdot \alpha +\varphi \alpha' - \varphi' \alpha + \beta\wedge\alpha' -\beta'\wedge\alpha - \alpha \lrcorner b' + \alpha' \lrcorner b \\
										&\phantom{=} - \tilde \beta \lrcorner a' + \tilde \beta' \lrcorner a \, .
								\end{split}
					\end{align}
				\end{subequations}
			Next, we obtain the explicit expression for the Dorfman derivative between two type IIA generalised vectors $V$ and $V'$. 
			By plugging~\eqref{redrules} into~\eqref{dorfM} we find:
					\begin{equation*}
						\begin{split}
							L_V V' =& \mathcal{L}_v v' + \left(\mathcal{L}_v \lambda' - \iota_{v^\prime} \dd \lambda\right) + \left(\iota_v \dd \omega_0' - \iota_{v^\prime} \dd \omega_0\right) \\
								& + \left(\mathcal{L}_v \omega_2^\prime - \iota_{v^\prime}\dd \omega_2 - \lambda' \wedge \dd \omega_0 + \omega_0' \dd \lambda\right) \\
								& + \left( \mathcal{L}_v \omega_4^\prime - \iota_{v^\prime}\dd \omega_4 - \lambda' \wedge \dd \omega_2 + \omega_2^\prime \wedge \dd \lambda\right) \\
								& + \left(\mathcal{L}_v \omega_6^\prime - \lambda' \wedge \dd \omega_4 + \omega_4' \wedge \dd \lambda\right) \\
								& + \left( \mathcal{L}_v \sigma' - \iota_{v^\prime}\dd \sigma + \omega_0^\prime \dd \omega_4 - \omega_2^\prime \wedge \dd \omega_2 + \omega_4' \wedge \dd \omega_0\right) \\
								& + \left(\mathcal{L}_v \tau' + j \sigma' \wedge \dd \lambda + \lambda^\prime \otimes \dd \sigma + \dd \omega_0 \otimes \omega_6^\prime + j \omega_4^\prime \wedge \dd \omega_2 - j \omega_2^\prime \wedge \dd \omega_4 \right) \, . 
						\end{split}
					\end{equation*}
			This expression can be cast in the more compact form given in~\eqref{dorfIIA}.
			
			Also the Dorfman derivative acting on a section of the adjoint bundle can be get by the M-theory one using the reduction rules~\eqref{redrules}.
			This reads,
					\begin{equation}
						\begin{split}
							L_V R  =& \left(-\frac{2}{3}\mathcal{L}_{v}\varphi+\frac{1}{2} \alpha_1 \lrcorner \mathrm{d} \omega_0	-\frac{2}{3}\beta \lrcorner \mathrm{d} \lambda+\frac{2}{3}\tilde \beta \lrcorner \mathrm{d} \sigma-\frac{1}{3} \alpha_3 \lrcorner \mathrm{d} \omega_2-\frac{1}{3} \alpha_5 \lrcorner \mathrm{d} \omega_4 \right) \\[1mm]
								&+ \left(\mathcal{L}_{v}\varphi-\frac{3}{2} \alpha_1 \lrcorner \mathrm{d} \omega_0+\beta \lrcorner \mathrm{d} \lambda-\tilde \beta \lrcorner \mathrm{d} \sigma+\frac{1}{2} \alpha_3 \lrcorner \mathrm{d} \omega_2+\frac{1}{2} \alpha_5 \lrcorner \mathrm{d} \omega_4	\right ) \\[1mm] 
								&+ \left( \mathcal{L}_{v} r  -[j\alpha \lrcorner j s(\mathrm{d} \omega)]_{-1\otimes 1} +j \beta \lrcorner j \mathrm{d} \lambda -j \tilde \beta \lrcorner j \mathrm{d} \sigma \right. \\
								& \phantom{++} \left. + \mathds{1} \left( -\frac{1}{3} \mathcal{L}_{v} \varphi	 +\frac{1}{2}[\alpha \lrcorner s(\mathrm{d}\omega)]_0  + \tilde \beta \lrcorner \mathrm{d} \sigma \right)\right ) \\[1mm]
								& + \left(\mathcal{L}_{v}\beta - \alpha_3 \lrcorner \mathrm{d} \omega_0 + \alpha_5 \lrcorner \mathrm{d} \omega_2\right ) \\[1mm]
				& + \left(\mathcal{L}_{v}b-r\cdot  \mathrm{d} \lambda	- \frac{1}{3}\varphi \mathrm{d} \lambda+\alpha_1 \lrcorner \mathrm{d} \omega_2+\alpha_3 \lrcorner \mathrm{d} \omega_4	\right ) \\[1mm]
				& + \left(\mathcal{L}_{v} \tilde \beta \right )  + \left(\mathcal{L}_{v} \tilde b 	+r\cdot \mathrm{d} \sigma	+\frac{1}{3} \varphi \mathrm{d} \sigma+[ \alpha \wedge s( \mathrm{d} \omega ) ]_6\right ) \\[1mm]
				& + \left( \mathcal{L}_{v}\alpha  -\tilde \beta \lrcorner \mathrm{d} \omega  +(\alpha_3 + \alpha_5 ) \lrcorner \mathrm{d} \lambda	\right ) \\[1mm]
				& + \left(\mathcal{L}_{v} \alpha	+r\cdot \mathrm{d} \omega -\alpha \lrcorner \mathrm{d} \sigma +(\alpha_1 +\alpha_3) \wedge \mathrm{d} \lambda+b \wedge ( \mathrm{d} \omega_0 + \mathrm{d} \omega_2) \right. \\
				& \phantom{++} \left. + \beta \lrcorner ( \mathrm{d} \omega_2 + \mathrm{d} \omega_4) -\varphi \mathrm{d} \omega_0 -\frac{1}{3} \varphi \mathrm{d} \omega_2+\frac{1}{3} \varphi \mathrm{d} \omega_4 \right) \, .
						\end{split}
					\end{equation}

			As in M-theory, the presence in type IIA of non-trivial gauge potentials leads to the definition of a twisted generalised tangent bundle whose sections are related to~\eqref{app:genvec} by the twist~\eqref{eq:twistC_short}. 
			In order to derive the explicit form of the twist we need to exponentiate the $\E_{7(7)}$ adjoint action on a generalised vector~\eqref{IIAadjvecCompact} with $l=\varphi = r = \beta =\tilde\beta= \alpha= 0$. 
			This corresponds to exponentiating a nilpotent sub-algebra of the $\mathfrak{e}_{7(7)}$ algebra, comprising precisely the form potentials of type IIA supergravity. 
			We find that the series expansion
					\begin{equation}
						V' = e^{R}\cdot V \ \equiv\ V + R \cdot V + \tfrac{1}{2} R \cdot(R \cdot V) + \ldots 
					\end{equation}
			truncates at fifth order, and is given by
					\begin{align*}
							v' 		&= v \, , \\
							\lambda' 	&= \lambda - \iota_v b \, , \\
							\sigma' 	&= \sigma + \iota_v \tilde b -\left[\mathcal{B}^{(1)} \wedge s(a)\wedge \omega + \mathcal{B}^{(2)}\wedge s(a) \wedge \iota_v a \right]_5 + a_1 \wedge a_3 \wedge \left(\lambda -\tfrac 13 \iota_v b \right) \, ,\\ 
							\begin{split}
							\tau' 		&= \tau + j \tilde b \wedge \left(\lambda - \tfrac 12 \iota_v b \right) - j s(a) \wedge \left(\mathcal{B}^{(1)}\wedge\omega + \mathcal{B}^{(2)}\wedge (\iota_v a + \lambda \wedge a)+ \mathcal{B}^{(3)}\wedge a\wedge \iota_v b \right) \\
									& \phantom{=} - j b \wedge \left(\sigma + \tfrac 12 \iota_v \tilde b - \mathcal{B}^{(2)}\wedge s(a) \wedge \omega - \mathcal{B}^{(3)}\wedge s(a)\wedge \iota_v a + \tfrac 13 a_1 \wedge a_3 \wedge \left(\lambda-\tfrac 14 \iota_v b\right)\right)\, , 
							\end{split}
							\\
							\omega' 	&= e^b \wedge\omega + \mathcal{B}^{(1)}\wedge (\iota_v a + \lambda \wedge a) + \mathcal{B}^{(2)}\wedge a\wedge \iota_v b \ ,
					\end{align*}
			where we introduced the shorthand notation,
					\begin{align}
						\mathcal{B}^{(1)} &= \frac{e^b-1}{b} = 1 + \tfrac{1}{2} b + \tfrac{1}{3!}b \wedge b + \ldots \, , \\
						\mathcal{B}^{(2)} &= \frac{e^b-1-b}{b\wedge b} = \tfrac{1}{2} + \tfrac{1}{3!} b + \tfrac{1}{4!}b\wedge b + \ldots \, , \\
						\mathcal{B}^{(3)} &= \frac{e^b-1-b- \tfrac 12 b \wedge b}{b\wedge b\wedge b} = \tfrac{1}{3!} + \tfrac{1}{4!}b + \tfrac{1}{5!}b \wedge b + \ldots\, .
					\end{align}

			We can also reduce to type IIA the bundle $N \subset S^2 E$ given in~\eqref{MthNbundle}. 
			In terms of bundles on $M_6$, we obtain 
					\begin{equation}\label{Nbundle}
						N \simeq \mathbb{R} \oplus \Lambda^4T^* \oplus \Lambda^{\mathrm{odd}}T^* \oplus \Lambda^6T^* \oplus (T^* \otimes \Lambda^5 T^*) \oplus (\Lambda^2T^* \oplus \Lambda^6T^* \oplus \Lambda^\mathrm{odd}T^*)\otimes \Lambda^6T^* \, .
					\end{equation}
			The full $N$ bundle in type IIA is described as a similar set of exact sequences to those in M-theory~\eqref{eq:N-sequences}. 
			Again, these provide us with a natural projection onto a smaller bundle $N'$, which is isomorphic to
					\begin{equation}\label{NprimeBundle}
						N' \simeq \mathbb{R} \oplus \Lambda^4T^* \oplus \Lambda^{\rm odd} T^* \, ,
					\end{equation}
			(note that this also includes $\Lambda^5 T^*$ and thus it is not just the reduction of the M-theory $N'$ bundle given in~\eqref{MthN'}).
			Given a basis $\{\hat E_A\}$, $A = 1,\ldots, 56$, for the generalised tangent bundle $E$, a section $S$ of $N'$ has the form
					\begin{equation}
						S = S^{AB}\hat E_A \otimes_{N'} \hat E_B \, ,
					\end{equation}
			where $S^{AB}$ are functions on the manifold and the map 
					\begin{equation}
						\otimes_{N'}: E \otimes E \longrightarrow N'
					\end{equation} 
			is defined as
					\begin{equation}\label{N'prod_IIA}
						\begin{split}
							V \otimes_{N'} V' =& v\lrcorner \lambda' + v' \lrcorner \lambda \\
										& + v \lrcorner \sigma' + v' \lrcorner \sigma + [\omega \wedge s(\omega')]_4 \\
										& + v \lrcorner \omega' + \lambda\wedge \omega' + v' \lrcorner \omega + \lambda' \wedge \omega \, .
						\end{split}
					\end{equation}
			As for~\eqref{N'prod_Mth}, this is the $\E_{7(7)}\times\RR^+$ covariant projection to $N$ further projected onto $N'$.
			\subsection{The split frame}\label{splitfr_MtoIIA}
				As discussed in section~\ref{gen_frame_metric}, a convenient way to compute the generalised metric is starting from the conformal split frame, namely a specific choice of frame on the generalised tangent bundle~\eqref{IIAtangbung}. 
				Here we derive the type IIA conformal split frame by reducing the M-theory one given in~\cite{Coimbra:2011ky}. 
				The latter reads
						\begin{equation}\label{eq:geom-basis}
							\begin{split}
							\mathcal{E}_{{\mathrm{M}} \, \hat{a}} &= e^{\Delta_{\mathrm{M}}} \Big( \hat{e}_{\hat{a}} + i_{\hat{e}_{\hat{a}}} A + i_{\hat{e}_{\hat{a}}} \tilde A + \tfrac{1}{2} A \wedge i_{\hat{e}_{\hat{a}}} A \\
														&\phantom{= e^{\Delta_{\mathrm{M}}} \Big( \hat{e}_{\hat{a}}} + j A \wedge i_{\hat{e}_{\hat{a}}}\tilde{A} + \tfrac{1}{6} j A \wedge A \wedge \iota_{\hat{e}_{\hat{a}}} A \Big) \, , \\
 							\mathcal{E}_{\mathrm{M}}^{\hat{a}\hat{b}} &= e^{\Delta_{\mathrm{M}}} \left( e^{\hat{a}\hat{b}} + A \wedge e^{\hat{a}\hat{b}} - j \tilde{A}\wedge e^{\hat{a}\hat{b}} + \tfrac{1}{2}j A \wedge A \wedge e^{\hat{a}\hat{b}} \right)\, , \\
							\mathcal{E}_{\mathrm{M}}^{\hat{a}_1\dots \hat{a}_5} &= e^{\Delta_{\mathrm{M}}} \left( e^{\hat{a}_1\dots \hat{a}_5} + j A \wedge e^{\hat{a}_1\dots \hat{a}_5} \right)\, , \\
 							\mathcal{E}_{\mathrm{M}}^{\hat{a},\hat{a}_1\dots \hat{a}_7} &= e^{\Delta_{\mathrm{M}}}\, e^{\hat{a},\hat{a}_1\dots \hat{a}_7}\, ,
							\end{split}
						\end{equation}
				where $\Delta_{\mathrm{M}}$ is the M-theory warp factor and $A$ and $\tilde A$ are the three- and six-form potentials of M-theory. 
				$\hat{e}_{\hat{a}}$ is a frame for $T M_7$, $e_{\hat{a}} $ is the dual one and $e^{\hat{a}_1\ldots \hat{a}_p}= e^{\hat{a}_1}\wedge \cdots \wedge e^{\hat{a}_p}$, and $e^{\hat{a},\hat{a}_1\ldots \hat{a}_7} = e^{\hat{a}}\otimes e^{\hat{a}_1\ldots \hat{a}_7}$. 
				The index $\hat{a}$ goes from 1 to 7 and, not to clutter the notation, we omitted the subscript M on $\hat{e}_{\hat{a}} $ and $e_{\hat{a}}$.

				In reducing to type IIA, we decompose the M-theory potentials as 
						\begin{align}
							A &= C_3 - B \wedge \dd z\, , \\
							\tilde{A} &= \tilde{B} - \tfrac{1}{2}C_5 \wedge C_1 + (C_5 - \tfrac{1}{2}B \wedge C_3) \wedge \dd z\, , 
						\end{align}
				where $z$ denotes again the circle direction along which we are reducing, and $B$, $\tilde{B}$ and $C_k$ are the IIA potentials. 
				As already pointed out, the IIA and M-theory warp factors are related by
						\begin{equation}
							\Delta_{\mathrm{M}} \,=\, \Delta - \phi/3 \ . 
						\end{equation}
				To reduce the split frame~\eqref{eq:geom-basis}, we also need to decompose the seven-dimensional indices as $\hat{a} = (a, z)$ with $a=1, \ldots, 6$ and write the seven-dimensional frames as 
					\begin{align}
								\hat{e}_{{\mathrm{M}}\ \hat{a}} = 	\begin{cases} 
																e^{\phi/3}\left(\hat{e}_a + C_a \partial_z\right) \, , \\
 																e^{-2\phi/3}\partial_z \, ,
															\end{cases}
							& &
								e_{\mathrm{M}}^{\hat{a}} =		\begin{cases} 
																e^{-\phi/3}e^a \, ,\\
																e^{2\phi/3}\left(\dd z - C_1\right)\ ,
															\end{cases}
						\end{align}
				where $\hat{e}_a$ and $e^a$ are basis for the IIA frame bundles and $C_a$ denotes the components of the one-form $C_1$. 
				The reduction gives
						\begin{equation*}
							\{ \hat{E}_A \} = \{ \hat{\mathcal{E}}_a \} \cup \{\mathcal{E}^a\} \cup \{ \mathcal{E}^{a_1 \ldots a_5} \} \cup \{ \mathcal{E}^{a,a_1\ldots a_6} \} \cup \{ \mathcal{E} \} \cup \{ \mathcal{E}^{a_1a_2} \} \cup \{ \mathcal{E}^{a_1\ldots a_4} \} \cup \{ \mathcal{E}^{a_1 \ldots a_6} \} \, ,
						\end{equation*} 
				with
						\begin{equation}\label{splitframe_explicit}
							\begin{split}
								\hat{\mathcal{E}}_a &= e^{\Delta} \left(\hat e_a + \iota_{\hat e_a}B + e^{-B}\wedge\iota_{\hat e_a}(C_1 + C_3 + C_5) + \iota_{\hat e_a}\tilde{B} + j\tilde{B} \wedge \iota_{\hat e_a}B \right. \\
												& \phantom{= e^{\Delta}} - \tfrac{1}{2} C_1 \wedge \iota_{\hat e_a}C_5 +\tfrac{1}{2} C_3 \wedge \iota_{\hat e_a} C_3 -\tfrac{1}{2} C_5 \wedge \iota_{\hat e_a} C_1 - \tfrac{1}{2} j C_5 \wedge \iota_{\hat e_a}C_3 \\
												& \phantom{= e^{\Delta}} \left. + \tfrac{1}{2}j B\wedge C_3 \wedge \iota_{\hat e_a}C_3 -\tfrac{1}{2}j B\wedge C_5 \wedge \iota_{\hat e_a}C_1 -\tfrac{1}{2}j B\wedge C_1 \wedge \iota_{\hat e_a}C_5 \right) \, , \\[2mm]
								\mathcal{E}^a &= e^{\Delta} \big(e^a - e^{-B}\wedge (C_1+ C_3 + C_5) \wedge e^a + j\tilde{B} \wedge e^a - C_3 \wedge C_1 \wedge e^a \\
											& \phantom{=e^{\Delta}} - j B \wedge C_3 \wedge C_1\wedge e^a + \tfrac{1}{2}j C_1 \wedge C_5\wedge e^a - \tfrac{1}{2} j C_3 \wedge C_3\wedge e^a \\
											& \phantom{=e^{\Delta}} + \tfrac{1}{2}j C_5 \wedge C_1\wedge e^a \big)\, , \\[2mm]
								\mathcal{E}^{a_1 \ldots a_5} &= e^{\Delta-2\phi}\left(e^{a_1 \ldots a_5} + jB \wedge e^{a_1 \ldots a_5} \right)\, , \\[2mm]
								\mathcal{E}^{a,a_1\ldots a_6} &= e^{\Delta-2\phi}\left(e^{a,a_1 \ldots a_6} \right)\, , \\[2mm]
								\mathcal{E} &= e^{\Delta-\phi} \left( e^{-B} - C_5 - j B\wedge C_5 \right)\ , \\[2mm]
								\mathcal{E}^{a_1 a_2} &= e^{\Delta-\phi} \left( e^{-B}\wedge e^{a_1 a_2} + C_3\wedge e^{a_1 a_2} - jC_5 \wedge e^{a_1 a_2} + jB \wedge C_3\wedge e^{a_1 a_2} \right)\, , \\[2mm]
								\mathcal{E}^{a_1 \ldots a_4} &= e^{\Delta-\phi}\left(e^{-B}\wedge e^{a_1 \ldots a_4} - C_1 \wedge e^{a_1 \ldots a_4} \right. \\
											 & \phantom{=e^{\Delta}}\left. + jC_3 \wedge e^{a_1 \ldots a_4} -jB\wedge C_1 \wedge e^{a_1\ldots a_4} \right) , \\[2mm]
								\mathcal{E}^{a_1 \ldots a_6} &= e^{\Delta-\phi}\left(e^{a_1 \ldots a_6} - j C_1 \wedge e^{a_1\ldots a_6} \right)\, .
							\end{split}
						\end{equation}
				These expressions can be summarised in the twist given in~\eqref{twist_splitfr}.
			%
			%
		\section{Twisted bundle and gauge transformations}\label{appsec:EGGgauge}
				In this section -- closely following~\cite{oscar1} -- we show how one can derive the patching conditions~\eqref{patchingIIA} of the generalised tangent bundle starting from the supergravity gauge transformations.
				We will refer to (massive) type IIA generalised geometry, however, the line of reasoning holds in general. 
				The key requirement will be that the generalised vector generates the diffeomorphism and gauge transformations that act on the supergravity fields. 
				We include the Romans mass in our computation, the massless case simply follows by setting $m=0$.

				We start imposing that in each chart $U$ covering the manifold $M_6$, a generalised vector $V$ generates a diffeomorphism and gauge transformation of the type IIA supergravity potentials:
						\begin{equation}\label{eq:gauge-trans-by-V-E1}
							\begin{split}
								\delta_{V} B &= \mathcal{L}_v B - \dd \lambda \, , \\
								\delta_{V} C &= \mathcal{L}_v C - e^B \wedge (\dd \omega - m \lambda) \, , \\
								\delta_{V} \tilde{B} &= \mathcal{L}_v \tilde{B} - (\dd \sigma + m \,\omega_6) - \tfrac{1}{2} \big[e^B \wedge (\dd \omega - m \lambda) \wedge s(C) \big]_6 \,,\\
							\end{split}
						\end{equation}
				where all the fields are defined on $U$.
				In these expressions, the diffeomorphism along a generic vector $v$ is generated by the ordinary Lie derivative $\mathcal{L}_v$, while the remaining terms correspond to the supergravity gauge transformation.

				We next require that the generalised diffeomorphism~\eqref{eq:gauge-trans-by-V-E1} be globally well-defined. 
				This means that on the intersection of a patch $U_\alpha$ with another patch $U_\beta$, the new field configuration defined by~\eqref{eq:gauge-trans-by-V-E1} is patched in the same way as the original one, so as to preserve the global structure (which cannot be changed by an infinitesimal transformation). 
				The patching conditions for the gauge potentials on $U_{\alpha} \cap U_{\beta}$ are given by the gauge transformation of the supergravity fields. 
				At the linearised level, these read
						\begin{equation}\label{eq:lin-gauge-E1}
							\begin{split}
								B_{(\alpha)} &= B_{(\beta)} + \dd \Lambda_{(\alpha\beta)} \, ,\\
								C_{(\alpha)} &= C_{(\beta)} + e^{B_{(\beta)} } \wedge (\dd \Omega_{(\alpha\beta)} -m \Lambda_{(\alpha\beta)})\, , \\
								\tilde{B}_{(\alpha)} &= \tilde{B}_{(\beta)} + \dd \tilde \Lambda_{(\alpha\beta)} + m \Omega_{6(\alpha\beta)} + \tfrac{1}{2} \left[e^{B_{(\beta)}}\wedge (\dd \Omega_{(\alpha\beta)} -m\Lambda_{(\alpha\beta)})\wedge s(C_{(\beta)}) \right]_6 \, ,
							\end{split}
						\end{equation}
				where the labels $(\alpha)$ and $(\beta)$ indicate fields on $U_\alpha$ and $U_\beta$, respectively, while $(\alpha\beta)$ denotes a field defined just on $U_{\alpha} \cap U_{\beta}$. 
				Note that these gauge transformations have the opposite signs with respect to those in~\eqref{eq:gauge-trans-by-V-E1}, as that equation describes an active transformation which shifts the field configuration to a physically equivalent one; contrastingly, equation~\eqref{eq:lin-gauge-E1} describes a patching relation needed to define the fields on the whole manifold, similar to coordinate invariance in general relativity, which is a passive transformation.

				From \eqref{eq:lin-gauge-E1} we construct the corresponding finite transformation.
				Its form is not uniquely determined, since it depends on the order one chooses for the exponentiation of the infinitesimal transformations. 
				We choose to exponentiate first the action of the RR transformation with parameter $\Omega$, then the NSNS transformation by $ \Lambda$ and finally the one by $\tilde \Lambda$. 
				This gives:
						\begin{equation}\label{eq:finite-gauge-E1}
							\begin{split}
								B_{(\alpha)} &= B_{(\beta)} + \dd \Lambda_{(\alpha\beta)} \, ,\\
								C_{(\alpha)} &= C_{(\beta)} + e^{B_{(\beta)} + \dd \Lambda_{(\alpha\beta)}} \wedge \dd \Omega_{(\alpha\beta)} - m e^{B_{(\beta)} } \wedge\Lambda_{(\alpha\beta)} \wedge \left( \tfrac{e^{\dd \Lambda}-1}{\dd \Lambda} \right)_{(\alpha\beta)} \, , \\
								\tilde{B}_{(\alpha)} &= \tilde{B}_{(\beta)} + \dd \tilde \Lambda_{(\alpha\beta)} + m \Omega_{6 \ (\alpha\beta)} + \tfrac{1}{2} m \Lambda_{(\alpha\beta)} \wedge \Big[ e^{-B_{(\beta)} } \wedge \left(\tfrac{ e^{-\dd \Lambda}-1}{\dd \Lambda} \right)_{(\alpha\beta)} \wedge C_{(\beta)} \Big]_5 \\
												&\phantom{=} + \tfrac{1}{2} \Big[ \dd \Omega_{(\alpha\beta)} \wedge e^{B_{(\beta)} + \dd \Lambda_{(\alpha\beta)}} \wedge s( C_{(\beta)} ) - m\, \dd \Omega_{(\alpha\beta)} \wedge \Lambda_{(\alpha\beta)} \wedge \left( \tfrac{e^{\dd \Lambda}-1}{\dd \Lambda} \right)_{(\alpha\beta)} \Big]_6 \,,
							\end{split}
						\end{equation}
				where we used the shorthand notation
						\begin{equation}
							\tfrac{e^{\pm\dd \Lambda} - 1}{\dd \Lambda} = \pm 1 + \tfrac{1}{2} \dd \Lambda \pm \tfrac{1}{3!} \dd \Lambda \wedge \dd \Lambda\, + \dots \, .
						\end{equation}

				Imposing that the new field configurations \eqref{eq:gauge-trans-by-V-E1} in two overlapping patches $U_\alpha$ and $U_\beta$ are still related in the intersection $U_\alpha \cap U_\beta$ by the transformation \eqref{eq:finite-gauge-E1}, and working to first order in the components of $V$, we obtain
					\begin{equation}\label{eq:well-defined}
						\begin{split}
							\delta_{V_{(\alpha)}} B_{(\alpha)} &= \delta_{V_{(\beta)}} B_{(\beta)}\,, \\
 							\delta_{V_{(\alpha)}} C_{(\alpha)} &= \delta_{V_{(\beta)}} C_{(\beta)} + e^{ B_{(\beta)} +\dd \Lambda} \wedge \delta_{V_{(\beta)}} B_{(\beta)} \wedge \dd \Omega \\
													&\phantom{=} - m e^{ B_{(\beta)} } \wedge \delta_{V_{(\beta)}} B_{(\beta)} \wedge \Lambda \wedge \big( \tfrac{e^{\dd \Lambda}-1}{\dd \Lambda} \big) \, , \\
							\delta_{V_{(\alpha)}} \tilde{B}_{(\alpha)} & = \delta_{V_{(\beta)}} \tilde{B}_{(\beta)} \\
														&\phantom{=}+ \tfrac{1}{2} m \Lambda \wedge \Big[ e^{-B_{(\beta)} } \wedge \big(\tfrac{ e^{-\dd \Lambda}-1}{\dd \Lambda} \big) \wedge ( \delta_{V_{(\beta)}} C_{(\beta)} - \delta_{V_{(\beta)}} B_{(\beta)} \wedge C_{(\beta)}) \Big]_5 \\
															&\phantom{=} + \tfrac{1}{2} \Big[ \dd \Omega \wedge e^{B_{(\beta)} + \dd \Lambda} \wedge\Big( s( \delta_{V_{(\beta)}} C_{(\beta)}) + \delta_{V_{(\beta)}} B_{(\beta)} \wedge s(C_{(\beta)})\Big) \Big]_6 \, ,
						\end{split}
					\end{equation}
				where for ease of notation we are omitting the label $(\alpha\beta)$ on $\Lambda$, $\tilde\Lambda$ and $\Omega$.
				This equation can be solved to give relations between the components of $V_{(\alpha)}$ and $V_{(\beta)}$. Also requiring that these relations are linear in $V_{(\alpha)}$ and $V_{(\beta)}$, we obtain the following patching rules for the generalised vector: 
						\begin{equation*}
							\begin{split}
								v_{(\alpha)}&= v_{(\beta)} \, , \\
								\lambda_{(\alpha)} &= \lambda_{(\beta)} + \iota_{v_{(\beta)}} \dd \Lambda \ , \\
								\sigma_{(\alpha)} &= \sigma_{(\beta)} +\iota_{v_{(\beta)}}(\dd\tilde\Lambda+m\Omega_6) + \dd\Omega_{0} \wedge \dd\Omega_{2} \wedge (\lambda_{(\beta)} + \iota_{ v_{(\beta)}}\dd\Lambda) \\ 
											& \phantom{=} - \big[ s(\dd\Omega)\wedge \big( e^{-\dd\Lambda} \wedge\omega_{(\beta)} + m \big(\tfrac{e^{-\dd\Lambda}-1}{\dd\Lambda}\big)\wedge (\iota_{v_{(\beta)}} \Lambda + \lambda_{(\beta)} \wedge \Lambda)\big)\big]_5 \\
											&\phantom{=} + \big[ \tfrac{1}{2}s(\dd\Omega)\wedge \iota_{v_{(\beta)}}\dd\Omega \big]_5 \\
											&\phantom{=} - \big[m \big(\tfrac{e^{-\dd\Lambda}-1}{\dd\Lambda}\big) \wedge\Lambda\wedge \omega_{(\beta)} + m^2 \big(\tfrac{e^{-\dd\Lambda}-1+ \dd\Lambda}{\dd\Lambda\wedge \dd\Lambda}\big)\wedge \Lambda \wedge \iota_{v_{(\beta)}} \Lambda \big]_5 \ , \\ 
								\omega_{(\alpha)} &= e^{-\dd\Lambda} \wedge\omega_{(\beta)} + \iota_{v_{(\beta)}}\dd\Omega + ( \lambda_{(\beta)} + \iota_{ v_{(\beta)}}\dd\Lambda)\wedge \dd\Omega \\ 
												&\phantom{=} + m \big(\tfrac{e^{-\dd\Lambda}-1}{\dd\Lambda}\big)\wedge (\iota_{v_{(\beta)}} \Lambda + \lambda_{(\beta)} \wedge \Lambda) + m \big(\tfrac{e^{-\dd\Lambda}-1+ \dd\Lambda}{\dd\Lambda\wedge \dd\Lambda}\big)\wedge \Lambda\wedge \iota_{v_{(\beta)}} \dd\Lambda \ .
							\end{split}
						\end{equation*}
				Setting $m=0$, these terms match precisely those following from~\eqref{patchingIIA} for the patching of the twisted generalised tangent space relevant to massless type IIA. 
				Keeping \hbox{$m \neq 0$}, we recover the corresponding terms of equation~\eqref{patching_m}. 

				Note however that by this procedure, one can construct the full twisted bundle $E$ only for compactifications on manifolds $M_d$ of dimension $d \leq 5$. 
 				Indeed, one can directly deduce the patching of the differential form parts of the generalised vector (which form a section of the bundle $E'$ in~\eqref{IIAexten}), but not the dual graviton charge, as there is no known treatment of the (non-linear) gauge transformations of the dual graviton field in an arbitrary background. 
				One can nevertheless infer the transformation of the $\tau$ component of the generalised vector by insisting that the patching is an $\E_{d+1 (d+1)}$ adjoint action. 
				In particular, for $m=0$ this yields:
						\begin{equation*}
							\begin{split}
							\tau_{(\alpha)} =& \tau_{(\beta)} + j \dd\Lambda\wedge \sigma_{(\beta)} + j\dd\tilde\Lambda\wedge (\lambda_{(\beta)} + \iota_{ v_{(\beta)}}\dd\Lambda) \\
										& -js(\dd\Omega)\wedge\big(e^{-\dd\Lambda}\wedge\omega_{(\beta)} + \tfrac{1}{2}\iota_{ v_{(\beta)}}\dd\Omega + \tfrac{1}{2} (\lambda_{(\beta)} + \iota_{ v_{(\beta)}}\dd\Lambda) \wedge \dd\Omega \big)\, .
							\end{split}
						\end{equation*}
					%
	
			%
			%
		\section{\texorpdfstring{Exceptional tangent bundle as extension of $\rmO(d,d)$ generalised geometry}{Exceptional bundle as extension of \rmO(d,d) generalised geometry}}\label{appsec:EGGodd}
			As for the previous section, we stick with type IIA for concreteness, but the way of proceeding can be easily extended to type IIB.
			
			In the formulae for exceptional generalised geometry for (massive) type IIA, one can identify combinations of terms familiar from Hitchin's generalised geometry~\cite{hitch1,gualtphd}.
			We devote this section to showing how the exceptional generalised tangent space can be formulated as an extension of that introduced by Hitchin, by $\rmO(d,d)\times\RR^+$ tensor bundles. 
			This clarifies how exceptional geometry constructions like the Dorfman derivative~\eqref{dorfIIAm}, are built out of objects and operators naturally associated to these $\rmO(d,d)\times\RR^+$ generalised geometric bundles.

			Recall that Hitchin's generalised tangent space~\cite{hitch1,gualtphd}, which we denote by $E'$, has the structure of the extension~\eqref{gentanext},
					\begin{equation}\label{eq:HitchinE}
						\begin{tikzcd}
							0 \arrow{r} &T^*M \arrow{r}{i} & E' \arrow{r}{\pi}& TM \arrow[bend left=50, color = red!60]{l}{B} \arrow{r}& 0 \ .
						\end{tikzcd}
					\end{equation}
			The supergravity $B$-field (red arrow in the~\eqref{eq:HitchinE}) provides a splitting of the sequence and thus an isomorphism
					\begin{equation}\label{eq:E'-isom}
						E' \cong T \oplus T^*\,.
					\end{equation}
			As in~\cite{Coimbra:2011nw}, we will view $E'$ as an $\rmO(d,d)\times\RR^+$ vector bundle with zero $\RR^+$-weight. We normalise the $\RR^+$ weight by fixing the line bundle $L \cong \Lambda^d T^*$ to have unit weight. 
			The spinor bundles associated to $E'$ with weight $\frac{1}{2}$, denoted $S^\pm(E')_{\frac{1}{2}}$, can then be represented as local polyforms
					\begin{equation}\label{eq:SE'-isom}
						S^\pm(E')_{\frac{1}{2}} \cong \Lambda^{\rm even/odd} T^*\,,
					\end{equation}
			while (in six dimensions) there is also an isomorphism
					\begin{equation}\label{eq:E'L-isom}
						E' \otimes L \cong \Lambda^5 T^* \oplus (T^* \otimes \Lambda^6 T^*)\,.
					\end{equation}
			The bundles $S^\pm(E')_{\frac12}$ and $E' \otimes L$ are themselves naturally formed from extensions, and the isomorphisms~\eqref{eq:SE'-isom} and~\eqref{eq:E'L-isom} are also provided by the supergravity $B$ field.

			The (massive) type IIA exceptional generalised tangent space $E$ then fits into the exact sequences
					\begin{equation}\label{eq:Odd-E}
						\begin{tikzcd}
							0 \arrow{r} &S^+(E')_{\frac{1}{2}} \arrow{r} &E'' \arrow{r}{\pi'} & E' \arrow{r} & 0 \, , \\
							0 \arrow{r} & E' \otimes L \arrow{r} &E \arrow{r} & E'' \arrow{r} & 0 \, .
						\end{tikzcd}
					\end{equation}
			These give us a mapping
					\begin{equation}\label{eq:Odd-anchor}
						\begin{tikzcd}[row sep=tiny]
							\pi' : E \arrow{r} & E' \\
							V \arrow[mapsto]{r} & X = v + \lambda\,,
						\end{tikzcd}
					\end{equation}
			which serves as an analogue of the anchor map when viewing the exceptional generalised tangent space $E$ as an extension of $E'$. 

			Some useful $\rmO(d,d)\times\RR^+$ covariant maps can be defined as follows. 
			First, given a section $\tilde{b} \in L$, one has the mapping
					\begin{equation}\label{eq:density-map}
						\begin{tikzcd}[row sep=tiny]
							\tilde{b} : E' \arrow{r}& E' \otimes L \\
							v + \lambda \arrow[mapsto]{r} & i_v \tilde{b} - \lambda \otimes \tilde{b} \, .
						\end{tikzcd}
					\end{equation}
			There is also a natural derivative
					\begin{equation}\label{eq:d-sigma-map}
						\begin{tikzcd}[row sep=tiny]
							\mathrm{der} : E' \otimes L \arrow{r}& L \\
							\tilde{X} = \sigma + \tau \arrow[mapsto]{r} & \langle \mathrm{der} , \tilde{X} \rangle = \dd \sigma \, ,
						\end{tikzcd}
					\end{equation}
			which is the analogue of the (covariant) divergence of a vector density in Riemannian geometry, and a covariant pairing of spinors of opposite chirality
					\begin{equation}\label{eq:Odd-spinor-bilinear}
						\begin{aligned}
							\langle (\dots), \Gamma^{(1)} (\dots) \rangle : S^+ (E')_{\frac{1}{2}} \otimes S^- (E')_{\frac{1}{2}} &\longrightarrow E' \otimes L \\
							\langle \omega , \Gamma^{(1)} \theta \rangle = - [s(\omega) \wedge \theta ]_5 - [j s(\omega)\wedge& \theta ]_{1,6}\,.
						\end{aligned}
					\end{equation}
			The supergravity fields\footnote{%
				In this appendix we use the $A$-basis for the RR fields as we wish for the $B$ field to appear purely in the twisting of the $\rmO(d,d)$ bundles in~\eqref{eq:Odd-E} and not in defining the isomorphism~\eqref{eq:Odd-E-isom}.%
				}
			$A$ and $\tilde{B}$ are naturally collections of local sections of $S^- (E')_{\frac{1}{2}}$ and $L$ respectively, patched by the relevant supergravity gauge transformations. 
			These provide splittings of the sequences~\eqref{eq:Odd-E} and thus an isomorphism 
					\begin{equation}\label{eq:Odd-E-isom}
						\begin{aligned}
							E &\cong E' \oplus S^+(E')_{\frac12} \oplus (E' \otimes L) \\
							V &\mapsto \tilde{X} + \tilde{\omega} + \tilde{\hat{X}}
						\end{aligned}
					\end{equation}
			which is given explicitly in terms of the maps above as
					\begin{equation}\label{eq:Odd-twist}
						\begin{aligned}
							\tilde{X} &= X \\
							\tilde{\omega} &= \omega - X \cdot A \\
							\tilde{\hat{X}} &= \tilde{X} - \tilde{B} \cdot X - \langle \omega - \tfrac12 X \cdot A, \Gamma^{(1)} A \rangle
						\end{aligned}
					\end{equation}
			where $X\cdot A$ is the Clifford product. 

			Let us now show how to write the massless type IIA Dorfman derivative~\eqref{dorfIIA} in terms of natural operations in $\rmO(d,d)\times\RR^+$ generalised geometry. 
			Denote by $X = \pi' (V) = v + \lambda$ and $X' = \pi'(V') = v' + \lambda'$ the projections of the generalised vectors $V$ and $V'$ onto $E'$ using the mapping~\eqref{eq:Odd-anchor}. 
			The vector and one-form parts of~\eqref{dorfIIA} correspond to the $\rmO(d,d)$ Dorfman derivative $L_X X' = \mathcal{L}_v v' + \left(\mathcal{L}_v \lambda' - \iota_{v^\prime} \mathrm{d}\lambda\right)$~\eqref{adjDorf}, so that one has $\pi' (L_V V') = L_{\pi' (V)} \pi' (V')$. 
			This is reminiscent of the situation for the usual anchor map $\pi : E \rightarrow TM$, which satisfies $\pi (L_V V') = \mathcal{L}_{\pi(V)} \pi(V')$ so that the Dorfman derivative descends to the Lie derivative. 
			Here, the mapping $\pi'$ preserves the Dorfman derivative structure. 
			We remark that the map $\pi'$ and the Dorfman derivatives can be viewed as providing a generalisation of the notion of an algebroid, where one replaces the tangent bundle with Hitchin's generalised tangent bundle.

			The poly-forms $\omega$ and $\omega'$ are local sections of the $\rmO(6,6)$ spinor bundle $S^+(E')_{\frac{1}{2}}$, and these are treated in an $\rmO(6,6)$-covariant way in~\eqref{dorfIIA}. 
			Indeed, $L_X \omega' = (\mathcal{L}_v + \dd \lambda \wedge)\omega'$ is a spinorial Lie derivative in $\rmO(d,d)$ generalised geometry, while $(\iota_v + \lambda\wedge)\dd\omega$ is the Clifford action of the $\rmO(6,6)$ generalised vector $X$ on $\dd \omega$. 

			In six dimensions, the last two parts $\sigma$ and $\tau$ form a local section $\tilde{X}$ of $E' \otimes L$ as in~\eqref{eq:E'L-isom}. 
			We see that the $\rmO(d,d)\times\RR^+$ Dorfman derivative $L_X \tilde{X}' = \mathcal{L}_v \sigma' + \mathcal{L}_v \tau' + j \sigma' \wedge \dd \lambda$ accounts for some of the terms involving these in $L_V V'$. 	
			From~\eqref{eq:d-sigma-map}, one can write $\dd \sigma = \langle \mathrm{der} , X \rangle$, a section of $L$, which can act on $X'$ via the map~\eqref{eq:density-map}, to give $\langle \mathrm{der} , X \rangle (X') = i_{v'} \dd \sigma - \lambda'\otimes\dd\sigma$. 
			Finally, the exterior derivative gives the natural $\rmO(d,d)\times\RR^+$ Dirac operator $S^+(E')_{\frac{1}{2}} \rightarrow S^-(E')_{\frac{1}{2}}$ and the pairing between $\omega'$ and $\dd\omega$ in the first and second lines of \eqref{dorfIIA} is the $O(6,6)\times\RR^+$ invariant given in~\eqref{eq:Odd-spinor-bilinear}.

			Putting all of this together, we can write the Dorfman derivative in terms of $\rmO(d,d)\times\RR^+$ objects as
					\begin{equation}\label{eq:Odd-Edd-Dorfman}
							L_V V' = L_X X' + (L_X \omega' - X' \cdot \dd \omega) + (L_X \tilde{X}' - \langle \mathrm{der} , \tilde{X} \rangle (X') - \langle \omega' , \Gamma^{(1)} \dd \omega \rangle)\,.
					\end{equation}
			This can be easily enhanced to include the mass terms in~\eqref{dorfIIAm}. 
			The mass can be viewed as a local section of the spinor bundle $S^+(E')_{\frac12}\cong \Lambda^{\rm even} T^*$ and we can write the massive version of~\eqref{eq:Odd-Edd-Dorfman} as
					\begin{equation}\label{eq:massive-Odd-Edd-Dorfman}
						\begin{split}
							L_V V' &= L_X X' + \Big[ L_X \omega' - X' \cdot (\dd\omega - X\cdot m)\Big] \\
		 						& \phantom{=}+ \Big[ L_X \tilde{X}' - \big(\langle \mathrm{der} , \tilde{X} \rangle + \langle \omega , m \rangle\big)(X') - \langle \omega' , \Gamma^{(1)} (\dd \omega - X\cdot m)\rangle \Big]\,.
						\end{split}
					\end{equation}
			Finally, we remark that the projected generalised metric appearing in~\eqref{eq:GB-metric} is formalised by the construction of this appendix as $\mathcal{H}^{-1} \in S^2(E')$, which is the image of the exceptional generalised metric $G^{-1} \in S^2(E)$ in the anchor-like mapping $\pi' : E \rightarrow E'$ from~\eqref{eq:Odd-anchor}. 
			This is much like the first line of~\eqref{invG_comp_1}, where $e^{2\Delta} g^{-1} \in S^2(TM)$ is seen to be the image of $G^{-1}$ in the anchor map $\pi : E \rightarrow TM$.
			%
			%
		\section{Generalised geometry for type IIB}\label{appsec:EGGIIB}
			We have seen in the main text that the generalised tangent bundle has sections transforming in the fundamental of $\E_{d+1(d+1)} \times \RR^+$.
			On a $d$-dimensional manifold it reads
					\begin{equation*}
						\begin{split}
							E 	& \cong TM\oplus T^{*}M\oplus(T^{*}M\oplus\Lambda^{3}T^{*}M\oplus\Lambda^{5}T^{*}M)\oplus\Lambda^{5}T^{*}M\oplus(T^{*}M\otimes\Lambda^{6}T^{*}M)\\
 								& \cong TM\oplus(T^{*}M\otimes S)\oplus\Lambda^{3}T^{*}M\oplus(\Lambda^{5}T^{*}M\otimes S)\oplus(T^{*}M\otimes\Lambda^{6}T^{*}M)\, ,
						\end{split}
					\end{equation*}
			where $S$ transforms as a doublet of $\SL{2}$.
			Formally it is defined by the series of short exact sequences,
					\begin{equation}\label{extIIB}
						\begin{tikzcd}[row sep=tiny]
							0 \arrow{r} & T^*M \arrow{r} & E''' \arrow{r} &TM \arrow{r} &0 	\, , \\
							0 \arrow{r} &\Lambda^{\mathrm{odd}} T^*M \arrow{r} &E'' \arrow{r} &E''' \arrow{r} &0 	\, , \\
							0 \arrow{r} &\Lambda^5 T^*M \arrow{r} &E' \arrow{r} &E'' \arrow{r} &0 \, . \\
							0 \arrow{r} &T^*M \otimes \Lambda^6 T^*M \arrow{r} &E \arrow{r} &E' \arrow{r} &0 \, .
						\end{tikzcd}
					\end{equation}
			analogously to the type IIA case.
			The sequences are split by maps living in the adjoint bundle,
					\begin{equation}
						\begin{split}
							\adj \tilde{F} = & \RR \oplus(TM\otimes T^{*}M) \oplus(S\otimes S^{*})_0 \oplus(S\otimes\Lambda^{2}TM) \oplus(S\otimes\Lambda^{2}T^{*}M) \\
										& \oplus\Lambda^{4}TM\oplus\Lambda^{4}T^{*}M \oplus(S \otimes\Lambda^{6}TM)\oplus(S\otimes\Lambda^{6}T^{*}M) \, ,
						\end{split}
					\end{equation}
			where the subscript on $(S\otimes S^*)_0$ denotes the traceless part. 
			We write adjoint sections as in~\eqref{eq:IIB-adj},
					\begin{equation}
						R=l+r+a+\beta^{i}+B^{i}+\gamma+C+\tilde{\alpha}^{i}+\tilde{a}^{i} \, .
					\end{equation}

			Taking $\{\hat{e}_{a}\}$ to be a basis for $TM$ with a dual basis $\{e^{a}\}$ on $T^{*}M$, one can define a natural $\mathfrak{gl}{d}$ action on tensors.
			
			The $\mathfrak{e}_{d+1(d+1)}$ subalgebra is generated by setting $l=r_{\phantom{a}a}^{a}/(8-d)$. This fixes the weight of generalised tensors under the $\mathbb{R}^+$ factor, so that a scalar of weight $k$ is a section of $(\det T^{*}M)^{k/(8-d)}$
					\begin{equation}\label{RplusIIB}
						\mathbf{1}_k \in \Gamma\Bigl( (\det T^{*}M)^{k/(8-d)} \Bigr).
					\end{equation}
			We define the adjoint action of $R\in\Gamma(\adj \tilde{F})$ on $V\in \Gamma(E)$ to be $V^{\prime} =R \cdot V$. 
			Then, the components of $V^{\prime}$ are
				\begin{equation}\label{eq:IIB_adjoint}
						\begin{split}
							v^{\prime} & =lv+r\cdot v+\gamma\lrcorner\rho+\epsilon_{ij}\beta^{i}\lrcorner\lambda^{j}+\epsilon_{ij}\tilde{\alpha}^{i}\lrcorner\sigma^{j},\\
							\lambda^{\prime i} & =l\lambda^{i}+r\cdot\lambda^{i}+a_{\phantom{i}j}^{i}\lambda^{j}-\gamma\lrcorner\sigma^{i}+v\lrcorner B^{i}+\beta^{i}\lrcorner\rho-\tilde{\alpha}^{i}\lrcorner\tau,\\
							\rho^{\prime} & =l\rho+r\cdot\rho+v\lrcorner C+\epsilon_{ij}\beta^{i}\lrcorner\sigma^{j}+\epsilon_{ij}\lambda^{i}\wedge B^{j}+\gamma\lrcorner\tau,\\
							\sigma^{\prime i} & =l\sigma^{i}+r\cdot\sigma^{i}+a_{\phantom{i}j}^{i}\sigma^{j}-C\wedge\lambda^{i}+\rho\wedge B^{i}+\beta^{i}\lrcorner\tau+v\lrcorner\tilde{a}^{i},\\
							\tau^{\prime} & =l\tau+r\cdot\tau+\epsilon_{ij}j\lambda^{i}\wedge\tilde{a}^{j}-j\rho\wedge C-\epsilon_{ij}j\sigma^{i}\wedge B^{j}.
						\end{split}
					\end{equation}
			In addition, one defines the adjoint action of $R$ on $R'$ to be the commutator $R''=[R,R']$.
			As consequence, the components of $R''$ are
					\begin{equation}\label{eq:ad-ad-IIB}
						\begin{split}
							l^{\prime\prime} 								& =\tfrac{1}{2}(\gamma\lrcorner C^{\prime}-\gamma^{\prime}\lrcorner C)+\tfrac{1}{4}\epsilon_{kl}(\beta^{k}\lrcorner B^{\prime l}-\beta^{\prime k}\lrcorner B^{l})+\tfrac{3}{4}\epsilon_{ij}(\tilde{\alpha}^{i}\lrcorner\tilde{a}^{\prime j}-\tilde{\alpha}^{\prime i}\lrcorner\tilde{a}^{j})\, ,\\
							r^{\prime\prime} 						& =(r\cdot r^{\prime}-r^{\prime}\cdot r)+\epsilon_{ij}(j\beta^{i}\lrcorner jB^{\prime j}-j\beta^{\prime i}\lrcorner jB^{j})-\tfrac{1}{4}\id\epsilon_{kl}(\beta^{k}\lrcorner B^{\prime l}-\beta^{\prime k}\lrcorner B^{l})\\
 											& \phantom{=} +(j\gamma\lrcorner jC^{\prime}-j\gamma^{\prime}\lrcorner jC)-\tfrac{1}{2}\id(\gamma\lrcorner C^{\prime}-\gamma^{\prime}\lrcorner C)\\
 											& \phantom{=} +\epsilon_{ij}(j\tilde{\alpha}^{i}\lrcorner j\tilde{a}^{\prime j}-j\tilde{\alpha}^{\prime i}\lrcorner j\tilde{a}^{j})-\tfrac{3}{4}\epsilon_{ij}(\tilde{\alpha}^{i}\lrcorner\tilde{a}^{\prime j}-\tilde{\alpha}^{\prime i}\lrcorner\tilde{a}^{j})\, ,\\
							a_{\phantom{\prime\prime i}j}^{\prime\prime i} 	& =(a\cdot a^{\prime}-a^{\prime}\cdot a)_{\phantom{i}j}^{i}+\epsilon_{jk}(\beta^{i}\lrcorner B^{\prime k}-\beta^{\prime i}\lrcorner B^{k})-\tfrac{1}{2}\delta_{\phantom{i}j}^{i}\epsilon_{kl}(\beta^{k}\lrcorner B^{\prime l}-\beta^{\prime k}\lrcorner B^{l})\\
											& \phantom{=} +\epsilon_{jk}(\tilde{\alpha}^{i}\lrcorner\tilde{a}^{\prime k}-\tilde{\alpha}^{\prime i}\lrcorner\tilde{a}^{k})-\tfrac{1}{2}\delta_{\phantom{i}j}^{i}\epsilon_{kl}(\tilde{\alpha}^{k}\lrcorner\tilde{a}^{\prime l}-\tilde{\alpha}^{\prime k}\lrcorner\tilde{a}^{l})\, ,\\
							\beta^{\prime\prime i} 					& =(r\cdot\beta^{\prime i}-r^{\prime}\cdot\beta^{i})+(a\cdot\beta^{\prime}-a^{\prime}\cdot\beta)^{i}-(\gamma\lrcorner B^{\prime i}-\gamma^{\prime}\lrcorner B^{i})-(\tilde{\alpha}^{i}\lrcorner C^{\prime}-\tilde{\alpha}^{\prime i}\lrcorner C)\, ,\\
							B^{\prime\prime i} 						& =(r\cdot B^{\prime i}-r^{\prime}\cdot B^{i})+(a\cdot B^{\prime}-a^{\prime}\cdot B)^{i}+(\beta^{i}\lrcorner C^{\prime}-\beta^{\prime i}\lrcorner C)-(\gamma\lrcorner\tilde{a}^{\prime i}-\gamma^{\prime}\lrcorner\tilde{a}^{i})\, ,\\
							\gamma^{\prime\prime} 					& =(r\cdot\gamma^{\prime}-r^{\prime}\cdot\gamma)+\epsilon_{ij}\beta^{i}\wedge\beta^{\prime j}+\epsilon_{ij}(\tilde{\alpha}^{i}\lrcorner B^{\prime j}-\tilde{\alpha}^{\prime i}\lrcorner B^{j})\, ,\\
							C^{\prime\prime} 						& =(r\cdot C^{\prime}-r^{\prime}\cdot C)-\epsilon_{ij}B^{i}\wedge B^{\prime j}+\epsilon_{ij}(\beta^{i}\lrcorner\tilde{a}^{\prime j}-\beta^{\prime i}\lrcorner\tilde{a}^{j})\, ,\\
							\tilde{\alpha}^{\prime\prime i} 				& =(r\cdot\tilde{\alpha}^{\prime i}-r^{\prime}\cdot\tilde{\alpha}^{i})+(a\cdot\tilde{\alpha}^{\prime}-a^{\prime}\cdot\tilde{\alpha})^{i}-(\beta^{i}\wedge\gamma^{\prime}-\beta^{\prime i}\wedge\gamma)\, ,\\
							\tilde{a}^{\prime\prime i} 					& =(r\cdot\tilde{a}^{\prime i}-r^{\prime}\cdot\tilde{a}^{i})+(a\cdot\tilde{a}^{\prime}-a^{\prime}\cdot\tilde{a})^{i}+(B^{i}\wedge C^{\prime}-B^{\prime i}\wedge C) \, .
						\end{split}
					\end{equation}

			The generalised Lie derivative is defined in~\eqref{dorfMadj}. 
			The twisted Dorfman derivative on a generic tensor can be written as
					\begin{equation}\label{eq:twisted_untwisted_lie_der}
							\mathbb{L}_{\tilde V} \tilde{\alpha} = \mathcal{L}_{\tilde v} \tilde{\alpha} - R_{\mathbb{L}_{\tilde V}} \cdot \tilde{\alpha}\, ,
					\end{equation}
			where $R_{\mathbb{L}_{\tilde V}}$ is the adjoint element 
					\begin{equation}\label{eq:tensor_r}
						R_{\mathbb{L}_{\tilde V}} = \dd \tilde{\lambda}^i -\iota_{\tilde v} F^i + \dd \tilde{\rho} - \iota_{\tilde v} F - \epsilon_{ij} \tilde{\lambda}^i \wedge F^j + \dd \tilde{\sigma}^i +\tilde{\lambda}^i \wedge F - \tilde{\rho} \wedge F^i \, . 
					\end{equation}
			Thus, its action on a generic adjoint element $R$ reads
					\begin{equation}\label{eq:IIB_Dorf_adjoint}
						\begin{split}
							L_{V}R = & (\mathcal{L}_{v} l + \tfrac{1}{2}\gamma\lrcorner \dd\rho + \tfrac{1}{4}\epsilon_{kl}\beta^{k}\lrcorner\dd\lambda^{l}+\tfrac{3}{4}\epsilon_{kl}\tilde{\alpha}^{k}\lrcorner\dd\sigma^{l})\\
 								& +(\mathcal{L}_{v}r+j\gamma\lrcorner j\dd\rho-\tfrac{1}{2}\id\gamma\lrcorner\dd\rho+\epsilon_{ij}j\beta^{i}\lrcorner j\dd\lambda^{j}-\tfrac{1}{4}\id\epsilon_{kl}\beta^{k}\lrcorner\dd\lambda^{l}\\
								& \phantom{{}+({}}+\epsilon_{ij}j\tilde{\alpha}^{i}\lrcorner j\dd\sigma^{j}-\tfrac{3}{4}\id\epsilon_{kl}\tilde{\alpha}^{k}\lrcorner\dd\sigma^{l})\\
 								& +(\mathcal{L}_{v}a_{\phantom{i}j}^{i}+\epsilon_{jk}\beta^{i}\lrcorner\dd\lambda^{k}-\tfrac{1}{2}\delta_{\phantom{i}j}^{i}\epsilon_{kl}\beta^{k}\lrcorner\dd\lambda^{l}+\epsilon_{jk}\tilde{\alpha}^{i}\lrcorner\dd\sigma^{k}-\tfrac{1}{2}\delta_{\phantom{i}j}^{i}\epsilon_{kl}\tilde{\alpha}^{k}\lrcorner\dd\sigma^{l})\\
 								& +(\mathcal{L}_{v}\beta^{i}-\gamma\lrcorner\dd\lambda^{i}-\tilde{\alpha}^{i}\lrcorner\dd\rho)\\
 								& +(\mathcal{L}_{v}B^{i}+r\cdot\dd\lambda^{i}+a_{\phantom{i}j}^{i}\dd\lambda^{j}+\beta^{i}\lrcorner\dd\rho-\gamma\lrcorner\dd\sigma^{i})\\
 								& +(\mathcal{L}_{v}\gamma+\epsilon_{ij}\tilde{\alpha}^{i}\lrcorner\dd\lambda^{j})\\
 								& +(\mathcal{L}_{v}C+r\cdot\dd\rho+\epsilon_{ij}\dd\lambda^{i}\wedge B^{j}+\epsilon_{ij}\beta^{i}\lrcorner\dd\sigma^{j})+(\mathcal{L}_{v}\tilde{\alpha}^{i})\\
 								& +(\mathcal{L}_{v}\tilde{a}^{i}+r\cdot\dd\sigma^{i}+a_{\phantom{i}j}^{i}\dd\sigma^{j}-\dd\lambda^{i}\wedge C+B^{i}\wedge\dd\rho)\, .
						\end{split}
					\end{equation}

			For $\E_{5(5)}$, we also need the vector bundle transforming in the $\mathbf{10}_{2}$ representation of $\mathrm{Spin}{5,5}\times\RR^{+}$. 
			We define this bundle as
					\begin{equation}
						N\simeq S\oplus\Lambda^{2}T^{*}M\oplus S\otimes\Lambda^{4}T^{*}M.
					\end{equation}
			We write sections of the $N$ bundle as
					\begin{equation}
						Q=m^{i}+n+p^{i},
					\end{equation}
			where $m^{i}\in\Gamma(S)$, $n\in\Gamma(\Lambda^{2}T^{*}M)$ and $p^{i}\in\Gamma(S\otimes\Lambda^{4}T^{*}M)$. 
			We define the adjoint action of $R\in\Gamma(\adj \tilde{F})$ on $Q\in\Gamma(N)$ to be $Q'=R\cdot Q$, with components
					\begin{equation}\label{Q_adj_IIB}
						\begin{split}
							m'^{i} & =2lm^{i}+a_{\phantom{i}j}^{i}m^j+\beta^{i}\lrcorner n-\gamma\lrcorner p^{i},\\
							n' & =2ln+r\cdot n+\epsilon_{ij}\beta^{i}\lrcorner p^{j}+\epsilon_{ij}m^{i}B^{j},\\
							p'^{i} & =2lp^{i}+r\cdot p^{i}+a_{\phantom{i}j}^{i}p^{j}+B^{i}\wedge n-m^{i}C.
						\end{split}
					\end{equation}
			Using $\mathrm{16}^{c} \times \mathrm{10}\rightarrow\mathrm{16}$, we define a projection to $E$ as
					\begin{equation}
						\times_E : E^{*}\otimes N\rightarrow E \, .
					\end{equation}
			Explicitly, as a section of $E$, this allows us to define 
					\begin{equation}\label{e5_proj_IIB}
						\dd Q \coloneqq \partial\times_E Q = \dd m^{i}+\dd n \, .
					\end{equation}

			Analogously to the $\rmO(d,d)$ $\eta$ metric, in $\E_{d(d)}$ geometry one has several invariants.
			
			The quadratic invariant for $\E_{5(5)}$ is
					\begin{equation}\label{eq:IIB_quadratic}
						\zeta(Q,Q)=\epsilon_{ij}m^i p^j -\tfrac{1}{2} n\wedge n\, .
					\end{equation}
			The cubic invariant for $\E_{6(6)}$ is
					\begin{equation}\label{eq:IIB_cubic}
						c(V,V,V)=-\tfrac{1}{2}(\imath_{v}\rho\wedge\rho+\epsilon_{ij}\rho\wedge\lambda^{i}\wedge\lambda^{j}-2\epsilon_{ij}\imath_{v}\lambda^{i}\sigma^{j}) \, .
					\end{equation}
			Finally, the symplectic invariant for $\E_{7(7)}$ is
					\begin{equation}\label{eq:IIB_symplectic}
						s(V,V^{\prime})=-\tfrac{1}{4}\bigl((\iota_{v}\tau^{\prime}-\iota_{v^{\prime}}\tau)+\epsilon_{ij}(\lambda^{i}\wedge\sigma^{\prime j}-\lambda^{\prime i}\wedge\sigma^{j})-\rho\wedge\rho^{\prime}\bigr) \, .	
					\end{equation}
			The $\mathfrak{e}_{d+1(d+1)}$ Killing form is given by the trace operator
					\begin{equation}\label{eq:IIB_Killing}
						\begin{split}
							\tr(R,R^{\prime}) =& \tfrac{1}{2}\Bigl(\tfrac{1}{8-d}\tr(r)\tr(r^{\prime})+\tr(rr^{\prime})+\tr(aa^{\prime})+\gamma\lrcorner C^{\prime}+\gamma^{\prime}\lrcorner C \\
 											& \phantom{\frac{1}{2}\biggl[} +\epsilon_{ij}(\beta^{i}\lrcorner B^{\prime j}+\beta^{\prime i}\lrcorner B^{j}) +\epsilon_{ij}(\tilde{\alpha}^{i}\lrcorner\tilde{a}^{\prime j}+\tilde{\alpha}^{\prime i}\lrcorner\tilde{a}^{j})\Bigr)\, .
						\end{split}
					\end{equation}

		\section{Type IIB parallelisation on $S^3$}\label{IIBonS3}

In this appendix, we present a parallelisation of the type IIB generalised tangent bundle on $S^3$ which satisfies an $\SO(4)$ gauge algebra.
 A consistent truncation of type IIB supergravity on $S^3$ down to $\SO(4)$ maximal supergravity in seven dimensions has recently been worked out in~\cite{Malek:2015hma} adopting an exceptional field theory approach. This was related to the $S^3$ reduction of massless type IIA by an external automorphism of $\SL(5)$ exchanging the ${\bf 10}\subset{\bf 15}$ and the ${\bf 10'}\subset {\bf 40'}$ representations. 
 Here we show that this type IIB truncation can also be understood in terms of generalised parallelisations.

The type IIB generalised tangent bundle $E$ on a three-dimensional manifold $M_3$ is
\begin{equation}
\label{gb}
E \cong T \oplus T^* \oplus T^* \oplus \Lambda^3 T^*\, ,
\end{equation}
and has structure group $\E_{4(4)} \times \RR^+\cong \SL\left(5,\mathbb{R}\right)\times \RR^+$.
A generalised vector transforms in the $\mathbf{10}_1$ representation and can be written as
\begin{equation*}
V = v + \lambda + \rho + \zeta\ ,
\end{equation*}
where $v\in T$, $\lambda \in T^*$, $\rho \in T^*$, and $\zeta\in \Lambda^3 T^*$.
The relevant Dorfman derivative can be obtained by truncating to three dimensions the type IIB, five-dimensional Dorfman derivative given in~\cite{spheres,AshmoreECY}. This yields
\begin{equation}
\label{dorfb3}
\begin{split}
L_V V' =& \mathcal{L}_v v' + \left(\mathcal{L}_v \lambda' - \iota_{v^\prime} \mathrm{d}\lambda \right) \\
& + \left(\mathcal{L}_v \rho' - \iota_{v^\prime} \mathrm{d}\rho \right) + \left(\mathcal{L}_v \zeta' + \mathrm{d}\lambda \wedge \rho'\right)\, .
\end{split}
\end{equation}

As in the type IIA example discussed in section~\ref{S3and7dsugra}, we decompose the generalised frame $\hat{E}_{IJ}$, $I,J =1,\ldots, 5$, under $\SL\left(4,\mathbb{R}\right)$ as $\{E_{ij} , E_{i5}\}$, with  $i=1,\ldots,4$.
Then we define a generalised parallelisation on $S^3$ as
\begin{equation}
\label{glpb}
\hat{E}_{IJ} = \begin{cases}\  \hat{E}_{ij} = v_{ij} + \rho_{ij} + \iota_{v_{ij}}\!\! \rg{B}\, ,\\
\hat{E}_{i5} = R \dd y_i + y_i \rg{\vol}_3 + R \dd y_i \wedge \rg{B} \, , \end{cases}
\end{equation}
with
\begin{equation}
\rho_{ij} \,=\, \rg{*}(R^2 \dd y_i \wedge\dd y_j)   \,=\, R\, \epsilon_{ijkl}\, y^k\mathrm{d}y^l \, .
\end{equation}
Here, $\hat{E}_{ij}$ parallelises $T \oplus T^*$, that is the part of the generalised tangent bundle common to type IIA, while $\hat{E}_i$ is a parallelisation on the complement $T^* \oplus \Lambda^3 T^*$.
As in section~\ref{S3and7dsugra}, the background two-form potential $\rg{B}$ is chosen such that 
\begin{equation}
\rg{H} \ =\, \dd\! \rg{B}\, \ = \, \frac{2}{R}\rg{\vol}_3\ 
\end{equation}
(we could also have twisted by a background RR two-form potential $\rg{C_2}$).

Evaluating the Dorfman derivative on the frame~\eqref{glpb}, we obtain
\begin{equation}
\begin{split}
\hat{E}_{kl} &= 2R^{-1}\big(\delta_{i[k}\hat{E}_{l]j} -  \delta_{j[k}\hat{E}_{l]i} \big)\ ,\\[1mm]
\hat{E}_{k5}&= -2R^{-1}\delta_{k[i}\hat{E}_{j]5} \ , \\[1mm]
\hat{E}_{kl} &=0 \, , \\[1mm]
\hat{E}_{k5}&= 0 \, ,
\end{split}
\end{equation}
which corresponds to an $SO(4)$ frame algebra.\footnote{This is the same algebra satisfied by the massive IIA generalised parallelisation on $S^3$ discussed in section~\ref{massive_algebras} (cf.~eq.~\eqref{massive_algebra_S3}) -- however in that case the parallelisation fails to be an $\SL(5)$ frame.} This is consistent with the $\SO(4)$ gauging of $D=7$ maximal supergravity originally discussed in~\cite{Samtleben:2005bp}. To see this, it is convenient to dualise $\hat E_{ij}$ to $\widetilde{E}_{ij} = \frac{1}{2} \epsilon_{ij}{}^{kl}\hat{E}_{kl}$. Also renaming $\widetilde{E}_{i5}= \hat{E}_{i5}$, the frame algebra now reads
\begin{equation}
L_{\widetilde{E}_{II'}} \widetilde{E}_{JJ'} \, =\, -X_{[II'][JJ']}{}^{[KK']} \widetilde{E}_{KK'}\ ,
\end{equation}
with
\begin{align}\label{strconst_m_S3}
& & X_{[II'][JJ']}{}^{[KK']} = -4 \, \epsilon_{5II'L[J}w^{L[K}\delta_{J']}^{K']}\ ,& &
w^{IJ} = \frac{1}{2R} \mathrm{diag}\big(1,1,1,1,0\big)\ , & & 
\end{align}
which matches the embedding tensor given in~\cite{Samtleben:2005bp} for the $\SO(4)$ gauging. 
		\nocite{*}
		\bibliography{./Resources/Bibliography/Bibliography}
		\listoffigures
	\abstractpage
\end{document}